\documentclass[PGL,biber]{nowfnt} 

\usepackage[utf8]{inputenc}

\usepackage{tablefootnote}
\usepackage{multirow}


\makeatletter
\@ifundefined{lhs2tex.lhs2tex.sty.read}%
  {\@namedef{lhs2tex.lhs2tex.sty.read}{}%
   \newcommand\SkipToFmtEnd{}%
   \newcommand\EndFmtInput{}%
   \long\def\SkipToFmtEnd#1\EndFmtInput{}%
  }\SkipToFmtEnd

\newcommand\ReadOnlyOnce[1]{\@ifundefined{#1}{\@namedef{#1}{}}\SkipToFmtEnd}
\usepackage{amstext}
\usepackage{amssymb}
\usepackage{stmaryrd}
\DeclareFontFamily{OT1}{cmtex}{}
\DeclareFontShape{OT1}{cmtex}{m}{n}
  {<5><6><7><8>cmtex8
   <9>cmtex9
   <10><10.95><12><14.4><17.28><20.74><24.88>cmtex10}{}
\DeclareFontShape{OT1}{cmtex}{m}{it}
  {<-> ssub * cmtt/m/it}{}

\DeclareFontShape{OT1}{cmtt}{bx}{n}
  {<5><6><7><8>cmtt8
   <9>cmbtt9
   <10><10.95><12><14.4><17.28><20.74><24.88>cmbtt10}{}
\DeclareFontShape{OT1}{cmtex}{bx}{n}
  {<-> ssub * cmtt/bx/n}{}

\newcommand{\Conid}[1]{\mathit{#1}}
\newcommand{\Varid}[1]{\mathit{#1}}
\newcommand{\anonymous}{\kern0.06em \vbox{\hrule\@width.5em}}

\renewcommand{\leq}{\leqslant}

\usepackage{polytable}

\@ifundefined{mathindent}%
  {\newdimen\mathindent\mathindent\leftmargini}%
  {}%

\def\resethooks{%
  \global\let\SaveRestoreHook\empty
  \global\let\ColumnHook\empty}
\newcommand*{\savecolumns}[1][default]%
  {\g@addto@macro\SaveRestoreHook{\savecolumns[#1]}}
\newcommand*{\restorecolumns}[1][default]%
  {\g@addto@macro\SaveRestoreHook{\restorecolumns[#1]}}
\newcommand*{\aligncolumn}[2]%
  {\g@addto@macro\ColumnHook{\column{#1}{#2}}}

\resethooks

\newcommand{\onelinecommentchars}{\quad-{}- }
\newcommand{\commentbeginchars}{\enskip\{-}
\newcommand{\commentendchars}{-\}\enskip}

\newcommand{\visiblecomments}{%
  \let\onelinecomment=\onelinecommentchars
  \let\commentbegin=\commentbeginchars
  \let\commentend=\commentendchars}

\newcommand{\invisiblecomments}{%
  \let\onelinecomment=\empty
  \let\commentbegin=\empty
  \let\commentend=\empty}

\visiblecomments

\newlength{\blanklineskip}
\newcommand{\hsindent}[1]{\quad}
\let\hspre\empty
\let\hspost\empty

\EndFmtInput
\makeatother
\ReadOnlyOnce{jfpcompat.fmt}%
\makeatletter
\def\@authortable{%
  \leavevmode \hbox \bgroup $\col@sep\tabcolsep
  \let\d@llarbegin\begingroup
  \let\d@llarend\endgroup
  \let\\\author@tabcrone
  \ignorespaces
  \@tabarray}

\makeatother
\EndFmtInput
\ReadOnlyOnce{polycode.fmt}%
\makeatletter

\newcommand{\hsnewpar}[1]%
  {{\parskip=0pt\parindent=0pt\par\vskip #1\noindent}}

\newcommand{\hscodestyle}{}

\newcommand{\sethscode}[1]%
  {\expandafter\let\expandafter\hscode\csname #1\endcsname
   \expandafter\let\expandafter\endhscode\csname end#1\endcsname}

  {\par\noindent
   \advance\leftskip\mathindent
   \hscodestyle
   \let\\=\@normalcr
   \let\hspre\(\let\hspost\)%
   \pboxed}%
  {\endpboxed\)%
   \par\noindent
   \ignorespacesafterend}

  {\hsnewpar\abovedisplayskip
   \advance\leftskip\mathindent
   \hscodestyle
   \let\hspre\(\let\hspost\)%
   \pboxed}%
  {\endpboxed%
   \hsnewpar\belowdisplayskip
   \ignorespacesafterend}

  {\hsnewpar\abovedisplayskip
   \advance\leftskip\mathindent
   \hscodestyle
   \let\\=\@normalcr
   \(\pboxed}%
  {\endpboxed\)%
   \hsnewpar\belowdisplayskip
   \ignorespacesafterend}

\newcommand{\plainhs}{\sethscode{plainhscode}}

\plainhs

  {\hsnewpar\abovedisplayskip
   \advance\leftskip\mathindent
   \hscodestyle
   \let\\=\@normalcr
   \(\parray}%
  {\endparray\)%
   \hsnewpar\belowdisplayskip
   \ignorespacesafterend}

  {\parray}{\endparray}

  {\(\parray}{\endparray\)}

\def\codeframewidth{\arrayrulewidth}
\RequirePackage{calc}

  {\parskip=\abovedisplayskip\par\noindent
   \hscodestyle
   \arrayrulewidth=\codeframewidth
   \tabular{@{}|p{\linewidth-2\arraycolsep-2\arrayrulewidth-2pt}|@{}}%
   \hline\framedhslinecorrect\\{-1.5ex}%
   \let\endoflinesave=\\
   \let\\=\@normalcr
   \(\pboxed}%
  {\endpboxed\)%
   \framedhslinecorrect\endoflinesave{.5ex}\hline
   \endtabular
   \parskip=\belowdisplayskip\par\noindent
   \ignorespacesafterend}

\newcommand{\framedhslinecorrect}[2]%
  {#1[#2]}

  {\(\def\column##1##2{}%
   \let\>\undefined\let\<\undefined\let\\\undefined
   \newcommand\>[1][]{}\newcommand\<[1][]{}\newcommand\\[1][]{}%
   \def\fromto##1##2##3{##3}%
   }{\) }%

  {\let\orighscode=\hscode
   \let\origendhscode=\endhscode
   \def\endhscode{\def\hscode{\endgroup\def\@currenvir{hscode}\\}\begingroup}
   \orighscode\def\hscode{\endgroup\def\@currenvir{hscode}}}%
  {\origendhscode
   \global\let\hscode=\orighscode
   \global\let\endhscode=\origendhscode}%

\makeatother
\EndFmtInput

\newcommand{\lio}{{\textbf{LIO}}}
\newcommand{\FG}{\ensuremath{\lambda}^{dF\mkern-2muG}}
\newcommand{\CG}{\ensuremath{\lambda}^{{dC\mkern-2muG}}}

\newcommand{\llangle}{\langle\mkern-5mu\langle}
\newcommand{\rrangle}{\rangle\mkern-5mu\rangle}

\newcommand{\ceqN}[1]{\ensuremath{\shortdownarrow\approx}_{ #1 }}

\newcommand{\flows}{\sqsubseteq} \newcommand{\lub}{\sqcup}

\newcommand{\greenBox}[1]{\colorbox[RGB]{214,250,219}{#1}}

\newcommand{\update}[3]{\ensuremath{#1 [ #2 \mapsto #3 ]}}

\newcommand{\srule}[1]{{[}\textsc{#1}{]}}

\usepackage{booktabs}   
\usepackage{subcaption} 

\usepackage{framed}
\usepackage{mathrsfs}
\usepackage{color}
\usepackage{xcolor}
\usepackage{xspace}
\newcommand{\Blue}[1]{{\color{blue} #1}}
\newcommand{\Red}[1]{{\color{red} #1}}

\usepackage{placeins}
\newcommand{\rulesep}{\unskip\ \vrule\ }

\newif\ifediting
\editingtrue

\ifediting
\usepackage[colorinlistoftodos,prependcaption,textsize=tiny,disable]{todonotes}
\usepackage{xargs}
\newcommandx{\unsure}[2][1=]{\todo[linecolor=red,backgroundcolor=red!25,bordercolor=red,#1]{#2}}
\newcommandx{\change}[2][1=]{\todo[linecolor=blue,backgroundcolor=blue!25,bordercolor=blue,#1]{#2}}
\newcommandx{\info}[2][1=]{\todo[linecolor=green,backgroundcolor=green!25,bordercolor=green,#1]{#2}}
\newcommandx{\mv}[2][1=]{\todo[linecolor=red,backgroundcolor=red!25,bordercolor=red,#1]{MV: #2}}
\newcommandx{\ar}[2][1=]{\todo[linecolor=yellow,backgroundcolor=yellow!25,bordercolor=red,#1]{AR: #2}}

\usepackage{mathpartir}
\usepackage{subcaption}
\DeclareCaptionFormat{caption-with-line}{#1#2#3\hrulefill}

\usepackage{mathtools}

\usepackage{tikz}
\usetikzlibrary{shapes,arrows,positioning}
\usepackage{tikz-cd}

\DeclarePairedDelimiter\abs{\lvert}{\rvert}%

\newtheorem{thm}{Theorem}

\newtheorem{property}{Property}

\newtheorem{cor}{Corollary}
\newtheorem{defn}{Definition}

\usepackage{enumitem}
\title{From Fine- to Coarse-Grained Dynamic Information Flow Control and Back}

\subtitle{A tutorial on dynamic Information Flow}

\maintitleauthorlist{
Marco Vassena \\
CISPA Helmholtz Center for Information Security \\
marco.vassena@cispa.saarland
\and
Alejandro Russo \\
Chalmers University of Technology \\
russo@chalmers.se
\and
Deepak Garg \\
Max Planck Institute for Software Systems \\
dg@mpi-sws.org
\and
Vineet Rajani \\
Max Planck Institute for Security and Privacy \\
vineet.rajani@csp.mpg.de
\and
Deian Stefan \\
University of California, San Diego \\
deian@cs.ucsd.edu
}

\issuesetup
{%
 copyrightowner={},
 volume        = xx,
 issue         = xx,
 pubyear       = 2018,
 isbn          = xxx-x-xxxxx-xxx-x,
 eisbn         = xxx-x-xxxxx-xxx-x,
 doi           = 10.1561/XXXXXXXXX,
 firstpage     = 1, 
 lastpage      = 18
 }

\author[1]{Vassena,Marco}
\author[2]{Russo,Alejandro}
\author[3]{Garg, Deepak}
\author[4]{Rajani, Vineet}
\author[5]{Stefan, Deian}

\affil[1]{CISPA Helmholtz Center for Information Security; marco.vassena@cispa.saarland>}
\affil[2]{Chalmers University of Technology; russo@chalmers.se}
\affil[3]{Max Planck Institute for Software Systems; dg@mpi-sws.org}
\affil[4]{Max Planck Institute for Security and Privacy; vineet.rajani@csp.mpg.de}
\affil[5]{University of California, San Diego; deian@cs.ucsd.edu}

\articledatabox{\nowfntstandardcitation}

\newenvironment{CompactItemize}%
  {\begin{list}{$\; \; \; \; \; \; \ \  \triangleright$}%
   { \leftmargin=12pt \itemsep=2pt \topsep=2pt
     \parsep=0pt \partopsep=0pt}}%
  {\end{list}}

\PassOptionsToPackage{hyphens}{url}
\usepackage{hyperref}

\usepackage{xurl}
\begin{document}


\begin{abstract}

%
This tutorial provides a complete and homogeneous account of the
latest advances in fine- and coarse-grained dynamic information-flow
control (IFC) security.
Since the 70’s, the programming language and the operating system
communities proposed different IFC approaches.
IFC operating systems track information flows in a coarse-grained
fashion, at the granularity of a process.
In contrast, traditional language-based approaches to IFC are
fine-grained: they track information flows at the granularity of
program variables.
For decades, researchers believed coarse-grained IFC to be strictly
less permissive than fine-grained IFC---coarse grained IFC systems
seem inherently less precise because they track less information—--and
so granularity appeared to be a fundamental feature of IFC systems.
We show that the granularity of the tracking system does \emph{not}
fundamentally restrict how precise or permissive dynamic IFC systems
can be.
%
%
To this end, we mechanize two mostly standard languages, one with a
fine-grained dynamic IFC system and the other with a coarse-grained
dynamic IFC system, and prove a semantics-preserving translation from
each language to the other.
In addition, we derive the standard security property of
non-interference of each language from that of the other, via our
verified translation.
%
%

%
These translations stand to have important implications on the
usability of IFC approaches.
The coarse- to fine-grained direction can be used to remove the label
annotation burden that fine-grained systems impose on developers,
while the fine- to coarse-grained translation shows that
coarse-grained systems---which are easier to design and
implement---can track information as precisely as fine-grained systems
and provides an algorithm for automatically retrofitting legacy
applications to run on existing coarse-grained systems.
\end{abstract}

\maketitle

\listoftodos
\chapter{Introduction}

Dynamic \emph{information-flow control} (IFC) is a principled approach to
protecting the confidentiality and integrity of data in software systems.
Conceptually, dynamic IFC systems are very simple---they associate
\emph{security} levels or \emph{labels} with every bit of data in the system to
subsequently track and restrict the flow of labeled data throughout the system,
e.g., to enforce a security property such as
\emph{non-interference}~\citep{Goguen:Meseguer:Noninterference}.
In practice, dynamic IFC implementations are considerably more
complex---the \emph{granularity} of the tracking system alone has
important implications for the usage of IFC technology.
Indeed, until somewhat
recently~\citep{roy2009laminar,stefan:2017:flexible}, granularity was
the main distinguishing factor between dynamic IFC operating systems
and programming languages.
Most IFC operating systems (e.g.,~\cite{Efstathopoulos:2005, Zeldovich:2006,
  krohn:flume}) are \emph{coarse-grained}, i.e., they track and enforce
information flow at the granularity of a process or thread.
Conversely, most programming languages with dynamic IFC
(e.g.,~\cite{Austin:Flanagan:PLAS09, zdancewic:thesis, hedin2014jsflow,
  hritcu2013all, yang2012language}) track the flow of information in a more
\emph{fine-grained} fashion, e.g., at the granularity of program variables and
references.

Dynamic coarse-grained IFC systems in the style of
LIO~\citep{stefan:2017:flexible, stefan:lio, stefan:2012:addressing,
  heule:2015:ifc-inside,buiras2015hlio, VASSENA2017} have several advantages
over dynamic fine-grained IFC systems.
%
%
Such coarse-grained systems are often easier to design and implement---they
inherently track less information.
For example, LIO protects against control-flow-based \emph{implicit flows} by
tracking information at a coarse-grained level---to branch on secrets, LIO
programs must first taint the context where secrets are going to be observed.
Finally, coarse-grained systems often require considerably fewer programmer
annotations---unlike fine-grained ones.
More specifically, developers often only need a single label-annotation to
protect everything in the scope of a thread or process responsible to handle
sensitive data.

Unfortunately, these advantages of coarse-grained systems give up on the many
benefits of fine-grained ones.
For instance, one main drawback of coarse-grained systems is that it requires
developers to compartmentalize their application in order to avoid both false
alarms and the \emph{label creep} problem, i.e., wherein the program gets too
``tainted'' to do anything useful.
To this end, coarse-grained systems often create special abstractions
(e.g., event processes~\citep{Efstathopoulos:2005},
gates~\citep{Zeldovich:2006}, and security
regions~\citep{roy2009laminar}) that compensate for the conservative
approximations of the coarse-grained tracking approach.
Furthermore, fine-grained systems do not impose the burden of focusing on
avoiding the label creep problem on developers.
By tracking information at fine granularity, such systems are seemingly more
flexible and do not suffer from false alarms and label creep
issues~\citep{Austin:Flanagan:PLAS09} as coarse-grained systems do.
Indeed, fine-grained systems such as JSFlow~\citep{hedin2014jsflow} can often be
used to secure existing, legacy applications; they only require developers to
properly annotate the application.

This paper removes the division between fine- and coarse-grained dynamic IFC
systems and the belief that they are fundamentally different.
In particular, we show that \emph{dynamic} fine-grained and coarse-grained IFC are equally
expressive.
Our work is inspired by the recent work of~\citet{RajaniBRG17, Rajani:2018}, who
prove similar results for \emph{static} fine-grained and coarse-grained IFC
systems.
Specifically, they establish a semantics- and type-preserving translation from a
coarse-grained IFC type system to a fine-grained one and vice-versa.
We complete the picture by showing a similar result for dynamic IFC systems that
additionally allow \emph{introspection on labels} at run-time.
While label introspection is meaningless in a static IFC system, in a dynamic
IFC system this feature is key to both writing practical applications and
mitigating the label creep problem~\citep{stefan:2017:flexible}.

Using the Agda proof assistant~\parencite{Norell09,Bove09}, we
formalize a traditional fine-grained system (in the style of
\cite{Austin:Flanagan:PLAS09}) extended with label introspection
primitives, as well as a coarse-grained system (in the style of
\cite{stefan:2017:flexible}).
We then define and formalize modular semantics-preserving translations between
them.
Our translations are macro-expressible in the sense
of~\citet{Felleisen:1991}, i.e., they can be expressed as a pure
source program rewriting.

We show that a translation from fine- to coarse-grained is possible when the
coarse-grained system is equipped with a primitive that limits the scope of
tainting (e.g., when reading sensitive data).
In practice, this is not an imposing requirement since most coarse-grained
systems rely on such primitives for compartmentalization.
For example,~\citet{stefan:2017:flexible, stefan:2012:addressing}, provide
\ensuremath{\mathbf{toLabeled}(\cdot \mkern1mu)} blocks and threads for precisely this purpose.
Dually, we show that the translation from coarse- to fine-grained is possible
when the fine-grained system has a primitive \ensuremath{\mathbf{taint}(\cdot \mkern1mu)} that relaxes precision to
keep the \emph{program counter label} synchronized when translating a program to
the coarse-grained language.
While this primitive is largely necessary for us to establish the coarse- to
fine-grained translation, extending existing fine-grained systems with it is
both secure and trivial.

The implications of our results are multi-fold.
The fine- to coarse-grained translation formally confirms an old OS-community
hypothesis that it is possible to restructure a system into smaller compartments
to address the label creep problem---indeed our translation is a (naive)
algorithm for doing so.
This translation also allows running legacy fine-grained IFC compatible
applications atop coarse-grained systems like LIO.
Dually, the coarse- to fine-grained translation allows developers building new
applications in a fine-grained system to avoid the annotation burden of the
fine-grained system by writing some of the code in the coarse-grained system and
compiling it automatically to the fine-grained system with our translation.
The technical contributions of this paper are:
\begin{itemize}
\item A pair of semantics-preserving translations between traditional
  dynamic fine-grained and coarse-grained IFC systems equipped with
  label introspection and flow-insensitive references
  (Theorems~\ref{thm:fg2cg-correct} and
  \ref{thm:cg2fg-forcing-correct}).
\item Two different proofs of \emph{termination-insensitive}
  non-interference (TINI) for each calculus: one is derived directly
  in the usual way (Theorems~\ref{thm:fg-tini} and \ref{thm:cg-tini}),
  while the other is recovered via our verified translation
  (Theorems~\ref{thm:recovery-fg2cg} and \ref{thm:recovery-cg2fg}).
\item Mechanized Agda proofs of our results (\textasciitilde 4,000
  LOC).
%
\end{itemize}

This monograph is based on our conference paper~\parencite{Vassena19} and
extended with:

\begin{itemize}
\item A tutorial-style introduction to fine- and coarse-grained
  dynamic IFC, which (i) illustrates their specific features and
  (apparent) differences through examples, and (ii) supplements our
  proof artifacts with general explanations of the proof techniques
  used.

\item \emph{Flow-sensitive} references, a key feature for boosting the
  permissiveness of dynamic IFC
  systems~\parencite{Austin:Flanagan:PLAS09}.
  We extend both fine- and coarse-grained language with flow-sensitive
  references (Sections~\ref{sec:fg-fs-refs} and \ref{sec:cg-fs-refs}),
  adapt their security proofs (Theorems~\ref{thm:fg-tini-bij} and
  \ref{thm:cg-tini-bij}), and the verified translations to each other.

\item A discussion and analysis of our extended proof artifact
  (\textasciitilde 6,900 LOC)\footnote{The extended artifact is
    available at \href{https://hub.docker.com/r/marcovassena/granularity-ftpl}{\url{https://hub.docker.com/r/marcovassena/granularity-ftpl}} and supersedes the artifact archived with
    the conference paper.}.  Our analysis highlights that the security
  proofs for fine-grained languages are between 43\% and 74\% longer
  than for coarse-grained languages. These empirical results confirm
  the general sense that coarse-grained IFC languages are easier to
  prove secure than fine-grained languages.

\end{itemize}

This tutorial is organized as follows.
Our dynamic fine- and coarse-grained IFC calculi are introduced in
Sections \ref{sec:fg-calculus} and \ref{sec:cg-calculus}, and then
extended with flow-sensitive references in
Sections~\ref{sec:fg-fs-refs} and \ref{sec:cg-fs-refs}, respectively.
We prove the soundness guarantees (i.e., termination-insensitive
non-interference) of the original languages
(Sections~\ref{subsec:fg-security} and \ref{subsec:cg-security}), and
of the extended languages
(Sections~\ref{subsec:fg-fs-refs-security} and
\ref{subsec:cg-fs-refs-sec}).
In Section~\ref{sec:cg-discussion}, we discuss our mechanized proof
artifacts and compare the security proofs of the two calculi, before
and after the extension.
In Section \ref{sec:fg2cg}, we present the fine- to coarse-grained
translation and a proof of non-interference for the fine-grained
calculus recovered from non-interference of the other calculus through
our verified translation. Section \ref{sec:cg2fg} presents similar
results in the other direction.
Related work is described in Section \ref{sec:related} and Section
\ref{sec:conclusions} concludes the paper.



\chapter{Fine-Grained IFC Calculus}
\label{sec:fg-calculus}

\begin{figure}[!t]
\begin{align*}
\text{Types: } & \ensuremath{\tau\Coloneqq\mathbf{unit}\;{\mid}\;\tau_{1}\to \tau_{2}\;{\mid}\;\tau_{1}\mathbin{+}\tau_{2}\;{\mid}\;\tau_{1}\;\times\;\tau_{2}\;{\mid}\;\mathscr{L}\;{\mid}\;\mathbf{Ref}\;\tau}
\\
\text{Labels: } & \ensuremath{{\ell},\Varid{pc}\;\in\;\mathscr{L}}
\\
\text{Addresses: } & \ensuremath{\Varid{n}\;\in\;\mathbb{N}}
\\
\text{Environments: } & \ensuremath{\theta\;\in\;\Conid{Var}\;\rightharpoonup\;\Conid{Value}}
\\
\text{Raw Values: } &  \ensuremath{{r}\Coloneqq()\;{\mid}\;(\Varid{x}.\Varid{e},\theta)\;{\mid}\;\mathbf{inl}({v}\mkern1mu)\;{\mid}\;\mathbf{inr}({v}\mkern1mu)\;{\mid}\;({v}_{1},{v}_{2})\;{\mid}\;{\ell}\;{\mid}\;{\Varid{n}}_{{\ell}}}
\\
\text{Values: } & \ensuremath{{v}\Coloneqq{{r}}^{{\ell}}}
\\
\text{Expression: } & \ensuremath{\Varid{e}\Coloneqq\Varid{x}\;{\mid}\lambda \Varid{x}.\Varid{e}\;{\mid}\;\Varid{e}_{1}\;\Varid{e}_{2}\;{\mid}\;()\;{\mid}\;{\ell}\;{\mid}\;(\Varid{e}_{1},\Varid{e}_{2})\;{\mid}\;\mathbf{fst}(\Varid{e}\mkern1mu)\;{\mid}\;\mathbf{snd}(\Varid{e}\mkern1mu)} \\
       & \ensuremath{{\mid}\;\mathbf{inl}(\Varid{e}\mkern1mu)\;{\mid}\;\mathbf{inr}(\Varid{e}\mkern1mu)\;{\mid}\;\mathbf{case}(\Varid{e},\Varid{x}.\Varid{e}_{1},\Varid{x}.\Varid{e}_{2}\mkern1mu)} \\
       & \ensuremath{{\mid}\;\mathbf{getLabel}\;{\mid}\;\mathbf{labelOf}(\Varid{e}\mkern1mu)\;{\mid}\;\Varid{e}_{1}\;\flows^{?}\;\Varid{e}_{2}\;{\mid}\;\mathbf{taint}(\Varid{e}_{1},\Varid{e}_{2}\mkern1mu)} \\
       & \ensuremath{{\mid}\;\mathbf{new}(\Varid{e}\mkern1mu)\;{\mid}\mathbin{!}\Varid{e}\;{\mid}\;\Varid{e}_{1}\mathbin{:=}\Varid{e}_{2}\;{\mid}\;\mathbf{labelOfRef}(\Varid{e}\mkern1mu)}
\\
\text{Type System: } & \ensuremath{\Gamma\vdash\Varid{e}\mathbin{:}\tau}
\\
\text{Configurations: } & \ensuremath{\Varid{c}\Coloneqq\langle\Sigma,\Varid{e}\rangle}
\\
\text{Stores: } & \ensuremath{\Sigma\;\in\;({\ell}\mathbin{:}\Conid{Label})\to \Conid{Memory}\;{\ell}} \\
\text{Memory }\ensuremath{{\ell}} \text{: } & \ensuremath{\Conid{M}\Coloneqq[\mskip1.5mu \mskip1.5mu]\;{\mid}\;{r}\mathbin{:}\Conid{M}}
\end{align*}
\caption{Syntax of \ensuremath{\FG}.}
\label{fig:fg-syntax}
\end{figure}

In order to compare in a rigorous way fine- and coarse-grained dynamic
IFC techniques, we formally define the operational semantics of two
\ensuremath{\lambda}-calculi that respectively perform fine- and coarse-grained
IFC dynamically.
Figure \ref{fig:fg-syntax} shows the syntax of the dynamic
fine-grained IFC calculus \ensuremath{\FG}, which is inspired by
\citet{Austin:Flanagan:PLAS09} and extended with a standard (security
unaware) type system \ensuremath{\Gamma\vdash\Varid{e}\mathbin{:}\tau} (omitted), sum and product
data types and security labels \ensuremath{{\ell}\;\in\;\mathscr{L}} that form a lattice
\ensuremath{(\mathscr{L},\flows)}.\footnote{ The lattice is arbitrary and
  fixed. In examples we will often use the two point lattice \ensuremath{\{\mskip1.5mu \Blue{\Conid{L}},\Red{\Conid{H}}\mskip1.5mu\}}, which only disallows secret to public flow of information,
  i.e., \ensuremath{\Red{\Conid{H}}\;\not\flows\;\Blue{\Conid{L}}}.}
\mv[inline]{Tutorial style? maybe we should introduce briefly the lattice and main properties in a general background section.}
In order to capture flows of information precisely at run-time, the
\ensuremath{\FG}-calculus features \emph{intrinsically labeled} values, written
\ensuremath{{{r}}^{{\ell}}}, meaning that raw value \ensuremath{{r}} has security level \ensuremath{{\ell}}.
Compound values, e.g., pairs and sums, carry labels to tag the
security level of each component, for example a pair
containing a secret and a public boolean would be written \ensuremath{({\mathbf{true}}^{\Red{\Conid{H}}},{\mathbf{false}}^{\Blue{\Conid{L}}})}.\footnote{We define the
  boolean type \ensuremath{\mathbf{bool}\mathrel{=}\mathbf{unit}\mathbin{+}\mathbf{unit}}, boolean values as raw values, i.e.,
  \ensuremath{\mathbf{true}\mathrel{=}\mathbf{inl}({()}^{\Blue{\Conid{L}}}\mkern1mu)}, \ensuremath{\mathbf{false}\mathrel{=}\mathbf{inr}({()}^{\Blue{\Conid{L}}}\mkern1mu)} and \ensuremath{\mathbf{if}\;\Varid{e}\;\mathbf{then}\;\Varid{e}_{1}\;\mathbf{else}\;\Varid{e}_{2}\mathrel{=}\mathbf{case}\;\Varid{e}\;\anonymous .\Varid{e}_{1}\;\anonymous .\Varid{e}_{2}}.}
%
%
Functional values are closures \ensuremath{(\Varid{x}.\Varid{e},\theta)}, where \ensuremath{\Varid{x}} is the
variable that binds the argument in the body of the function \ensuremath{\Varid{e}} and
all other free variables are mapped to some labeled value in the
environment \ensuremath{\theta}.
The \ensuremath{\FG}-calculus features a labeled partitioned store, i.e., \ensuremath{\Sigma\;\in\;({\ell}\mathbin{:}\mathscr{L})\to \Conid{Memory}\;{\ell}}, where \ensuremath{\Conid{Memory}\;{\ell}} is the memory that
contains values at security level \ensuremath{{\ell}}.
Each reference carries an additional label annotation that records the
label of the memory it refers to---reference \ensuremath{{\Varid{n}}_{{\ell}}} points to
the \ensuremath{\Varid{n}}-th cell of the \ensuremath{{\ell}}-labeled memory, i.e., \ensuremath{\Sigma({\ell})}.
Notice that this label has nothing to do with the \emph{intrinsic}
label that decorates the reference itself.
For example, a reference \ensuremath{{({\Varid{n}}_{\Red{\Conid{H}}})}^{\Blue{\Conid{L}}}} represents a secret
reference in a public context, whereas \ensuremath{{({\Varid{n}}_{\Blue{\Conid{L}}})}^{\Red{\Conid{H}}}}
represents a public reference in a secret context.
Notice that there is no order invariant between those labels---in the
latter case, the IFC runtime monitor prevents writing data to the
reference to avoid \emph{implicit flows}.
A program can create, read and write a labeled reference via
constructs \ensuremath{\mathbf{new}(\Varid{e}\mkern1mu)}, \ensuremath{\mathbin{!}\Varid{e}} and \ensuremath{\Varid{e}_{1}\mathbin{:=}\Varid{e}_{2}} and inspect its subscripted
label with the primitive \ensuremath{\mathbf{labelOfRef}(\cdot \mkern1mu)}.


\section{Dynamics}

\label{subsec:fg-sem}
\begin{figure}[p]
\begin{mathpar}
\inferrule[(Var)]
{}
{\ensuremath{\langle\Sigma,\Varid{x}\rangle\Downarrow^{\theta}_{\Varid{pc}}\langle\Sigma,\theta(\Varid{x})\;\lub\;\Varid{pc}\rangle}}
\and
\inferrule[(Unit)]
{}
{\ensuremath{\langle\Sigma,()\rangle\Downarrow^{\theta}_{\Varid{pc}}\langle\Sigma,{()}^{\Varid{pc}}\rangle}}
\and
\inferrule[(Label)]
{}
{\ensuremath{\langle\Sigma,{\ell}\rangle\Downarrow^{\theta}_{\Varid{pc}}\langle\Sigma,{{\ell}}^{\Varid{pc}}\rangle}}
\and
\inferrule[(Fun)]
{}
{\ensuremath{\langle\Sigma,\lambda \Varid{x}.\Varid{e}\rangle\Downarrow^{\theta}_{\Varid{pc}}\langle\Sigma,{(\Varid{x}.\Varid{e},\theta)}^{\Varid{pc}}\rangle}}
\and
\inferrule[(App)]
{\ensuremath{\langle\Sigma,\Varid{e}_{1}\rangle\Downarrow^{\theta}_{\Varid{pc}}\langle\Sigma',{(\Varid{x}.\Varid{e},\theta')}^{{\ell}}\rangle} \\
 \ensuremath{\langle\Sigma',\Varid{e}_{2}\rangle\Downarrow^{\theta}_{\Varid{pc}}\langle\Sigma'',{v}_{2}\rangle} \\
 \ensuremath{\langle\Sigma'',\Varid{e}\rangle\Downarrow^{\update{\theta'}{\Varid{x}}{{v}_{2}}}_{\Varid{pc}\;\lub\;{\ell}}\langle\Sigma''',{v}\rangle} }
{\ensuremath{\langle\Sigma,\Varid{e}_{1}\;\Varid{e}_{2}\rangle\Downarrow^{\theta}_{\Varid{pc}}\langle\Sigma''',{v}\rangle}}
\and
\inferrule[(Inl)]
{\ensuremath{\langle\Sigma,\Varid{e}\rangle\Downarrow^{\theta}_{\Varid{pc}}\langle\Sigma',{v}\rangle}}
{\ensuremath{\langle\Sigma,\mathbf{inl}(\Varid{e}\mkern1mu)\rangle\Downarrow^{\theta}_{\Varid{pc}}\langle\Sigma',{\mathbf{inl}({v}\mkern1mu)}^{\Varid{pc}}\rangle}}
\and
\inferrule[(Inr)]
{\ensuremath{\langle\Sigma,\Varid{e}\rangle\Downarrow^{\theta}_{\Varid{pc}}\langle\Sigma',{v}\rangle}}
{\ensuremath{\langle\Sigma,\mathbf{inr}(\Varid{e}\mkern1mu)\rangle\Downarrow^{\theta}_{\Varid{pc}}\langle\Sigma',{\mathbf{inr}({v}\mkern1mu)}^{\Varid{pc}}\rangle}}
\and
\inferrule[(Case$_1$)]
{\ensuremath{\langle\Sigma,\Varid{e}\rangle\Downarrow^{\theta}_{\Varid{pc}}\langle\Sigma',{\mathbf{inl}({v}_{1}\mkern1mu)}^{{\ell}}\rangle} \\ \ensuremath{\langle\Sigma',\Varid{e}_{1}\rangle\Downarrow^{\update{\theta}{\Varid{x}}{{v}_{1}}}_{\Varid{pc}\;\lub\;{\ell}}\langle\Sigma'',{v}\rangle}}
{\ensuremath{\langle\Sigma,\mathbf{case}(\Varid{e},\Varid{x}.\Varid{e}_{1},\Varid{x}.\Varid{e}_{2}\mkern1mu)\rangle\Downarrow^{\theta}_{\Varid{pc}}\langle\Sigma'',{v}\rangle}}
\and
\inferrule[(Case$_2$)]
{\ensuremath{\langle\Sigma,\Varid{e}\rangle\Downarrow^{\theta}_{\Varid{pc}}\langle\Sigma',{\mathbf{inr}({v}_{2}\mkern1mu)}^{{\ell}}\rangle} \\ \ensuremath{\langle\Sigma',\Varid{e}_{2}\rangle\Downarrow^{\update{\theta}{\Varid{x}}{{v}_{2}}}_{\Varid{pc}\;\lub\;{\ell}}\langle\Sigma'',{v}\rangle}}
{\ensuremath{\langle\Sigma,\mathbf{case}(\Varid{e},\Varid{x}.\Varid{e}_{1},\Varid{x}.\Varid{e}_{2}\mkern1mu)\rangle\Downarrow^{\theta}_{\Varid{pc}}\langle\Sigma'',{v}\rangle}}
\and
\inferrule[(Pair)]
{\ensuremath{\langle\Sigma,\Varid{e}_{1}\rangle\Downarrow^{\theta}_{\Varid{pc}}\langle\Sigma',{v}_{1}\rangle} \\ \ensuremath{\langle\Sigma',\Varid{e}_{2}\rangle\Downarrow^{\theta}_{\Varid{pc}}\langle\Sigma'',{v}_{2}\rangle}}
{\ensuremath{\langle\Sigma,(\Varid{e}_{1},\Varid{e}_{2})\rangle\Downarrow^{\theta}_{\Varid{pc}}\langle\Sigma'',{({v}_{1},{v}_{2})}^{\Varid{pc}}\rangle}}
\and
\inferrule[(Fst)]
{\ensuremath{\langle\Sigma,\Varid{e}\rangle\Downarrow^{\theta}_{\Varid{pc}}\langle\Sigma',{({v}_{1},{v}_{2})}^{{\ell}}\rangle}}
{\ensuremath{\langle\Sigma,\mathbf{fst}(\Varid{e}\mkern1mu)\rangle\Downarrow^{\theta}_{\Varid{pc}}\langle\Sigma',{v}_{1}\;\lub\;{\ell}\rangle}}
\and
\inferrule[(Snd)]
{\ensuremath{\langle\Sigma,\Varid{e}\rangle\Downarrow^{\theta}_{\Varid{pc}}\langle\Sigma',{({v}_{1},{v}_{2})}^{{\ell}}\rangle}}
{\ensuremath{\langle\Sigma,\mathbf{snd}(\Varid{e}\mkern1mu)\rangle\Downarrow^{\theta}_{\Varid{pc}}\langle\Sigma',{v}_{2}\;\lub\;{\ell}\rangle}}

\end{mathpar}
\caption{Big-step semantics for \ensuremath{\FG} (part I).}
\label{fig:fg-semantics}
\end{figure}

\begin{figure}[t]
\begin{mathpar}
\inferrule[(LabelOf)]
{\ensuremath{\langle\Sigma,\Varid{e}\rangle\Downarrow^{\theta}_{\Varid{pc}}\langle\Sigma',{{r}}^{{\ell}}\rangle}}
{\ensuremath{\langle\Sigma,\mathbf{labelOf}(\Varid{e}\mkern1mu)\rangle\Downarrow^{\theta}_{\Varid{pc}}\langle\Sigma',{{\ell}}^{{\ell}}\rangle}}
\and
\inferrule[(GetLabel)]
{}
{\ensuremath{\langle\Sigma,\mathbf{getLabel}\rangle\Downarrow^{\theta}_{\Varid{pc}}\langle\Sigma'',{\Varid{pc}}^{\Varid{pc}}\rangle}}
\and
\inferrule[(\ensuremath{\flows^{?}}-T)]
{\ensuremath{\langle\Sigma,\Varid{e}_{1}\rangle\Downarrow^{\theta}_{\Varid{pc}}\langle\Sigma',{{\ell}_{1}}^{{\ell}_{1}'}\rangle} \\
 \ensuremath{\langle\Sigma',\Varid{e}_{2}\rangle\Downarrow^{\theta}_{\Varid{pc}}\langle\Sigma'',{{\ell}_{2}}^{{\ell}_{2}'}\rangle} \\
 \ensuremath{{\ell}_{1}\;\flows\;{\ell}_{2}}
}
{\ensuremath{\langle\Sigma,\Varid{e}_{1}\;\flows^{?}\;\Varid{e}_{2}\rangle\Downarrow^{\theta}_{\Varid{pc}}\langle\Sigma'',{\mathbf{inl}({()}^{\Varid{pc}}\mkern1mu)}^{{\ell}_{1}'\;\lub\;{\ell}_{2}'}\rangle}}
\and
\inferrule[(\ensuremath{\flows^{?}}-F)]
{\ensuremath{\langle\Sigma,\Varid{e}_{1}\rangle\Downarrow^{\theta}_{\Varid{pc}}\langle\Sigma',{{\ell}_{1}}^{{\ell}_{1}'}\rangle} \\
 \ensuremath{\langle\Sigma',\Varid{e}_{2}\rangle\Downarrow^{\theta}_{\Varid{pc}}\langle\Sigma'',{{\ell}_{2}}^{{\ell}_{2}'}\rangle} \\
 \ensuremath{{\ell}_{1}\;\not\flows\;{\ell}_{2}}
}
{\ensuremath{\langle\Sigma,\Varid{e}_{1}\;\flows^{?}\;\Varid{e}_{2}\rangle\Downarrow^{\theta}_{\Varid{pc}}\langle\Sigma'',{\mathbf{inr}({()}^{\Varid{pc}}\mkern1mu)}^{{\ell}_{1}'\;\lub\;{\ell}_{2}'}\rangle}}
\and
\inferrule[(Taint)]
{\ensuremath{\langle\Sigma,\Varid{e}_{1}\rangle\Downarrow^{\theta}_{\Varid{pc}}\langle\Sigma',{{\ell}}^{{\ell}'}\rangle} \\
 \ensuremath{{\ell}'\;\flows\;{\ell}}  \\
 \ensuremath{\langle\Sigma',\Varid{e}_{2}\rangle\Downarrow^{\theta}_{{\ell}}\langle\Sigma'',{v}\rangle}}
{\ensuremath{\langle\Sigma,\mathbf{taint}(\Varid{e}_{1},\Varid{e}_{2}\mkern1mu)\rangle\Downarrow^{\theta}_{\Varid{pc}}\langle\Sigma'',{v}\rangle}}
\end{mathpar}
\caption{Big-step semantics for \ensuremath{\FG} (part II).}
\label{fig:fg-semantics2}
\end{figure}

The operational semantics of \ensuremath{\FG} includes a security
monitor that propagates the label annotations of input values during
program execution and assigns security labels to the result
accordingly.
The monitor prevents information leakage by stopping the execution of
potentially leaky programs, which is reflected in the semantics by not
providing reduction rules for the cases that may cause insecure
information flow.\footnote{In this work, we ignore leaks that exploit
  program termination and prove \emph{termination insensitive}
  non-interference for \ensuremath{\FG} (Theorem~\ref{thm:fg-tini}).}
The relation \ensuremath{\langle\Sigma,\Varid{e}\rangle\Downarrow^{\theta}_{\Varid{pc}}\langle\Sigma',{v}\rangle} denotes
the evaluation of program \ensuremath{\Varid{e}} with initial store \ensuremath{\Sigma} that
terminates with labeled value \ensuremath{{v}} and final store \ensuremath{\Sigma'}.
The environment \ensuremath{\theta} stores the input values of the program and is
extended with intermediate results during function application and
case analysis.
The subscript \ensuremath{\Varid{pc}} is the \emph{program counter} label
\citep{Sabelfeld:Myers:JSAC}--- it is a label that represents the
security level of the context in which the expression is evaluated.
The semantics employs the program counter label to (i)
propagate and assign labels to values computed by a program and (ii)
prevent implicit flow leaks that exploit the control flow and the
store (explained below).
In particular, when a program produces a value, the monitor tags the
raw value with the program counter label in order to record the
security level of the context in which it was computed.
For this reason all the introduction rules for ground and compound
types (\srule{Unit,Label,Fun,Inl,Inr,Pair}) assign security level \ensuremath{\Varid{pc}}
to the result.
Other than that, these rules are fairly standard---we simply note that
rule \srule{Fun} creates a closure by capturing the current
environment \ensuremath{\theta}.

When the control flow of a program \emph{depends} on some intermediate
value, the program counter label is joined with the value's label so that the label
of the final result will be tainted with the result of the
intermediate value.
For instance, consider case analysis, i.e., \ensuremath{\mathbf{case}\;\Varid{e}\;\Varid{x}.\Varid{e}_{1}\;\Varid{x}.\Varid{e}_{2}}.
Rules \srule{Case$_1$} and \srule{Case$_2$} evaluate the scrutinee \ensuremath{\Varid{e}}
to a value (either \ensuremath{{\mathbf{inl}({v}\mkern1mu)}^{{\ell}}} or \ensuremath{{\mathbf{inr}({v}\mkern1mu)}^{{\ell}}}), add the
value to the environment, i.e., \ensuremath{\update{\theta}{\Varid{x}}{{v}}}, and then
evaluate the appropriate branch with a program counter label tainted
with \ensuremath{{v}}'s security label, i.e., \ensuremath{\Varid{pc}\;\lub\;{\ell}}.
As a result, the monitor tracks data dependencies across control flow
constructs through the label of the result.
%
Function application follows the same principle.
In rule \srule{App}, since the first premise evaluates the function to
some closure \ensuremath{(\Varid{x}.\Varid{e},\theta')} at security level \ensuremath{{\ell}}, the third premise
evaluates the body with program counter label raised to \ensuremath{\Varid{pc}\;\lub\;{\ell}}.
The evaluation strategy is call-by-value: it evaluates the argument
before the body in the second premise and binds the corresponding
variable to its value in the environment of the closure, i.e.,
\ensuremath{\update{\theta'}{\Varid{x}}{{v}_{2}}}.
Notice that the security level of the argument is irrelevant at this
stage and that this is beneficial to not over-tainting the result:
if the function never uses its argument then the label of the result
depends exclusively on the program counter label, e.g., \ensuremath{(\lambda \Varid{x}.())\;\Varid{y}\;\Downarrow^{[\mskip1.5mu \Varid{y}\ \mapsto\ {\mathbf{true}}^{\Red{\Conid{H}}}\mskip1.5mu]}_{\Blue{\Conid{L}}}\;{()}^{\Blue{\Conid{L}}}}.
%
%
The elimination rules for variables and pairs taint the label of the
corresponding value with the program counter label for security
reasons.
In rules \srule{Var,Fst,Snd} the notation, \ensuremath{{v}\;\lub\;{\ell}'} upgrades the
label of \ensuremath{{v}} with \ensuremath{{\ell}'}---it is a shorthand for \ensuremath{{{r}}^{{\ell}\;\lub\;{\ell}'}}
with \ensuremath{{v}\mathrel{=}{{r}}^{{\ell}}}.
Intuitively, public values must be considered secret when the program
counter is secret, for example \ensuremath{\Varid{x}\;\Downarrow^{[\mskip1.5mu \Varid{x}\ \mapsto\ {()}^{\Blue{\Conid{L}}}\mskip1.5mu]}_{\Red{\Conid{H}}}\;{()}^{\Red{\Conid{H}}}}.

\subsection{Label Introspection}
\label{subsec:fg-label-introspection}
The \ensuremath{\FG}-calculus features primitives for label introspection, namely
\ensuremath{\mathbf{getLabel}}, \ensuremath{\mathbf{labelOf}(\cdot \mkern1mu)} and \ensuremath{\flows^{?}}---see Figure \ref{fig:fg-semantics2}.
These operations allow to respectively retrieve the current program
counter label, obtain the label annotations of values, and compare two
labels (inspecting labels at run-time is useful for controlling and
mitigating the label creep problem).

Enabling label introspection raises the question of what label should
be assigned to the label itself (in \ensuremath{\FG} every value, including all
label values, must be annotated with a label).
As a matter of fact, labels can be used to encode secret information
and thus careless label introspection may open the doors to
information leakage \citep{stefan:2017:flexible}.
Notice that in \ensuremath{\FG}, the label annotation on the result is computed by
the semantics together with the result and thus it is as
sensitive as the result itself (the label annotation on a value
depends on the sensitivity of all values affecting the
\emph{control-flow} of the program up to the point where the result is
computed).
This motivates the design choice to protect each projected label with
the label itself, i.e., \ensuremath{{{\ell}}^{{\ell}}} and \ensuremath{{\Varid{pc}}^{\Varid{pc}}} in rules
\srule{GetLabel} and \srule{LabelOf} in Figure \ref{fig:fg-semantics2}.
We remark that this choice is consistent with previous work on coarse-grained IFC languages \citep{Buiras:2014, stefan:2017:flexible}, but novel in the
context of fine grained IFC.
%
%

Finally, primitive \ensuremath{\mathbf{taint}(\Varid{e}_{1},\Varid{e}_{2}\mkern1mu)} temporarily raises the program
counter label to the label given by the first argument in order to
evaluate the second argument.
The fine-to-coarse translation in Section \ref{sec:fg2cg}
uses \ensuremath{\mathbf{taint}(\cdot \mkern1mu)} to loosen the precision of \ensuremath{\FG} in a controlled way and
match the \emph{coarse} approximation of our coarse-grained IFC
calculus (\ensuremath{\CG}) by upgrading the labels of intermediate values
systematically.
In rule \srule{Taint}, the constraint \ensuremath{{\ell}'\;\flows\;{\ell}} ensures that
the label of the nested context \ensuremath{{\ell}} is at least as sensitive as the
program counter label \ensuremath{\Varid{pc}}.
In particular, this constraint ensures that the operational semantics
have Property~\ref{prop:fg-pc-raise} (\emph{``the label of the result
  of any \ensuremath{\FG} program is always at least as sensitive as the program
  counter label''}) even with rule \srule{Taint}.

\begin{property}
\label{prop:fg-pc-raise}
If \ensuremath{\langle\Sigma,\Varid{e}\rangle\Downarrow^{\theta}_{\Varid{pc}}\langle\Sigma',{{r}}^{{\ell}}\rangle} then \ensuremath{\Varid{pc}\;\flows\;{\ell}}.
\end{property}
\begin{proof} By induction on the given evaluation derivation.
\end{proof}


\subsection{References}
\label{subsec:fg-fi-ref}

\begin{figure}[t]
\begin{mathpar}
\inferrule[(New)]
{\ensuremath{\langle\Sigma,\Varid{e}\rangle\Downarrow^{\theta}_{\Varid{pc}}\langle\Sigma',{{r}}^{{\ell}}\rangle} \\ 
 \ensuremath{\Varid{n}\mathrel{=}\abs{\Sigma'({\ell})}} }
{\ensuremath{\langle\Sigma,\mathbf{new}(\Varid{e}\mkern1mu)\rangle\Downarrow^{\theta}_{\Varid{pc}}\langle\update{\Sigma'}{{\ell}}{\update{\Sigma'({\ell})}{\Varid{n}}{{r}}},{({\Varid{n}}_{{\ell}})}^{\Varid{pc}}\rangle}}
\and
\inferrule[(Read)]
{\ensuremath{\langle\Sigma,\Varid{e}\rangle\Downarrow^{\theta}_{\Varid{pc}}\langle\Sigma',{{\Varid{n}}_{{\ell}}}^{{\ell}'}\rangle} \\ \ensuremath{\Sigma'({\ell}){[}\Varid{n}{]}\mathrel{=}{r}}}
{\ensuremath{\langle\Sigma,\mathbin{!}\Varid{e}\rangle\Downarrow^{\theta}_{\Varid{pc}}\langle\Sigma',{{r}}^{{\ell}\;\lub\;{\ell}'}\rangle}}
\and
\inferrule[(Write)]
{\ensuremath{\langle\Sigma,\Varid{e}_{1}\rangle\Downarrow^{\theta}_{\Varid{pc}}\langle\Sigma',{{\Varid{n}}_{{\ell}}}^{{\ell}_{1}}\rangle} \\ \ensuremath{{\ell}_{1}\;\flows\;{\ell}} \\ 
 \ensuremath{\langle\Sigma',\Varid{e}_{2}\rangle\Downarrow^{\theta}_{\Varid{pc}}\langle\Sigma'',{{r}}^{{\ell}_{2}}\rangle} \\ \ensuremath{{\ell}_{2}\;\flows\;{\ell}}}
{\ensuremath{\langle\Sigma,\Varid{e}_{1}\mathbin{:=}\Varid{e}_{2}\rangle\Downarrow^{\theta}_{\Varid{pc}}\langle\update{\Sigma''}{{\ell}}{\update{\Sigma''({\ell})}{\Varid{n}}{{r}}},{}^{\Varid{pc}}\rangle}}
\and
\inferrule[(LabelOfRef)]
{\ensuremath{\langle\Sigma,\Varid{e}\rangle\Downarrow^{\theta}_{\Varid{pc}}\langle\Sigma',{{\Varid{n}}_{{\ell}}}^{{\ell}'}\rangle}}
{\ensuremath{\langle\Sigma,\mathbf{labelOfRef}(\Varid{e}\mkern1mu)\rangle\Downarrow^{\theta}_{\Varid{pc}}\langle\Sigma',{{\ell}}^{{\ell}\;\lub\;{\ell}'}\rangle}}
\end{mathpar}
\caption{Big-step semantics for \ensuremath{\FG} (references).}
\label{fig:fg-memory}
\end{figure}

We now extend the semantics presented earlier with primitives that
inspect, access and modify the labeled store via labeled
references. See Figure \ref{fig:fg-memory}.
Rule \srule{New} creates a reference \ensuremath{{\Varid{n}}_{{\ell}}}, labeled with the
security level of the initial content, i.e., label \ensuremath{{\ell}}, in the \ensuremath{{\ell}}-labeled
memory \ensuremath{\Sigma({\ell})} and updates the memory store
accordingly.\footnote{\ensuremath{\abs{\Conid{M}}} denotes the length of memory
  \ensuremath{\Conid{M}}---memory indices start at 0.}
Since the security level of the reference is as sensitive as the
content, which is at least as sensitive as the program counter label
by Property \ref{prop:fg-pc-raise} (\ensuremath{\Varid{pc}\;\flows\;{\ell}}) this operation
does not leak information via \emph{implicit flows}.
When reading the content of reference \ensuremath{{\Varid{n}}_{{\ell}}} at security level
\ensuremath{{\ell}'}, rule \srule{Read} retrieves the corresponding raw value from the
\ensuremath{\Varid{n}}-th cell of the \ensuremath{{\ell}}-labeled memory, i.e.,
\ensuremath{\Sigma'({\ell}){[}\Varid{n}{]}\mathrel{=}{r}} and upgrades its label to \ensuremath{{\ell}\;\lub\;{\ell}'} since the
decision to read from that particular reference depends on information
at security level \ensuremath{{\ell}'}.
When writing to a reference the monitor performs security checks to
avoid leaks via explicit or implicit flows.
Rule \srule{Write} achieves this by evaluating the reference, i.e.,
\ensuremath{{({\Varid{n}}_{{\ell}})}^{{\ell}_{1}}} and replacing its content with the value of the
second argument, i.e., \ensuremath{{{r}}^{{\ell}_{2}}}, under the conditions that the
decision of ``which'' reference to update does not depend on data more
sensitive than the reference itself, i.e., \ensuremath{{\ell}_{1}\;\flows\;{\ell}} (not checking
this would leak via an \emph{implicit flow})\footnote{ Notice that \ensuremath{\Varid{pc}\;\flows\;{\ell}_{1}} by Property \ref{prop:fg-pc-raise}, thus \ensuremath{\Varid{pc}\;\flows\;{\ell}_{1}\;\flows\;{\ell}} by transitivity. An \emph{implicit flow} would occur if
  a reference is updated in a \emph{high branch}, i.e., depending on
  the secret, e.g., \ensuremath{\mathbf{let}\;\Varid{x}\mathrel{=}\mathbf{new}(\mathrm{0}\mkern1mu)\;\mathbf{in}\;\mathbf{if}\;\Varid{secret}\;\mathbf{then}\;\Varid{x}\mathbin{:=}\mathrm{1}\;\mathbf{else}\;()}.}, and that the new content is no more sensitive than the
reference itself, i.e., \ensuremath{{\ell}_{2}\;\flows\;{\ell}} (not checking this would leak
sensitive information to a less sensitive reference via an
\emph{explicit flow}).
Lastly, rule \srule{LabelOfRef} retrieves the label of the reference
and protects it with the label itself (as explained before) and taints
it with the security level of the reference, i.e., \ensuremath{{{\ell}}^{{\ell}\;\lub\;{\ell}'}}
to avoid leaks.
Intuitively, the label of the reference, i.e., \ensuremath{{\ell}}, depends also on
data at security level \ensuremath{{\ell}'} as seen in the premise.
%

\paragraph{Other Extensions.}
We consider \ensuremath{\FG} equipped with references as sufficient foundation to
study the relationship between fine-grained and coarse-grained IFC.
We remark that extending it with other side-effects such as file
operations, or other IO-operations would not change our claims in
Section \ref{sec:fg2cg} and \ref{sec:cg2fg}.
The main reason for this is that, typically, handling such effects would be
done at the same granularity in both IFC enforcements.
For instance, when adding file operations, both fine- (e.g.,
\cite{DBLP:conf/aplas/BrobergDS13}) and coarse-grained (e.g.,
\cite{Russo+:Haskell08,stefan:lio,Efstathopoulos:2005,krohn:flume})
enforcements are likely to assign a single \emph{flow-insensitive}
(i.e., immutable) label to each file in order to denote the
sensitivity of its content.
Then, those features could be handled \emph{flow-insensitively} in
both systems (e.g.,
\cite{Pottier:Simonet:POPL02,jif,stefan:lio,Vassena:2016}), in a
manner similar to what we have just shown for references in \ensuremath{\FG}.
  Importantly, fine- and coarse-grained IFC are equally expressive
  also when extended with \emph{flow-sensitive} (i.e., mutable)
  labels.
  Unlike flow-insensitive labels, these labels can \emph{change}
  during program execution to reflect the current sensitivity of the
  content of various resources (e.g., references, files etc.).
  %
  %
  %
  In order to show that fine- and coarse-grained IFC equally support
  flow-sensitive features, we follow the same approach described
  above.
  In particular, we add \emph{flow-sensitive}
  references~\parencite{Austin:Flanagan:PLAS09} to \ensuremath{\FG}
  (Section~\ref{sec:fg-fs-refs}) and \ensuremath{\CG}
  (Section~\ref{sec:cg-fs-refs}) and then complete our pair of
  verified semantics- and security-preserving translations from one
  language to the other (Sections~\ref{sec:fg2cg} and \ref{sec:cg2fg}).

\section{Security}
\label{subsec:fg-security}

\begin{figure}[t]

\begin{mathpar}
\inferrule[(Value\ensuremath{{}_{\Blue{\Conid{L}}}})\textsl{}]
{\ensuremath{{\ell}\;\flows\;\Blue{\Conid{L}}} \\ \ensuremath{{r}_{1}\approx_{\Blue{\Conid{L}}}{r}_{2}}}
{\ensuremath{{{r}_{1}}^{{\ell}}\approx_{\Blue{\Conid{L}}}{{r}_{2}}^{{\ell}}}}
\and
\inferrule[(Value\ensuremath{{}_{\Red{\Conid{H}}}})]
{\ensuremath{{\ell}_{1}\;\not\flows\;\Blue{\Conid{L}}} \\ \ensuremath{{\ell}_{2}\;\not\flows\;\Blue{\Conid{L}}}}
{\ensuremath{{{r}_{1}}^{{\ell}_{1}}\approx_{\Blue{\Conid{L}}}{{r}_{2}}^{{\ell}_{2}}}}
\and
\inferrule[(Unit)]
{}
{\ensuremath{()\approx_{\Blue{\Conid{L}}}()}}
\and
\inferrule[(Label)]
{}
{\ensuremath{{\ell}\approx_{\Blue{\Conid{L}}}{\ell}}}
\and
\inferrule[(Closure)]
{\ensuremath{\Varid{e}_{1}\equiv_{\alpha}\Varid{e}_{2}} \\ \ensuremath{\theta_{1}\approx_{\Blue{\Conid{L}}}\theta_{2}}}
{\ensuremath{(\Varid{e}_{1},\theta_{1})\approx_{\Blue{\Conid{L}}}(\Varid{e}_{2},\theta_{2})}}
\and
\inferrule[(Inl)]
{\ensuremath{{v}_{1}\approx_{\Blue{\Conid{L}}}{v}_{2}}}
{\ensuremath{\mathbf{inl}({v}_{1}\mkern1mu)\approx_{\Blue{\Conid{L}}}\mathbf{inl}({v}_{2}\mkern1mu)}}
\and
\inferrule[(Inr)]
{\ensuremath{{v}_{1}\approx_{\Blue{\Conid{L}}}{v}_{2}}}
{\ensuremath{\mathbf{inr}({v}_{1}\mkern1mu)\approx_{\Blue{\Conid{L}}}\mathbf{inr}({v}_{2}\mkern1mu)}}
\and
\inferrule[(Pair)]
{\ensuremath{{v}_{1}\approx_{\Blue{\Conid{L}}}{v}_{1}'} \\ \ensuremath{{v}_{2}\approx_{\Blue{\Conid{L}}}{v}_{2}'}}
{\ensuremath{({v}_{1},{v}_{2})\approx_{\Blue{\Conid{L}}}({v}_{1}',{v}_{2}')}}
\and
\inferrule[(Ref\ensuremath{{}_{\Blue{\Conid{L}}}})]
{\ensuremath{{\ell}\;\flows\;\Blue{\Conid{L}}}}
{\ensuremath{{\Varid{n}}_{{\ell}}\approx_{\Blue{\Conid{L}}}{\Varid{n}}_{{\ell}}}}
\and
\inferrule[(Ref\ensuremath{{}_{\Red{\Conid{H}}}})]
{\ensuremath{{\ell}_{1}\;\not\flows\;\Blue{\Conid{L}}} \\ \ensuremath{{\ell}_{2}\;\not\flows\;\Blue{\Conid{L}}}}
{\ensuremath{{\Varid{n}_{1}}_{{\ell}_{1}}\approx_{\Blue{\Conid{L}}}{\Varid{n}_{2}}_{{\ell}_{2}}}}
\end{mathpar}



\caption{\ensuremath{\Blue{\Conid{L}}}-equivalence for \ensuremath{\FG} values and raw values.}
\label{fig:fg-low-equiv}
\end{figure}
We now prove that \ensuremath{\FG} is secure, i.e., it satisfies \emph{termination
  insensitive non-interference} (TINI)
\citep{Goguen:Meseguer:Noninterference,Volpano:1997:ECF:794197.795081}.
Intuitively, the security condition says that no terminating \ensuremath{\FG}
program leaks information, i.e., changing secret inputs does not
produce any publicly visible effect.
The proof technique is standard and based on the notion of
\ensuremath{\Blue{\Conid{L}}}-equivalence, written \ensuremath{{v}_{1}\approx_{\Blue{\Conid{L}}}{v}_{2}}, which relates values (and
similarly raw values, environments, stores and configurations) that
are indistinguishable for an attacker at security level \ensuremath{\Blue{\Conid{L}}}.
For clarity we use the 2-points lattice, assume that secret data is
labeled with \ensuremath{\Red{\Conid{H}}} and that the attacker can only observe data at
security level \ensuremath{\Blue{\Conid{L}}}. Our mechanized proofs are parametric in the
lattice and in the security level of the attacker.

\subsection{\ensuremath{\Blue{\Conid{L}}}-Equivalence}
\ensuremath{\Blue{\Conid{L}}}-equivalence for values and raw-values is defined formally by mutual induction in Figure \ref{fig:fg-low-equiv}.
Rule \srule{Value\ensuremath{{}_{\Blue{\Conid{L}}}}} relates observable values, i.e.,
raw values labeled below the security level of the attacker.
These values have the \emph{same} observable label (\ensuremath{{\ell}\;\flows\;\Blue{\Conid{L}}})
and related raw values, i.e., \ensuremath{{r}_{1}\approx_{\Blue{\Conid{L}}}{r}_{2}}.
Rule \srule{Value\ensuremath{{}_{\Red{\Conid{H}}}}} relates non-observable values, which may
have different labels not below the attacker level, i.e., \ensuremath{{\ell}_{1}\;\not\flows\;\Blue{\Conid{L}}} and \ensuremath{{\ell}_{2}\;\not\flows\;\Blue{\Conid{L}}}. In this case, the raw values
can be arbitrary.
Raw values are \ensuremath{\Blue{\Conid{L}}}-equivalent when they consist of the same ground
value (i.e., rules \srule{Unit} and \srule{Label}), or are
homomorphically related for compound
values. 
For example, for the sum type the relation requires that both values
are either a left or a right injection through rules \srule{Inl} and
\srule{Inr}.
Closures are related by rule \srule{Closure}, if they contain the
\emph{same} function (up to \ensuremath{\alpha}-renaming)\footnote{Symbol \ensuremath{{\equiv }_{\alpha}} denotes \ensuremath{\alpha}-equivalence. In our mechanized proofs we
  use De Bruijn indexes and syntactic equivalence.} and
\ensuremath{\Blue{\Conid{L}}}-equivalent environments, i.e., the environments are
\ensuremath{\Blue{\Conid{L}}}-equivalent pointwise. Formally, \ensuremath{\theta_{1}\approx_{\Blue{\Conid{L}}}\theta_{2}} iff \ensuremath{dom(\theta_{1})\equiv dom(\theta_{2})} and \ensuremath{\forall\Varid{x}.\theta_{1}(\Varid{x})\approx_{\Blue{\Conid{L}}}\theta_{2}(\Varid{x})}.
We define \ensuremath{\Blue{\Conid{L}}}-equivalence for stores pointwise, i.e., \ensuremath{\Sigma_1\approx_{\Blue{\Conid{L}}}\Sigma_2} iff for all labels \ensuremath{{\ell}\;\in\;\mathscr{L}}, \ensuremath{\Sigma_1({\ell})\approx_{\Blue{\Conid{L}}}\Sigma_2({\ell})}.
Memory \ensuremath{\Blue{\Conid{L}}}-equivalence relates arbitrary \ensuremath{{\ell}}-labeled memories if \ensuremath{{\ell}\;\not\flows\;\Blue{\Conid{L}}}, and pointwise otherwise, i.e., \ensuremath{\Conid{M}_1\approx_{\Blue{\Conid{L}}}\Conid{M}_2} iff \ensuremath{\Conid{M}_1} and
\ensuremath{\Conid{M}_2} are memories labeled with \ensuremath{{\ell}\;\flows\;\Blue{\Conid{L}}}, \ensuremath{\abs{\Conid{M}_1}\mathrel{=}\abs{\Conid{M}_2}}
and for all \ensuremath{\Varid{n}\;\in\;\{\mskip1.5mu \mathrm{0}\mathinner{\ldotp\ldotp}\abs{\Conid{M}_1}\mathbin{-}\mathrm{1}\mskip1.5mu\}}, \ensuremath{\Conid{M}_1{[}\Varid{n}{]}\approx_{\Blue{\Conid{L}}}\Conid{M}_2{[}\Varid{n}{]}}.
Similarly, \ensuremath{\Blue{\Conid{L}}}-equivalence relates any two secret references through
rule \srule{Ref\ensuremath{{}_{\Red{\Conid{H}}}}}, but requires the same label and address
for public references in rule \srule{Ref\ensuremath{{}_{\Blue{\Conid{L}}}}}.
We naturally lift \ensuremath{\Blue{\Conid{L}}}-equivalence to initial configurations, i.e.,
\ensuremath{\Varid{c}_{1}\approx_{\Blue{\Conid{L}}}\Varid{c}_{2}} iff \ensuremath{\Varid{c}_{1}\mathrel{=}\langle\Sigma_1,\Varid{e}_{1}\rangle}, \ensuremath{\Varid{c}_{2}\mathrel{=}\langle\Sigma_2,\Varid{e}_{2}\rangle},
\ensuremath{\Sigma_1\approx_{\Blue{\Conid{L}}}\Sigma_2} and \ensuremath{\Varid{e}_{1}\equiv_{\alpha}\Varid{e}_{2}}, and final
configurations, i.e., \ensuremath{\Varid{c}_{1}'\approx_{\Blue{\Conid{L}}}\Varid{c}_{2}'} iff \ensuremath{\Varid{c}_{1}'\mathrel{=}\langle\Sigma_{1}',{v}_{1}\rangle}, \ensuremath{\Varid{c}_{2}'\mathrel{=}\langle\Sigma_{2}',{v}_{2}\rangle} and \ensuremath{\Sigma_{1}'\approx_{\Blue{\Conid{L}}}\Sigma_{2}'} and \ensuremath{{v}_{1}\approx_{\Blue{\Conid{L}}}{v}_{2}}.
The \ensuremath{\Blue{\Conid{L}}}-equivalence relation defined above is \emph{reflexive},
\emph{symmetric}, and \emph{transitive}.

\begin{property}
\label{prop:fg-leq}
Let $x$, $y$, $z$ range over labeled values, raw values, environments,
labeled memories, stores, and configurations:
\begin{enumerate}
\item \label{prop:fg-leq-refl} \textbf{Reflexivity.} For all \ensuremath{\Varid{x}}, \ensuremath{\Varid{x}\approx_{\Blue{\Conid{L}}}\Varid{x}}.
\item \label{prop:fg-leq-sym} \textbf{Symmetricity.} For all \ensuremath{\Varid{x}} and \ensuremath{\Varid{y}}, if \ensuremath{\Varid{x}\approx_{\Blue{\Conid{L}}}\Varid{y}}, then \ensuremath{\Varid{y}\approx_{\Blue{\Conid{L}}}\Varid{x}}.
\item \label{prop:fg-leq-trans} \textbf{Transitivity.} For all \ensuremath{\Varid{x}}, \ensuremath{\Varid{y}}, \ensuremath{\Varid{z}}, if \ensuremath{\Varid{x}\approx_{\Blue{\Conid{L}}}\Varid{y}} and \ensuremath{\Varid{y}\approx_{\Blue{\Conid{L}}}\Varid{z}}, then \ensuremath{\Varid{x}\approx_{\Blue{\Conid{L}}}\Varid{z}}.
\end{enumerate}
\end{property}
These properties simplify the security analysis of \ensuremath{\FG}: they let us
reason about \ensuremath{\Blue{\Conid{L}}}-equivalent terms using \emph{commutative diagrams}.
\mv[inline]{Is there a good reference for introducing these diagrams
  and using them for IFC?}
%
%
For example, consider the \emph{Square Commutative Diagram for Stores}
outlined in Figure~\ref{fig:fg-square-store}.
In the diagram, the arrows connect \ensuremath{\Blue{\Conid{L}}}-equivalent stores (e.g., the
arrow from \ensuremath{\Sigma_1} to \ensuremath{\Sigma_{1}'} indicates that \ensuremath{\Sigma_1\approx_{\Blue{\Conid{L}}}\Sigma_{1}'}).
The diagram provides a visual representation of
Lemma~\ref{lemma:fg-square-store}: solid arrows represent the
assumptions of the lemma and the dashed arrow represents the
conclusion.
To prove the lemma, we have to show that the diagram \emph{commutes},
i.e., we need to construct a path from \ensuremath{\Sigma_{1}'} to \ensuremath{\Sigma_{2}'} using
the solid arrows.
Thanks to Property~\ref{prop:fg-leq}, we can derive additional arrows
to construct this path.
For example, we can reverse arrows using \emph{symmetricity} (e.g.,
\ensuremath{\Sigma_{1}'\approx_{\Blue{\Conid{L}}}\Sigma_1} from \ensuremath{\Sigma_1\approx_{\Blue{\Conid{L}}}\Sigma_{1}'}) and \emph{transitivity}
let us compose consecutive arrows (e.g., \ensuremath{\Sigma_1\approx_{\Blue{\Conid{L}}}\Sigma_{2}'} from
\ensuremath{\Sigma_1\approx_{\Blue{\Conid{L}}}\Sigma_2} and \ensuremath{\Sigma_2\approx_{\Blue{\Conid{L}}}\Sigma_{2}'}).

%
%
\begin{figure}[t]
\begin{mathpar}
\begin{tikzcd}
  \Sigma_1 \arrow[r , "\ensuremath{\approx_{\Blue{\Conid{L}}}}"] \arrow[d , swap, "\ensuremath{\approx_{\Blue{\Conid{L}}}}"]  & \Sigma_1' \arrow[d, dashed, "\ensuremath{\approx_{\Blue{\Conid{L}}}}"] \\
  \Sigma_2 \arrow[r , swap, "\ensuremath{\approx_{\Blue{\Conid{L}}}}"] & \Sigma_2'
\end{tikzcd}
\end{mathpar}
\caption{Square commutative diagram for stores
  (Lemma~\ref{lemma:fg-square-store}).}
\label{fig:fg-square-store}
\end{figure}

\begin{lemma}[Square Commutative Diagram for Stores]
If \ensuremath{\Sigma_1\approx_{\Blue{\Conid{L}}}\Sigma_{1}'}, \ensuremath{\Sigma_1\approx_{\Blue{\Conid{L}}}\Sigma_2}, \ensuremath{\Sigma_2\approx_{\Blue{\Conid{L}}}\Sigma_{2}'}, then \ensuremath{\Sigma_{1}'\approx_{\Blue{\Conid{L}}}\Sigma_{2}'}.
\label{lemma:fg-square-store}
\end{lemma}
\begin{proof}
%
  We show that the square diagram commutes using \emph{symmetricity}
  (Property~\ref{prop:fg-leq}.2) and \emph{transitivity}
  (Property~\ref{prop:fg-leq}.3) to draw the red arrows.
\begin{mathpar}
\begin{tikzcd}
  \Sigma_1 \arrow[r , "\ensuremath{\approx_{\Blue{\Conid{L}}}}"] \arrow[d , swap, "\ensuremath{\approx_{\Blue{\Conid{L}}}}"] \arrow[dr, red]  & \Sigma_1' \arrow[l, red, bend left=25] \arrow[d, red] \\
  \Sigma_2 \arrow[r , swap, "\ensuremath{\approx_{\Blue{\Conid{L}}}}"] & \Sigma_2'
\end{tikzcd}
\end{mathpar}
\end{proof}

\subsection{Termination-Insensitive Non-Interference}
The security monitor of \ensuremath{\FG} enforces termination-insensitive
non-interference.
Intuitively, this property guarantees that \emph{terminating} programs
do not leak secret data into public values and memories of the store.
%
%
More formally, a program satisfies non-interference if, given
indistinguishable inputs (initial stores and environments), then it
produces outputs (final stores and values) that are also
indistinguishable to the attacker.
%
%
%
%
We prove this result through two key lemmas: \emph{store confinement}
and \emph{\ensuremath{\Blue{\Conid{L}}}-equivalence preservation}.
In the following, we give a high-level overview of these lemmas and
their proof, focusing on the general proof technique and illustrative
cases.
We refer to our mechanized proof scripts for complete proofs.

\paragraph{Store Confinement.}
At a high-level, store confinement ensures that programs cannot leak
secret data \emph{implicitly} through observable side-effects in the
labeled store.
%
%
Intuitively, the side-effects of programs running in secret contexts
must be \emph{confined} to secret memories in the labeled store to
enforce security---programs that do otherwise may leak and are
therefore conservatively aborted by the security monitor.
%
%
%
This lemma holds for \ensuremath{\FG} precisely because the constraints in rules
\srule{New} and \srule{Write} only allow programs to write memories
labeled above the program counter label.
In particular, these constraints prevent programs running in secret
contexts from writing public memories, which remain unchanged and thus
indistinguishable to the attacker.
%

%
%
\begin{lemma}[Store Confinement]
\label{lemma:fg-confinement}
For all configurations \ensuremath{\Varid{c}\mathrel{=}\langle\Sigma,\Varid{e}\rangle}, \ensuremath{\Varid{c'}\mathrel{=}\langle\Sigma',{v}\rangle}, program
counter labels \ensuremath{\Varid{pc}\;\not\flows\;\Blue{\Conid{L}}}, if \ensuremath{\Varid{c}\;\Downarrow^{\theta}_{\Varid{pc}}\;\Varid{c'}}, then
\ensuremath{\Sigma\approx_{\Blue{\Conid{L}}}\Sigma'}.
\end{lemma}
\begin{proof}
  The proof is by induction on the big-step reduction, using
  \emph{reflexivity} (Property~\ref{prop:fg-leq}.1) in the base cases
  and \emph{transitivity} (Property~\ref{prop:fg-leq}.3) in the
  inductive cases.
  In the inductive cases (e.g., \srule{Case$_1$}), we observe that the
  program counter label of the nested computations is always \emph{at
  least as sensitive} as the initial program counter \ensuremath{\Varid{pc}}, and
  therefore above the attacker's label \ensuremath{\Blue{\Conid{L}}}, i.e., if \ensuremath{\Varid{pc}\;\not\flows\;\Blue{\Conid{L}}}, then \ensuremath{\Varid{pc}\;\lub\;{\ell}\;\not\flows\;\Blue{\Conid{L}}} for any label \ensuremath{{\ell}}.
  %
  The interesting cases are those that change the store, i.e., cases
  \srule{New} and \srule{Write}, where we use
  Property~\ref{prop:fg-pc-raise} to show that these rules can modify
  only secret memories.
  %
  For example, in case \srule{New}, the program creates a reference in
  a secret context (\ensuremath{\Varid{pc}\;\not\flows\;\Blue{\Conid{L}}}).
  First, the program computes a value labeled \ensuremath{{\ell}} above the attacker's
  label \ensuremath{\Blue{\Conid{L}}}, i.e., \ensuremath{{\ell}\;\not\flows\;\Blue{\Conid{L}}} by
  Property~\ref{prop:fg-pc-raise}, and then allocates it in the
  corresponding secret memory also labeled \ensuremath{{\ell}}.
  Since \ensuremath{{\ell}\;\not\flows\;\Blue{\Conid{L}}}, the original and the extended memory are
  indistinguishable by the attacker and so are the initial and final
  stores.
\end{proof}

\paragraph{\ensuremath{\Blue{\Conid{L}}}-equivalence Preservation.}
Termination-insensitive non-interference ensures that programs that
receive \ensuremath{\Blue{\Conid{L}}}-equivalent inputs produce outputs that are also
\ensuremath{\Blue{\Conid{L}}}-equivalent, i.e., terminating programs \emph{preserve}
\ensuremath{\Blue{\Conid{L}}}-equivalence.
Importantly, programs must preserve \ensuremath{\Blue{\Conid{L}}}-equivalence regardless of the
sensitivity of the context in which they are executed.
Therefore, we consider \ensuremath{\Blue{\Conid{L}}}-equivalence preservation in public and
secret contexts \emph{separately}.
Then, we combine these individual results to prove \ensuremath{\Blue{\Conid{L}}}-equivalence
preservation in arbitrary contexts, i.e., termination-insensitive
non-interference.
%
%
%
More precisely, we prove two preservation lemmas: the first relates
executions in secret contexts (\ensuremath{\Varid{pc}\;\not\flows\;\Blue{\Conid{L}}}) involving
\emph{arbitrary expressions}, while the other relates executions
of the \emph{same expression} in public contexts (\ensuremath{\Varid{pc}\;\flows\;\Blue{\Conid{L}}}).
%

%
The first lemma ensures that programs cannot leak secret data
implicitly through the \emph{program control-flow}.
%
%
For example, consider the program \ensuremath{\mathbf{if}\; s\;\mathbf{then}\;\Varid{e}_{1}\;\mathbf{else}\;\Varid{e}_{2}}, which
branches on a secret boolean \ensuremath{ s}.
Depending on the value of \ensuremath{ s}, this program evaluates either
expression \ensuremath{\Varid{e}_{1}} or \ensuremath{\Varid{e}_{2}}, which may reveal the value of the secret
through secret-dependent store updates (e.g., \ensuremath{\mathbf{if}\; s\;\mathbf{then}\;\Varid{p}\mathbin{:=}\mathbf{true}\;\mathbf{else}\;()} for some public reference \ensuremath{\Varid{p}}) or results (e.g., \ensuremath{\mathbf{if}\; s\;\mathbf{then}\;\mathbf{true}\;\mathbf{else}\;\mathbf{false}}).
%
%
This lemma ensures that even these programs cannot leak secrets
through the final result or observable changes to the stores.
Formally, we have to show that programs preserve \ensuremath{\Blue{\Conid{L}}}-equivalence in
secret contexts even if they evaluate different expressions, i.e., if
\ensuremath{\Sigma_1\approx_{\Blue{\Conid{L}}}\Sigma_2}, \ensuremath{\langle\Sigma_1,\Varid{e}_{1}\rangle\Downarrow^{\theta_{1}}_{\Red{\Conid{H}}}\;\Varid{c}_{1}} and \ensuremath{\langle\Sigma_2,\Varid{e}_{2}\rangle\Downarrow^{\theta_{2}}_{\Red{\Conid{H}}}\;\Varid{c}_{2}}, then \ensuremath{\Varid{c}_{1}\approx_{\Blue{\Conid{L}}}\Varid{c}_{2}}.
How should we prove this lemma?
At first, we might try to prove it \emph{directly} by induction on
the reduction steps.
However, this approach is not practical: since these executions
involve \emph{arbitrary} expressions, we would have to reason about a
large number of completely unrelated reduction steps!
%
%
%
Instead, we observe that these programs are executed in secret
contexts (\ensuremath{\Varid{pc}\mathrel{=}\Red{\Conid{H}}}), so they are restricted by the security monitor
to avoid leaks.
From these restrictions, we establish two program \emph{invariants} to
show that the final stores and values are \ensuremath{\Blue{\Conid{L}}}-equivalent: in secret
contexts, programs can (i) modify \emph{only} secret memories
(\emph{Store Confinement}), and (ii) produce results labeled secret
(Property~\ref{prop:fg-pc-raise}).
Notice that these invariants hold for individual executions: we still
need to combine them to relate the final configurations of the two
executions.
To do that, we reason separately about the final stores and values.
The final stores are related via a \emph{Square Commutative Diagram}
(Fig.~\ref{fig:fg-square-store}), while the secret results are
trivially \ensuremath{\Blue{\Conid{L}}}-equivalent by rule \srule{Value\ensuremath{{}_{\Red{\Conid{H}}}}}
(Fig.~\ref{fig:fg-low-equiv}).
%
%
%
%

\begin{lemma}[\ensuremath{\Blue{\Conid{L}}}-Equivalence Preservation in Secret Contexts]
\label{lemma:fg-leq-preservation-secret}
For all program counter labels \ensuremath{\Varid{pc}_{1}\;\not\flows\;\Blue{\Conid{L}}} and \ensuremath{\Varid{pc}_{2}\;\not\flows\;\Blue{\Conid{L}}}, and \emph{arbitrary} expressions \ensuremath{\Varid{e}_{1}} and \ensuremath{\Varid{e}_{2}}, if
\ensuremath{\Sigma_1\approx_{\Blue{\Conid{L}}}\Sigma_2}, \ensuremath{\langle\Sigma_1,\Varid{e}_{1}\rangle\Downarrow^{\theta_{1}}_{\Varid{pc}_{1}}\;\Varid{c}_{1}}, and \ensuremath{\langle\Sigma_2,\Varid{e}_{2}\rangle\Downarrow^{\theta_{2}}_{\Varid{pc}_{2}}\;\Varid{c}_{2}}, then \ensuremath{\Varid{c}_{1}\approx_{\Blue{\Conid{L}}}\Varid{c}_{2}}.
\end{lemma}
\begin{proof}
  Assume \ensuremath{\Varid{pc}_{1}\;\not\flows\;\Blue{\Conid{L}}}, \ensuremath{\Varid{pc}_{2}\;\not\flows\;\Blue{\Conid{L}}}, \ensuremath{\Sigma_1\approx_{\Blue{\Conid{L}}}\Sigma_2},
  and let the final configurations be \ensuremath{\Varid{c}_{1}\mathrel{=}\langle\Sigma_{1}',{v}_{1}\rangle} and \ensuremath{\Varid{c}_{2}\mathrel{=}\langle\Sigma_{2}',{v}_{2}\rangle}.
  First, we apply \emph{Store Confinement}
  (Lemma~\ref{lemma:fg-confinement}) to \ensuremath{\langle\Sigma_1,\Varid{e}_{1}\rangle\Downarrow^{\theta_{1}}_{\Varid{pc}_{1}}\langle\Sigma_{1}',{v}_{1}\rangle} and \ensuremath{\langle\Sigma_2,\Varid{e}_{2}\rangle\Downarrow^{\theta_{2}}_{\Varid{pc}_{2}}\langle\Sigma_{2}',{v}_{2}\rangle} and obtain \ensuremath{\Sigma_1\approx_{\Blue{\Conid{L}}}\Sigma_{1}'} and \ensuremath{\Sigma_2\approx_{\Blue{\Conid{L}}}\Sigma_{2}'}, respectively.
  Then, we construct the \emph{Square Commutative Diagram for Stores}
  (Lemma~\ref{lemma:fg-square-store}) using \ensuremath{\Sigma_1\approx_{\Blue{\Conid{L}}}\Sigma_2}, \ensuremath{\Sigma_{1}'\approx_{\Blue{\Conid{L}}}\Sigma_1} \ensuremath{\Sigma_2\approx_{\Blue{\Conid{L}}}\Sigma_{2}'}, and obtain
  \ensuremath{\Sigma_{1}'\approx_{\Blue{\Conid{L}}}\Sigma_{2}'}.
  To show that the values \ensuremath{{v}_{1}\mathrel{=}{{r}_{1}}^{{\ell}_{1}}} and \ensuremath{{v}_{2}\mathrel{=}{{r}_{2}}^{{\ell}_{2}}} are
  \ensuremath{\Blue{\Conid{L}}}-equivalent, it suffices to show that they are labeled
  \emph{secret}.
  %
  %
  Since \ensuremath{\Varid{pc}_{1}\;\not\flows\;\Blue{\Conid{L}}} and \ensuremath{\Varid{pc}_{2}\;\not\flows\;\Blue{\Conid{L}}} by assumption,
  \ensuremath{\Varid{pc}_{1}\;\flows\;{\ell}_{1}} and \ensuremath{\Varid{pc}_{2}\;\flows\;{\ell}_{2}} by
  Property~\ref{prop:fg-pc-raise}, we have \ensuremath{{\ell}_{1}\;\not\flows\;\Blue{\Conid{L}}} and \ensuremath{{\ell}_{2}\;\not\flows\;\Blue{\Conid{L}}}, and thus \ensuremath{{v}_{1}\mathrel{=}{{r}_{1}}^{{\ell}_{1}}\approx_{\Blue{\Conid{L}}}{{r}_{2}}^{{\ell}_{2}}\mathrel{=}{v}_{2}} by
  rule \srule{Value\ensuremath{{}_{\Red{\Conid{H}}}}}.
  Since \ensuremath{\Sigma_{1}'\approx_{\Blue{\Conid{L}}}\Sigma_{2}'} and \ensuremath{{v}_{1}\approx_{\Blue{\Conid{L}}}{v}_{2}}, we have \ensuremath{\Varid{c}_{1}\mathrel{=}\langle\Sigma_{1}',{v}_{1}\rangle\approx_{\Blue{\Conid{L}}}\langle\Sigma_{2}',{v}_{2}\rangle\mathrel{=}\Varid{c}_{2}}, as desired.\looseness=-1
\end{proof}
%

We now turn our attention to \ensuremath{\Blue{\Conid{L}}}-equivalence preservation in
\emph{public contexts}.
%
Unlike \ensuremath{\Blue{\Conid{L}}}-equivalence preservation in secret contexts, we must
inspect the program executions and examine their specific reduction
steps to prove this lemma.
This is because these executions occur in a public context (\ensuremath{\Varid{pc}\;\flows\;\Blue{\Conid{L}}}), therefore their side-effects and final results may be
observable by the attacker and so we have to verify that no leaks
occur in each case.
Luckily, this task is simplified by the fact that the initial
configurations are \ensuremath{\Blue{\Conid{L}}}-equivalent and thus contain \ensuremath{\alpha}-equivalent
expressions.
Intuitively, this means that the two programs are \emph{synchronized}
and proceed in \emph{lock-step}.
In other words, their reductions are almost identical, i.e., the
programs perform the same operation on \ensuremath{\Blue{\Conid{L}}}-equivalent inputs and so
produce \ensuremath{\Blue{\Conid{L}}}-equivalent outputs.
This makes the proof of this lemma straightforward in most cases.
%
%
%
%
However, reductions that involve control-flow expressions (e.g.,
\ensuremath{\mathbf{case}(\Varid{e},\Varid{x}.\Varid{e}_{1},\Varid{x}.\Varid{e}_{2}\mkern1mu)}) are more complicated.
In general, these programs may follow different paths and evaluate
different expressions: how can they preserve \ensuremath{\Blue{\Conid{L}}}-equivalence?
Intuitively, since the programs considered in this lemma are initially
synchronized, they can start following different paths \emph{only}
after branching on secret data.
%
%
As a result, their program counter label gets tainted and the programs
enter a \emph{secret context}, in which public side-effects are not
allowed and results are guaranteed to be labeled secret, as proved in
the previous lemma.
%

\begin{lemma}[\ensuremath{\Blue{\Conid{L}}}-Equivalence Preservation in Public Contexts]
\label{lemma:fg-leq-preservation-public}
For all program counter labels \ensuremath{\Varid{pc}\;\flows\;\Blue{\Conid{L}}}, if \ensuremath{\Varid{c}_{1}\approx_{\Blue{\Conid{L}}}\Varid{c}_{2}}, \ensuremath{\theta_{1}\approx_{\Blue{\Conid{L}}}\theta_{2}},
\ensuremath{\Varid{c}_{1}\;\Downarrow^{\theta_{1}}_{\Varid{pc}}\;\Varid{c}_{1}'}, and \ensuremath{\Varid{c}_{2}\;\Downarrow^{\theta_{2}}_{\Varid{pc}}\;\Varid{c}_{2}'},
then \ensuremath{\Varid{c}_{1}'\approx_{\Blue{\Conid{L}}}\Varid{c}_{2}'}.
\end{lemma}
\begin{proof}
  Let \ensuremath{\Varid{c}_{1}\mathrel{=}\langle\Sigma_1,\Varid{e}_{1}\rangle} and \ensuremath{\Varid{c}_{2}\mathrel{=}\langle\Sigma_2,\Varid{e}_{2}\rangle} and proceed by
  induction on the big-step reductions.
  Since \ensuremath{\Varid{c}_{1}\approx_{\Blue{\Conid{L}}}\Varid{c}_{2}}, we know that the initial stores are
  \ensuremath{\Blue{\Conid{L}}}-equivalent, i.e., \ensuremath{\Sigma_1\approx_{\Blue{\Conid{L}}}\Sigma_2}, and the big-step
  reductions evaluate \ensuremath{\alpha}-equivalent expressions, i.e., \ensuremath{\Varid{e}_{1}\;{\equiv }_{\alpha}\;\Varid{e}_{2}}.
  As a result, the expressions step following the same rule in most
  cases and thus produce \ensuremath{\Blue{\Conid{L}}}-equivalent values and stores.
  %
  %

  %
The interesting cases are those that influence the control-flow of the
program, where the two executions may deviate from each other, i.e., case
\srule{Case$_1$}, \srule{Case$_2$}, and \srule{App}.
  In these cases, the security label \ensuremath{{\ell}} of the scrutinee determines
whether the two executions stay in a public context (\ensuremath{{\ell}\;\flows\;\Blue{\Conid{L}}})
and remain \emph{synchronized} on the same path, or not (\ensuremath{{\ell}\;\not\flows\;\Blue{\Conid{L}}}). 
  For example, consider expression \ensuremath{\mathbf{case}(\Varid{e},\Varid{x}.\Varid{e}_{1},\Varid{x}.\Varid{e}_{2}\mkern1mu)}, which steps
  through rule \srule{Case$_1$} in the first reduction and either
  through rule \srule{Case$_1$} or \srule{Case$_2$} in the second
  reduction (the opposite cases are symmetric).
  If both rules step through rule \srule{Case$_1$}, then the
  scrutinees are both left injections, i.e., \ensuremath{{\mathbf{inl}({v}_{1}\mkern1mu)}^{{\ell}_{1}}\approx_{\Blue{\Conid{L}}}{\mathbf{inl}({v}_{2}\mkern1mu)}^{{\ell}_{2}}} by induction hypothesis.
  Then, we perform case analysis on the \ensuremath{\Blue{\Conid{L}}}-equivalence judgment and
  have two sub-cases: \srule{Value\ensuremath{{}_{\Blue{\Conid{L}}}}} and \srule{Value\ensuremath{{}_{\Red{\Conid{H}}}}}.
  In the first case, both scrutinees are public, i.e., \ensuremath{{\ell}_{1}\mathrel{=}{\ell}_{2}\;\flows\;\Blue{\Conid{L}}} and \ensuremath{{v}_{1}\approx_{\Blue{\Conid{L}}}{v}_{2}} by rule \srule{Value\ensuremath{{}_{\Blue{\Conid{L}}}}}, and
  the proof follows by induction.
  In the other case, both scrutinees are secret, i.e., \ensuremath{{\ell}_{1}\;\not\flows\;\Blue{\Conid{L}}} and \ensuremath{{\ell}_{2}\;\not\flows\;\Blue{\Conid{L}}} by rule \srule{Value\ensuremath{{}_{\Red{\Conid{H}}}}}, and the
  program enters a secret context and we apply \emph{\ensuremath{\Blue{\Conid{L}}}-equivalence
  Preservation in Secret Contexts}
  (Lemma~\ref{lemma:fg-leq-preservation-secret}).
  Finally, if \ensuremath{\mathbf{case}(\Varid{e},\Varid{x}.\Varid{e}_{1},\Varid{x}.\Varid{e}_{2}\mkern1mu)} steps through different rules
  (e.g., \srule{Case$_1$} and \srule{Case$_2$}), then the scrutinees
  \emph{must} be secret and thus we also apply \emph{\ensuremath{\Blue{\Conid{L}}}-equivalence
  Preservation in Secret Contexts}.
  In particular, it is impossible for \ensuremath{\Blue{\Conid{L}}}-equivalent public scrutinees
  to have different injection.
  To see that, assume \ensuremath{{\mathbf{inl}({v}_{1}\mkern1mu)}^{{\ell}_{1}}\approx_{\Blue{\Conid{L}}}{\mathbf{inr}({v}_{2}\mkern1mu)}^{{\ell}_{2}}} and \ensuremath{{\ell}_{1}\mathrel{=}{\ell}_{2}\;\flows\;\Blue{\Conid{L}}} by rule \srule{Value\ensuremath{{}_{\Blue{\Conid{L}}}}}.
  Then, we also have a proof that the \emph{raw values} are
  \ensuremath{\Blue{\Conid{L}}}-equivalent, i.e., \ensuremath{\mathbf{inl}({v}_{1}\mkern1mu)\approx_{\Blue{\Conid{L}}}\mathbf{inr}({v}_{2}\mkern1mu)}.
  But this is impossible: raw values of sum type can be related only
  if they are the same injection (i.e., rules \srule{Inl} and
  \srule{Inr} in Figure~\ref{fig:fg-low-equiv}).
\end{proof}

Finally, we combine the \ensuremath{\Blue{\Conid{L}}}-equivalence preservation lemmas from
above and prove \emph{termination-insensitive} \emph{non-interference}
(TINI) for \ensuremath{\FG}.



\begin{thm}[\ensuremath{\FG}-TINI]
\label{thm:fg-tini}
If \ensuremath{\Varid{c}_{1}\;\Downarrow^{\theta_{1}}_{\Varid{pc}}\;\Varid{c}_{1}'}, \ensuremath{\Varid{c}_{2}\;\Downarrow^{\theta_{2}}_{\Varid{pc}}\;\Varid{c}_{2}'}, \ensuremath{\theta_{1}\approx_{\Blue{\Conid{L}}}\theta_{2}}
and \ensuremath{\Varid{c}_{1}\approx_{\Blue{\Conid{L}}}\Varid{c}_{2}} then \ensuremath{\Varid{c}_{1}'\approx_{\Blue{\Conid{L}}}\Varid{c}_{2}'}.
\end{thm}
\begin{proof}
  By case analysis over the program counter label \ensuremath{\Varid{pc}}.  If \ensuremath{\Varid{pc}\;\flows\;\Blue{\Conid{L}}}, we apply \emph{\ensuremath{\Blue{\Conid{L}}}-equivalence Preservation in
    Public Contexts}
  (Lemma~\ref{lemma:fg-leq-preservation-public}). If \ensuremath{\Varid{pc}\;\not\flows\;\Blue{\Conid{L}}}, we apply \emph{\ensuremath{\Blue{\Conid{L}}}-equivalence Preservation in Secret
    Contexts} (Lemma~\ref{lemma:fg-leq-preservation-secret}).
\end{proof}

\section{Flow-Sensitive References}
\label{sec:fg-fs-refs}
This section extends \ensuremath{\FG} with \emph{flow-sensitive} references \parencite{Austin:Flanagan:PLAS09}, an
  important feature to boost the permissiveness of IFC systems.
These references differ slightly from the labeled references presented in Section~\ref{subsec:fg-fi-ref}, which are instead \emph{flow-insensitive}.
The key difference between them lies in the way the IFC system treats
their label.
In particular, the label of flow-insensitive references is
\emph{immutable}, i.e., when a program creates a reference, the IFC
monitor assigns it a label, which remains fixed throughout the
execution of the program.
In contrast, the label of flow-sensitive references is \emph{mutable}.
Intuitively, the label of these references can change to reflect the
sensitivity of the data that they currently store.
This change is completely transparent to the program: when the program
writes some data to a flow-sensitive reference, the IFC monitor simply
replaces the label of the reference with the label of the new content.
However, if added naively, mutable labels open a new channel for
implicitly leaking data, therefore, the IFC monitor changes the labels
of flow-sensitive references only as long as this operation does not
leak information.
In this section, we formally discuss the differences of flow-sensitive
references and their subtleties for security in the context of \ensuremath{\FG}.
First, we show that flow-sensitive references are more permissive than
flow-insensitive references with an example.

\begin{example}
\label{example:fg-fs-refs}
Consider the following program, which creates a new reference, initially containing some public data \ensuremath{\Varid{p}}, overwrites it with secret data \ensuremath{ s}, and finally reads it back from the reference.
\begin{center}
\setlength{\parindent}{0pt}
\setlength{\mathindent}{0pt}
\begin{hscode}\SaveRestoreHook
\column{B}{@{}>{\hspre}l<{\hspost}@{}}%
\column{7}{@{}>{\hspre}l<{\hspost}@{}}%
\column{E}{@{}>{\hspre}l<{\hspost}@{}}%
\>[B]{}\mathbf{let}\;{}\<[7]%
\>[7]{}{r}\mathrel{=}\mathbf{new}(\Varid{p}\mkern1mu)\;\mathbf{in}{}\<[E]%
\\
\>[7]{}{r}\mathbin{:=} s;{}\<[E]%
\\
\>[7]{}\mathbin{!}{r}{}\<[E]%
\ColumnHook
\end{hscode}\resethooks
\end{center}
The execution of the \ensuremath{\FG} program above with program counter label public, i.e.,
\ensuremath{\Varid{pc}\mathrel{=}\Blue{\Conid{L}}}, environment \ensuremath{\theta\mathrel{=}[\mskip1.5mu \Varid{p}\ \mapsto\ {\mathbf{true}}^{\Blue{\Conid{L}}}, s\ \mapsto\ {\mathbf{false}}^{\Red{\Conid{H}}}\mskip1.5mu]}, and empty store \ensuremath{\Sigma\mathrel{=}\lambda {\ell}.[\mskip1.5mu \mskip1.5mu]} is aborted by the IFC monitor with
flow-insensitive references from Section~\ref{subsec:fg-fi-ref}.
%
%
Rule \srule{New} (Fig.~\ref{fig:fg-memory}) assigns the \emph{fixed} public label \ensuremath{\Blue{\Conid{L}}} to the reference, which then causes rule \srule{Write} to fail, as
the program tries to write secret data (i.e., \ensuremath{{\mathbf{false}}^{\Red{\Conid{H}}}}) into a public reference (i.e., \ensuremath{{\mathrm{0}}_{\Blue{\Conid{L}}}}).
In particular, the last premise of rule \srule{Write} does not hold (\ensuremath{\Red{\Conid{H}}\;\not\flows\;\Blue{\Conid{L}}}) and thus the program gets stuck.
Had the monitor not aborted the execution, the program would have leaked the secret value of \ensuremath{ s}.
Specifically, the program would have written the \emph{raw} value of \ensuremath{ s}, i.e., \ensuremath{\mathbf{false}}, into the public \ensuremath{\Blue{\Conid{L}}}-labeled memory, from where it would be extracted by the last instruction through rule \srule{Read} and labeled with \ensuremath{\Blue{\Conid{L}}}, i.e., \ensuremath{{\mathbf{false}}^{\Blue{\Conid{L}}}}, causing the leak.
In contrast, the IFC monitor presented in this section and extended with flow-sensitive references does \emph{not} abort the program.
The extended monitor allows the program to write secret data into the
public reference, but after doing so, it \emph{upgrades} the label of
the reference from \ensuremath{\Blue{\Conid{L}}} to \ensuremath{\Red{\Conid{H}}} to indicate that it now contains
secret data.
As a result, when the program reads back the reference in the last
instruction, the result gets tainted with \ensuremath{\Red{\Conid{H}}}, i.e., \ensuremath{{\mathbf{false}}^{\Red{\Conid{H}}}},
thus eliminating the leak.
\end{example}

In the following, we extend \ensuremath{\FG} with flow-sensitive references
(§~\ref{subsec:fg-fs-refs-syntax}), define their operational semantics
(§~\ref{subsec:fg-fs-refs-dynamics}), and finally reestablish
non-interference (§~\ref{subsec:fg-fs-refs-security}).

\subsection{Syntax}
\label{subsec:fg-fs-refs-syntax}
\begin{figure}[!t]
\centering
\begin{subfigure}{.9\textwidth}
\begin{align*}
\text{Flow-sensitivity tags: }      & \ensuremath{ s\Coloneqq\Conid{I}\;{\mid}\;\Conid{S}} \\
\text{Types: }                & \ensuremath{\tau\Coloneqq\cdots\;{\mid}\;\mathbf{Ref}\; s\;\tau} \\
\text{Raw Values: }           & \ensuremath{{r}\Coloneqq\cdots\;{\mid}\;\Varid{n}} \\
\text{Heaps: } & \ensuremath{\mu\Coloneqq[\mskip1.5mu \mskip1.5mu]\;{\mid}\;{{r}}^{{\ell}}\mathbin{:}\mu} \\
\text{Configurations: }        & \ensuremath{\Varid{c}\Coloneqq\langle\Sigma,\mu,\Varid{e}\rangle}
\end{align*}
\captionsetup{format=caption-with-line}
\caption{Syntax.}
\label{fig:fs-fg-refs-syntax}
\end{subfigure}%
\\
\begin{subfigure}{.9\textwidth}
\begin{mathpar}
\inferrule[(New-FS)]
{\ensuremath{\langle\Sigma,\mu,\Varid{e}\rangle\Downarrow^{\theta}_{\Varid{pc}}\langle\Sigma',\mu',{v}\rangle} \\
 \ensuremath{\Varid{n}\mathrel{=}\abs{\mu'}} \\
 \ensuremath{\mu''\mathrel{=}\update{\mu'}{\Varid{n}}{{v}}} }
{\ensuremath{\langle\Sigma,\mu,\mathbf{new}(\Varid{e}\mkern1mu)\rangle\Downarrow^{\theta}_{\Varid{pc}}\langle\Sigma',\mu'',{\Varid{n}}^{\Varid{pc}}\rangle}}
\and
\inferrule[(Read-FS)]
{\ensuremath{\langle\Sigma,\mu,\Varid{e}\rangle\Downarrow^{\theta}_{\Varid{pc}}\langle\Sigma',\mu',{\Varid{n}}^{{\ell}}\rangle} \\ \ensuremath{\mu'{[}\Varid{n}{]}\mathrel{=}{{r}}^{{\ell}'}}}
{\ensuremath{\langle\Sigma,\mu,\mathbin{!}\Varid{e}\rangle\Downarrow^{\theta}_{\Varid{pc}}\langle\Sigma',\mu',{{r}}^{{\ell}\;\lub\;{\ell}'}\rangle}}
\and
\inferrule[(LabelOfRef-FS)]
{\ensuremath{\langle\Sigma,\mu,\Varid{e}\rangle\Downarrow^{\theta}_{\Varid{pc}}\langle\Sigma',\mu',{\Varid{n}}^{{\ell}_{1}}\rangle} \\ \ensuremath{\mu'{[}\Varid{n}{]}\mathrel{=}{{r}}^{{\ell}_{2}}}}
{\ensuremath{\langle\Sigma,\mu,\mathbf{labelOfRef}(\Varid{e}\mkern1mu)\rangle\Downarrow^{\theta}_{\Varid{pc}}\langle\Sigma',\mu',{{\ell}_{2}}^{{\ell}_{1}\;\lub\;{\ell}_{2}}\rangle}}
\and
\inferrule[(Write-FS)]
{\ensuremath{\langle\Sigma,\mu,\Varid{e}_{1}\rangle\Downarrow^{\theta}_{\Varid{pc}}\langle\Sigma',\mu',{\Varid{n}}^{{\ell}}\rangle}  \\
 \ensuremath{\langle\Sigma',\mu',\Varid{e}_{2}\rangle\Downarrow^{\theta}_{\Varid{pc}}\langle\Sigma'',\mu'',{{r}_{2}}^{{\ell}_{2}}\rangle} \\
 \ensuremath{\mu''{[}\Varid{n}{]}\mathrel{=}{{r}_{1}}^{{\ell}_{1}}} \\
 \ensuremath{\greenBox{\ensuremath{{\ell}\;\flows\;{\ell}_{1}}}} \\
 \ensuremath{\mu'''\mathrel{=}\update{\mu''}{\Varid{n}}{{{r}_{2}}^{{\ell}_{2}\;\lub\;{\ell}}}} }
{\ensuremath{\langle\Sigma,\mu,\Varid{e}_{1}\mathbin{:=}\Varid{e}_{2}\rangle\Downarrow^{\theta}_{\Varid{pc}}\langle\Sigma'',\mu''',{()}^{\Varid{pc}}\rangle}}
\end{mathpar}
\captionsetup{format=caption-with-line}
\caption{Dynamics. The \greenBox{shaded constraint} is the \emph{no-sensitive upgrade} security check.}
\label{fig:fs-fg-refs-dynamics}
\end{subfigure}
\caption{\ensuremath{\FG} extended with flow-sensitive references.}
\label{fig:fs-fg-refs}
\end{figure}

We introduce the syntax of \ensuremath{\FG} extended with flow-sensitive
references in Figure \ref{fig:fs-fg-refs-syntax}, where we omit the
constructs and the syntactic categories that remain unchanged.
First, we annotate the type of references with flow-sensitivity tags, e.g., \ensuremath{\mathbf{Ref}\; s\;\tau}, where the tag \ensuremath{ s} allows programs to distinguish between flow-insensitive (\ensuremath{ s\mathrel{=}\Conid{I}}) and flow-sensitive (\ensuremath{ s\mathrel{=}\Conid{S}}) references.
We do not introduce new constructs to create, write, and read flow-sensitive references.
Instead, we reuse the same constructs introduced for flow-sensitive references (i.e., \ensuremath{\mathbf{new}(\Varid{e}\mkern1mu)}, \ensuremath{\Varid{e}_{1}\mathbin{:=}\Varid{e}_{2}}, \ensuremath{\mathbin{!}\Varid{e}}, and \ensuremath{\mathbf{labelOfRef}(\Varid{e}\mkern1mu)} from Fig.~\ref{fig:fg-syntax})\footnote{We do not bother to introduce separate constructs because the translations
  of these constructs given in Sections~\ref{sec:fg2cg} and \ref{sec:cg2fg} are the same for both kinds of references. } and rely on the type-level sensitivity tag to determine which kind of reference is used.\footnote{In this presentation 
   we assume that terms are implicitly well-typed. In our mechanized proofs, terms are explicitly and intrinsically well-typed.}
Raw values for flow-sensitive references are plain addresses \ensuremath{\Varid{n}\;\in\;\mathbb{N}}.
Since the label of flow-sensitive references is \emph{mutable} (as explained above), these addresses
are \emph{not} annotated with a label like flow-insensitive references (e.g., \ensuremath{{\Varid{n}}_{{\ell}}}) and hence do not point into a labeled memory in the partitioned store.
Instead, we store both the content and the label of flow-sensitive references in the new, linear heap \ensuremath{\mu}, which can contain data at different security levels.
Specifically, a flow-sensitive reference \ensuremath{\Varid{n}} of type \ensuremath{\mathbf{Ref}\;\Conid{S}\;\tau} points to the \ensuremath{\Varid{n}}-th cell of the heap \ensuremath{\mu}, i.e., \ensuremath{\mu{[}\Varid{n}{]}\mathrel{=}{{r}}^{{\ell}}}, for some raw value \ensuremath{{r}} of type \ensuremath{\tau} at security level \ensuremath{{\ell}}, which also represents the label of the reference.
Finally, we extend program configurations with a heap \ensuremath{\mu}, e.g., \ensuremath{\Varid{c}\mathrel{=}\langle\Sigma,\mu,\Varid{e}\rangle}.

\subsection{Dynamics}
\label{subsec:fg-fs-refs-dynamics}

Figure~\ref{fig:fs-fg-refs-dynamics} gives the semantics rules for the
\ensuremath{\FG} constructs that operate on flow-sensitive references (the rules
in Figure~\ref{fig:fg-semantics}-\ref{fig:fg-memory} are adapted for configurations extended with the heap in
the obvious way).
Rule \srule{New-FS} allocates a new flow-sensitive reference
in the heap at fresh address \ensuremath{\Varid{n}\mathrel{=}\abs{\mu'}}, where value \ensuremath{{v}} gets stored, i.e., \ensuremath{\update{\mu'}{\Varid{n}}{{v}}}.
Similarly to rule \srule{New} (Fig.~\ref{fig:fg-memory}), this rule does not leak data through implicit flows because value \ensuremath{{v}} has security level at least equal to the program counter label.\footnote{In particular, Property~\ref{prop:fg-pc-raise} holds also for \ensuremath{\FG} extended with flow-sensitive references.}
Rule \srule{Read-FS} is also similar to rule \srule{Read}.
The rule reads reference \ensuremath{\Varid{n}} at security level \ensuremath{{\ell}} from the heap
\ensuremath{\mu'}, i.e., \ensuremath{\mu'{[}\Varid{n}{]}\mathrel{=}{{r}}^{{\ell}'}}, and upgrades the label of \ensuremath{{r}} to \ensuremath{{\ell}\;\lub\;{\ell}'} to reflect the fact that reading this particular
reference depends on information at security level \ensuremath{{\ell}}.
Rule \srule{LabelOfRef-FS} retrieves the label of reference \ensuremath{\Varid{n}} by reading the corresponding cell in the heap, i.e., \ensuremath{\mu'{[}\Varid{n}{]}\mathrel{=}{{r}}^{{\ell}_{2}}}, and protects it with the label itself and taints it with the security level of the reference, i.e., \ensuremath{{{\ell}_{2}}^{{\ell}_{1}\;\lub\;{\ell}_{2}}}.  
Although the label of values in the heap represent also the label of the reference, we cannot obtain exactly the label of the reference by reading the reference and then projecting the label of the value, i.e., in general \ensuremath{\mathbf{labelOfRef}(\Varid{e}\mkern1mu)\not\equiv \mathbf{labelOf}(\mathbin{!}\Varid{e}\mkern1mu)} for a flow-sensitive reference \ensuremath{\Varid{e}}.
To see this, suppose \ensuremath{\Varid{e}} evaluates to reference \ensuremath{{\Varid{n}}^{{\ell}_{1}}} such that \ensuremath{\mu{[}\Varid{n}{]}\mathrel{=}{{r}}^{{\ell}_{2}}} for some raw value \ensuremath{{r}}.
Then, expression \ensuremath{\mathbf{labelOfRef}(\Varid{e}\mkern1mu)} results in \ensuremath{{{\ell}_{2}}^{{\ell}_{1}\;\lub\;{\ell}_{2}}} through rule \srule{LabelOfRef-FS}, which is in general different from \ensuremath{{({\ell}_{1}\;\lub\;{\ell}_{2})}^{{\ell}_{1}\;\lub\;{\ell}_{2}}} obtained from \ensuremath{\mathbf{labelOf}(\mathbin{!}\Varid{e}\mkern1mu)} through rule \srule{LabelOf} (Fig.~\ref{fig:fg-semantics2}) applied to rule \srule{Read-FS}.
Hence, we need to define \ensuremath{\mathbf{labelOfRef}(\cdot \mkern1mu)} as a primitive construct of the calculus.
%

%
Rule \srule{Write-FS} updates flow-sensitive references but, in contrast to rule \srule{Write}, it also updates the label of the (flow-sensitive) reference.
In particular, the rule updates reference \ensuremath{\Varid{n}} at security level \ensuremath{{\ell}} by writing value \ensuremath{{{r}_{2}}^{{\ell}_{2}}} tainted with \ensuremath{{\ell}} in the
 \ensuremath{\Varid{n}}-th cell of the heap, i.e., \ensuremath{\update{\mu''}{\Varid{n}}{{{r}_{2}}^{{\ell}_{2}\;\lub\;{\ell}}}}.
Notice that the label of the updated reference, i.e., \ensuremath{{\ell}_{2}\;\lub\;{\ell}}, depends \emph{only} on the security level of the data written into the reference and of the reference that gets updated---this label can be completely different from the label of the reference before the update, i.e., \ensuremath{{\ell}_{1}} from \ensuremath{\mu''{[}\Varid{n}{]}\mathrel{=}{{r}_{1}}^{{\ell}_{1}}}.
For example, the rule allows turning a public
reference into a secret one (i.e., changing the label of the reference from \ensuremath{\Blue{\Conid{L}}} to \ensuremath{\Red{\Conid{H}}}) by writing secret data into it (like in Example~\ref{example:fg-fs-refs}).
Perhaps more surprisingly, the rule also allows changing the label in the opposite direction (i.e., from \ensuremath{\Red{\Conid{H}}} to \ensuremath{\Blue{\Conid{L}}}).
For example, it is possible to make a secret reference public by
overwriting its content with public data (e.g., swap \ensuremath{\Varid{p}} and \ensuremath{ s} in
Example~\ref{example:fg-fs-refs}).
However, rule \srule{Write-FS} is not completely unrestricted: it includes a security check known as \emph{no-sensitive upgrade} (NSU) to avoid leaking information through \emph{implicit flows}~\citep{Austin:Flanagan:PLAS09}.
Intuitively, this check (\greenBox{shaded} in Fig.~\ref{fig:fs-fg-refs-dynamics}) prohibits updates in which the decision to update a particular reference depends on information that is \emph{more sensitive} than the label of the reference itself.
The following example shows why this check is needed to avoid leaks.

\begin{example}
\label{example:fg-fs-refs-leak}
Consider the following program, which branches on secret data, updates an initially public (flow-sensitive) reference in one branch, and then reads back the reference.

\begin{center}
\setlength{\parindent}{0pt}
\setlength{\mathindent}{0pt}
\begin{hscode}\SaveRestoreHook
\column{B}{@{}>{\hspre}l<{\hspost}@{}}%
\column{7}{@{}>{\hspre}l<{\hspost}@{}}%
\column{E}{@{}>{\hspre}l<{\hspost}@{}}%
\>[B]{}\mathbf{let}\;{}\<[7]%
\>[7]{}{r}\mathrel{=}\mathbf{new}(\Varid{p}\mkern1mu)\;\mathbf{in}{}\<[E]%
\\
\>[7]{}\mathbf{if}\; s\;\mathbf{then}\;{r}\mathbin{:=} s\;\mathbf{else}\;();{}\<[E]%
\\
\>[7]{}\mathbin{!}{r}{}\<[E]%
\ColumnHook
\end{hscode}\resethooks
\end{center}
Without the no-sensitive upgrade check described above, this program \emph{leaks} information through an implicit flow.
Formally, we show that the program above breaks non-interference (Theorem~\ref{thm:fg-tini}).
Consider two executions of the program with program counter label public, i.e., \ensuremath{\Varid{pc}\mathrel{=}\Blue{\Conid{L}}}, and with two \ensuremath{\Blue{\Conid{L}}}-equivalent input environments, e.g., \ensuremath{\theta_{1}\mathrel{=}[\mskip1.5mu \Varid{p}\ \mapsto\ {\mathbf{true}}^{\Blue{\Conid{L}}}, s\ \mapsto\ {\mathbf{false}}^{\Red{\Conid{H}}}\mskip1.5mu]\approx_{\Blue{\Conid{L}}}[\mskip1.5mu \Varid{p}\ \mapsto\ {\mathbf{true}}^{\Blue{\Conid{L}}}, s\ \mapsto\ {\mathbf{true}}^{\Red{\Conid{H}}}\mskip1.5mu]\mathrel{=}\theta_{2}}.
In the first execution with environment \ensuremath{\theta_{1}}, reference \ensuremath{{r}} is \emph{not} updated, and thus the result of the program is public value \ensuremath{\Varid{p}\mathrel{=}{\mathbf{true}}^{\Blue{\Conid{L}}}}.
Instead, in the second execution with environment \ensuremath{\theta_{2}}, the program updates reference \ensuremath{{r}} (rule \srule{Write-FS} without the \greenBox{NSU check}) and thus we obtain the secret value \ensuremath{ s\mathrel{=}{\mathbf{true}}^{\Red{\Conid{H}}}} as result.
However, this breaks Theorem~\ref{thm:fg-tini}: the input environments are \ensuremath{\Blue{\Conid{L}}}-equivalent, i.e., \ensuremath{\theta_{1}\approx_{\Blue{\Conid{L}}}\theta_{2}}, but
the results of the program are not, i.e., \ensuremath{\Varid{p}\mathrel{=}{\mathbf{true}}^{\Blue{\Conid{L}}}\not\approx_{\Blue{\Conid{L}}}{\mathbf{true}}^{\Red{\Conid{H}}}\mathrel{=} s}, because \ensuremath{\Blue{\Conid{L}}\;\flows\;\Blue{\Conid{L}}}, but \ensuremath{\Red{\Conid{H}}\;\not\flows\;\Blue{\Conid{L}}} (neither rule \srule{Value\ensuremath{{}_{\Red{\Conid{H}}}}} nor \srule{Value\ensuremath{{}_{\Blue{\Conid{L}}}}} from Fig.~\ref{fig:fg-low-equiv} apply).\footnote{The mismatch in the label of the results can also be propagated to raw values through the label introspection primitives. We refer to \citep{Austin:Flanagan:PLAS09} for a more elaborated example that leaks directly through raw values. }

\end{example}

To avoid such leaks, we include the no-sensitive upgrade check in rule \srule{Write-FS} through the constraint \ensuremath{{\ell}\;\flows\;{\ell}_{1}}, which ensures that the decision of updating
reference \ensuremath{\Varid{n}} depends on information at some security level \ensuremath{{\ell}}
below the current label of the reference, i.e., \ensuremath{{\ell}_{1}} from \ensuremath{\mu''{[}\Varid{n}{]}\mathrel{=}{{r}_{1}}^{{\ell}_{1}}}.
This constraint causes the IFC monitor to abort the second execution of the program above (\ensuremath{\Red{\Conid{H}}\;\not\flows\;\Blue{\Conid{L}}}), which gets stuck and thus satisfies (trivially) Theorem~\ref{thm:fg-tini}.

\subsection{Security}
\label{subsec:fg-fs-refs-security}
\ensuremath{\FG} extended with flow-sensitive references is also secure, i.e., it satisfies termination insensitive non-interference.
However, the theorem for the extended calculus is more complicated
than Theorem~\ref{thm:fg-tini}: it requires a relaxed notion of
\ensuremath{\Blue{\Conid{L}}}-equivalence up to a \emph{bijection}, which relates observable
flow-sensitive references with different heap addresses, and a
side-condition on program configurations to rule out dangling
references~\citep{Banerjee:2005}.
In this section, we make these differences rigorous and we reestablish
non-interference.

\paragraph{The need for Bijections.}
Since the label of flow-sensitive references can change throughout
program execution, our semantics allocates cells for both public and
secret references in a single, unlabeled heap.
An unfortunate consequence of this design choice is that allocations
made in secret contexts can influence the addresses of public
references allocated in the rest of the program, which get shifted by
the number of secret references previously allocated.
This can be problematic for security because attackers can exploit
these deterministic side-effects to leak information through the
address of public references~\citep{Hedin:Sands:CSFW06}.
However, the addresses considered in this work are \emph{opaque},
i.e., programs cannot inspect and compare them, and so attackers
cannot use them to leak secrets.
To show that, we extend the \ensuremath{\Blue{\Conid{L}}}-equivalence relation from
Section~\ref{subsec:fg-security} with a
\emph{bijection}~\citep{Banerjee:2005}, a one-to-one correspondence
between heap addresses, which we use to relate the addresses of
references allocated in public contexts.
%

%
\begin{defn}[Bijection]
A bijection \ensuremath{\beta\mathbin{:}\Conid{Addr}\;\rightharpoonup\;\Conid{Addr}} is a bijective finite partial function between heap addresses. Formally, for all addresses \ensuremath{\Varid{n}} and \ensuremath{\Varid{n'}},
\ensuremath{\beta(\Varid{n})\mathrel{=}\Varid{n'}\Longleftrightarrow\beta^{-1}(\Varid{n'})\mathrel{=}\Varid{n}}, where \ensuremath{\beta^{-1}} is the inverse function of \ensuremath{\beta}.
\end{defn}
%
\paragraph{Notation and Auxiliary Definitions.}
In the following, we treat partial functions as sets of input-output pairs and for clarity we write \ensuremath{(\Varid{n}_{1},\Varid{n}_{2})\;\in\;\beta} for \ensuremath{\beta(\Varid{n}_{1})\mathrel{=}\Varid{n}_{2}}, i.e., if the partial function \ensuremath{\beta} is defined on input \ensuremath{\Varid{n}_{1}} and has output \ensuremath{\Varid{n}_{2}}.
We say that a bijection \ensuremath{\beta'} \emph{extends} another bijection \ensuremath{\beta}, written \ensuremath{\beta\;\subseteq\;\beta'}, if and only if for all input-output pairs \ensuremath{(\Varid{n}_{1},\Varid{n}_{2})}, \ensuremath{(\Varid{n}_{1},\Varid{n}_{2})\;\in\;\beta} implies \ensuremath{(\Varid{n}_{1},\Varid{n}_{2})\;\in\;\beta'}.
Bijection composition is written \ensuremath{\beta_{2}\;\circ\;\beta_{1}\mathrel{=}\{\mskip1.5mu (\Varid{n}_{1},\Varid{n}_{3})\;{\mid}\;(\Varid{n}_{1},\Varid{n}_{2})\;\in\;\beta_{1}\land(\Varid{n}_{2},\Varid{n}_{3})\;\in\;\beta_{2}\mskip1.5mu\}} and the inverse operator as \ensuremath{\beta^{-1}\mathrel{=}\{\mskip1.5mu (\Varid{n}_{2},\Varid{n}_{1})\;{\mid}\;(\Varid{n}_{1},\Varid{n}_{2})\;\in\;\beta\mskip1.5mu\}}.
We also define the domain and range of bijections in the obvious way, i.e., \ensuremath{dom(\beta)\mathrel{=}\{\mskip1.5mu \Varid{n}_{1}\;{\mid}\;(\Varid{n}_{1},\Varid{n}_{2})\;\in\;\beta\mskip1.5mu\}}, and \ensuremath{rng(\beta)\mathrel{=}\{\mskip1.5mu \Varid{n}_{2}\;{\mid}\;(\Varid{n}_{1},\Varid{n}_{2})\;\in\;\beta\mskip1.5mu\}}.
Furthermore, we write \ensuremath{{\iota}_{\Varid{n}}} for the \emph{finite}, partial identity bijection defined up to \ensuremath{\Varid{n}},
i.e., \ensuremath{{\iota}_{\Varid{n}}\mathrel{=}\{\mskip1.5mu (\Varid{n'},\Varid{n'})\;{\mid}\;\Varid{n'}\;\in\;\{\mskip1.5mu \mathrm{0},\ldots,\Varid{n}\mathbin{-}\mathrm{1}\mskip1.5mu\}\mskip1.5mu\}}.
Identity bijections satisfy the following laws.
\begin{property}[Identity Laws]
\label{prop:identity-laws}
For all bijections \ensuremath{\beta} and \ensuremath{\Varid{n}\;\in\;\mathbb{N}}:
\begin{enumerate}
\item \textbf{Inverse Identity.} \ensuremath{{\iota}_{\Varid{n}}^{-1}\equiv {\iota}_{\Varid{n}}}.
\item \textbf{Absorb Left.} If \ensuremath{rng(\beta)\;\subseteq\;dom({\iota}_{\Varid{n}})}, then \ensuremath{{\iota}_{\Varid{n}}\;\circ\;\beta\equiv \beta}.
\item \textbf{Absorb Right.} If \ensuremath{dom(\beta)\;\subseteq\;rng({\iota}_{\Varid{n}})}, then \ensuremath{\beta\;\circ\;{\iota}_{\Varid{n}}\equiv \beta}.

\end{enumerate}
\end{property}


%
\paragraph{\ensuremath{\Blue{\Conid{L}}}-Equivalence up to Bijection.}
Formally, we redefine \ensuremath{\Blue{\Conid{L}}}-equivalence as a relation \ensuremath{\approx_{\Blue{\Conid{L}}}^{\beta}}\ \ensuremath{\subseteq\;\Conid{Value}\;\times\;\Conid{Value}} and write \ensuremath{{v}_{1}\approx_{\Blue{\Conid{L}}}^{\beta}{v}_{2}} to indicate that values \ensuremath{{v}_{1}} and
\ensuremath{{v}_{2}} are indistinguishable to an attacker at security level \ensuremath{\Blue{\Conid{L}}} up to
bijection \ensuremath{\beta}, which relates the heap addresses of corresponding
observable flow-sensitive references in \ensuremath{{v}_{1}} and \ensuremath{{v}_{2}}.
We extend the \ensuremath{\Blue{\Conid{L}}}-equivalence relation for all other syntactic
categories (raw values, memories, and stores) with a bijection in the
same way.
Besides the bijection \ensuremath{\beta}, which we only add to \ensuremath{\approx_{\Blue{\Conid{L}}}}, all the rules from Figure~\ref{fig:fg-low-equiv} remain unchanged.
In particular, rules \srule{Ref\ensuremath{{}_{\Blue{\Conid{L}}}}} and \srule{Ref\ensuremath{{}_{\Red{\Conid{H}}}}}
for flow-insensitive references simply ignore the bijection.
Since we keep labeled memories partitioned in the store and isolated
from the heap, the problem of mismatching addresses does not arise for
flow-insensitive references.
Intuitively, the \emph{address space} of public memories is completely
independent from the address space of secret memories, and thus
allocations made in secret contexts cannot influence the memory
addresses of public references.
%
As a result, rule \srule{Ref\ensuremath{{}_{\Blue{\Conid{L}}}}} guarantees that
\ensuremath{\Blue{\Conid{L}}}-equivalent public references have exactly the same \emph{memory
  address}.
In contrast, heap allocations made in secret contexts can shift the \emph{heap addresses} of observable flow-sensitive references (as explained above), which then may not be necessarily the same.
Therefore, we relate observable flow-sensitive references with
possibly different addresses through the following new rule:
\begin{mathpar}
\inferrule[(Ref-FS)]
{\ensuremath{(\Varid{n}_{1},\Varid{n}_{2})\;\in\;\beta}}
{\ensuremath{\Varid{n}_{1}\approx_{\Blue{\Conid{L}}}^{\beta}\Varid{n}_{2}}}
\end{mathpar}
Rule \srule{Ref-FS} relates references with arbitrary heap addresses
\ensuremath{\Varid{n}_{1}} and \ensuremath{\Varid{n}_{2}}, as long as their addresses are ``matched'' by the
bijection \ensuremath{\beta}, i.e., \ensuremath{(\Varid{n}_{1},\Varid{n}_{2})\;\in\;\beta}.
Similarly, we define heap \ensuremath{\Blue{\Conid{L}}}-equivalence by relating heap cells that are at corresponding addresses according to the bijection.
\begin{defn}[Heap \ensuremath{\Blue{\Conid{L}}}-equivalence]
\label{def:fg-heap-equivalence}
Two heaps \ensuremath{\mu_{1}} and
\ensuremath{\mu_{2}} are \ensuremath{\Blue{\Conid{L}}}-equivalent up to bijection \ensuremath{\beta}, written \ensuremath{\mu_{1}\approx_{\Blue{\Conid{L}}}^{\beta}\mu_{2}}, if and only if:
\begin{enumerate}
\item \ensuremath{dom(\beta)\;\subseteq\;\{\mskip1.5mu \mathrm{0},\ldots,\abs{\mu_{1}}\mathbin{-}\mathrm{1}\mskip1.5mu\}},
\item \ensuremath{rng(\beta)\;\subseteq\;\{\mskip1.5mu \mathrm{0},\ldots,\abs{\mu_{2}}\mathbin{-}\mathrm{1}\mskip1.5mu\}}, and
\item For all addresses \ensuremath{\Varid{n}_{1}} and \ensuremath{\Varid{n}_{2}}, if \ensuremath{(\Varid{n}_{1},\Varid{n}_{2})\;\in\;\beta}, then \ensuremath{\mu_{1}{[}\Varid{n}_{1}{]}\approx_{\Blue{\Conid{L}}}^{\beta}\mu_{2}{[}\Varid{n}_{2}{]}}.
\end{enumerate}
\end{defn}

In the definition above, the first two conditions require that  the domain and the range of \ensuremath{\beta} are contained in the address space of \ensuremath{\mu_{1}} and \ensuremath{\mu_{2}}, respectively.\footnote{Heap addresses start at 0 just like memory addresses.}
These side-conditions ensure that the heap addresses considered in the
third condition are \emph{valid}, i.e., they point to some cell in
each heap.
The definition of \ensuremath{\Blue{\Conid{L}}}-equivalence for heaps is complicated by the fact that heaps can contain both public and secret cells, i.e., cells allocated in public and secret contexts, respectively.
%
In particular, the \ensuremath{\Blue{\Conid{L}}}-equivalence relation needs to relate the values contained in corresponding public cells, which are at possibly different addresses in the heaps.\footnote{Secret cells in each heap are simply disregarded as they may not necessarily have a counterpart in the other heap.}
This task is simplified by the bijection \ensuremath{\beta}, which identifies exactly the addresses of corresponding public cells.
Therefore, the third condition above ensures that the values contained in the cells at
addresses related by the bijection \ensuremath{\beta}, i.e., \ensuremath{(\Varid{n}_{1},\Varid{n}_{2})\;\in\;\beta}, are themselves \ensuremath{\Blue{\Conid{L}}}-equivalent, i.e., \ensuremath{\mu_{1}{[}\Varid{n}_{1}{]}\approx_{\Blue{\Conid{L}}}^{\beta}\mu_{2}{[}\Varid{n}_{2}{]}}.
Finally, we extend \ensuremath{\Blue{\Conid{L}}}-equivalence for configurations  with a bijection and additionally require their heaps to be related. %
Formally, \ensuremath{\Varid{c}_{1}\approx_{\Blue{\Conid{L}}}^{\beta}\Varid{c}_{2}} iff \ensuremath{\Varid{c}_{1}\mathrel{=}\langle\Sigma_1,\mu_{1},{v}_{1}\rangle}, \ensuremath{\Varid{c}_{2}\mathrel{=}\langle\Sigma_2,\mu_{2},{v}_{2}\rangle}, and all their components are related, i.e.,  \ensuremath{\Sigma_1\approx_{\Blue{\Conid{L}}}^{\beta}\Sigma_2}, \ensuremath{\mu_{1}\approx_{\Blue{\Conid{L}}}^{\beta}\mu_{2}}, and \ensuremath{{v}_{1}\approx_{\Blue{\Conid{L}}}^{\beta}{v}_{2}}.

Next, we discuss some side-conditions over program configurations and inputs that are needed for reasoning with \ensuremath{\Blue{\Conid{L}}}-equivalence relations up to bijection.

\paragraph{Valid References.}
\begin{figure}[t]
\centering
\begin{subfigure}{.9\linewidth}
\begin{mathpar}
\inferrule[(Valid-Value)]
{\ensuremath{\Varid{n}\vdash \mathbf{Valid}({r})}}
{\ensuremath{\Varid{n}\vdash \mathbf{Valid}({{r}}^{{\ell}})}}
\and
\inferrule[(Valid-FS-Ref)]
{\ensuremath{\Varid{n'}\;<\;\Varid{n}}}
{\ensuremath{\Varid{n}\vdash \mathbf{Valid}(\Varid{n'})}}
\and
\inferrule[(Valid-FI-Ref)]
{}
{\ensuremath{\Varid{n}\vdash \mathbf{Valid}({\Varid{n'}}_{{\ell}})}}
\and
\inferrule[(Valid-Closure)]
{\ensuremath{\Varid{n}\vdash \mathbf{Valid}(\theta)}}
{\ensuremath{\Varid{n}\vdash \mathbf{Valid}((\Varid{x}.\Varid{e},\theta))}}
\end{mathpar}
\captionsetup{format=caption-with-line}
\caption{Judgment for values \ensuremath{\Varid{n}\vdash \mathbf{Valid}({v})} and raw values \ensuremath{\Varid{n}\vdash \mathbf{Valid}({r})} (selected cases).}
\label{fig:fg-valid-values}
\end{subfigure}
\\

%
\begin{subfigure}{.9\linewidth}
\begin{mathpar}
\inferrule[(Valid-Inputs)]
{\ensuremath{\Varid{c}\mathrel{=}\langle\Sigma,\mu,\Varid{e}\rangle} \\ \ensuremath{\Varid{n}\mathrel{=}\abs{\mu}} \\
 \ensuremath{\Varid{n}\vdash \mathbf{Valid}(\Sigma)} \\
 \ensuremath{\Varid{n}\vdash \mathbf{Valid}(\mu)} \\
 \ensuremath{\Varid{n}\vdash \mathbf{Valid}(\theta)} }
{\ensuremath{\vdash \mathbf{Valid}(\Varid{c},\theta)}}
\and
\inferrule[(Valid-Outputs)]
{\ensuremath{\Varid{c}\mathrel{=}\langle\Sigma,\mu,{v}\rangle} \\ \ensuremath{\Varid{n}\mathrel{=}\abs{\mu}} \\
 \ensuremath{\Varid{n}\vdash \mathbf{Valid}(\Sigma)} \\
 \ensuremath{\Varid{n}\vdash \mathbf{Valid}(\mu)} \\
 \ensuremath{\Varid{n}\vdash \mathbf{Valid}({v})}}
{\ensuremath{\vdash \mathbf{Valid}(\Varid{c})}}
\end{mathpar}
\captionsetup{format=caption-with-line}
\caption{Judgments for valid program inputs and outputs in \ensuremath{\FG}.}
\label{fig:fg-valid-conf}
\end{subfigure}
\caption{Some valid judgments for \ensuremath{\FG}.}
\label{fig:fg-valid}
\end{figure}
In Section~\ref{subsec:fg-security}, we proved that the
\ensuremath{\Blue{\Conid{L}}}-equivalence relation is reflexive, symmetric, and transitive, and
showed how these properties helped us structure and simplify the proof
of non-interference.
%
%
Does the \ensuremath{\Blue{\Conid{L}}}-equivalence extended with bijection also enjoy these
properties?
Unfortunately, reflexivity does not hold \emph{unconditionally}
anymore for \ensuremath{\Blue{\Conid{L}}}-equivalence relations up to bijection.
The culprit of the problem is in the definition of \ensuremath{\Blue{\Conid{L}}}-equivalence for heaps.
In this case, proving reflexivity requires showing that any
heap is \ensuremath{\Blue{\Conid{L}}}-equivalent to itself up to the identity bijection defined over its address space, i.e., \ensuremath{\forall\mu.\mu\;\approx_{\Blue{\Conid{L}}}^{{\iota}_{\abs{\mu}}}\;\mu}.
Technically, this property does not hold in general for any \emph{arbitrary} heap \ensuremath{\mu}.
To see this, consider for example the ill-formed heap \ensuremath{\mu'\mathrel{=}{\mathrm{1}}^{\Blue{\Conid{L}}}\mathbin{:}[\mskip1.5mu \mskip1.5mu]}, which contains the \emph{dangling} flow-sensitive reference \ensuremath{{\mathrm{1}}^{\Blue{\Conid{L}}}}.
In order to prove \ensuremath{\mu'\;\approx_{\Blue{\Conid{L}}}^{{\iota}_{\mathrm{1}}}\;\mu'}, we must show that this dangling reference is \ensuremath{\Blue{\Conid{L}}}-equivalent to itself up to bijection \ensuremath{{\iota}_{\mathrm{1}}} (Def.~\ref{def:fg-heap-equivalence}.3), i.e.,  \ensuremath{{\mathrm{1}}^{\Blue{\Conid{L}}}\;\approx_{\Blue{\Conid{L}}}^{{\iota}_{\mathrm{1}}}\;{\mathrm{1}}^{\Blue{\Conid{L}}}}, which by rule  \srule{Ref-FS} would require a proof for the false statement \ensuremath{(\mathrm{1},\mathrm{1})\;\in\;{\iota}_{\mathrm{1}}\mathrel{=}\{\mskip1.5mu (\mathrm{0},\mathrm{0})\mskip1.5mu\}}.
One may be tempted to repair reflexivity by choosing a sufficiently large domain for the identity bijection, so that also dangling references can be related (e.g., for \ensuremath{\mu'} we could pick \ensuremath{{\iota}_{\mathrm{2}}} and prove \ensuremath{{\mathrm{1}}^{\Blue{\Conid{L}}}\;\approx_{\Blue{\Conid{L}}}^{{\iota}_{\mathrm{2}}}\;{\mathrm{1}}^{\Blue{\Conid{L}}}}).
Unfortunately, this is not possible without breaking the first and the second condition of Definition~\ref{def:fg-heap-equivalence}.
For any heap \ensuremath{\mu}, the identity bijection \ensuremath{{\iota}_{\abs{\mu}}} has the largest domain and range that satisfies the side-conditions \ensuremath{dom({\iota}_{\abs{\mu}})\;\subseteq\;\{\mskip1.5mu \mathrm{0}\mathinner{\ldotp\ldotp}\abs{\mu}\mathbin{-}\mathrm{1}\mskip1.5mu\}} (Def.~\ref{def:fg-heap-equivalence}.1) and \ensuremath{rng({\iota}_{\abs{\mu}})\;\subseteq\;\{\mskip1.5mu \mathrm{0}\mathinner{\ldotp\ldotp}\abs{\mu}\mathbin{-}\mathrm{1}\mskip1.5mu\}} (Def.~\ref{def:fg-heap-equivalence}.2).
The solution to this technical nuisance is to simply prove a weaker version of reflexivity, which holds only for \emph{well-formed} heaps (and similarly values, stores, memories, and configurations) that contain only \emph{valid} (i.e., not dangling) references.
Luckily, this restriction is inconsequential in our setting because references are always valid in \ensuremath{\FG}.
Intuitively, references are \emph{unforgeable}: programs can only create fresh, valid references through the \ensuremath{\mathbf{new}(\cdot \mkern1mu)} construct.
After creation references simply remain valid throughout program
execution: they can be accessed, passed around, or stored, but never
deallocated (we simply provide no construct to do so) or tempered.
We now formalize this insight in the form of a \emph{valid judgment}, which rules out dangling reference from all the categories of the calculus together with the proof that this judgment is an \emph{invariant} of the operational semantics of \ensuremath{\FG}.

\paragraph{Valid Judgment.}
Figure~\ref{fig:fg-valid} defines a valid judgment for  \ensuremath{\FG} values and configurations.
These judgments ensure that all flow-sensitive references values of a
program configuration are valid (i.e., not dangling) in a heap of a
given size.
For values (Fig.~\ref{fig:fg-valid-values}), this judgment takes the form of \ensuremath{\Varid{n}\vdash \mathbf{Valid}({v})}, which indicates that all heap addresses contained in value \ensuremath{{v}} are valid in a heap of size \ensuremath{\Varid{n}}.
This judgment is mutually defined with similar judgments for raw values, i.e., \ensuremath{\Varid{n}\vdash \mathbf{Valid}({r})}, and environments \ensuremath{\Varid{n}\vdash \mathbf{Valid}(\theta)}.
Importantly, rule \srule{Valid-FS-Ref} disallows \emph{dangling
   references}: a flow-sensitive reference is valid if and only if its
heap address \ensuremath{\Varid{n'}} is strictly smaller than the size \ensuremath{\Varid{n}} of the heap.
Notice that this judgment holds unconditionally for \emph{flow-insensitive}
values through rule \srule{Valid-FI-Ref}.
Intuitively, these references contain memory addresses,
which are not affected by the technical issue described above.
Most of the remaining rules are fairly straightforward:
they simply require that all sub-values are homogeneously valid in heaps of the same size.
For example, values \ensuremath{{{r}}^{{\ell}}} is valid only if and only if the raw
value \ensuremath{{r}} is also valid (rule \srule{Valid-Value}) and closures \ensuremath{(\Varid{x}.\Varid{e},\theta)} are valid if and only if the value environment \ensuremath{\theta} is also
valid (rule \srule{Valid-Closure}).\footnote{Since expressions do not
  contain heap addresses or values (they belong to
  distinct syntactic categories), we do not need to define a valid
  judgment for expressions.}
Environments \ensuremath{\theta} are valid if they contain only valid values, i.e., \ensuremath{\Varid{n}\vdash \mathbf{Valid}(\theta)} iff \ensuremath{\forall\Varid{x}\;\in\;dom(\theta).\Varid{n}\vdash \mathbf{Valid}(\theta(\Varid{x}))}.
Similarly, valid stores \ensuremath{\Sigma} contain only valid memories, i.e., \ensuremath{\Varid{n}\vdash \mathbf{Valid}(\Sigma)} iff \ensuremath{\forall{\ell}\;\in\;\mathscr{L}.\Varid{n}\vdash \mathbf{Valid}(\Sigma({\ell}))}, and so on for memories and heaps, i.e., \ensuremath{\Varid{n}\vdash \mathbf{Valid}(\Conid{M})} iff \ensuremath{\forall\Varid{n}\;\in\;\{\mskip1.5mu \mathrm{0},\ldots,\abs{\Conid{M}}\mathbin{-}\mathrm{1}\mskip1.5mu\}.\Varid{n}\vdash \mathbf{Valid}(\Conid{M}{[}\Varid{n}{]})} and \ensuremath{\Varid{n}\vdash \mathbf{Valid}(\mu)} iff \ensuremath{\forall\Varid{n}\;\in\;\{\mskip1.5mu \mathrm{0},\ldots,\abs{\mu}\mathbin{-}\mathrm{1}\mskip1.5mu\}.\Varid{n}\vdash \mathbf{Valid}(\mu{[}\Varid{n}{]})}, respectively.
Finally, the judgments for initial and final configurations
(Fig.~\ref{fig:fg-valid-conf}) instantiate the parameter \ensuremath{\Varid{n}} of the
judgments above with the size of the heap.
For example, a configuration \ensuremath{\Varid{c}\mathrel{=}\langle\Sigma,\mu,\Varid{e}\rangle} and input environment \ensuremath{\theta} are valid through rule \srule{Valid-Inputs} if and only if all the components of the configuration and the value environment are valid in the heap of size \ensuremath{\Varid{n}\mathrel{=}\abs{\mu}}, i.e., \ensuremath{\vdash \mathbf{Valid}(\Varid{c},\theta)} iff \ensuremath{\Varid{n}\vdash \mathbf{Valid}(\Sigma)}, \ensuremath{\Varid{n}\vdash \mathbf{Valid}(\mu)}, and \ensuremath{\Varid{n}\vdash \mathbf{Valid}(\theta)}.
Rule \srule{Valid-Final} is for final configurations and is similar.

Before showing that this valid judgment is an invariant of the
semantics of \ensuremath{\FG}, we prove two simple helping lemmas.
The first lemma shows that values that are valid in a heap of a
certain size, are also valid in any larger heap, while the second
lemma shows that heaps can only grow larger in the semantics of \ensuremath{\FG}.

\begin{lemma}[Valid Weakening]
\label{lemma:fg-valid-wken}
For all values \ensuremath{{v}}, environments \ensuremath{\theta}, raw values \ensuremath{{r}} and natural numbers \ensuremath{\Varid{n}} and \ensuremath{\Varid{n'}}:
\begin{enumerate}
\item If \ensuremath{\Varid{n}\vdash \mathbf{Valid}({v})} and \ensuremath{\Varid{n}\;\leq\;\Varid{n'}}, then \ensuremath{\Varid{n'}\vdash \mathbf{Valid}({v})},
\item If \ensuremath{\Varid{n}\vdash \mathbf{Valid}({r})} and \ensuremath{\Varid{n}\;\leq\;\Varid{n'}}, then \ensuremath{\Varid{n'}\vdash \mathbf{Valid}({r})}
\item If \ensuremath{\Varid{n}\vdash \mathbf{Valid}(\theta)} and \ensuremath{\Varid{n}\;\leq\;\Varid{n'}}, then \ensuremath{\Varid{n'}\vdash \mathbf{Valid}(\theta)}
\end{enumerate}
\end{lemma}
\begin{proof} By mutual induction over the judgments and using
transitivity of \ensuremath{\leq} for the case \srule{Valid-FS-Ref}.
\end{proof}

\begin{lemma}[Heaps Only Grow]
\label{lemma:fg-heap-grow}
Let \ensuremath{\Varid{c}\mathrel{=}\langle\Sigma,\mu,\Varid{e}\rangle}, \ensuremath{\Varid{c'}\mathrel{=}\langle\Sigma',\mu',{v}\rangle}, if \ensuremath{\Varid{c}\;\Downarrow^{\theta}_{\Varid{pc}}\;\Varid{c'}}, then \ensuremath{\abs{\mu}\;\leq\;\abs{\mu'}}.
\end{lemma}
\begin{proof}
  By induction on the big-step reduction. In particular, notice that
  no rule deallocates cells from the heap, which can only grow through
  rule \srule{New-FS}.
\end{proof}

\begin{property}[Valid Invariant]
\label{prop:fg-valid-invariant}
If \ensuremath{\Varid{c}\;\Downarrow^{\theta}_{\Varid{pc}}\;\Varid{c'}} and \ensuremath{\vdash \mathbf{Valid}(\Varid{c},\theta)}, then \ensuremath{\vdash \mathbf{Valid}(\Varid{c'})}.
\end{property}
\begin{proof}
  By induction on the big-step reduction and using Lemmas
  \ref{lemma:fg-valid-wken} and \ref{lemma:fg-heap-grow}.
\end{proof}
%
%
We now reconsider the properties of \ensuremath{\Blue{\Conid{L}}}-equivalence up to bijection.
\begin{property}[]
\label{prop:fg-low-equiv-props-with-bij}
For all values, raw values, environments, memories, and stores \ensuremath{\Varid{x}}, \ensuremath{\Varid{y}}, \ensuremath{\Varid{z}}, and \ensuremath{\Varid{n}\;\in\;\mathbb{N}}:
\begin{enumerate}
\item \textbf{Restricted Reflexivity.} If \ensuremath{\Varid{n}\vdash \mathbf{Valid}(\Varid{x})}, then \ensuremath{\Varid{x}\;\approx_{\Blue{\Conid{L}}}^{{\iota}_{\Varid{n}}}\;\Varid{x}}.
\item \textbf{Symmetricity.} If \ensuremath{\Varid{x}\approx_{\Blue{\Conid{L}}}^{\beta}\Varid{y}}, then \ensuremath{\Varid{y}\;\approx_{\Blue{\Conid{L}}}^{\beta^{-1}}\;\Varid{x}}.
\item \textbf{Transitivity.} If \ensuremath{\Varid{x}\;\approx_{\Blue{\Conid{L}}}^{\beta_{1}}\;\Varid{y}} and \ensuremath{\Varid{y}\;\approx_{\Blue{\Conid{L}}}^{\beta_{2}}\;\Varid{z}}, then \ensuremath{\Varid{x}\;\approx_{\Blue{\Conid{L}}}^{\beta_{2}\;\circ\;\beta_{1}}\;\Varid{z}}.
\item \textbf{Weakening.} If \ensuremath{\Varid{x}\approx_{\Blue{\Conid{L}}}^{\beta}\Varid{y}} and \ensuremath{\beta\;\subseteq\;\beta'}, then \ensuremath{\Varid{x}\;\approx_{\Blue{\Conid{L}}}^{\beta'}\;\Varid{y}}.
\end{enumerate}
\end{property}
First, \emph{reflexivity} up to a \emph{identity bijection} \ensuremath{{\iota}_{\Varid{n}}} is restricted to program constructs that are valid in a heap of
size \ensuremath{\Varid{n}}, as explained above.
In a relation \ensuremath{\Varid{x}\;\approx_{\Blue{\Conid{L}}}^{\beta}\;\Varid{y}}, the bijection \ensuremath{\beta} maps the heap
addresses of \ensuremath{\Varid{x}} to the corresponding heap addresses of \ensuremath{\Varid{y}}.
Then, in order to obtain the mapping from \ensuremath{\Varid{y}} to \ensuremath{\Varid{x}}, the bijection
must be inverted in the symmetric relation, i.e., \ensuremath{\Varid{y}\;\approx_{\Blue{\Conid{L}}}^{\beta^{-1}}\;\Varid{x}}.
Similarly, when composing \ensuremath{\Varid{x}\;\approx_{\Blue{\Conid{L}}}^{\beta_{1}}\;\Varid{y}} and \ensuremath{\Varid{y}\;\approx_{\Blue{\Conid{L}}}^{\beta_{2}}\;\Varid{z}}
through \emph{transitivity}, the bijections must also be composed,
i.e., \ensuremath{\Varid{x}\;\approx_{\Blue{\Conid{L}}}^{\beta_{2}\;\circ\;\beta_{1}}\;\Varid{z}}, so that the composed bijection
\ensuremath{\beta_{2}\;\circ\;\beta_{1}} first maps the addresses of \ensuremath{\Varid{x}} to the
corresponding addresses of \ensuremath{\Varid{y}} through \ensuremath{\beta_{1}}, and then the addresses
of \ensuremath{\Varid{y}} to those of \ensuremath{\Varid{z}} through \ensuremath{\beta_{2}}.
Lastly, \emph{weakening} allows to relax a relation \ensuremath{\Varid{x}\;\approx_{\Blue{\Conid{L}}}^{\beta}\;\Varid{y}}
with an extended bijection \ensuremath{\beta'\;\supseteq\;\beta}.
All the properties above hold also for \ensuremath{\Blue{\Conid{L}}}-equivalence for heaps
(Def.~\ref{def:fg-heap-equivalence}), with the exception for
\emph{weakening}, due to the side conditions on the domain and range
of the bijection.

\begin{figure}[t]
\centering
\begin{tikzcd}
  \ensuremath{\mu_{1}} \arrow[r , "\ensuremath{\beta_{1}}"] \arrow[d , swap, "\ensuremath{\beta}"]  & \ensuremath{\mu_{1}'} \arrow[d, dashed, "\ensuremath{\beta'}\ \ensuremath{\mathrel{=}}\ \ensuremath{\beta_{2}\;\circ\;\beta\;\circ\;\beta_{1}^{-1}}"] \\
  \ensuremath{\mu_{2}} \arrow[r , swap, "\ensuremath{\beta_{2}}"] & \ensuremath{\mu_{2}'}
\end{tikzcd}
\caption{
Square commutative diagram for heaps
  (Lemma~\ref{lemma:fg-square-heaps}). Solid arrows represent
  the assumptions of the lemma and the dashed arrow represents the
  conclusion. If \ensuremath{\beta_{1}\mathrel{=}{\iota}_{\abs{\mu_{1}}}}, \ensuremath{\beta_{2}\mathrel{=}{\iota}_{\abs{\mu_{2}}}}, then \ensuremath{\beta'} is simply equal to \ensuremath{\beta}.
}
\label{fig:fg-square-heaps}
\end{figure}
We conclude this part with the \emph{square commutative digram} for
heaps, outlined in Figure~\ref{fig:fg-square-heaps}.\footnote{We omit the
  analogous square commutative diagram for stores.}
In the diagram, the arrows connect \ensuremath{\Blue{\Conid{L}}}-equivalent heaps up to a given
bijection, e.g., the arrow from \ensuremath{\mu_{1}} to \ensuremath{\mu_{1}'} labeled \ensuremath{\beta_{1}}
indicates that \ensuremath{\mu_{1}\;\approx_{\Blue{\Conid{L}}}^{\beta_{1}}\;\mu_{1}'}.
Compared to the square diagram without bijections from Figure
\ref{fig:fg-square-store}, this diagram is complicated by the fact
that the bijections must be composed and inverted appropriately, as
needed by \emph{symmetricity} and \emph{transitivity}, in order to
obtain the dashed arrow that completes the square.
However, when \ensuremath{\beta_{1}} and \ensuremath{\beta_{2}} are (appropriate) identity
bijections, the bijection between the final heaps can be simplified
and shown to be equal to the bijection between the initial heaps.

\begin{lemma}[Square Commutative Diagram for Heaps]
  If \ensuremath{\mu_{1}\approx_{\Blue{\Conid{L}}}^{\beta}\mu_{2}}, \ensuremath{\mu_{1}\;\approx_{\Blue{\Conid{L}}}^{\beta_{1}}\;\mu_{1}'}, and \ensuremath{\mu_{2}\;\approx_{\Blue{\Conid{L}}}^{\beta_{2}}\;\mu_{2}'}, then \ensuremath{\mu_{1}'\;\approx_{\Blue{\Conid{L}}}^{\beta'}\;\mu_{2}'}, where \ensuremath{\beta'\mathrel{=}\beta_{2}\;\circ\;\beta\;\circ\;\beta_{1}^{-1}}.  Furthermore, if \ensuremath{\beta_{1}\mathrel{=}{\iota}_{\abs{\mu_{1}}}}
  and \ensuremath{\beta_{2}\mathrel{=}{\iota}_{\abs{\mu_{2}}}}, then \ensuremath{\mu_{1}'\;\approx_{\Blue{\Conid{L}}}^{\beta}\;\mu_{2}'}.
\label{lemma:fg-square-heaps}
\end{lemma}
\begin{proof}
  By applying \emph{symmetricity}
  (Property~\ref{prop:fg-low-equiv-props-with-bij}.2) and
  \emph{transitivity}
  (Property~\ref{prop:fg-low-equiv-props-with-bij}.3) like in
  Lemma~\ref{lemma:fg-square-store}, we obtain directly \ensuremath{\mu_{1}'\;\approx_{\Blue{\Conid{L}}}^{\beta'}\;\mu_{2}'} and \ensuremath{\beta'\mathrel{=}\beta_{2}\;\circ\;\beta\;\circ\;\beta_{1}^{-1}}.
  If \ensuremath{\beta_{1}\mathrel{=}{\iota}_{\abs{\mu_{1}}}} and
  \ensuremath{\beta_{2}\mathrel{=}{\iota}_{\abs{\mu_{2}}}}, then \ensuremath{\beta'\mathrel{=}{\iota}_{\abs{\mu_{2}}}\;\circ\;\beta\;\circ\;{\iota}_{\abs{\mu_{1}}}^{-1}}.
  Using the \emph{identity laws} (Property~\ref{prop:identity-laws}),
  we show that \ensuremath{\beta'\mathrel{=}{\iota}_{\abs{\mu_{2}}}\;\circ\;\beta\;\circ\;{\iota}_{\abs{\mu_{1}}}^{-1}\mathrel{=}\beta} and thus we have \ensuremath{\mu_{1}'\;\approx_{\Blue{\Conid{L}}}^{\beta}\;\mu_{2}'}.
\end{proof}

\paragraph{Termination-Insensitive Non-Interference.}

We now turn our attention to the termination-insensitive
non-interference theorem for \ensuremath{\FG} extended with flow-sensitive references.
First, the theorem assumes that the initial configurations are
\ensuremath{\Blue{\Conid{L}}}-equivalent up to a bijection, which we use to account for
secret-dependent heap allocations and hence relate the heap addresses
of corresponding public flow-sensitive references, as explained above.
Then, the theorem guarantees that the final configurations are also
\ensuremath{\Blue{\Conid{L}}}-equivalent, but up to some \emph{extended bijection}.
Intuitively, programs may dynamically allocate new cells in the heap,
therefore the heap addresses contained in their final configurations
must be shown to be related with respect to the address space of the
final heaps.
Additionally, the theorem explicitly assumes that the initial program
configurations are valid and thus do not contain dangling references.
As explained before, this extra assumption is needed for technical
reasons and, besides having to explicitly propagate it as needed in
our mechanized proof scripts, it does not impact the security
guarantees of \ensuremath{\FG}.
The proof of this theorem is also structured on two key lemmas:
\emph{store and heap confinement} and \emph{\ensuremath{\Blue{\Conid{L}}}-equivalence
preservation}.
In the following, we focus on the changes needed to adapt these lemmas
for flow-sensitive references, heaps, and \ensuremath{\Blue{\Conid{L}}}-equivalence up to
bijection.
Confinement guarantees that programs running in a secret context
cannot leak secret data implicitly through the program state, i.e.,
the store and the heap.
Formally, we show that if the program counter label is above the
attacker's label (\ensuremath{\Varid{pc}\;\not\flows\;\Blue{\Conid{L}}}), then the initial store and heap
are \ensuremath{\Blue{\Conid{L}}}-equivalent to the final store and heap obtained at the end of
the execution, up to the \emph{identity bijection}.
In the lemma, \ensuremath{\Blue{\Conid{L}}}-equivalence is up to the identity bijection defined
over the address space of the initial heap.
Intuitively, this is because the references allocated in secret
contexts cannot be observed by the attacker, and therefore must be
disregarded by the bijection, which instead keeps track only of
observable references, i.e., those allocated in public contexts.

\begin{lemma}[Store and Heap Confinement]
\label{lemma:fg-confinement-bij}
For all configurations \ensuremath{\Varid{c}\mathrel{=}\langle\Sigma,\mu,\Varid{e}\rangle}, \ensuremath{\Varid{c'}\mathrel{=}\langle\Sigma',\mu',{v}\rangle}, program counter labels \ensuremath{\Varid{pc}\;\not\flows\;\Blue{\Conid{L}}}, if \ensuremath{\vdash \mathbf{Valid}(\Varid{c},\theta)} and \ensuremath{\Varid{c}\;\Downarrow^{\theta}_{\Varid{pc}}\;\Varid{c'}}, then \ensuremath{\Sigma\;\approx_{\Blue{\Conid{L}}}^{{\iota}_{\abs{\mu}}}\;\Sigma'} and \ensuremath{\mu\;\approx_{\Blue{\Conid{L}}}^{{\iota}_{\abs{\mu}}}\;\mu'}.
\end{lemma}
\begin{proof} By induction on the big-step reduction and using
the fact the semantics preserves valid references
(Property~\ref{prop:fg-valid-invariant}) to propagate the valid
judgment through the intermediate configurations.
In particular, we apply \emph{restricted reflexivity}
(Property~\ref{prop:fg-low-equiv-props-with-bij}.1) in the base cases,
and \emph{transitivity}
(Property~\ref{prop:fg-low-equiv-props-with-bij}.3) and the bijection
\emph{identity laws} (Property~\ref{prop:identity-laws}) in the
inductive cases.
Intuitively, when we apply transitivity, we need to show that the
composition of identity bijections defined over different heaps
results in the identity bijection over the initial heap required by
the lemma.
For example, to show that \ensuremath{\mu\;\approx_{\Blue{\Conid{L}}}^{{\iota}_{\abs{\mu}}}\;\mu''} through some intermediate heap \ensuremath{\mu'} such that \ensuremath{\mu\;\approx_{\Blue{\Conid{L}}}^{{\iota}_{\abs{\mu}}}\;\mu'} and \ensuremath{\mu'\;\approx_{\Blue{\Conid{L}}}^{{\iota}_{\abs{\mu'}}}\;\mu''}, we apply transitivity and  obtain \ensuremath{\mu\;\approx_{\Blue{\Conid{L}}}^{{\iota}_{\abs{\mu'}}\;\circ\;{\iota}_{\abs{\mu}}}\;\mu''}.
Since heaps only grow in size (Lemma~\ref{lemma:fg-heap-grow}),
we have that \ensuremath{\abs{\mu}\;\leq\;\abs{\mu'}}, thus \ensuremath{rng({\iota}_{\abs{\mu}})\;\subseteq\;dom({\iota}_{\abs{\mu'}})} and \ensuremath{{\iota}_{\abs{\mu'}}\;\circ\;{\iota}_{\abs{\mu}}\mathrel{=}{\iota}_{\abs{\mu}}} by
Property~\ref{prop:identity-laws}.2 and so we have \ensuremath{\mu\;\approx_{\Blue{\Conid{L}}}^{{\iota}_{\abs{\mu}}}\;\mu''}, as required.
\end{proof}

For \ensuremath{\Blue{\Conid{L}}}-equivalence preservation, we assume that the program inputs
are \ensuremath{\Blue{\Conid{L}}}-equivalent up to some bijection \ensuremath{\beta} and must show that the
final configurations are \ensuremath{\Blue{\Conid{L}}}-equivalent up to some extended bijection
\ensuremath{\beta'\;\supseteq\;\beta}.
Since we cannot predict, in general, how arbitrary programs allocate
references at run-time, in these lemmas we \emph{existentially
  quantify} the bijection \ensuremath{\beta'} that relates the heap addresses in
final configurations, but guarantee that the final bijection \ensuremath{\beta'}
is at least as large as the initial bijection \ensuremath{\beta}, i.e., \ensuremath{\beta\;\subseteq\;\beta'}.
This invariant is critical for the proof, where the inductive steps
require combining \ensuremath{\Blue{\Conid{L}}}-equivalent values and environments with
different bijections.
In particular, the fact that bijections only get extended, e.g., \ensuremath{\beta\;\subseteq\;\beta'}, lets us lift \ensuremath{\Blue{\Conid{L}}}-equivalence relations up to a small
bijection \ensuremath{\beta} into \ensuremath{\Blue{\Conid{L}}}-equivalence relations up to a larger
bijection \ensuremath{\beta'} via \emph{weakening}
(Property~\ref{prop:fg-low-equiv-props-with-bij}.4).
%

%
Like before, we prove two \ensuremath{\Blue{\Conid{L}}}-equivalence preservation lemmas, for
secret and public contexts, respectively, which we then combine in the
proof of termination-insensitive non-interference.
For preservation in secret contexts, we observe that the bijection
that relates the final configurations is \emph{exactly the same}
bijection that relates the initial configurations.\footnote{This is a
  special case of \ensuremath{\Blue{\Conid{L}}}-equivalence preservation, where we can predict
  the final bijection \ensuremath{\beta}, and therefore we omit the existential
  quantification.  Notice that in this case the invariant \ensuremath{\beta\;\subseteq\;\beta} is trivially satisfied.}
Since the program executions occur in secret contexts, their
allocations are secret-dependent and therefore not observable by the
attacker, which is reflected in the lemma by relating the initial and
final configurations with the same bijection.
The proof technique for this lemma is similar to that of the previous
\ensuremath{\Blue{\Conid{L}}}-equivalence preservation lemma.
We consider each program execution individually and
reason independently about program state (i.e., heap and store) and
values in the final configurations.
In particular, we apply \emph{store and heap confinement} to each
execution and relate the initial and final program states with the
\emph{identity bijection}, which are then combined in a square
commutative diagram (Figure~\ref{fig:fg-square-heaps}) to relate the
final configurations.

\begin{lemma}[\ensuremath{\Blue{\Conid{L}}}-Equivalence Preservation in Secret Contexts]
\label{lemma:fg-leq-preservation-secret-bij}
For all program counter labels \ensuremath{\Varid{pc}_{1}\;\not\flows\;\Blue{\Conid{L}}} and \ensuremath{\Varid{pc}_{2}\;\not\flows\;\Blue{\Conid{L}}}, valid inputs \ensuremath{\vdash \mathbf{Valid}(\Varid{c}_{1},\theta_{1})} and \ensuremath{\vdash \mathbf{Valid}(\Varid{c}_{2},\theta_{2})}, such that
\ensuremath{\Varid{c}_{1}\mathrel{=}\langle\Sigma_1,\mu_{1},\Varid{e}_{1}\rangle}, \ensuremath{\Varid{c}_{2}\mathrel{=}\langle\Sigma_2,\mu_{2},\Varid{e}_{2}\rangle}, \ensuremath{\Sigma_1\;\approx_{\Blue{\Conid{L}}}^{\beta}\;\Sigma_2} and \ensuremath{\mu_{1}\;\approx_{\Blue{\Conid{L}}}^{\beta}\;\mu_{2}}, if \ensuremath{\langle\Sigma_1,\mu_{1},\Varid{e}_{1}\rangle\Downarrow^{\theta_{1}}_{\Varid{pc}_{1}}\;\Varid{c}_{1}} and \ensuremath{\langle\Sigma_2,\mu_{2},\Varid{e}_{2}\rangle\Downarrow^{\theta_{2}}_{\Varid{pc}_{2}}\;\Varid{c}_{2}}, then \ensuremath{\Varid{c}_{1}\;\approx_{\Blue{\Conid{L}}}^{\beta}\;\Varid{c}_{2}}.
\end{lemma}
\begin{proof}
  Assume \ensuremath{\Varid{pc}_{1}\;\not\flows\;\Blue{\Conid{L}}}, \ensuremath{\Varid{pc}_{2}\;\not\flows\;\Blue{\Conid{L}}}, \ensuremath{\Sigma_1\;\approx_{\Blue{\Conid{L}}}^{\beta}\;\Sigma_2}, \ensuremath{\mu_{1}\;\approx_{\Blue{\Conid{L}}}^{\beta}\;\mu_{2}}, and let the final configurations be
  \ensuremath{\Varid{c}_{1}\mathrel{=}\langle\Sigma_{1}',\mu_{1}',{v}_{1}\rangle} and \ensuremath{\Varid{c}_{2}\mathrel{=}\langle\Sigma_{2}',\mu_{2}',{v}_{2}\rangle}.
  First, we apply \emph{Store and Heap Confinement}
  (Lemma~\ref{lemma:fg-confinement-bij}) to \ensuremath{\langle\Sigma_1,\mu_{1},\Varid{e}_{1}\rangle\Downarrow^{\theta_{1}}_{\Varid{pc}_{1}}\langle\Sigma_{1}',\mu_{1}',{v}_{1}\rangle} and \ensuremath{\langle\Sigma_2,\mu_{2},\Varid{e}_{2}\rangle\Downarrow^{\theta_{2}}_{\Varid{pc}_{2}}\langle\Sigma_{2}',\mu_{2},{v}_{2}\rangle} and obtain \ensuremath{\mu_{1}\;\approx_{\Blue{\Conid{L}}}^{{\iota}_{\abs{\mu_{1}}}}\;\mu_{1}'} and \ensuremath{\mu_{2}\;\approx_{\Blue{\Conid{L}}}^{{\iota}_{\abs{\mu_{2}}}}\;\mu_{2}'},
  respectively.
  %
  %
  Then, we construct the \emph{Square Commutative Diagram for Heaps}
  (Lemma~\ref{lemma:fg-square-heaps}) using \ensuremath{\mu_{1}\;\approx_{\Blue{\Conid{L}}}^{\beta}\;\mu_{2}},
  \ensuremath{\mu_{1}\;\approx_{\Blue{\Conid{L}}}^{{\iota}_{\abs{\mu_{1}}}}\;\mu_{1}'} and \ensuremath{\mu_{2}\;\approx_{\Blue{\Conid{L}}}^{{\iota}_{\abs{\mu_{2}}}}\;\mu_{2}'}, and obtain \ensuremath{\mu_{1}'\;\approx_{\Blue{\Conid{L}}}^{\beta}\;\mu_{2}'}.
  We derive \ensuremath{\Sigma_{1}'\;\approx_{\Blue{\Conid{L}}}^{\beta}\;\Sigma_{2}'} with a similar construction
  and prove \ensuremath{{v}_{1}\;\approx_{\Blue{\Conid{L}}}^{\beta}\;{v}_{2}} like in
  Lemma~\ref{lemma:fg-leq-preservation-secret}.
  %
  %
  %
   %
  %
\end{proof}

The proof of \ensuremath{\Blue{\Conid{L}}}-equivalence preservation in public contexts also follows
the same structure of the proof without heaps.
In particular, the proof is by simultaneous induction on the two
big-step reductions and relies on \emph{\ensuremath{\Blue{\Conid{L}}}-Equivalence Preservation
  in Secret Contexts} when the program executions deviate from each
other at a secret-dependent control-flow point.
Here, we discuss only the interesting case where the programs
allocate a flow-sensitive reference and we must extend
the bijection with a new pair of matching addresses.

\begin{lemma}[\ensuremath{\Blue{\Conid{L}}}-Equivalence Preservation in Public Contexts]
\label{lemma:fg-leq-preservation-public-bij}
For all program counter labels \ensuremath{\Varid{pc}\;\flows\;\Blue{\Conid{L}}}, valid inputs \ensuremath{\vdash \mathbf{Valid}(\Varid{c}_{1},\theta_{1})} and \ensuremath{\vdash \mathbf{Valid}(\Varid{c}_{2},\theta_{2})}, such that \ensuremath{\Varid{c}_{1}\;\approx_{\Blue{\Conid{L}}}^{\beta}\;\Varid{c}_{2}},
\ensuremath{\theta_{1}\;\approx_{\Blue{\Conid{L}}}^{\beta}\;\theta_{2}}, if \ensuremath{\Varid{c}_{1}\;\Downarrow^{\theta_{1}}_{\Varid{pc}}\;\Varid{c}_{1}'}, and \ensuremath{\Varid{c}_{2}\;\Downarrow^{\theta_{2}}_{\Varid{pc}}\;\Varid{c}_{2}'}, then there exists an extended bijection \ensuremath{\beta'\;\supseteq\;\beta}, such that \ensuremath{\Varid{c}_{1}'\;\approx_{\Blue{\Conid{L}}}^{\beta'}\;\Varid{c}_{2}'}.
\end{lemma}
\begin{proof}
  From \ensuremath{\Blue{\Conid{L}}}-equivalence, i.e., \ensuremath{\Varid{c}_{1}\approx_{\Blue{\Conid{L}}}^{\beta}\Varid{c}_{2}}, the program expressions are
  \ensuremath{\alpha}-equivalent, therefore the two reductions step following the
  same rule and produce \ensuremath{\Blue{\Conid{L}}}-equivalent results and perform
  \ensuremath{\Blue{\Conid{L}}}-equivalent operations on the program store and heap.
  In particular, for rule \srule{New-FS}, the programs allocate
  \ensuremath{\Blue{\Conid{L}}}-equivalent heap cells, e.g., \ensuremath{{v}_{1}\;\approx_{\Blue{\Conid{L}}}^{\beta'}\;{v}_{2}} for some
  extended bijection \ensuremath{\beta'\;\supseteq\;\beta} by induction hypothesis, at
  fresh addresses \ensuremath{\Varid{n}_{1}\mathrel{=}\abs{\mu_{1}'}} and \ensuremath{\Varid{n}_{2}\mathrel{=}\abs{\mu_{2}'}} of related heaps
  \ensuremath{\mu_{1}'\;\approx_{\Blue{\Conid{L}}}^{\beta'}\;\mu_{2}'}, which  remain related under the
  \emph{extended} bijection \ensuremath{\beta''\mathrel{=}\beta' \cup \{\mskip1.5mu (\Varid{n}_{1},\Varid{n}_{2})\mskip1.5mu\}\;\supseteq\;\beta} by transitivity, i.e., \ensuremath{\update{\mu_{1}}{\Varid{n}_{1}}{{v}_{1}}\;\approx_{\Blue{\Conid{L}}}^{\beta''}\;\update{\mu_{2}}{\Varid{n}_{2}}{{v}_{2}}}, and so produce related references \ensuremath{{\Varid{n}_{1}}^{\Blue{\Conid{L}}}\;\approx_{\Blue{\Conid{L}}}^{\beta''}\;{\Varid{n}_{2}}^{\Blue{\Conid{L}}}} by rule \srule{Value\ensuremath{{}_{\Blue{\Conid{L}}}}} applied to rule
  \srule{Ref-FS}.
\end{proof}

Lastly, we combine the \ensuremath{\Blue{\Conid{L}}}-equivalence preservation lemmas for public
and secret contexts and prove termination-insensitive non-interference
for \ensuremath{\FG} extended with flow-sensitive references.

\begin{thm}[\ensuremath{\FG}-TINI with Bijections]
\label{thm:fg-tini-bij}
For all program counter labels \ensuremath{\Varid{pc}} and valid inputs \ensuremath{\vdash \mathbf{Valid}(\Varid{c}_{1},\theta_{1})} and \ensuremath{\vdash \mathbf{Valid}(\Varid{c}_{2},\theta_{2})}, such that \ensuremath{\Varid{c}_{1}\;\approx_{\Blue{\Conid{L}}}^{\beta}\;\Varid{c}_{2}},
\ensuremath{\theta_{1}\;\approx_{\Blue{\Conid{L}}}^{\beta}\;\theta_{2}}, if \ensuremath{\Varid{c}_{1}\;\Downarrow^{\theta_{1}}_{\Varid{pc}}\;\Varid{c}_{1}'}, and \ensuremath{\Varid{c}_{2}\;\Downarrow^{\theta_{2}}_{\Varid{pc}}\;\Varid{c}_{2}'}, then there exists an extended bijection \ensuremath{\beta'\;\supseteq\;\beta}, such that \ensuremath{\Varid{c}_{1}'\;\approx_{\Blue{\Conid{L}}}^{\beta'}\;\Varid{c}_{2}'}.
\end{thm}

\paragraph{Conclusion.}
Dynamic language-based fine-grained IFC, of which \ensuremath{\FG} is just a
particular instance, represents an intuitive approach to tracking
information flows in programs.
Programmers annotate input values with labels that represent their
sensitivity and a label-aware instrumented security monitor propagates
those labels during execution and computes the result of the program
together with a conservative approximation of its sensitivity.
The next section describes an IFC monitor that tracks information
flows at \emph{coarse} granularity.



\chapter{Coarse-Grained IFC Calculus}
\label{sec:cg-calculus}

\begin{figure}[t]
\begin{align*}
\text{Types: } & \ensuremath{\tau\Coloneqq\mathbf{unit}\;{\mid}\;\tau_{1}\to \tau_{2}\;{\mid}\;\tau_{1}\mathbin{+}\tau_{2}\;{\mid}\;\tau_{1}\;\times\;\tau_{2}\;{\mid}\;\mathscr{L}} \\ & \ensuremath{{\mid}\;\mathbf{LIO}\xspace\;\tau\;{\mid}\;\mathbf{Labeled}\;\tau\;{\mid}\;\mathbf{Ref}\;\tau}
\\
\text{Labels: } & \ensuremath{{\ell},\Varid{pc}\;\in\;\mathscr{L}}
\\
\text{Addresses: } & \ensuremath{\Varid{n}\;\in\;\mathbb{N}} \\
\text{Environments: } & \ensuremath{\theta\;\in\;\Conid{Var}\;\rightharpoonup\;\Conid{Value}}
\\
\text{Values: } & \ensuremath{{v}\Coloneqq()\;{\mid}\;(\Varid{x}.\Varid{e},\theta)\;{\mid}\;\mathbf{inl}({v}\mkern1mu)\;{\mid}\;\mathbf{inr}({v}\mkern1mu)\;{\mid}\;({v}_{1},{v}_{2})\;{\mid}\;{\ell}}\
\\ & \ensuremath{{\mid}\;\mathbf{Labeled}\;{\ell}\;{v}\;{\mid}\;(\Varid{t},\theta)\;{\mid}\;{\Varid{n}}_{{\ell}}}
\\
\text{Expressions: } & \ensuremath{\Varid{e}\Coloneqq\Varid{x}\;{\mid}\lambda \Varid{x}.\Varid{e}\;{\mid}\;\Varid{e}_{1}\;\Varid{e}_{2}\;{\mid}\;()\;{\mid}\;{\ell}\;{\mid}\;(\Varid{e}_{1},\Varid{e}_{2})\;{\mid}\;\mathbf{fst}(\Varid{e}\mkern1mu)\;{\mid}\;\mathbf{snd}(\Varid{e}\mkern1mu)}
\\ & \ensuremath{{\mid}\;\mathbf{inl}(\Varid{e}_{1}\mkern1mu)\;{\mid}\;\mathbf{inr}(\Varid{e}_{2}\mkern1mu)\;{\mid}\;\mathbf{case}(\Varid{e},\Varid{x}.\Varid{e}_{1},\Varid{x}.\Varid{e}_{2}\mkern1mu)\;{\mid}\;\Varid{e}_{1}\;\flows^{?}\;\Varid{e}_{2}\;{\mid}\;\Varid{t}}
\\
\text{Thunks: } & \ensuremath{\Varid{t}\Coloneqq\mathbf{return}(\Varid{e}\mkern1mu)\;{\mid}\;\mathbf{bind}(\Varid{e},\Varid{x}.\Varid{e}\mkern1mu)\;{\mid}\;\mathbf{unlabel}(\Varid{e}\mkern1mu)}
\\ & \ensuremath{{\mid}\;\mathbf{toLabeled}(\Varid{e}\mkern1mu)\;{\mid}\;\mathbf{labelOf}(\Varid{e}\mkern1mu)\;{\mid}\;\mathbf{getLabel}\;{\mid}\;\mathbf{taint}(\Varid{e}\mkern1mu)}
\\ & \ensuremath{{\mid}\;\mathbf{new}(\Varid{e}\mkern1mu)\;{\mid}\mathbin{!}\Varid{e}\;{\mid}\;\Varid{e}_{1}\mathbin{:=}\Varid{e}_{2}\;{\mid}\;\mathbf{labelOfRef}(\Varid{e}\mkern1mu)}
\\
\text{Type System: } & \ensuremath{\Gamma\vdash\Varid{e}\mathbin{:}\tau}
\\
\text{Configurations: } & \ensuremath{\Varid{c}\Coloneqq\langle\Sigma,\Varid{pc},\Varid{e}\rangle}
\\
\text{Stores: } & \ensuremath{\Sigma\;\in\;({\ell}\mathbin{:}\Conid{Label})\to \Conid{Memory}\;{\ell}}
\\
\text{Memory }\ensuremath{{\ell}}\text{: } & \ensuremath{\Conid{M}\Coloneqq[\mskip1.5mu \mskip1.5mu]\;{\mid}\;{v}\mathbin{:}\Conid{M}}
\end{align*}
\caption{Syntax of \ensuremath{\CG}.}
\label{fig:cg-syntax}
\end{figure}

One of the drawbacks of dynamic fine-grained IFC is that the programming model
requires all input values to be explicitly and fully annotated with
their security labels.
Imagine a program with many inputs and highly structured data: it
quickly becomes cumbersome, if not impossible, for the programmer to
specify all the labels.
The label of some inputs may be sensitive (e.g., passwords, pin codes,
etc.), but the sensitivity of the rest may probably be irrelevant for
the computation, yet a programmer must come up with appropriate labels
for them as well.
The programmer is then torn between two opposing risks:
over-approximating the actual sensitivity can negatively affect
execution (the monitor might stop secure programs),
under-approximating the sensitivity can endanger security.
Even worse, specifying many labels manually is error-prone and
assigning the wrong security label to a piece of sensitive data can be
catastrophic for security and completely defeat the purpose of IFC.
Dynamic coarse-grained IFC represents an attractive alternative that
requires fewer annotations, in particular it allows the programmer to
label only the inputs that need to be protected.

%

\paragraph{Syntax.}
Figure \ref{fig:cg-syntax} shows the syntax of \ensuremath{\CG}, a standard
simply-typed \ensuremath{\lambda}-calculus extended with security primitives for
dynamic coarse-grained IFC, inspired by \citet{stefan:lio} and adapted
to use call-by-value instead of call-by-name to match \ensuremath{\FG}.
The \ensuremath{\CG}-calculus features both standard (unlabeled) values and
\emph{explicitly labeled} values. For example, \ensuremath{\mathbf{Labeled}\;\Red{\Conid{H}}\;\mathbf{true}}
represents a secret boolean value of type \ensuremath{\mathbf{Labeled}\;\mathbf{bool}}.\footnote{As
  in \ensuremath{\FG}, we define \ensuremath{\mathbf{bool}\mathrel{=}\mathbf{unit}\mathbin{+}\mathbf{unit}} and \ensuremath{\mathbf{if}\;\Varid{e}\;\mathbf{then}\;\Varid{e}_{1}\;\mathbf{else}\;\Varid{e}_{2}\mathrel{=}\mathbf{case}(\Varid{e},\anonymous .\Varid{e}_{1},\anonymous .\Varid{e}_{2}\mkern1mu)}.  Unlike \ensuremath{\FG} values, \ensuremath{\CG} values are not
  intrinsically labeled, thus we encode boolean constants simply as
  \ensuremath{\mathbf{true}\mathrel{=}\mathbf{inl}(()\mkern1mu)} and \ensuremath{\mathbf{false}\mathrel{=}\mathbf{inr}(()\mkern1mu)}.}
The type constructor \ensuremath{\mathbf{LIO}\xspace} encapsulates a security state monad, whose
state consists of a labeled store and the program counter label.
In addition to standard \ensuremath{\mathbf{return}(\cdot \mkern1mu)} and \ensuremath{\mathbf{bind}(\cdot \mkern1mu)} constructs, the monad
provides primitives that regulate the creation and the inspection of
labeled values, i.e., \ensuremath{\mathbf{toLabeled}(\cdot \mkern1mu)}, \ensuremath{\mathbf{unlabel}(\cdot \mkern1mu)} and \ensuremath{\mathbf{labelOf}(\cdot \mkern1mu)}, and the
interaction with the labeled store, allowing the creation, reading and
writing of labeled references \ensuremath{{\Varid{n}}_{{\ell}}} through the constructs
\ensuremath{\mathbf{new}(\Varid{e}\mkern1mu)}, \ensuremath{\mathbin{!}\Varid{e}}, \ensuremath{\Varid{e}_{1}\mathbin{:=}\Varid{e}_{2}}, respectively.\footnote{We extend \ensuremath{\CG}
  with \emph{flow-sensitive} references and \emph{heaps} in
  Section~\ref{sec:cg-fs-refs}.}
%
%
The primitives of the \ensuremath{\mathbf{LIO}\xspace} monad are listed in a separate
sub-category of expressions called \emph{thunk}.
Intuitively, a thunk is just a description of a stateful computation,
which only the top-level security monitor can execute---a \emph{thunk
  closure}, i.e., \ensuremath{(\Varid{t},\theta)}, provides a way to suspend nested
computations.

\section{Dynamics}

\begin{figure}[t]
\centering
\begin{mathpar}
\inferrule[(Thunk)]
{}
{\ensuremath{\Varid{t}\;\Downarrow^{\theta}\;(\Varid{t},\theta)}}
\and
\inferrule[(Fun)]
{}
{\ensuremath{\lambda \Varid{x}.\Varid{e}\;\Downarrow^{\theta}\;(\Varid{x}.\Varid{e},\theta)}}
\and
\inferrule[(Var)]
{}
{\ensuremath{\Varid{x}\;\Downarrow^{\theta}\;\theta(\Varid{x})}}
\and
\inferrule[(App)]
{\ensuremath{\Varid{e}_{1}\;\Downarrow^{\theta}\;(\Varid{x}.\Varid{e},\theta')} \\ \ensuremath{\Varid{e}_{2}\;\Downarrow^{\theta}\;{v}_{2}} \\ \ensuremath{\Varid{e}\;\Downarrow^{\update{\theta'}{\Varid{x}}{{v}_{2}}}\;{v}}}
{\ensuremath{\Varid{e}_{1}\;\Varid{e}_{2}\;\Downarrow^{\theta}\;{v}}}
\and
\inferrule[(Case$_1$)]
{\ensuremath{\Varid{e}_{1}\;\Downarrow^{\theta}\;\mathbf{inl}({v}_{1}\mkern1mu)} \\ \ensuremath{\Varid{e}_{1}\;\Downarrow^{\update{\theta}{\Varid{x}}{{v}_{1}}}\;{v}} }
{\ensuremath{\mathbf{case}(\Varid{e},\Varid{x}.\Varid{e}_{1},\Varid{x}.\Varid{e}_{2}\mkern1mu)\;\Downarrow^{\theta}\;{v}}}
\and
\inferrule[(Case$_2$)]
{\ensuremath{\Varid{e}_{1}\;\Downarrow^{\theta}\;\mathbf{inr}({v}_{2}\mkern1mu)} \\ \ensuremath{\Varid{e}_{2}\;\Downarrow^{\update{\theta}{\Varid{x}}{{v}_{2}}}\;{v}} }
{\ensuremath{\mathbf{case}(\Varid{e},\Varid{x}.\Varid{e}_{1},\Varid{x}.\Varid{e}_{2}\mkern1mu)\;\Downarrow^{\theta}\;{v}}}
\end{mathpar}
\caption{Pure semantics: \ensuremath{\Varid{e}\;\Downarrow^{\theta}\;{v}} (selected rules). }
\label{fig:cg-pure}
\end{figure}
\begin{figure}[p]
\centering
\begin{subfigure}[t]{\linewidth}
\begin{mathpar}
\inferrule[(Force)]
{\ensuremath{\Varid{e}\;\Downarrow^{\theta}\;(\Varid{t},\theta')} \\
 \ensuremath{\langle\Sigma,\Varid{pc},\Varid{t}\rangle\Downarrow^{\theta'}\langle\Sigma',\Varid{pc}',{v}\rangle}}
{\ensuremath{\langle\Sigma,\Varid{pc},\Varid{e}\rangle\Downarrow^{\theta}\langle\Sigma',\Varid{pc}',{v}\rangle}}
\end{mathpar}
\captionsetup{format=caption-with-line}
\caption{Forcing semantics: \ensuremath{\langle\Sigma,\Varid{pc},\Varid{e}\rangle\Downarrow^{\theta}\langle\Sigma',\Varid{pc}',{v}\rangle}.}
\label{fig:cg-forcing}
\end{subfigure}

\begin{subfigure}[t]{\linewidth}
\centering
\begin{mathpar}
\inferrule[(Return)]
{\ensuremath{\Varid{e}\;\Downarrow^{\theta}\;{v}}}
{\ensuremath{\langle\Sigma,\Varid{pc},\mathbf{return}(\Varid{e}\mkern1mu)\rangle\Downarrow^{\theta}\langle\Sigma,\Varid{pc},{v}\rangle}}
\and
\inferrule[(Bind)]
{\ensuremath{\langle\Sigma,\Varid{pc},\Varid{e}_{1}\rangle\Downarrow^{\theta}\langle\Sigma',\Varid{pc}',{v}_{1}\rangle} \\
 \ensuremath{\langle\Sigma',\Varid{pc}',\Varid{e}_{2}\rangle\Downarrow^{\update{\theta}{\Varid{x}}{{v}_{1}}}\langle\Sigma'',\Varid{pc''},{v}\rangle}}
{\ensuremath{\langle\Sigma,\Varid{pc},\mathbf{bind}(\Varid{e}_{1},\Varid{x}.\Varid{e}_{2}\mkern1mu)\rangle\Downarrow^{\theta}\langle\Sigma'',\Varid{pc''},{v}\rangle}}
\and
\inferrule[(ToLabeled)]
{\ensuremath{\langle\Sigma,\Varid{pc},\Varid{e}\rangle\Downarrow^{\theta}\langle\Sigma',\Varid{pc}',{v}\rangle}}
{\ensuremath{\langle\Sigma,\Varid{pc},\mathbf{toLabeled}(\Varid{e}\mkern1mu)\rangle\Downarrow^{\theta}\langle\Sigma',\Varid{pc},\mathbf{Labeled}\;\Varid{pc}'\;{v}\rangle}}
\and
\inferrule[(Unlabel)]
{\ensuremath{\Varid{e}\;\Downarrow^{\theta}\;\mathbf{Labeled}\;{\ell}\;{v}} }
{\ensuremath{\langle\Sigma,\Varid{pc},\mathbf{unlabel}(\Varid{e}\mkern1mu)\rangle\Downarrow^{\theta}\langle\Sigma,\Varid{pc}\;\lub\;{\ell},{v}\rangle}}
\and
\inferrule[(LabelOf)]
{\ensuremath{\Varid{e}\;\Downarrow^{\theta}\;\mathbf{Labeled}\;{\ell}\;{v}}}
{\ensuremath{\langle\Sigma,\Varid{pc},\mathbf{labelOf}(\Varid{e}\mkern1mu)\rangle\Downarrow^{\theta}\langle\Sigma,\Varid{pc}\;\lub\;{\ell},{\ell}\rangle}}
\and
\inferrule[(GetLabel)]
{}
{\ensuremath{\langle\Sigma,\Varid{pc},\mathbf{getLabel}\rangle\Downarrow^{\theta}\langle\Sigma,\Varid{pc},\Varid{pc}\rangle}}
\and
\inferrule[(Taint)]
{\ensuremath{\Varid{e}\;\Downarrow^{\theta}\;{\ell}}}
{\ensuremath{\langle\Sigma,\Varid{pc},\mathbf{taint}(\Varid{e}\mkern1mu)\rangle\Downarrow^{\theta}\langle\Sigma,\Varid{pc}\;\lub\;{\ell},()\rangle}}
\end{mathpar}
\captionsetup{format=caption-with-line}
\caption{Thunk semantics: \ensuremath{\langle\Sigma,\Varid{pc},\Varid{t}\rangle\Downarrow^{\theta}\langle\Sigma',\Varid{pc}',{v}\rangle}.}
\label{fig:cg-thunk}
\end{subfigure}

\caption{Big-step semantics for \ensuremath{\CG}.}
\label{fig:cg-semantics}
\end{figure}

In order to track information flows dynamically at coarse granularity,
\ensuremath{\CG} employs a technique called \emph{floating-label}, which was
originally developed for IFC operating systems (e.g.,
\cite{Zeldovich:2006,zeldovich:dstar}) and that was later applied in a
language-based setting.
In this technique, throughout a program's execution, the program
counter label \emph{floats} above the label of any value observed
during program execution and thus represents (an upper bound on) the
sensitivity of all the values that are not explicitly labeled.
For this reason, \ensuremath{\CG} stores the program counter label in the program
configuration, so that the primitives of the \ensuremath{\mathbf{LIO}\xspace} monad can control
it explicitly.
In technical terms the program counter is said to be
\emph{flow-sensitive}, i.e., it may assume different values in the
final configuration depending on the control flow of the
program.\footnote{In contrast, we consider \ensuremath{\FG}'s program counter
  \emph{flow-insensitive} because it is part of the evaluation
  judgment and its value changes only inside nested judgments.}

Like \ensuremath{\FG}, the operational semantics of \ensuremath{\CG} consists of a security
monitor that fully evaluates secure programs but prevents the
execution of insecure programs and similarly enforces
\emph{termination-insensitive} non-interference
(Theorem~\ref{thm:cg-tini}).
The big-step operational semantics of \ensuremath{\CG} is structured in two parts:
(i) a straightforward call-by-value side-effect-free semantics for
pure expressions (Fig.~\ref{fig:cg-pure}), and (ii) a top-level
security monitor for monadic programs (Fig.~\ref{fig:cg-semantics}).
The semantics of the security monitor is further split into two
mutually recursive reduction relations, one for arbitrary expressions
(Fig.~\ref{fig:cg-forcing}) and one specific to thunks
(Fig.~\ref{fig:cg-thunk}).
These constitute the \emph{forcing} semantics of the monad, which
reduces a thunk to a pure value and perform side-effects.
In particular, given the initial store \ensuremath{\Sigma}, program counter label \ensuremath{\Varid{pc}},
expression \ensuremath{\Varid{e}} of type \ensuremath{\mathbf{LIO}\xspace\;\tau} for some type \ensuremath{\tau} and input values
\ensuremath{\theta} (which may or may not be labeled), the monitor executes the
program, i.e., \ensuremath{\langle\Sigma,\Varid{pc},\Varid{e}\rangle\Downarrow^{\theta}\langle\Sigma',\Varid{pc}',{v}\rangle}
and gives an updated store \ensuremath{\Sigma'}, updated program counter \ensuremath{\Varid{pc}'} and
a final value \ensuremath{{v}} of type \ensuremath{\tau}, which also might not be labeled.
The execution starts with rule \srule{Force}, which reduces the pure
expression to a thunk closure, i.e., \ensuremath{(\Varid{t},\theta')} and then forces the
thunk \ensuremath{\Varid{t}} in its environment \ensuremath{\theta'} with the thunk semantics.
The pure semantics is fairly standard---we report some selected rules
in Fig.~\ref{fig:cg-pure} for comparison with \ensuremath{\FG}.
A pure reduction, written \ensuremath{\Varid{e}\;\Downarrow^{\theta}\;{v}}, evaluates an expression
\ensuremath{\Varid{e}} with an appropriate environment \ensuremath{\theta} to a pure value \ensuremath{{v}}.
%
%
Notice that, unlike \ensuremath{\FG}, all reduction rules of the pure semantics
ignore security, even those that affect the control flow of the
program, like rules \srule{App}, \srule{Case$_1$}, and
\srule{Case$_2$}: they do not feature the program counter label or
label annotations.
This is because these reductions are \emph{pure}---they cannot perform
side-effects and so leak sensitive data---and thus are inherently
secure and need not to be monitored~\parencite{VASSENA2017}.\footnote{
  The strict separation between side-effect-free and side-effectful
  code is a distinctive feature of coarse-grained IFC, which is
  crucial to lightweight approaches to enforce security via software
  libraries~\parencite{Russo+:Haskell08,buiras2015hlio,Russo:2015,stefan:2012:addressing}}
For example, since pure programs do not have access to the store, they
cannot leak through \emph{implicit flows}, which are then a concern
only for monadic programs.
How does the semantics prevent pure programs from performing
side-effects, without getting the program stuck and raising a false
alarm?
If the pure evaluation reaches a side-effectful computation, i.e.,
thunk \ensuremath{\Varid{t}}, it \emph{suspends} the computation by creating a thunk
closure that captures the current environment \ensuremath{\theta} (see rule
\srule{Thunk}).\footnote{Notice that \emph{type preservation} for the
  pure semantics preserves types \emph{exactly} i.e., if \ensuremath{\Gamma\vdash\Varid{e}\mathbin{:}\tau}, \ensuremath{\Varid{e}\;\Downarrow^{\theta}\;{v}} and \ensuremath{\vdash\theta\mathbin{:}\Gamma}, then \ensuremath{\vdash{v}\mathbin{:}\tau},
  which reflects the \emph{suspending} behavior for the monadic type
  \ensuremath{\mathbf{LIO}\xspace\;\tau}. In contrast, type preservation for the \emph{thunk} and
  \emph{forcing} semantics assumes that the expression (resp. thunk)
  has a monadic type, i.e., \ensuremath{\mathbf{LIO}\xspace\;\tau} for some type \ensuremath{\tau}, and
  guarantees that the final value has type \ensuremath{\tau}. }
Notice that \emph{thunk closures} and \emph{function closures} are
distinct values created by different rules, \srule{Thunk} and
\srule{Fun} respectively.\footnote{It would have also been possible to
  define thunk values in terms of function closures using explicit
  suspension and an opaque constructor wrapper, e.g., \ensuremath{\mathbf{LIO}\xspace\;(\anonymous .\Varid{t},\theta)}.  }
Function application succeeds only when the function evaluates to a
function closure (rule \srule{App}).
In the thunk semantics, rule \srule{Return} evaluates a pure value
embedded in the monad via \ensuremath{\mathbf{return}(\cdot \mkern1mu)} and leaves the state unchanged,
while rule \srule{Bind} executes the first computation with the
forcing semantics, binds the result in the environment i.e.,
\ensuremath{\update{\theta}{\Varid{x}}{{v}_{1}}}, passes it on to the second computation
together with the updated state and returns the final result and
state.
Rule \srule{Unlabel} is interesting.
Following the \emph{floating-label} principle, it returns the value
wrapped inside the labeled value, i.e., \ensuremath{{v}}, and raises the program
counter with its label, i.e., \ensuremath{\Varid{pc}\;\lub\;{\ell}}, to reflect the fact that
new data at security level \ensuremath{{\ell}} is now in scope.
Floating-label based coarse-grained IFC systems like \lio~ suffer from
the \emph{label creep} problem, which occurs when the program counter
gets over-tainted, e.g., because too many secrets have been unlabeled, to
the point that no useful further computation can be performed.
Primitive \ensuremath{\mathbf{toLabeled}(\cdot \mkern1mu)} provides a mechanism to address this problem by
(i) creating a separate context where some sensitive computation can
take place and (ii) restoring the original program counter label
afterwards.
Rule \srule{ToLabeled} formalizes this idea.
Notice that the result of the nested sensitive computation, i.e., \ensuremath{{v}},
cannot be simply returned to the lower context---that would be a leak,
so \ensuremath{\mathbf{toLabeled}(\cdot \mkern1mu)} wraps that piece of information in a labeled value
protected by the final program counter of the sensitive computation,
i.e., \ensuremath{\mathbf{Labeled}\;\Varid{pc}'\;{v}}.\footnote{\citet{stefan:2017:flexible} have
  proposed an alternative flow-insensitive primitive, i.e.,
  \ensuremath{\mathbf{toLabeled}({\ell},\Varid{e}\mkern1mu)}, which labels the result with the user-assigned
  label \ensuremath{{\ell}}. The semantics of \ensuremath{\FG} forced us to use \ensuremath{\mathbf{toLabeled}(\Varid{e}\mkern1mu)}.
  \label{footnote:flow-insensitive-api}}
Furthermore, notice that \ensuremath{\Varid{pc}'}, the label that tags the result \ensuremath{{v}}, is
as sensitive as the result itself because the final program counter
depends on all the \ensuremath{\mathbf{unlabel}(\cdot \mkern1mu)} operations performed to compute the
result.
This motivates why primitive \ensuremath{\mathbf{labelOf}(\cdot \mkern1mu)} does not simply project the
label from a labeled value, but additionally taints the program
counter with the label itself in rule \srule{LabelOf}--a label in a
labeled value has sensitivity equal to the label itself, thus the
program counter label rises to accommodate reading new sensitive data.
Lastly, rule \srule{GetLabel} returns the value of the program
counter, which does not rise (because \ensuremath{\Varid{pc}\;\lub\;\Varid{pc}\mathrel{=}\Varid{pc}}), and rule
\srule{Taint} simply taints the program counter with the given label
and returns unit (this primitive matches the functionality of \ensuremath{\mathbf{taint}(\cdot \mkern1mu)}
in \ensuremath{\FG}).
Note that, in \ensuremath{\CG}, \ensuremath{\mathbf{taint}(\cdot \mkern1mu)} takes \emph{only} the label with which the
program counter must be tainted whereas, in \ensuremath{\FG}, it additionally
requires the expression that must be evaluated in the tainted
environment.
This difference highlights the \emph{flow-sensitive} nature of the
program counter label in \ensuremath{\CG}.


%

\subsection{References}
\begin{figure}[t]
\begin{mathpar}
\inferrule[(New)]
{\ensuremath{\Varid{e}\;\Downarrow^{\theta}\;\mathbf{Labeled}\;{\ell}\;{v}} \\ \ensuremath{\Varid{pc}\;\flows\;{\ell}} \\ \ensuremath{\Varid{n}\mathrel{=}\abs{\Sigma({\ell})}}}
{\ensuremath{\langle\Sigma,\Varid{pc},\mathbf{new}(\Varid{e}\mkern1mu)\rangle\Downarrow^{\theta}\langle\update{\Sigma}{{\ell}}{\update{\Sigma({\ell})}{\Varid{n}}{{v}}},\Varid{pc},{\Varid{n}}_{{\ell}}\rangle}}
\and
\inferrule[(Read)]
{\ensuremath{\Varid{e}\;\Downarrow^{\theta}\;{\Varid{n}}_{{\ell}}} \\ \ensuremath{\Sigma({\ell}){[}\Varid{n}{]}\mathrel{=}{v}}}
{\ensuremath{\langle\Sigma,\Varid{pc},\mathbin{!}\Varid{e}\rangle\Downarrow^{\theta}\langle\Sigma,\Varid{pc}\;\lub\;{\ell},{v}\rangle}}
\and
\inferrule[(Write)]
{\ensuremath{\Varid{e}_{1}\;\Downarrow^{\theta}\;{\Varid{n}}_{{\ell}_{1}}} \\ \ensuremath{\Varid{e}_{2}\;\Downarrow^{\theta}\;\mathbf{Labeled}\;{\ell}_{2}\;{v}} \\ \ensuremath{{\ell}_{2}\;\flows\;{\ell}_{1}} \\ \ensuremath{\Varid{pc}\;\flows\;{\ell}_{1}}}
{\ensuremath{\langle\Sigma,\Varid{pc},\Varid{e}_{1}\mathbin{:=}\Varid{e}_{2}\rangle\Downarrow^{\theta}\langle\update{\Sigma}{{\ell}_{1}}{\update{\Sigma({\ell}_{1})}{\Varid{n}}{{v}}},\Varid{pc},()\rangle}}
\and
\inferrule[(LabelOfRef)]
{\ensuremath{\Varid{e}\;\Downarrow^{\theta}\;{\Varid{n}}_{{\ell}}}}
{\ensuremath{\langle\Sigma,\Varid{pc},\mathbf{labelOfRef}(\Varid{e}\mkern1mu)\rangle\Downarrow^{\theta}\langle\Sigma,\Varid{pc}\;\lub\;{\ell},{\ell}\rangle}}
\end{mathpar}
\caption{Big-step semantics for \ensuremath{\CG} (references).}
\label{fig:cg-memory}
\end{figure}
\ensuremath{\CG} features \emph{flow-insensitive} labeled references similar to
\ensuremath{\FG} and allows programs to create, read, update and inspect the label
inside the \ensuremath{\mathbf{LIO}\xspace} monad (see Figure \ref{fig:cg-memory}).
The API of these primitives takes explicitly labeled values as
arguments, by making explicit at the type level, the tagging that
occurs in memory, which was left implicit in previous work
\parencite{stefan:2017:flexible}.
Rule \srule{New} creates a reference labeled with the same label
annotation as that of the labeled value it receives as an argument, and
checks that \ensuremath{\Varid{pc}\;\flows\;{\ell}} in order to avoid implicit flows.
Rule \srule{Read} retrieves the content of the reference from the
\ensuremath{{\ell}}-labeled memory and returns it. Since this brings data at security
level \ensuremath{{\ell}} in scope, the program counter is tainted accordingly, i.e.,
\ensuremath{\Varid{pc}\;\lub\;{\ell}}.
%
%
Rule \srule{Write} performs security checks analogous to those in \ensuremath{\FG}
and updates the content of a given reference and rule
\srule{LabelOfRef} returns the label on a reference and taints the
context accordingly.
We conclude this section by noting that the forcing and the thunk
semantics of \ensuremath{\CG} satisfy Property \ref{prop:cg-pc-raise} (\emph{``the
  final value of the program counter label of any \ensuremath{\CG} program is
  always at least as sensitive as the initial value''}).

\begin{property}[]
\label{prop:cg-pc-raise}
\leavevmode
\begin{itemize}
\item If \ensuremath{\langle\Sigma,\Varid{pc},\Varid{e}\rangle\Downarrow^{\theta}\langle\Sigma',\Varid{pc}',{v}\rangle} then \ensuremath{\Varid{pc}\;\flows\;\Varid{pc}'}.
\item If \ensuremath{\langle\Sigma,\Varid{pc},\Varid{t}\rangle\Downarrow^{\theta}\langle\Sigma',\Varid{pc}',{v}\rangle} then
  \ensuremath{\Varid{pc}\;\flows\;\Varid{pc}'}.
\end{itemize}
\end{property}
\begin{proof} By mutual induction on the given evaluation derivations.
\end{proof}

\section{Security}
\label{subsec:cg-security}

\begin{figure}[t!]

\begin{mathpar}
\inferrule[(Labeled\ensuremath{{}_{\Blue{\Conid{L}}}})]
{\ensuremath{{\ell}\;\flows\;\Blue{\Conid{L}}} \\ \ensuremath{{v}_{1}\approx_{\Blue{\Conid{L}}}{v}_{2}}}
{\ensuremath{\mathbf{Labeled}\;{\ell}\;{v}_{1}\approx_{\Blue{\Conid{L}}}\mathbf{Labeled}\;{\ell}\;{v}_{2}}}
\and
\inferrule[(Labeled\ensuremath{{}_{\Red{\Conid{H}}}})]
{\ensuremath{{\ell}_{1}\;\not\flows\;\Blue{\Conid{L}}} \\ \ensuremath{{\ell}_{2}\;\not\flows\;\Blue{\Conid{L}}}}
{\ensuremath{\mathbf{Labeled}\;{\ell}_{1}\;{v}_{1}\approx_{\Blue{\Conid{L}}}\mathbf{Labeled}\;{\ell}_{2}\;{v}_{2}}}
\and
\inferrule[(Inl)]
{\ensuremath{{v}_{1}\approx_{\Blue{\Conid{L}}}{v}_{2}}}
{\ensuremath{\mathbf{inl}({v}_{1}\mkern1mu)\approx_{\Blue{\Conid{L}}}\mathbf{inl}({v}_{2}\mkern1mu)}}
\and
\inferrule[(Inr)]
{\ensuremath{{v}_{1}\approx_{\Blue{\Conid{L}}}{v}_{2}}}
{\ensuremath{\mathbf{inr}({v}_{1}\mkern1mu)\approx_{\Blue{\Conid{L}}}\mathbf{inr}({v}_{2}\mkern1mu)}}
\and
\inferrule[(F-Closure)]
{\ensuremath{\Varid{e}_{1}\equiv_{\alpha}\Varid{e}_{2}} \\ \ensuremath{\theta_{1}\approx_{\Blue{\Conid{L}}}\theta_{2}}}
{\ensuremath{(\Varid{e}_{1},\theta_{1})\approx_{\Blue{\Conid{L}}}(\Varid{e}_{2},\theta_{2})}}
\and
\inferrule[(T-Closure)]
{\ensuremath{\Varid{t}_{1}\equiv_{\alpha}\Varid{t}_{2}} \\ \ensuremath{\theta_{1}\approx_{\Blue{\Conid{L}}}\theta_{2}}}
{\ensuremath{(\Varid{t}_{1},\theta_{1})\approx_{\Blue{\Conid{L}}}(\Varid{t}_{2},\theta_{2})}}
\and
\inferrule[(Ref\ensuremath{{}_{\Blue{\Conid{L}}}})]
{\ensuremath{{\ell}\;\flows\;\Blue{\Conid{L}}}}
{\ensuremath{{\Varid{n}}^{{\ell}}\approx_{\Blue{\Conid{L}}}{\Varid{n}}^{{\ell}}}}
\and
\inferrule[(Ref\ensuremath{{}_{\Red{\Conid{H}}}})]
{\ensuremath{{\ell}_{1}\;\not\flows\;\Blue{\Conid{L}}} \\ \ensuremath{{\ell}_{2}\;\not\flows\;\Blue{\Conid{L}}}}
{\ensuremath{{\Varid{n}_{1}}^{{\ell}_{1}}\approx_{\Blue{\Conid{L}}}{\Varid{n}_{2}}^{{\ell}_{2}}}}
\and
\inferrule[(Pc\ensuremath{{}_{\Red{\Conid{H}}}})]
{\ensuremath{\Sigma_1\approx_{\Blue{\Conid{L}}}\Sigma_2} \\ \ensuremath{\Varid{pc}_{1}\;\not\flows\;\Blue{\Conid{L}}} \\ \ensuremath{\Varid{pc}_{2}\;\not\flows\;\Blue{\Conid{L}}}}
{\ensuremath{\langle\Sigma_1,\Varid{pc}_{1},{v}_{1}\rangle\approx_{\Blue{\Conid{L}}}\langle\Sigma_2,\Varid{pc}_{2},{v}_{2}\rangle}}
\and
\inferrule[(Pc\ensuremath{{}_{\Blue{\Conid{L}}}})]
{\ensuremath{\Sigma_1\approx_{\Blue{\Conid{L}}}\Sigma_2} \\ \ensuremath{\Varid{pc}\;\flows\;\Blue{\Conid{L}}} \\ \ensuremath{{v}_{1}\approx_{\Blue{\Conid{L}}}{v}_{2}}}
{\ensuremath{\langle\Sigma_1,\Varid{pc},{v}_{1}\rangle\approx_{\Blue{\Conid{L}}}\langle\Sigma_2,\Varid{pc},{v}_{2}\rangle}}
\end{mathpar}



\caption{\ensuremath{\Blue{\Conid{L}}}-equivalence for \ensuremath{\CG} values (selected rules) and
  configurations.}
\label{fig:cg-low-equiv}
\end{figure}

We now prove that \ensuremath{\CG} is secure, i.e., it satisfies
\emph{termination-insensitive non-interference}.
The meaning of the security condition is intuitively similar to that
presented in Section \ref{subsec:fg-security} for \ensuremath{\FG}--- when secret
inputs are changed, terminating programs do not produce any publicly
observable effect---and based on a similar indistinguishability
relation.

\subsection{\ensuremath{\Blue{\Conid{L}}}-Equivalence}
\label{subsec:cg-leq}
Figure \ref{fig:cg-low-equiv} presents the definition of
\ensuremath{\Blue{\Conid{L}}}-equivalence for the interesting cases only.
Firstly, \ensuremath{\Blue{\Conid{L}}}-equivalence for \ensuremath{\CG} labeled values relates public and
secret values analogously to \ensuremath{\FG} values.
Specifically, rule \srule{Labeled\ensuremath{{}_{\Blue{\Conid{L}}}}} relates public labeled
values that share the \emph{same} observable label, i.e., \ensuremath{{\ell}\;\flows\;\Blue{\Conid{L}}},
and contain related values, i.e., \ensuremath{{v}_{1}\approx_{\Blue{\Conid{L}}}{v}_{2}}, while rule
\srule{Labeled\ensuremath{{}_{\Red{\Conid{H}}}}} relates secret labeled values, with
arbitrary sensitivity labels not below \ensuremath{\Blue{\Conid{L}}}, i.e., \ensuremath{{\ell}_{1}\;\not\flows\;\Blue{\Conid{L}}} and
\ensuremath{{\ell}_{2}\;\not\flows\;\Blue{\Conid{L}}}, and contents.
Secondly, \ensuremath{\Blue{\Conid{L}}}-equivalence relates standard (unlabeled) values
homomorphically.
For example, values of the sum type are related only as follows: \ensuremath{\mathbf{inl}({v}_{1}\mkern1mu)\approx_{\Blue{\Conid{L}}}\mathbf{inl}({v}_{1}'\mkern1mu)} iff \ensuremath{{v}_{1}\approx_{\Blue{\Conid{L}}}{v}_{1}'} through rule \srule{Inl} and \ensuremath{\mathbf{inr}({v}_{2}\mkern1mu)\approx_{\Blue{\Conid{L}}}\mathbf{inr}({v}_{2}'\mkern1mu)} iff \ensuremath{{v}_{2}\approx_{\Blue{\Conid{L}}}{v}_{2}'} through rule \srule{Inr}, i.e., we do not
provide any rule to relate values with different injections, just like
in \ensuremath{\FG}.
Function and thunk closures are related by rules \srule{F-Closure} and
\srule{T-Closure}, respectively.
In the rules the function and the monadic computations are
\ensuremath{\alpha}-equivalent and their environments are related, i.e., \ensuremath{\theta_{1}\approx_{\Blue{\Conid{L}}}\theta_{2}} iff \ensuremath{dom(\theta_{1})\equiv dom(\theta_{2})} and \ensuremath{\forall\Varid{x}.\theta_{1}(\Varid{x})\approx_{\Blue{\Conid{L}}}\theta_{2}(\Varid{x})}.
Labeled references, memories and stores are related by
\ensuremath{\Blue{\Conid{L}}}-equivalence analogously to \ensuremath{\FG}.
Lastly, \ensuremath{\Blue{\Conid{L}}}-equivalence relates \emph{initial} configurations with
related stores, equal program counters and \ensuremath{\alpha}-equivalent
expressions (resp. thunks), i.e., \ensuremath{\Varid{c}_{1}\approx_{\Blue{\Conid{L}}}\Varid{c}_{2}} iff \ensuremath{\Varid{c}_{1}\mathrel{=}\langle\Sigma_1,\Varid{pc}_{1},\Varid{e}_{1}\rangle}, \ensuremath{\Varid{c}_{2}\mathrel{=}\langle\Sigma_2,\Varid{pc}_{2},\Varid{e}_{2}\rangle}, \ensuremath{\Sigma_1\approx_{\Blue{\Conid{L}}}\Sigma_2}, \ensuremath{\Varid{pc}_{1}\equiv \Varid{pc}_{2}}, and \ensuremath{\Varid{e}_{1}\equiv_{\alpha}\Varid{e}_{2}} (resp. \ensuremath{\Varid{t}_{1}\equiv_{\alpha}\Varid{t}_{2}} for thunks \ensuremath{\Varid{t}_{1}} and
\ensuremath{\Varid{t}_{2}}), and \emph{final} configurations with related stores and (i)
equal public program counter label, i.e., \ensuremath{\Varid{pc}\;\flows\;\Blue{\Conid{L}}}, and
related values through rule \srule{Pc\ensuremath{{}_{\Blue{\Conid{L}}}}}, or (ii) arbitrary
secret program counter labels, i.e., \ensuremath{\Varid{pc}_{1}\;\not\flows\;\Blue{\Conid{L}}} and \ensuremath{\Varid{pc}_{2}\;\not\flows\;\Blue{\Conid{L}}}, and arbitrary values through rule \srule{Pc\ensuremath{{}_{\Red{\Conid{H}}}}}.

\ensuremath{\Blue{\Conid{L}}}-equivalence for \ensuremath{\CG} is \emph{reflexive}, \emph{symmetric}, and
\emph{transitive}, similarly to \ensuremath{\FG} (Property~\ref{prop:fg-leq}), and
so these properties can be combine to derive a \emph{Square
  Commutative Diagram for \ensuremath{\CG} Stores} as well, analogous to
Figure~\ref{fig:fg-square-store}.

\subsection{Termination-Insensitive Non-Interference}
We now formally prove that also the security monitor of \ensuremath{\CG} is
secure, i.e., it enforces \emph{termination-insensitive
  non-interference} (TINI).
The proof technique is the same used for \ensuremath{\FG} and similarly based on
\emph{store confinement} and \emph{\ensuremath{\Blue{\Conid{L}}}-equivalence preservation}.
However, since the semantics of the security monitor of \ensuremath{\CG} is
defined by two mutually recursive relations, i.e., the forcing and the
thunk semantics, each lemma is stated as a pair of lemmas (one for
each semantics relation), which are then proved by mutual induction.
%
%

\paragraph{Store Confinement. }
Store confinement ensures that programs running in a secret context
cannot leak data implicitly through the labeled store.
To prevent these leaks, the security monitor of \ensuremath{\CG} aborts programs
that attempt to write public memories in secret contexts, as these
changes may depend on secret data and would be observable by the
attacker.
Rules \srule{New} and \srule{Write} enforce exactly this security
policy by allowing programs to write memories only if they are labeled
above the program counter label.

\begin{lemma}[Store Confinement]
  \leavevmode For all program counter labels \ensuremath{\Varid{pc}\;\not\flows\;\Blue{\Conid{L}}},
  initial configurations \ensuremath{\Varid{c}}, and final configurations \ensuremath{\Varid{c'}\mathrel{=}\langle\Sigma',\Varid{pc}',{v}\rangle}:
\begin{itemize}
\item If \ensuremath{\Varid{c}\mathrel{=}\langle\Sigma,\Varid{pc},\Varid{e}\rangle} and  \ensuremath{\Varid{c}\;\Downarrow^{\theta}\;\Varid{c'}}, then \ensuremath{\Sigma\approx_{\Blue{\Conid{L}}}\Sigma'}.
\item If \ensuremath{\Varid{c}\mathrel{=}\langle\Sigma,\Varid{pc},\Varid{t}\rangle} and  \ensuremath{\Varid{c}\;\Downarrow^{\theta}\;\Varid{c'}}, then \ensuremath{\Sigma\approx_{\Blue{\Conid{L}}}\Sigma'}.
\end{itemize}
\label{fig:cg-store-confinement}
\end{lemma}
\begin{proof}
  By mutual induction, using \emph{reflexivity} in most base cases and
  \emph{transitivity} in case \srule{Bind}, where the program counter
  label remains above the attacker's level in the intermediate
  configuration by Property~\ref{prop:cg-pc-raise}.
  In the base cases \srule{New} and \srule{Write}, the programs run
  with program counter label secret, i.e., \ensuremath{\Varid{pc}\;\not\flows\;\Blue{\Conid{L}}}, and
  write a secret memory labeled \ensuremath{{\ell}} above \ensuremath{\Varid{pc}}, i.e., \ensuremath{\Varid{pc}\;\flows\;{\ell}}, and thus also not observable by the attacker, i.e., \ensuremath{{\ell}\;\not\flows\;\Blue{\Conid{L}}}.
  %
\end{proof}

\paragraph{\ensuremath{\Blue{\Conid{L}}}-Equivalence Preservation.}
Next, we prove \ensuremath{\Blue{\Conid{L}}}-equivalence preservation in secret contexts, i.e.,
programs running with program counter label secret (\ensuremath{\Varid{pc}\;\not\flows\;\Blue{\Conid{L}}}) cannot leak secret data implicitly through their final value or
observable changes to the store.
%
%
In \ensuremath{\CG}, values are not intrinsically labeled like \ensuremath{\FG} values,
therefore the final values computed by these programs may not be
explicitly labeled.
If these values are unlabeled, how can we establish if they depend on
secret data, or if they are \ensuremath{\Blue{\Conid{L}}}-equivalent?
Luckily, we can safely approximate the sensitivity of these values
with the program counter label.
The program counter label always represents an upper bound over the
sensitivity of all data not explicitly labeled in a program, including
its final value.
This is precisely why \ensuremath{\CG} stores it in the program configuration, so
that the \ensuremath{\mathbf{LIO}\xspace} monad can control it through the floating-label
mechanism explained above.
Therefore, the fact that the program counter label is secret in a
final configuration indicates that the program has unlabeled secret
data and thus the final value \emph{may} depend on secrets.
This approximation is sound, but also very \emph{conservative}---in a
secret context, the result of a program must always be considered
secret even if the program has not actually used any secret data to
compute it---and central to the design of the \emph{coarse-grained}
IFC approach embodied by \ensuremath{\CG}.\footnote{Due to this conservative
  behavior, coarse-grained IFC languages like \ensuremath{\CG} may seem inherently
  limited, compared to fine-grained languages like \ensuremath{\FG}, which track
  data-dependencies more precisely, i.e., the result gets labeled
  secret \emph{only if} secret data has been used to compute its
  value.  In Section~\ref{sec:fg2cg}, we prove that \ensuremath{\CG} can track
  data-dependencies as precisely as \ensuremath{\FG}.}
The \ensuremath{\Blue{\Conid{L}}}-equivalence relation for final configurations reflects this
reading of the program counter label for unlabeled values.
Specifically, if both program counter labels are secret, then rule
\srule{Pc\ensuremath{{}_{\Red{\Conid{H}}}}} (Fig.~\ref{fig:cg-low-equiv}) simply ignores the
values in the configurations (because they \emph{may} depend on secret
data) and only requires the stores to be \ensuremath{\Blue{\Conid{L}}}-equivalent.\footnote{In
  particular, rule \srule{Pc\ensuremath{{}_{\Red{\Conid{H}}}}} accepts arbitrary final
  values just like rule \srule{Labeled\ensuremath{{}_{\Red{\Conid{H}}}}} for explicitly
  labeled values.}
Then, to prove \ensuremath{\Blue{\Conid{L}}}-equivalence preservation in secret contexts, we
simply observe that the program counter label can only increase during
program execution (Property~\ref{prop:cg-pc-raise}), thus programs
that start in secret contexts must also necessarily terminate in
secret contexts.
Therefore, we can apply rule \srule{Pc\ensuremath{{}_{\Red{\Conid{H}}}}} provided that the
final stores are \ensuremath{\Blue{\Conid{L}}}-equivalent, i.e., if we can show that programs
cannot leak implicitly through the public memories of the store.
To do that, we follow the same proof technique used for \ensuremath{\FG}, i.e., we
first prove that the final stores are \ensuremath{\Blue{\Conid{L}}}-equivalent to the initial
stores by applying \emph{store confinement}
(Lemma~\ref{fig:cg-store-confinement}) to each execution and then
derive \ensuremath{\Blue{\Conid{L}}}-equivalence of the final stores by constructing a
\emph{square commutative diagram} (Fig.~\ref{fig:fg-square-store}).

\begin{lemma}[\ensuremath{\Blue{\Conid{L}}}-Equivalence Preservation in Secret Contexts]
\label{lemma:cg-leq-preservation-secret}
For all program counter labels \ensuremath{\Varid{pc}_{1}\;\not\flows\;\Blue{\Conid{L}}} and \ensuremath{\Varid{pc}_{2}\;\not\flows\;\Blue{\Conid{L}}}, and stores \ensuremath{\Sigma_1\approx_{\Blue{\Conid{L}}}\Sigma_2}:
\begin{itemize}
\item If \ensuremath{\langle\Sigma_1,\Varid{pc}_{1},\Varid{e}_{1}\rangle\Downarrow^{\theta_{1}}\;\Varid{c}_{1}} and \ensuremath{\langle\Sigma_2,\Varid{pc}_{2},\Varid{e}_{2}\rangle\Downarrow^{\theta_{2}}\;\Varid{c}_{2}},
then \ensuremath{\Varid{c}_{1}\approx_{\Blue{\Conid{L}}}\Varid{c}_{2}}.
\item If \ensuremath{\langle\Sigma_1,\Varid{pc}_{1},\Varid{t}_{1}\rangle\Downarrow^{\theta_{1}}\;\Varid{c}_{1}} and \ensuremath{\langle\Sigma_2,\Varid{pc}_{2},\Varid{t}_{2}\rangle\Downarrow^{\theta_{2}}\;\Varid{c}_{2}},
then \ensuremath{\Varid{c}_{1}\approx_{\Blue{\Conid{L}}}\Varid{c}_{2}}.
\end{itemize}
\end{lemma}
\begin{proof}
Assume \ensuremath{\Varid{pc}_{1}\;\not\flows\;\Blue{\Conid{L}}}, \ensuremath{\Varid{pc}_{2}\;\not\flows\;\Blue{\Conid{L}}}, \ensuremath{\Sigma_1\approx_{\Blue{\Conid{L}}}\Sigma_2},
and let \ensuremath{\Varid{c}_{1}\mathrel{=}\langle\Sigma_{1}',\Varid{pc}_{1}',{v}_{1}\rangle} and \ensuremath{\Varid{c}_{2}\mathrel{=}\langle\Sigma_{2}',\Varid{pc}_{2}',{v}_{2}\rangle}.
First, we apply \emph{store confinement} (Lemma~\ref{fig:cg-store-confinement}) to each big-step reduction and obtain \ensuremath{\Sigma_1\approx_{\Blue{\Conid{L}}}\Sigma_{1}'} and \ensuremath{\Sigma_2\approx_{\Blue{\Conid{L}}}\Sigma_{2}'}.
These are then combined with the assumption \ensuremath{\Sigma_1\approx_{\Blue{\Conid{L}}}\Sigma_2} to form the commutative square diagram for \ensuremath{\CG} stores (Fig.~\ref{fig:fg-square-store}), which gives \ensuremath{\Sigma_{1}'\approx_{\Blue{\Conid{L}}}\Sigma_{2}'}.
Then, we observe that the program counter labels in the final
configurations are \emph{secret}, i.e., \ensuremath{\Varid{pc}_{1}'\;\not\flows\;\Blue{\Conid{L}}} and \ensuremath{\Varid{pc}_{2}'\;\not\flows\;\Blue{\Conid{L}}} by Property~\ref{prop:cg-pc-raise}, and therefore we
have \ensuremath{\Varid{c}_{1}\approx_{\Blue{\Conid{L}}}\Varid{c}_{2}} by rule \srule{Pc\ensuremath{{}_{\Red{\Conid{H}}}}}.
\end{proof}

Now, we consider \ensuremath{\Blue{\Conid{L}}}-equivalence preservation in public contexts.
Since \ensuremath{\CG} separates \emph{pure computations} from \emph{monadic
  computations} in different semantics judgments, we first need to
prove that \ensuremath{\Blue{\Conid{L}}}-equivalence is preserved also by the pure semantics.
Notice that this separation is not only beneficial for security, as
explained above, but simplifies the security analysis as well.
For example, pure computations cannot inspect secret data in a public
context, as this operation requires performing a side-effect.
%
%
%
%
This is because, in public contexts, secret data is \emph{explicitly}
protected through the \ensuremath{\mathbf{Labeled}} type, which prevents programs from
inspecting secrets directly as they must first \emph{extract} them via
\ensuremath{\mathbf{unlabel}(\cdot \mkern1mu)\mathbin{:}\mathbf{Labeled}\;\tau\to \mathbf{LIO}\xspace\;\tau}.
The monadic type of this thunk indicates that this computation may
perform side-effects (i.e., tainting the program counter label), which
crucially, can be performed only by the security monitor, in the thunk
semantics.\footnote{The pure semantics simply \emph{suspends} the
  evaluation of thunks via rule \srule{Thunk}.}
%
%
As a result, the execution of pure programs in public contexts cannot
depend on secret data---they simply cannot access secrets in the first
place---and so their (unlabeled) results are always indistinguishable
by the attacker.
%

\begin{lemma}[Pure \ensuremath{\Blue{\Conid{L}}}-Equivalence Preservation]
  For all expressions \ensuremath{\Varid{e}_{1}\;{\equiv }_{\alpha}\;\Varid{e}_{2}} and environments
  \ensuremath{\theta_{1}\approx_{\Blue{\Conid{L}}}\theta_{2}}, if \ensuremath{\Varid{e}_{1}\;\Downarrow^{\theta_{1}}\;{v}_{1}} and \ensuremath{\Varid{e}_{2}\;\Downarrow^{\theta_{2}}\;{v}_{2}}, then \ensuremath{{v}_{1}\approx_{\Blue{\Conid{L}}}{v}_{2}}.
\label{lemma:cg-pure-preservation}
\end{lemma}
\begin{proof}
  By induction on the big-step reductions, which \emph{must} always
  step according to the same rule.
  In the spurious cases, e.g., when expression \ensuremath{\mathbf{case}(\Varid{e},\Varid{x}.\Varid{e}_{1},\Varid{x}.\Varid{e}_{2}\mkern1mu)} steps through rules \srule{Case$_1$} and \srule{Case$_2$},
  we show a contradiction.
%
  In particular, assume that the executions follow different paths,
  e.g., we have reductions \ensuremath{\Varid{e}\;\Downarrow^{\theta_{1}}\;\mathbf{inl}({v}_{1}\mkern1mu)} and \ensuremath{\Varid{e}\;\Downarrow^{\theta_{2}}\;\mathbf{inr}({v}_{2}\mkern1mu)} for the scrutinee \ensuremath{\Varid{e}}.
  Then, we apply the induction hypothesis to these reductions and
  obtain a proof for \ensuremath{\mathbf{inl}({v}_{1}\mkern1mu)\approx_{\Blue{\Conid{L}}}\mathbf{inr}({v}_{2}\mkern1mu)}.
  But this is impossible: \ensuremath{\Blue{\Conid{L}}}-equivalence relates values of sum types
  only if they have the same injection, i.e., rules \srule{Inl} and
  \srule{Inr} in Figure~\ref{fig:cg-low-equiv}.
  Therefore, the two executions must evaluate the scrutinee to the
  same injection and follow the same path.
%
\end{proof}

Similarly, the fact that \ensuremath{\CG} requires programs to explicitly unlabel
secret data simplifies also the analysis of \emph{implicit flows} in monadic
computations.
This is because the control flow of \ensuremath{\CG} programs can only depend on
unlabeled data, whose sensitivity is \emph{coarsely} approximated by
the program counter label, as explained above.
Since the program counter label only gets tainted in response to
specific monadic actions (e.g., \ensuremath{\mathbf{unlabel}(\Varid{e}\mkern1mu)}), and not by regular
control-flow construct (e.g., \ensuremath{\mathbf{case}(\Varid{e},\Varid{x}.\Varid{e}_{1},\Varid{x}.\Varid{e}_{2}\mkern1mu)}), the evaluation of
pure expressions \emph{cannot} cause implicit information flows.
In particular, by encapsulating secret data in the \ensuremath{\mathbf{Labeled}} data
type, \ensuremath{\CG} makes all data dependencies---even those implicit in the
program control flow---\emph{explicit} through the program counter
label.
%
%
%
%
%

\begin{lemma}[\ensuremath{\Blue{\Conid{L}}}-Equivalence Preservation in Public Contexts]
\label{lemma:cg-leq-preservation-public}
For all public program counter labels \ensuremath{\Varid{pc}\;\flows\;\Blue{\Conid{L}}}, environments \ensuremath{\theta_{1}\approx_{\Blue{\Conid{L}}}\theta_{2}}, and stores \ensuremath{\Sigma_1\approx_{\Blue{\Conid{L}}}\Sigma_2}:
\begin{itemize}
\item If \ensuremath{\Varid{e}_{1}\;{\equiv }_{\alpha}\;\Varid{e}_{2}}, \ensuremath{\langle\Sigma_1,\Varid{pc},\Varid{e}_{1}\rangle\Downarrow^{\theta_{1}}\;\Varid{c}_{1}}, \ensuremath{\langle\Sigma_2,\Varid{pc},\Varid{e}_{2}\rangle\Downarrow^{\theta_{2}}\;\Varid{c}_{2}},
then \ensuremath{\Varid{c}_{1}\approx_{\Blue{\Conid{L}}}\Varid{c}_{2}};
\item If \ensuremath{\Varid{t}_{1}\;{\equiv }_{\alpha}\;\Varid{t}_{2}}, \ensuremath{\langle\Sigma_1,\Varid{pc},\Varid{t}_{1}\rangle\Downarrow^{\theta_{1}}\;\Varid{c}_{1}}, \ensuremath{\langle\Sigma_2,\Varid{pc},\Varid{t}_{2}\rangle\Downarrow^{\theta_{2}}\;\Varid{c}_{2}},
then \ensuremath{\Varid{c}_{1}\approx_{\Blue{\Conid{L}}}\Varid{c}_{2}}.
\end{itemize}
\end{lemma}
\begin{proof}
  The two lemmas are proved mutually, by simultaneous induction on the
  big-step reductions.  The \emph{forcing semantics} has only a single
  rule, i.e., \srule{Force} in Figure~\ref{fig:cg-forcing}. Therefore,
  for the first lemma, we simply apply \emph{Pure \ensuremath{\Blue{\Conid{L}}}-Equivalence
    Preservation} (Lemma~\ref{lemma:cg-pure-preservation}) to the pure
  reductions \ensuremath{\Varid{e}_{1}\;\Downarrow^{\theta_{1}}\;(\Varid{t}_{1},\theta_{1}')} and \ensuremath{\Varid{e}_{2}\;\Downarrow^{\theta_{2}}\;(\Varid{t}_{2},\theta_{2}')}, which gives \ensuremath{(\Varid{t}_{1},\theta_{1}')\approx_{\Blue{\Conid{L}}}(\Varid{t}_{2},\theta_{2}')}, i.e.,
  \ensuremath{\Varid{t}_{1}\;{\equiv }_{\alpha}\;\Varid{t}_{2}} and \ensuremath{\theta_{1}'\approx_{\Blue{\Conid{L}}}\theta_{2}'}, so we complete
  the proof by mutual induction with the lemma for thunks.
  For the second lemma, we observe that the thunks are
  \ensuremath{\alpha}-equivalent, i.e., \ensuremath{\Varid{t}_{1}\;{\equiv }_{\alpha}\;\Varid{t}_{2}}, therefore their
  reductions always step according to the same rule.
  For example, in case \srule{Unlabel}, the labeled values that get
  unlabeled are \ensuremath{\Blue{\Conid{L}}}-equivalent, i.e., \ensuremath{\mathbf{Labeled}\;{\ell}_{1}\;{v}_{1}\approx_{\Blue{\Conid{L}}}\mathbf{Labeled}\;{\ell}_{2}\;{v}_{2}}, by Lemma~\ref{lemma:cg-pure-preservation} applied to the pure
  reductions, and then we proceed by cases on the \ensuremath{\Blue{\Conid{L}}}-equivalence
  judgment.
  (Notice that the store does not change in rule \srule{Unlabel}, so
  the final stores are \ensuremath{\Blue{\Conid{L}}}-equivalent by assumption).
  In case \srule{Labeled\ensuremath{{}_{\Blue{\Conid{L}}}}}, the two values are labeled
  public, i.e., \ensuremath{{\ell}_{1}\mathrel{=}{\ell}_{2}\;\flows\;\Blue{\Conid{L}}}, and the values are related,
  i.e., \ensuremath{{v}_{1}\approx_{\Blue{\Conid{L}}}{v}_{2}}.
  Therefore, the program counter label in the final configurations
  remain observable by the attacker, i.e., \ensuremath{\Varid{pc}\;\lub\;{\ell}_{1}\;\flows\;\Blue{\Conid{L}}}
  since \ensuremath{\Varid{pc}\;\flows\;\Blue{\Conid{L}}} and \ensuremath{{\ell}_{1}\;\flows\;\Blue{\Conid{L}}}, and the final
  configurations are \ensuremath{\Blue{\Conid{L}}}-equivalent by rule \srule{Pc\ensuremath{{}_{\Blue{\Conid{L}}}}}.
  In case \srule{Labeled\ensuremath{{}_{\Red{\Conid{H}}}}}, the two values are labeled
  secret, i.e., \ensuremath{{\ell}_{1}\;\not\flows\;\Blue{\Conid{L}}} and \ensuremath{{\ell}_{2}\;\not\flows\;\Blue{\Conid{L}}}, and the
  program counter label in the final configurations are secret, i.e.,
  \ensuremath{\Varid{pc}\;\lub\;{\ell}_{1}\;\not\flows\;\Blue{\Conid{L}}} and \ensuremath{\Varid{pc}\;\lub\;{\ell}_{2}\;\not\flows\;\Blue{\Conid{L}}}, and so
  the configurations are \ensuremath{\Blue{\Conid{L}}}-equivalent by rule \srule{Pc\ensuremath{{}_{\Red{\Conid{H}}}}}.
\end{proof}

Finally, we prove termination-insensitive non-interference for \ensuremath{\CG} by
combining the \ensuremath{\Blue{\Conid{L}}}-equivalence preservation lemmas above.


\begin{thm}[\ensuremath{\CG}-TINI]
\label{thm:cg-tini}
If \ensuremath{\Varid{c}_{1}\;\Downarrow^{\theta_{1}}\;\Varid{c}_{1}'}, \ensuremath{\Varid{c}_{2}\;\Downarrow^{\theta_{2}}\;\Varid{c}_{2}'}, \ensuremath{\theta_{1}\approx_{\Blue{\Conid{L}}}\theta_{2}}
and \ensuremath{\Varid{c}_{1}\approx_{\Blue{\Conid{L}}}\Varid{c}_{2}} then \ensuremath{\Varid{c}_{1}'\approx_{\Blue{\Conid{L}}}\Varid{c}_{2}'}.
\end{thm}

\section{Flow-Sensitive References}
\label{sec:cg-fs-refs}
We now continue our exploration of coarse-grained IFC by adding
flow-sensitive references to \ensuremath{\CG} and then showing that the extended
language is secure.
This extension is in many ways analogous to the extension presented in
Section~\ref{sec:fg-fs-refs} for \ensuremath{\FG}.
Therefore, we focus mainly on the (apparently different) security
checks performed by the security monitor of \ensuremath{\CG} and how they relate
to those performed by the monitor of \ensuremath{\FG}.
In Section~\ref{subsec:cg-fs-refs-sec}, we formally establish the
security guarantees of \ensuremath{\CG} extended with flow-sensitive references.
Though the security analysis of \ensuremath{\CG} is also complicated by bijections
like \ensuremath{\FG}, we find that the separation between pure and monadic
computations in \ensuremath{\CG} limits the extra complexity of the analysis only
to the monadic fragment of the semantics.
In particular, this extension does not affect the pure fragment of the
semantics and so adapting this part of the analysis is
straightforward.
Finally, we leverage our mechanized proof scripts to compare and
analyze the security proofs of \ensuremath{\FG} and \ensuremath{\CG} and discuss our findings
in Section~\ref{sec:cg-discussion}.
%


\subsection{Syntax}

\begin{figure}[t!]
\centering
\begin{subfigure}{.9\textwidth}
\begin{align*}
\text{Types: }                \ensuremath{\tau} & \ensuremath{\Coloneqq\cdots\;{\mid}\;\mathbf{Ref}\; s\;\tau} \\
\text{Values: }               \ensuremath{{v}} & \ensuremath{\Coloneqq\cdots\;{\mid}\;\Varid{n}} \\
\text{Heap: } \ensuremath{\mu} & \ensuremath{\Coloneqq[\mskip1.5mu \mskip1.5mu]\;{\mid}\;\mathbf{Labeled}\;{\ell}\;{v}\mathbin{:}\mu} \\
\text{Configuration: }        \ensuremath{\Varid{c}} & \ensuremath{\Coloneqq\langle\Sigma,\mu,\Varid{pc},\Varid{e}\rangle}
\end{align*}
\captionsetup{format=caption-with-line}
\caption{Syntax.}
\label{fig:fs-cg-refs-syntax}
\end{subfigure}%
\\
\begin{subfigure}{.9\textwidth}
\begin{mathpar}
\inferrule[(New-FS)]
{\ensuremath{\Varid{e}\;\Downarrow^{\theta}\;\mathbf{Labeled}\;{\ell}\;{v}} \\ \ensuremath{\Varid{pc}\;\flows\;{\ell}} \\ \ensuremath{\Varid{n}\mathrel{=}\abs{\mu}} \\
 \ensuremath{\mu'\mathrel{=}\update{\mu}{\Varid{n}}{\mathbf{Labeled}\;{\ell}\;{v}}} }
{\ensuremath{\langle\Sigma,\mu,\Varid{pc},\mathbf{new}(\Varid{e}\mkern1mu)\rangle\Downarrow^{\theta}\langle\Sigma,\mu',\Varid{pc},\Varid{n}\rangle}}
\and
\inferrule[(Read-FS)]
{\ensuremath{\Varid{e}\;\Downarrow^{\theta}\;\Varid{n}} \\ \ensuremath{\mu{[}\Varid{n}{]}\mathrel{=}\mathbf{Labeled}\;{\ell}\;{v}}}
{\ensuremath{\langle\Sigma,\mu,\Varid{pc},\mathbin{!}\Varid{e}\rangle\Downarrow^{\theta}\langle\Sigma,\mu,\Varid{pc}\;\lub\;{\ell},{v}\rangle}}
\and
\inferrule[(LabelOfRef-FS)]
{\ensuremath{\Varid{e}\;\Downarrow^{\theta}\;\Varid{n}} \\
 \ensuremath{\mu{[}\Varid{n}{]}\mathrel{=}\mathbf{Labeled}\;{\ell}\;\anonymous }}
{\ensuremath{\langle\Sigma,\mu,\Varid{pc},\mathbf{labelOfRef}(\Varid{e}\mkern1mu)\rangle\Downarrow^{\theta}\langle\Sigma,\mu,\Varid{pc}\;\lub\;{\ell},{\ell}\rangle}}
\and
\inferrule[(Write-FS)]
{\ensuremath{\Varid{e}_{1}\;\Downarrow^{\theta}\;\Varid{n}} \\
 \ensuremath{\Varid{e}_{2}\;\Downarrow^{\theta}\;\mathbf{Labeled}\;{\ell}'\;{v}} \\
 \ensuremath{\mu{[}\Varid{n}{]}\mathrel{=}\mathbf{Labeled}\;{\ell}\;\anonymous } \\
 \ensuremath{\greenBox{\ensuremath{\Varid{pc}\;\flows\;{\ell}}}} \\
 \ensuremath{\mu'\mathrel{=}\update{\mu}{\Varid{n}}{\mathbf{Labeled}\;(\Varid{pc}\;\lub\;{\ell}')\;{v}}} }
  {\ensuremath{\langle\Sigma,\mu,\Varid{pc},\Varid{e}_{1}\mathbin{:=}\Varid{e}_{2}\rangle\Downarrow^{\theta}\langle\Sigma,\mu',\Varid{pc},()\rangle}}
\end{mathpar}
\captionsetup{format=caption-with-line}
\caption{Dynamics. The \greenBox{shaded constraint} corresponds to the
  \emph{no-sensitive upgrade} security check in \ensuremath{\FG}.}
\label{fig:fs-cg-refs-dynamics}
\end{subfigure}%

\caption{\ensuremath{\CG} extended with flow-sensitive references.}
\label{fig:fs-cg-refs}

\end{figure}
Figure~\ref{fig:fs-cg-refs} introduces the new syntactic constructs
and semantics rules for flow-sensitive references.
First, we annotate the reference type constructor with a tag \ensuremath{ s\;\in\;\{\mskip1.5mu \Conid{I},\Conid{S}\mskip1.5mu\}}, which indicates the sensitivity of a reference term.
A plain address \ensuremath{\Varid{n}\mathbin{:}\mathbf{Ref}\;\Conid{S}\;\tau} represents a flow-sensitive reference
pointing to a value of type \ensuremath{\tau}, stored in the \ensuremath{\Varid{n}}-th cell of the
heap.
Notice that these references are \emph{not} annotated with a label
representing the sensitivity of their content, like flow-insensitive
references.
This is because flow-sensitive references are, by design, allowed to
store data at different security levels, i.e., the label of these
references can \emph{change} throughout program execution.
Thus, the references themselves are \emph{unlabeled} and, instead, the
security monitor \emph{explicitly} labels their content in the heap
when they are created and updated (see below).\footnote{In contrast,
  flow-insensitive references are annotated with a \emph{fixed} label,
  e.g., \ensuremath{{\ell}} for \ensuremath{{\Varid{n}}_{{\ell}}\mathbin{:}\mathbf{Ref}\;\Conid{I}\;\tau}, which represents (an upper
  bound over) the sensitivity of its content.
  Since this label does not change, the content of the reference can
  be stored directly in the memory labeled \ensuremath{{\ell}}, without being
  explicitly labeled.}
To ensure that values in the heap are always labeled, heaps \ensuremath{\mu} are
syntactically defined as lists of explicitly labeled values \ensuremath{\mathbf{Labeled}\;{\ell}\;{v}}, whose label \ensuremath{{\ell}} represents the sensitivity of value \ensuremath{{v}} and, at
the same time, the label of any reference pointing to it.
The heap is stored in program configurations \ensuremath{\langle\Sigma,\mu,\Varid{pc},\Varid{e}\rangle},
where programs can access it through the same thunk constructs used
for flow-insensitive references, i.e., \ensuremath{\mathbf{new}(\Varid{e}\mkern1mu)}, \ensuremath{\Varid{e}_{1}\mathbin{:=}\Varid{e}_{2}}, \ensuremath{\mathbin{!}\Varid{e}}, and
\ensuremath{\mathbf{labelOfRef}(\Varid{e}\mkern1mu)}.
As we explain next, these operations must be regulated by the security
monitor of \ensuremath{\CG} to enforce security.

%

\subsection{Dynamics}
Figure~\ref{fig:fs-cg-refs} extends the thunk semantics with new rules
that allow programs to access the heap through flow-sensitive
references without leaking data.
Rule \srule{New-FS} creates a new reference by allocating a new cell
in the heap at fresh address \ensuremath{\Varid{n}\mathrel{=}\abs{\mu}}, which is initialized with
the given labeled value argument, i.e., \ensuremath{\update{\mu}{\Varid{n}}{\mathbf{Labeled}\;{\ell}\;{v}}}.
Importantly, the rule requires the label of the value to be above the
program counter label, i.e., \ensuremath{\Varid{pc}\;\flows\;{\ell}}, to avoid
leaks.\footnote{This restriction, known as \emph{no
    write-down}~\parencite{Bell1976}, is a core design principles of
  several static IFC libraries~\parencite{Russo:2015,VASSENA2017},
  which was not identified as such in dynamic IFC
  libraries~\parencite{stefan:lio,stefan:2017:flexible}.  }
Without this constraint, a program could create a public reference in
a secret context, which constitutes a leak.
Compare this rule with the corresponding rule for \ensuremath{\FG}, i.e., rule
\srule{New-FS} in Figure~\ref{fig:fs-fg-refs-dynamics}.
The security monitor of \ensuremath{\FG} does not check that the label of the
value is above the program counter label---in fact the rule does not
contain any security check at all!
Why must \ensuremath{\CG} include that check and instead \ensuremath{\FG} can skip it?
This is because that check is \emph{redundant} in \ensuremath{\FG}: the constraint
\ensuremath{\Varid{pc}\;\flows\;{\ell}} performed by the security monitor of \ensuremath{\CG} is an
invariant (Property~\ref{prop:fg-pc-raise}) of the semantics of \ensuremath{\FG}.
Intuitively, \ensuremath{\FG} programs cannot create public references in secret
contexts because the result of a computation is always labeled above
the current program counter label.
Since \ensuremath{\CG} does not enjoy this invariant, i.e., computations can
return values labeled public even in secret contexts, that check
\emph{must} be included in all the rules that perform write
side-effects (i.e., also in rules \srule{New}, \srule{Write} for
flow-insensitive references in Figure~\ref{fig:cg-memory}) to avoid
leaks.
Rule \srule{Read-FS} reads a flow-sensitive reference by retrieving
its content from the heap, i.e., \ensuremath{\mu{[}\Varid{n}{]}\mathrel{=}\mathbf{Labeled}\;{\ell}\;{v}}, and
returning its value \ensuremath{{v}}.
Importantly, the rule taints the program counter label, i.e., \ensuremath{\Varid{pc}\;\lub\;{\ell}}, to indicate that data at security level \ensuremath{{\ell}} is now in scope.
Rule \srule{LabelOfRef-FS} is similar, but returns the label of the
reference, i.e., the same label \ensuremath{{\ell}} that also annotates its content,
and so it taints the program counter label with the label itself,
i.e., \ensuremath{\Varid{pc}\;\lub\;{\ell}}.
%

%
%
Lastly, rule \srule{Write-FS} updates the content of a flow-sensitive
reference \ensuremath{\Varid{n}} with a new value \ensuremath{\mathbf{Labeled}\;{\ell}'\;{v}}, which replaces the
current value stored in the heap, i.e., \ensuremath{\mu{[}\Varid{n}{]}\mathrel{=}\mathbf{Labeled}\;{\ell}\;\anonymous }.
Notice that the new value can be less or \emph{even more} sensitive
than the old value, i.e., labels \ensuremath{{\ell}} and \ensuremath{{\ell}'} can be in any relation.
This is exactly why flow-sensitive references are, in general, more
permissive than flow-insensitive references also in \ensuremath{\CG}.\footnote{
  In particular, Example~\ref{example:fg-fs-refs} can be adapted to
  \ensuremath{\CG} as well, i.e., program \ensuremath{{r}\leftarrow \mathbf{new}(\Varid{p}\mkern1mu);{r}\mathbin{:=} s;\mathbin{!}{r}} is aborted
  with flow-insensitive references, but accepted with flow-sensitive
  references.}\looseness=-1
In the rule, the constraint \ensuremath{\Varid{pc}\;\flows\;{\ell}} is analogous to the NSU
check for \ensuremath{\FG}.
(For convenience, these equivalent checks are \greenBox{shaded} in
Figure \ref{fig:fs-fg-refs-dynamics} and
\ref{fig:fs-cg-refs-dynamics}).
Intuitively, this check allows a reference update only if the decision
to update \emph{that} reference depends on data less sensitive than
the reference itself.\footnote{\ensuremath{\CG} would be insecure without this
  constraint. In particular, programs could leak data through
  secret-dependent reference updates, similarly to
  Example~\ref{example:fg-fs-refs-leak}.  }
Though the NSU checks in \ensuremath{\FG} and \ensuremath{\CG} protect against the same type
of implicit data leaks, they involve different labels, i.e., the
intrinsic label of the reference in \ensuremath{\FG} and the program counter label
in \ensuremath{\CG}.
In \ensuremath{\FG}, the sensitivity of the reference is represented by its
intrinsic label, which can then be used directly in the NSU check.
%
In contrast, \ensuremath{\CG} does not feature intrinsic labels, therefore, we use
the program counter label as a coarse, but \emph{sound}, approximation
of the sensitivity of the reference, hence the NSU check \ensuremath{\Varid{pc}\;\flows\;{\ell}}.
%
%
Notice that when the new value is written in the heap, the rule taints
its label with the program counter label, i.e., \ensuremath{\update{\mu}{\Varid{n}}{\mathbf{Labeled}\;(\Varid{pc}\;\lub\;{\ell}')\;{v}}}, for the same reason.
Intuitively, the sensitivity of the content is determined by the new
value, which is explicitly labeled \ensuremath{{\ell}'}, and by the sensitivity of the
(unlabeled) reference, approximated by the program counter label \ensuremath{\Varid{pc}},
i.e., the new content has sensitivity at most \ensuremath{\Varid{pc}\;\lub\;{\ell}'}.\footnote{In addition, by tainting the label \ensuremath{{\ell}'} with the program
  counter label \ensuremath{\Varid{pc}}, the rule automatically respects the \emph{no
    write-down} restriction, i.e., \ensuremath{\Varid{pc}\;\flows\;\Varid{pc}\;\lub\;{\ell}'}, hence no
  explicit security check is needed like in rules \srule{New},
  \srule{Write}, and \srule{New-FS}.}

\subsection{Security}
\begin{figure}[t]
\begin{mathpar}
\inferrule[(Valid-Inputs)]
{\ensuremath{\Varid{c}\mathrel{=}\langle\Sigma,\mu,\Varid{pc}\;\Varid{e}\rangle} \\ \ensuremath{\Varid{n}\mathrel{=}\abs{\mu}} \\
 \ensuremath{\Varid{n}\vdash \mathbf{Valid}(\Sigma)} \\
 \ensuremath{\Varid{n}\vdash \mathbf{Valid}(\mu)} \\
 \ensuremath{\Varid{n}\vdash \mathbf{Valid}(\theta)} }
{\ensuremath{\vdash \mathbf{Valid}(\Varid{c},\theta)}}
\and
\inferrule[(Valid-Outputs)]
{\ensuremath{\Varid{c}\mathrel{=}\langle\Sigma,\mu,\Varid{pc},{v}\rangle} \\ \ensuremath{\Varid{n}\mathrel{=}\abs{\mu}} \\
 \ensuremath{\Varid{n}\vdash \mathbf{Valid}(\Sigma)} \\
 \ensuremath{\Varid{n}\vdash \mathbf{Valid}(\mu)} \\
 \ensuremath{\Varid{n}\vdash \mathbf{Valid}({v})}}
{\ensuremath{\vdash \mathbf{Valid}(\Varid{c})}}
\end{mathpar}
\caption{Judgments for valid program inputs and outputs in \ensuremath{\CG}.}
\label{fig:cg-valid}
\end{figure}

\label{subsec:cg-fs-refs-sec}
The security analysis of \ensuremath{\CG} extended with flow-sensitive references
is very much similar to the analysis presented in
Section~\ref{subsec:fg-fs-refs-security} for \ensuremath{\FG}.
In particular, the security monitor of \ensuremath{\CG} allocates both public and
secret data in the same linear heap and so secret-dependent
allocations can influence the addresses of subsequent public
references, just like in \ensuremath{\FG}.
Although this can create a dependency between concrete heap addresses
and secret data, security is not at stake because references are
\emph{opaque} in \ensuremath{\CG} as well.
To formally show that, we need to adapt the \ensuremath{\Blue{\Conid{L}}}-equivalence relation,
and consequently the security analysis, to use a \emph{bijection} to
reason about corresponding references with secret-dependent, yet
indistinguishable, heap addresses.

\paragraph{\ensuremath{\Blue{\Conid{L}}}-Equivalence up to Bijection.}
%



%
Formally, we redefine \ensuremath{\Blue{\Conid{L}}}-equivalence as a relation \ensuremath{\approx_{\Blue{\Conid{L}}}^{\beta}}\ \ensuremath{\subseteq\;\Conid{Value}\;\times\;\Conid{Value}} and add the bijection \ensuremath{\beta} to \ensuremath{\approx_{\Blue{\Conid{L}}}} in all the rules
previously defined for \ensuremath{\CG}.
These rules do not use the bijection, which is needed instead only in
the rules for flow-sensitive references and heaps.
In particular, two flow-sensitive references are indistinguishable
only if their heap addresses are matched by the bijection, i.e., we
add the following new rule:
\begin{mathpar}
\inferrule[(Ref-FS)]
{\ensuremath{(\Varid{n}_{1},\Varid{n}_{2})\;\in\;\beta}}
{\ensuremath{\Varid{n}_{1}\approx_{\Blue{\Conid{L}}}^{\beta}\Varid{n}_{2}}}
\end{mathpar}
Similarly, in the definition of \ensuremath{\Blue{\Conid{L}}}-equivalence for heaps, the
addresses related by the bijection identify corresponding heap cells,
which must be recursively related, as they can be read through related
references.\footnote{Notice that these addresses are valid in the heap
  thanks to the first two side-conditions, which ensure that the
  domain and range of the bijection are compatible with the address
  space of the heaps. See also the explanation of
  Definition~\ref{def:fg-heap-equivalence}.}

\begin{defn}[Heap \ensuremath{\Blue{\Conid{L}}}-equivalence]
\label{def:cg-heap-equivalence}
Two heaps \ensuremath{\mu_{1}} and
\ensuremath{\mu_{2}} are \ensuremath{\Blue{\Conid{L}}}-equivalent up to bijection \ensuremath{\beta}, written \ensuremath{\mu_{1}\approx_{\Blue{\Conid{L}}}^{\beta}\mu_{2}}, if and only if:
\begin{enumerate}
\item \ensuremath{dom(\beta)\;\subseteq\;\{\mskip1.5mu \mathrm{0},\ldots,\abs{\mu_{1}}\mathbin{-}\mathrm{1}\mskip1.5mu\}},
\item \ensuremath{rng(\beta)\;\subseteq\;\{\mskip1.5mu \mathrm{0},\ldots,\abs{\mu_{2}}\mathbin{-}\mathrm{1}\mskip1.5mu\}}, and
\item For all addresses \ensuremath{\Varid{n}_{1}} and \ensuremath{\Varid{n}_{2}}, if \ensuremath{(\Varid{n}_{1},\Varid{n}_{2})\;\in\;\beta}, then \ensuremath{\mu_{1}{[}\Varid{n}_{1}{]}\approx_{\Blue{\Conid{L}}}^{\beta}\mu_{2}{[}\Varid{n}_{2}{]}}.
\end{enumerate}
\end{defn}

In the definition above, only the rules for labeled values can relate heap
cells \ensuremath{\mu_{1}{[}\Varid{n}_{1}{]}\approx_{\Blue{\Conid{L}}}^{\beta}\mu_{2}{[}\Varid{n}_{2}{]}}.
This is because heaps are syntactically defined as a list of
explicitly labeled values in Figure~\ref{fig:fs-cg-refs-syntax}, i.e.,
\ensuremath{\mu_{1}{[}\Varid{n}_{1}{]}\mathrel{=}\mathbf{Labeled}\;{\ell}_{1}\;{v}_{1}} for some label \ensuremath{{\ell}_{1}} and value \ensuremath{{v}_{1}} and
similarly \ensuremath{\mu_{2}{[}\Varid{n}_{2}{]}\mathrel{=}\mathbf{Labeled}\;{\ell}_{2}\;{v}_{2}}, and the only rules that give
\ensuremath{\mathbf{Labeled}\;{\ell}_{1}\;{v}_{1}\approx_{\Blue{\Conid{L}}}^{\beta}\mathbf{Labeled}\;{\ell}_{2}\;{v}_{2}} are \srule{Labeled\ensuremath{{}_{\Blue{\Conid{L}}}}} and
\srule{Labeled\ensuremath{{}_{\Red{\Conid{H}}}}} in Figure~\ref{fig:cg-low-equiv}.

%
%

Similarly to \ensuremath{\FG}, the \ensuremath{\Blue{\Conid{L}}}-equivalence relation up to bijection for
\ensuremath{\CG} satisfies \emph{restricted reflexivity}, \emph{symmetricity},
\emph{transitivity}, and \emph{weakening} (i.e.,
Property~\ref{prop:fg-low-equiv-props-with-bij}), and so we can
construct square commutative diagrams for stores and heaps (i.e.,
Lemma~\ref{lemma:fg-square-heaps}) in \ensuremath{\CG} as well.
Notice that reflexivity is restricted only to \emph{valid} program
constructs, i.e., values, environments, memories, stores, heaps, and
configurations that are \emph{free} of dangling flow-sensitive
references, just like it is in \ensuremath{\FG} and for the same technical
reasons.
These side conditions are formalized by straightforward judgments,
e.g., \ensuremath{\Varid{n}\vdash \mathbf{Valid}(\theta)} and \ensuremath{\Varid{n}\vdash \mathbf{Valid}({v})} indicating that the references
contained in value \ensuremath{{v}} and environment \ensuremath{\theta} are valid in a heap of
size \ensuremath{\Varid{n}}, which are mutually and recursively defined like in
Figure~\ref{fig:fg-valid}.
In these judgments, the size parameter \ensuremath{\Varid{n}} is instantiated by the
top-level judgment for program inputs and outputs with the size of the
current heap \ensuremath{\mu}, i.e., \ensuremath{\Varid{n}\mathrel{=}\abs{\mu}} as shown in
Figure~\ref{fig:cg-valid}.
The fact that the security lemmas of \ensuremath{\CG} assume valid configurations
does not actually weaken the security guarantees of the language:
these judgments are naturally preserved by the pure and monadic
semantics of \ensuremath{\CG}.

\begin{property}[Valid Invariant]
\leavevmode
\label{prop:cg-valid-invariant}
\begin{enumerate}
\item If \ensuremath{\Varid{e}\;\Downarrow^{\theta}\;{v}} and  \ensuremath{\Varid{n}\vdash \mathbf{Valid}(\theta)}, then \ensuremath{\Varid{n}\vdash \mathbf{Valid}({v})}.
\item If \ensuremath{\Varid{c}\;\Downarrow^{\theta}\;\Varid{c'}} and \ensuremath{\vdash \mathbf{Valid}(\Varid{c},\theta)}, then \ensuremath{\vdash \mathbf{Valid}(\Varid{c'})}.
\end{enumerate}
\end{property}

\paragraph{Termination-Insensitive Non-Interference.}
We now prove that \ensuremath{\CG} extended with flow-sensitive references
enforces termination-insensitive non-interference.
This result is also based on two lemmas, i.e., \emph{store and heap
  confinement} and \emph{\ensuremath{\Blue{\Conid{L}}}-equivalence preservation}, which are
complicated by the fact that \ensuremath{\Blue{\Conid{L}}}-equivalence is defined up to a
bijection, similarly to \ensuremath{\FG}.
These lemmas rely on bijections to relate flow-sensitive references
allocated in public contexts: these references are observable by the
attacker, but their heap addresses may differ due to previous,
secret-dependent allocations.
In contrast, references allocated in secret contexts cannot be observed
by the attacker and must be ignored by the bijection.
For example, in the store and heap confinement lemma, we show that
initial and final stores and heaps are \ensuremath{\Blue{\Conid{L}}}-equivalent, for programs
executed in secret contexts.
Which bijection should we use to relate the addresses of
flow-sensitive references in this lemma?
Intuitively, the addresses of references already allocated in the
initial configuration remain unchanged in the final configuration
(because references cannot be tempered in \ensuremath{\CG}), so they can be
related by the \emph{identity bijection}.
%
%
Furthermore, since references allocated in a secret context are not
observable by the attacker, the lemma shows \ensuremath{\Blue{\Conid{L}}}-equivalence up to a
identity bijection that ignores any new heap allocation performed by
the program, i.e., up to the identity bijection \emph{restricted to
  the domain of the initial heap}.

\begin{lemma}[Store and Heap Confinement]
\label{lemma:cg-confinement-bij}
\leavevmode
For all program counter labels \ensuremath{\Varid{pc}\;\not\flows\;\Blue{\Conid{L}}}, valid initial
configurations and environments \ensuremath{\vdash \mathbf{Valid}(\Varid{c},\theta)}, and final
configurations \ensuremath{\Varid{c'}\mathrel{=}\langle\Sigma',\mu',\Varid{pc}',{v}\rangle}:
\begin{itemize}
\item If \ensuremath{\Varid{c}\mathrel{=}\langle\Sigma,\mu,\Varid{pc},\Varid{e}\rangle} and  \ensuremath{\Varid{c}\;\Downarrow^{\theta}\;\Varid{c'}}, then \ensuremath{\Sigma\;\approx_{\Blue{\Conid{L}}}^{{\iota}_{\abs{\mu}}}\;\Sigma'} and \ensuremath{\mu\;\approx_{\Blue{\Conid{L}}}^{{\iota}_{\abs{\mu}}}\;\mu'}.
\item If \ensuremath{\Varid{c}\mathrel{=}\langle\Sigma,\mu,\Varid{pc},\Varid{t}\rangle} and  \ensuremath{\Varid{c}\;\Downarrow^{\theta}\;\Varid{c'}}, then \ensuremath{\Sigma\;\approx_{\Blue{\Conid{L}}}^{{\iota}_{\abs{\mu}}}\;\Sigma'} and \ensuremath{\mu\;\approx_{\Blue{\Conid{L}}}^{{\iota}_{\abs{\mu}}}\;\mu'}.
\end{itemize}
\label{fig:cg-store-heap-confinement}
\end{lemma}
\begin{proof} Analogous to Lemma~\ref{lemma:fg-confinement-bij}.
%
  %
  Since \ensuremath{\CG} encapsulates side-effects in a monad, we need to
  explicitly propagate the valid judgment only in very few cases
  compared to \ensuremath{\FG}. In particular, we only apply
  Property~\ref{prop:cg-valid-invariant}.1 in rule \srule{Force} and
  Property~\ref{prop:cg-valid-invariant}.2 in rule \srule{Bind}.
\end{proof}

Next, we prove \ensuremath{\Blue{\Conid{L}}}-equivalence preservation in \emph{secret contexts}.
Since the programs considered in this lemma run in a secret context,
the addresses of the flow-sensitive references allocated during these
executions must be ignored by the bijection, i.e., the final
configurations must be \ensuremath{\Blue{\Conid{L}}}-equivalent up to the \emph{same bijection}
used for the initial configurations.
We prove that by constructing a square commutative diagram for heaps
and stores, similar to Figure~\ref{fig:fg-square-heaps}.
(We explain the construction for heaps, stores are treated
analogously).\footnote{Heaps and stores are very often treated
  homogeneously in proofs, therefore our mechanized proof scripts
  pairs them up in a \emph{program state} helper data structure to
  shorten some proofs.}
In the figure, heaps \ensuremath{\mu_{1}} and \ensuremath{\mu_{2}} are from the initial
configurations of the lemma, while \ensuremath{\mu_{1}'} and \ensuremath{\mu_{2}'} are the heaps
from the final configurations.
Since the initial configurations are \ensuremath{\Blue{\Conid{L}}}-equivalent, we have the
vertical edge on the left, i.e., \ensuremath{\mu_{1}\approx_{\Blue{\Conid{L}}}^{\beta}\mu_{2}}, by assumption.
Instead, the horizontal edges of the square are obtained by applying
store and heap confinement to each individual program execution.
In particular, the lemma shows that the initial and final heaps in
these executions are \ensuremath{\Blue{\Conid{L}}}-equivalent up to appropriate identity
bijections, i.e., \ensuremath{\mu_{1}\;\approx_{\Blue{\Conid{L}}}^{{\iota}_{\abs{\mu_{1}}}}\;\mu_{1}'} and \ensuremath{\mu_{2}\;\approx_{\Blue{\Conid{L}}}^{{\iota}_{\abs{\mu_{2}}}}\;\mu_{2}'}, which ignore new heap allocations,
as explained above.
Therefore, the bijection that relates the final heaps from the vertical
edge on the right of the square can be simplified.
Specifically, the identity bijections \emph{cancel out} in the
composition with the bijection from the initial \ensuremath{\Blue{\Conid{L}}}-equivalence
relation, i.e., \ensuremath{{\iota}_{\abs{\mu_{2}}}\;\circ\;\beta\;\circ\;{\iota}_{\abs{\mu_{1}}}^{-1}\mathrel{=}\beta}, and so the final heaps remain related up to the
same bijection \ensuremath{\beta} used for the initial heaps, i.e., \ensuremath{\mu_{1}'\approx_{\Blue{\Conid{L}}}^{\beta}\mu_{2}'}.
In the following, we use dot-notation to access individual elements of
a configuration, e.g., we write \ensuremath{\Varid{c}.\Varid{pc}} to extract the program counter \ensuremath{\Varid{pc}}
from the configuration \ensuremath{\Varid{c}\mathrel{=}\langle\anonymous ,\anonymous ,\Varid{pc},\anonymous \rangle}.

\begin{lemma}[\ensuremath{\Blue{\Conid{L}}}-Equivalence Preservation in Secret Contexts]
\leavevmode
For all valid inputs \ensuremath{\vdash \mathbf{Valid}(\Varid{c}_{1},\theta_{1})} and \ensuremath{\vdash \mathbf{Valid}(\Varid{c}_{2},\theta_{2})} and
bijections \ensuremath{\beta}, such that \ensuremath{\Varid{c}_{1}.\Varid{pc}\;\not\flows\;\Blue{\Conid{L}}}, \ensuremath{\Varid{c}_{2}.\Varid{pc}\;\not\flows\;\Blue{\Conid{L}}}, and \ensuremath{\Varid{c}_{1}\approx_{\Blue{\Conid{L}}}^{\beta}\Varid{c}_{2}}, if \ensuremath{\Varid{c}_{1}\;\Downarrow^{\theta_{1}}\;\Varid{c}_{1}'} and \ensuremath{\Varid{c}_{2}\;\Downarrow^{\theta_{2}}\;\Varid{c}_{2}'}, then \ensuremath{\Varid{c}_{1}'\approx_{\Blue{\Conid{L}}}^{\beta}\Varid{c}_{2}'}.
\label{lemma:cg-leq-preservation-secret-bij}
\end{lemma}

We now adapt the proof of \ensuremath{\Blue{\Conid{L}}}-equivalence in \emph{public contexts}.
The references allocated by the program considered in this lemma are
observable by the attacker and so we have to construct an appropriate
bijection to relate their addresses in the final configurations.
However, since these allocations occur at run time and may depend on
the program inputs, we cannot, in general, predict the exact bijection
needed to relate them.
Therefore, the final bijection is \emph{existentially quantified} in
the lemma, so that the right bijection can be precisely constructed,
depending on the program execution, in each case of the proof itself.
Unfortunately, existential quantification also complicates the proof,
which now involves reasoning about terms that are \ensuremath{\Blue{\Conid{L}}}-equivalent, but
up to arbitrary bijections, and so may not be combined together.
To solve this issue, the lemma additionally requires to prove an
ordering invariant about these bijections, i.e., that the final
bijection \emph{extends} the initial bijection.
Intuitively, this extra property solves the issue described above
because \ensuremath{\Blue{\Conid{L}}}-equivalent relations up to ``smaller'' bijections can be
lifted to ``larger'' bijections via \emph{weakening}
(Property~\ref{prop:fg-low-equiv-props-with-bij}.4), and so combined
together.
It is worth pointing out that this issue is not specific to \ensuremath{\CG}: the
security analysis of \ensuremath{\FG} presents the same issue, which we solve in
the same way in Section~\ref{lemma:fg-leq-preservation-public}.
However, only the analysis of the monadic fragment of the semantics of
\ensuremath{\CG} is affected by the issue described above: the final bijection does not
need to be quantified when reasoning about the pure fragment of the
semantics.
%
This is because pure reductions cannot allocate new references (they
do not even have access to a heap) and so the same bijection that
relates their input values can also relate their output values.
This simplifies our formal analysis, as previous lemmas and their
proofs are largely unaffected by the addition of bijections.
For example, to adapt Lemma~\ref{lemma:cg-pure-preservation}, i.e.,
\emph{Pure \ensuremath{\Blue{\Conid{L}}}-Equivalence Preservation}, for flow-sensitive
references, we \emph{only} need to add the bijection \ensuremath{\beta} to the
\ensuremath{\Blue{\Conid{L}}}-equivalent relations in the statement of lemma: the proof
requires no changes.

\begin{lemma}[Pure \ensuremath{\Blue{\Conid{L}}}-Equivalence Preservation]
  For all expressions \ensuremath{\Varid{e}_{1}\;{\equiv }_{\alpha}\;\Varid{e}_{2}}, bijections \ensuremath{\beta}, and
  environments \ensuremath{\theta_{1}\approx_{\Blue{\Conid{L}}}^{\beta}\theta_{2}}, if \ensuremath{\Varid{e}_{1}\;\Downarrow^{\theta_{1}}\;{v}_{1}} and \ensuremath{\Varid{e}_{2}\;\Downarrow^{\theta_{2}}\;{v}_{2}}, then \ensuremath{{v}_{1}\approx_{\Blue{\Conid{L}}}^{\beta}{v}_{2}}.
\label{lemma:cg-pure-preservation-bij}
\end{lemma}

The separation between pure and monadic semantics fragments of \ensuremath{\CG}
simplifies the \ensuremath{\Blue{\Conid{L}}}-equivalence preservation of monadic computations
as well.
This is because most of the rules of this fragment of the semantics
only include pure reductions and so intermediate \ensuremath{\Blue{\Conid{L}}}-equivalent
results can be combined directly, i.e., no weakening is required in
most cases.

\begin{lemma}[\ensuremath{\Blue{\Conid{L}}}-Equivalence Preservation in Public Contexts]
\leavevmode
For all valid initial inputs \ensuremath{\vdash \mathbf{Valid}(\Varid{c}_{1},\theta_{1})} and \ensuremath{\vdash \mathbf{Valid}(\Varid{c}_{2},\theta_{2})}
and bijections \ensuremath{\beta}, such that \ensuremath{\Varid{c}_{1}.\Varid{pc}\mathrel{=}\Varid{c}_{2}.\Varid{pc}\;\flows\;\Blue{\Conid{L}}}, \ensuremath{\Varid{c}_{1}\approx_{\Blue{\Conid{L}}}^{\beta}\Varid{c}_{2}}, and \ensuremath{\theta_{1}\approx_{\Blue{\Conid{L}}}^{\beta}\theta_{2}}, if \ensuremath{\Varid{c}_{1}\;\Downarrow^{\theta_{1}}\;\Varid{c}_{1}'} and \ensuremath{\Varid{c}_{2}\;\Downarrow^{\theta_{2}}\;\Varid{c}_{2}'}, then there exists an extended bijection \ensuremath{\beta'\;\supseteq\;\beta} such that \ensuremath{\Varid{c}_{1}'\;\approx_{\Blue{\Conid{L}}}^{\beta'}\;\Varid{c}_{2}'}.
\label{lemma:cg-leq-preservation-public-bij}
\end{lemma}

Finally, we combine \ensuremath{\Blue{\Conid{L}}}-equivalence preservation in public and secret
contexts and prove \emph{termination-insensitive non-interference} for
\ensuremath{\CG} extended with flow-sensitive references.

\begin{thm}[\ensuremath{\CG}-TINI with Bijections]
\label{thm:cg-tini-bij}
For all valid inputs \ensuremath{\vdash \mathbf{Valid}(\Varid{c}_{1},\theta_{1})} and \ensuremath{\vdash \mathbf{Valid}(\Varid{c}_{2},\theta_{2})} and
bijections \ensuremath{\beta}, such that \ensuremath{\Varid{c}_{1}\;\approx_{\Blue{\Conid{L}}}^{\beta}\;\Varid{c}_{2}}, \ensuremath{\theta_{1}\;\approx_{\Blue{\Conid{L}}}^{\beta}\;\theta_{2}}, if \ensuremath{\Varid{c}_{1}\;\Downarrow^{\theta_{1}}\;\Varid{c}_{1}'}, and \ensuremath{\Varid{c}_{2}\;\Downarrow^{\theta_{2}}\;\Varid{c}_{2}'}, then
there exists an extended bijection \ensuremath{\beta'\;\supseteq\;\beta}, such that
\ensuremath{\Varid{c}_{1}'\;\approx_{\Blue{\Conid{L}}}^{\beta'}\;\Varid{c}_{2}'}.
\end{thm}

\paragraph{Conclusion.}
At this point, we have formalized two calculi---\ensuremath{\FG} and \ensuremath{\CG}---that perform dynamic IFC
at \emph{fine} and \emph{coarse} granularity, respectively.
While they have some similarities, i.e., they are both functional
languages that feature labeled annotated data, references and label
introspection primitives, and ensure a termination-insensitive
security condition, they also have striking differences.
First and foremost, they differ in the number of label
annotations---pervasive in \ensuremath{\FG} and optional in \ensuremath{\CG}---with
significant implications for the programming model and usability.
Second, they differ in the nature of the program counter,
\emph{flow-insensitive} in \ensuremath{\FG} and \emph{flow-sensitive} in \ensuremath{\CG}.
Third, they differ in the way they deal with side-effects---\ensuremath{\CG}
allows side-effectful computations exclusively inside the monad, while
\ensuremath{\FG} is \emph{impure}, i.e., any \ensuremath{\FG} expression can modify the state.
This difference affects the effort required to implement a system that
performs language-based fine- and coarse-grained dynamic IFC.
In fact, several coarse-grained IFC languages
\parencite{Schmitz:2018,buiras2015hlio,Jaskelioff:SME,Tsai:Russo:Hughes:CSF07,Russo+:Haskell08,Russo:2015}
have been implemented as an embedded domain specific language (EDSL)
in a Haskell library with little effort, exploiting the strict control
that the host language provides on side-effects.
Adapting an existing language to perform fine-grained IFC requires
major engineering effort, because several components (all the way from
the parser to the runtime system) must be adapted to be label-aware.
%

%
In the next section we discuss our verified artifacts of \ensuremath{\FG} and \ensuremath{\CG}
and compare their mechanized proofs of non-interference.
Then, in Section~\ref{sec:fg2cg} and \ref{sec:cg2fg} we show
that---despite their differences---these two calculi are, in fact,
equally expressive.


\chapter{Verified Artifacts}
\label{sec:cg-discussion}

\begin{table}[t]
\centering
\caption{The table reports the size (Agda LOC excluding blank lines
  and comments) of the proof scripts that formalize \ensuremath{\FG} and \ensuremath{\CG} and
  their security proofs with flow-insensitive references (\textsc{FI})
  and with also flow-sensitive references (\textsc{FI + FS}).}
\begin{tabular}{r | c c | c c}
& \multicolumn{2}{c|}{\textsc{FI}} & \multicolumn{2}{c}{\textsc{FI + FS}} \\
                         & \ensuremath{\FG} & \ensuremath{\CG}                  & \ensuremath{\FG} & \ensuremath{\CG} \\
\hline
\textsc{Types}            & 15  & 16                     & 16  & 17   \\
\textsc{Syntax}           & 82  & 85                     & 86  & 93   \\
\textsc{Semantics}        & 137 & 134                    & 164 & 162  \\
\textsc{Valid}            & -   & -                      & 230 & 198  \\
\textsc{\ensuremath{\Blue{\Conid{L}}}-Equivalence} & 143 & 138                    & 212 & 217  \\
\textsc{Security}             & 222 & 155                    & 409 & 235  \\
\hline
\textsc{Lattice}          & \multicolumn{2}{c|}{148} & \multicolumn{2}{c}{148} \\
\textsc{Store}            & \multicolumn{2}{c|}{\multirow{2}{*}{152}} & \multicolumn{2}{c}{112} \\
\textsc{Memory}           & \multicolumn{2}{c|}{}    & \multicolumn{2}{c}{153} \\
\textsc{Heap}             & \multicolumn{2}{c|}{-}   & \multicolumn{2}{c}{337} \\
\textsc{Bijection}        & \multicolumn{2}{c|}{-}   & \multicolumn{2}{c}{478} \\
\textsc{Other}            & \multicolumn{2}{c|}{66}  & \multicolumn{2}{c}{485} \\
\hline
\textsc{Total} & 965 & 894 & 2,830 & 2,635 \\
\end{tabular}
\label{table:fg-cg-loc}
\end{table}

\begin{table}[t]
\centering
\caption{The table reports the size of the main parts of the formal
  security analysis (\textsc{Security}) of \ensuremath{\FG} and \ensuremath{\CG} with and
  without flow-sensitive references.  \textsc{Confinement} refers to
  the store and heap confinement lemma, \textsc{Secret Context} and
  \textsc{Public Context} to \ensuremath{\Blue{\Conid{L}}}-equivalence preservation in secret
  and public contexts, respectively, and \textsc{Pure Preservation} to
  \ensuremath{\Blue{\Conid{L}}}-equivalence preservation for the pure semantics fragment of
  \ensuremath{\CG}.  }
\begin{tabular}{r | c c | c c}
& \multicolumn{2}{c|}{\textsc{FI}} & \multicolumn{2}{c}{\textsc{FI + FS}} \\
                                 & \ensuremath{\FG} & \ensuremath{\CG}                  & \ensuremath{\FG} & \ensuremath{\CG} \\
\hline
\textsc{Confinement}             & 51  & 24                     & 89  & 44   \\
\textsc{Secret Context}          & 13  & 23                     & 16  & 29   \\
\textsc{Pure Preservation}       & -   & 35                     & -   & 35   \\
\textsc{Public Context} & 138 & 46                     & 273 & 85   \\
\textsc{Tini}                    & 9   & 17                     & 10  & 21   \\
\textsc{Other}                   & 11  & 10                     & 21  & 21   \\
\hline
\textsc{Total}                   & 222 & 155                    & 409 & 235  \\
\end{tabular}
\label{table:fg-cg-loc-security}
\end{table}

We now discuss our verified artifacts, in which we model \ensuremath{\FG} and \ensuremath{\CG}
and provide machine-checked proofs of their security guarantees.
We have formalized these languages using
Agda~\parencite{Norell09,Bove09}, a dependently typed functional
language and an interactive proof assistant based on intuitionistic
type theory.
In our proof scripts, we embed the syntax, the type system, and the
semantics of \ensuremath{\FG} and \ensuremath{\CG} into Agda data types, where we leverage
dependent types to maintain additional assumptions about terms and
judgments.
For example, we use \emph{well-typed syntax} and typed DeBrujin
indexes to ensure that expressions, values, and environments are
intrinsically well-scoped and well-typed~\parencite{POPLMarkReloaded},
which in turn let us define \emph{type-preserving} big-step semantics
judgments.
In the following, we analyze our artifact and find that the security
proofs for \ensuremath{\FG} are longer (between 43\% and 74\%) than those for
\ensuremath{\CG}.
These empirical results confirm the general sense that coarse-grained
IFC languages are easier to analyze and prove secure than fine-grained
IFC languages.
%

\section{Artifact Analysis}
Table~\ref{table:fg-cg-loc} summarizes the size of our proof scripts.
In the table, we report the number of lines of Agda code needed to
formalize different parts of the fine- and coarse-grained languages
(\ensuremath{\FG} and \ensuremath{\CG}) featuring only flow-insensitive references
(\textsc{FI}) and also with the addition of flow-sensitive references
(\textsc{FI + FS}).\footnote{\textsc{FI} is the companion artifact of
  the conference version of this article~\parencite{Vassena19},
  available at
  \href{https://hub.docker.com/r/marcovassena/granularity}{https://hub.docker.com/r/marcovassena/granularity}.
  We developed artifact \textsc{FI + FS} by adding flow-sensitive
  references to artifact \textsc{FI}. The extended artifact is
  available at \href{https://hub.docker.com/r/marcovassena/granularity-ftpl}{https://hub.docker.com/r/marcovassena/granularity-ftpl}.}
The bottom part of the table lists parts of the formalization that are
shared and reused for both languages, where categories \textsc{Store},
\textsc{Memory} and \textsc{Heap} include also definitions and proofs
relative to their \ensuremath{\Blue{\Conid{L}}}-equivalence relations and valid
judgments.\footnote{We report only one number for \textsc{Store} and
  \textsc{Memory} because they are defined together in the same module
  in \textsc{FI}, but in separate modules in \textsc{FI + FS}. In the
  second artifact, \textsc{Memory} and \textsc{Heap} are also defined
  by instantiating a generic container data structure for labeled
  data, which is counted in \textsc{Other}.}
The line counts for basic definitions (i.e., \textsc{Types},
\textsc{Syntax}, \textsc{Semantics}, and \textsc{\ensuremath{\Blue{\Conid{L}}}-equivalence})
are roughly the same for the two languages, in both artifacts.
The only significant difference is in \textsc{Security}, which
represents the size of the scripts that state and prove the main
security lemmas and theorems of the languages.
In particular, we find that the security proofs in \ensuremath{\FG} are about 43\%
longer than in \ensuremath{\CG} in the artifact with only flow-insensitive
references (\textsc{FI}) and 74\% longer for the languages extended
with flow-sensitive references (\textsc{FI + FS}).
To understand better the reason behind this gap, we compare
\textsc{Security} in more detail in Table
\ref{table:fg-cg-loc-security}, which reports the size of the
following lemmas and theorems:
\begin{itemize}[label={$\triangleright$}]
  \setlength\itemsep{2pt}
  \setlength\topsep{2pt}
  \setlength\parsep{0pt}
  \setlength\partopsep{0pt}
\item \textsc{Store and Heap Confinement}
\item \textsc{\ensuremath{\Blue{\Conid{L}}}-Equivalence Preservation in Secret Context}
\item \textsc{Pure \ensuremath{\Blue{\Conid{L}}}-Equivalence Preservation}
\item \textsc{\ensuremath{\Blue{\Conid{L}}}-Equivalence Preservation in Public Context}
\item \textsc{Termination-Insensitive Non-Interference}
\end{itemize}
%

First of all, we observe that in \ensuremath{\FG} the \textsc{Confinement} lemma
is about twice as long as in \ensuremath{\CG}.
This is because side-effects can occur in every reduction rule in
\ensuremath{\FG}, while most constructs in \ensuremath{\CG} are side-effect free and so we
simply have to consider fewer cases in the second proof.
%
%
For the languages extended with flow-sensitive reference
(\textsc{FI + FS}), these proofs roughly \emph{double} in size, as they
additionally need to consider the validity of program configurations
in order to reason about \ensuremath{\Blue{\Conid{L}}}-equivalence up to bijection.
The size of \textsc{Secret Context} and \textsc{Tini} is modest in
both variants of each languages.
Compared to \ensuremath{\FG}, \ensuremath{\CG} requires \emph{twice} as many lines of code for
these results simply because for this language each result has to be
stated and proved \emph{twice}, once for the forcing semantics and
once for the thunk semantics.
\textsc{Public Context} is the lemma where we observe the greatest
gap in the number of lines needed in \ensuremath{\FG} and \ensuremath{\CG}.
In artifact \textsc{FI}, we find that \textsc{Public Context} for \ensuremath{\FG}
is 70\% longer than \textsc{Public Context} and \textsc{Pure
  Preservation} for \ensuremath{\CG} combined, and 128\% longer when considering
also flow-sensitive references (\textsc{FI + FS}).
Since all values are intrinsically labeled in \ensuremath{\FG}, those proofs have
to \emph{explicitly} (i) reason about the sensitivity of program
inputs and intermediate values in most cases of the proof, and (ii)
rule out implicit flows for control-flow constructs.
In contrast, most constructs in \ensuremath{\CG} are security unaware and
so evaluated by the pure fragment of the semantics, where the
security analysis is straightforward.
In particular, since pure reductions cannot perform side-effects or
inspect labeled data, no implicit flows can arise in the proof of
\textsc{Pure Preservation}, which follows by simple induction (35
LOC).
Only in the proof of \textsc{Public Context} for \ensuremath{\CG} we need to
reason about data flows explicitly using security labels, but even
there implicit flows are less problematic, thanks to the monadic
structure of computations.
The line counts for \textsc{Public Context} in \textsc{FI + FS} show
that the extra complexity of dealing with valid assumptions increases
the size of the proofs in both languages, which grows by 85\% for \ensuremath{\CG}
and 98\% for \ensuremath{\FG}.\footnote{Importantly though, the addition of
  flow-sensitive references did not require any change in the proof of
  \textsc{Pure Preservation}.}
Not only the proofs of \ensuremath{\FG} are longer than \ensuremath{\CG}, but they are also
more complicated because they often must (i) combine data related
up to different bijections, and (ii) propagate the assumption that
program configurations are valid in inductive cases.
This results in an increased number of calls to helper lemmas and
properties in \ensuremath{\FG}, compared to \ensuremath{\CG}.
For example, the proof for \textsc{Public Context} in \ensuremath{\FG} requires
\textbf{22} calls to \emph{weakening}
(Property~\ref{prop:fg-low-equiv-props-with-bij}.4) and \textbf{52}
calls to \emph{valid invariant}
(Property~\ref{prop:fg-valid-invariant}), while the same proof in \ensuremath{\CG}
requires only \textbf{4} calls in total.



\chapter{Fine- to Coarse-Grained Program Translation}
\label{sec:fg2cg}

This section presents a provably semantics-preserving program
translation from the fine-grained dynamic IFC calculus \ensuremath{\FG} to the
coarse-grained calculus \ensuremath{\CG}.
At a high level, the translation performs two tasks (i) it embeds
the \emph{intrinsic} label annotation of \ensuremath{\FG} values into an
\emph{explicitly} labeled \ensuremath{\CG} value via the \ensuremath{\mathbf{Labeled}} type constructor and
(ii) it restructures \ensuremath{\FG} \emph{side-effectful} expressions into
\emph{monadic operations} inside the \ensuremath{\mathbf{LIO}\xspace} monad.

\section{Types and Values}

\begin{figure}[t]
\begin{subfigure}[b]{0.61\textwidth}
\setlength{\parindent}{0pt}
\setlength{\mathindent}{6pt}
\begin{hscode}\SaveRestoreHook
\column{B}{@{}>{\hspre}l<{\hspost}@{}}%
\column{17}{@{}>{\hspre}l<{\hspost}@{}}%
\column{23}{@{}>{\hspre}l<{\hspost}@{}}%
\column{E}{@{}>{\hspre}l<{\hspost}@{}}%
\>[B]{}{\llangle}\mathbf{unit}{\rrangle}\mathrel{=}\mathbf{Labeled}\;\mathbf{unit}{}\<[E]%
\\
\>[B]{}{\llangle}\mathscr{L}{\rrangle}\mathrel{=}{}\<[17]%
\>[17]{}\mathbf{Labeled}\;\mathscr{L}{}\<[E]%
\\
\>[B]{}{\llangle}\tau_{1}\;\times\;\tau_{2}{\rrangle}\mathrel{=}\mathbf{Labeled}\;({\llangle}\tau_{1}{\rrangle}\times{\llangle}\tau_{2}{\rrangle}){}\<[E]%
\\
\>[B]{}{\llangle}\tau_{1}\mathbin{+}\tau_{2}{\rrangle}\mathrel{=}\mathbf{Labeled}\;({\llangle}\tau_{1}{\rrangle}\mathbin{+}{\llangle}\tau_{2}{\rrangle}){}\<[E]%
\\
\>[B]{}{\llangle}\tau_{1}\to \tau_{2}{\rrangle}\mathrel{=}{}\<[23]%
\>[23]{}\mathbf{Labeled}\;({\llangle}\tau_{1}{\rrangle}\to \mathbf{LIO}\xspace\;\thinspace\mkern-3mu{\llangle}\tau_{2}{\rrangle}){}\<[E]%
\\
\>[B]{}{\llangle}\mathbf{Ref}\; s\;\tau{\rrangle}\mathrel{=}\mathbf{Labeled}\;(\mathbf{Ref}\; s\;\thinspace\mkern-3mu{\llangle}\tau{\rrangle}){}\<[E]%
\ColumnHook
\end{hscode}\resethooks
\caption{Types.}
\label{fig:fg2cg-ty}
\end{subfigure}%
\rulesep
\begin{subfigure}[b]{0.37\textwidth}
\setlength{\parindent}{0pt}
\setlength{\mathindent}{6pt}
\begin{hscode}\SaveRestoreHook
\column{B}{@{}>{\hspre}l<{\hspost}@{}}%
\column{E}{@{}>{\hspre}l<{\hspost}@{}}%
\>[B]{}{\llangle}{{r}}^{{\ell}}{\rrangle}\mathrel{=}\mathbf{Labeled}\;{\ell}\;\thinspace\mkern-3mu{\llangle}{r}{\rrangle}{}\<[E]%
\\
\>[B]{}{\llangle}(){\rrangle}\mathrel{=}(){}\<[E]%
\\
\>[B]{}{\llangle}{\ell}{\rrangle}\mathrel{=}{\ell}{}\<[E]%
\\
\>[B]{}{\llangle}({v}_{1},{v}_{2}){\rrangle}\mathrel{=}({\llangle}{v}_{1}{\rrangle},{\llangle}{v}_{2}{\rrangle}){}\<[E]%
\\
\>[B]{}{\llangle}\mathbf{inl}({v}\mkern1mu){\rrangle}\mathrel{=}\mathbf{inl}({\llangle}{v}{\rrangle}\mkern1mu){}\<[E]%
\\
\>[B]{}{\llangle}\mathbf{inr}({v}\mkern1mu){\rrangle}\mathrel{=}\mathbf{inr}({\llangle}{v}{\rrangle}\mkern1mu){}\<[E]%
\\
\>[B]{}{\llangle}(\Varid{x}.\Varid{e},\theta){\rrangle}\mathrel{=}(\Varid{x}.{\llangle}\Varid{e}{\rrangle},{\llangle}\theta{\rrangle}){}\<[E]%
\\
\>[B]{}{\llangle}{\Varid{n}}_{{\ell}}{\rrangle}\mathrel{=}{\Varid{n}}_{{\ell}}{}\<[E]%
\\
\>[B]{}{\llangle}\Varid{n}{\rrangle}\mathrel{=}\Varid{n}{}\<[E]%
\ColumnHook
\end{hscode}\resethooks
\caption{Values.}
\label{fig:fg2cg-value}
\end{subfigure}%
\caption{Translation from \ensuremath{\FG} to \ensuremath{\CG}.}
\label{fig:fg2cg-ty-value}
\end{figure}

Our type-driven approach starts by formalizing this intuition in the
function \ensuremath{{\llangle}\cdot {\rrangle}}, which maps the \ensuremath{\FG} type \ensuremath{\tau} to the corresponding
\ensuremath{\CG} type \ensuremath{{\llangle}\tau{\rrangle}} (see Figure \ref{fig:fg2cg-ty}).
The function is defined by induction on types and recursively adds the
\ensuremath{\mathbf{Labeled}} type constructor to each existing \ensuremath{\FG} type constructor.
For the function type \ensuremath{\tau_{1}\to \tau_{2}}, the result is additionally
monadic, i.e., \ensuremath{{\llangle}\tau_{1}{\rrangle}\to \mathbf{LIO}\xspace\;\thinspace\mkern-3mu{\llangle}\tau_{2}{\rrangle}}. This is because the
function's body in \ensuremath{\FG} may have side-effects.
Furthermore, the translation for references types \ensuremath{\mathbf{Ref}\; s\;\tau}
preserves the sensitivity tag of the reference, i.e., \ensuremath{\mathbf{Ref}\; s\;\thinspace\mkern-3mu{\llangle}\tau{\rrangle}}.

The translation for values (Figure \ref{fig:fg2cg-value}) is
straightforward. Each \ensuremath{\FG} label tag becomes the label annotation in a
\ensuremath{\CG} labeled value. The translation is homomorphic in the constructors
on raw values.
The translation converts a \ensuremath{\FG} function closure into a \ensuremath{\CG} thunk
closure by translating the body of the function to a thunk, i.e., \ensuremath{{\llangle}\Varid{e}{\rrangle}} (see below), and translating the environment pointwise, i.e.,
\ensuremath{{\llangle}\theta{\rrangle}\mathrel{=}\lambda \Varid{x}.{\llangle}\theta(\Varid{x}){\rrangle}}.
Finally, the translation preserves the memory address and the label
for flow-insensitive references, i.e., \ensuremath{{\llangle}{\Varid{n}}_{{\ell}}{\rrangle}\mathrel{=}{\Varid{n}}_{{\ell}}},
and the heap address for flow-sensitive references, i.e., \ensuremath{{\llangle}\Varid{n}{\rrangle}\mathrel{=}\Varid{n}}.

\section{Expressions}
\begin{figure}[!p]
\begin{minipage}[t]{0.5\textwidth}
\setlength{\parindent}{0pt}
\setlength{\mathindent}{6pt}
\begin{hscode}\SaveRestoreHook
\column{B}{@{}>{\hspre}l<{\hspost}@{}}%
\column{3}{@{}>{\hspre}l<{\hspost}@{}}%
\column{5}{@{}>{\hspre}l<{\hspost}@{}}%
\column{E}{@{}>{\hspre}l<{\hspost}@{}}%
\>[B]{}{\llangle}(){\rrangle}\mathrel{=}\mathbf{toLabeled}(\mathbf{return}(()\mkern1mu)\mkern1mu){}\<[E]%
\\[\blanklineskip]%
\>[B]{}{\llangle}{\ell}{\rrangle}\mathrel{=}\mathbf{toLabeled}(\mathbf{return}({\ell}\mkern1mu)\mkern1mu){}\<[E]%
\\[\blanklineskip]%
\>[B]{}{\llangle}(\lambda \Varid{x}.\Varid{e}){\rrangle}\mathrel{=}{}\<[E]%
\\
\>[B]{}\hsindent{3}{}\<[3]%
\>[3]{}\mathbf{toLabeled}(\mathbf{return}(\lambda \Varid{x}.{\llangle}\Varid{e}{\rrangle}\mkern1mu)\mkern1mu){}\<[E]%
\\[\blanklineskip]%
\>[B]{}{\llangle}\mathbf{inl}(\Varid{e}\mkern1mu){\rrangle}\mathrel{=}\mathbf{toLabeled}(\;\mathbf{do}{}\<[E]%
\\
\>[B]{}\hsindent{3}{}\<[3]%
\>[3]{}\Varid{lv}\leftarrow {\llangle}\Varid{e}{\rrangle}{}\<[E]%
\\
\>[B]{}\hsindent{3}{}\<[3]%
\>[3]{}\mathbf{return}(\mathbf{inl}(\Varid{lv}\mkern1mu)\mkern1mu)\mkern1mu){}\<[E]%
\\[\blanklineskip]%
\>[B]{}{\llangle}\mathbf{inr}(\Varid{e}\mkern1mu){\rrangle}\mathrel{=}\mathbf{toLabeled}(\;\mathbf{do}{}\<[E]%
\\
\>[B]{}\hsindent{3}{}\<[3]%
\>[3]{}\Varid{lv}\leftarrow {\llangle}\Varid{e}{\rrangle}{}\<[E]%
\\
\>[B]{}\hsindent{3}{}\<[3]%
\>[3]{}\mathbf{return}(\mathbf{inr}(\Varid{lv}\mkern1mu)\mkern1mu)\mkern1mu){}\<[E]%
\\[\blanklineskip]%
\>[B]{}{\llangle}(\Varid{e}_{1},\Varid{e}_{2}){\rrangle}\mathrel{=}\mathbf{toLabeled}(\;\mathbf{do}{}\<[E]%
\\
\>[B]{}\hsindent{3}{}\<[3]%
\>[3]{}{\Varid{lv}}_{\mathrm{1}}\leftarrow {\llangle}\Varid{e}_{1}{\rrangle}{}\<[E]%
\\
\>[B]{}\hsindent{3}{}\<[3]%
\>[3]{}{\Varid{lv}}_{\mathrm{2}}\leftarrow {\llangle}\Varid{e}_{2}{\rrangle}{}\<[E]%
\\
\>[B]{}\hsindent{3}{}\<[3]%
\>[3]{}\mathbf{return}({\Varid{lv}}_{\mathrm{1}},{\Varid{lv}}_{\mathrm{2}}\mkern1mu)\mkern1mu){}\<[E]%
\\[\blanklineskip]%
\>[B]{}{\llangle}\Varid{x}{\rrangle}\mathrel{=}\mathbf{toLabeled}(\mathbf{unlabel}(\Varid{x}\mkern1mu)\mkern1mu){}\<[E]%
\\[\blanklineskip]%
\>[B]{}{\llangle}\Varid{e}_{1}\;\Varid{e}_{2}{\rrangle}\mathrel{=}\mathbf{toLabeled}(\;\mathbf{do}{}\<[E]%
\\
\>[B]{}\hsindent{5}{}\<[5]%
\>[5]{}\Varid{lv}_{1}\leftarrow {\llangle}\Varid{e}_{1}{\rrangle}{}\<[E]%
\\
\>[B]{}\hsindent{5}{}\<[5]%
\>[5]{}\Varid{lv}_{2}\leftarrow {\llangle}\Varid{e}_{2}{\rrangle}{}\<[E]%
\\
\>[B]{}\hsindent{5}{}\<[5]%
\>[5]{}{v}_{1}\leftarrow \mathbf{unlabel}(\Varid{lv}_{1}\mkern1mu){}\<[E]%
\\
\>[B]{}\hsindent{5}{}\<[5]%
\>[5]{}\Varid{lv}\leftarrow {v}_{1}\;\Varid{lv}_{2}{}\<[E]%
\\
\>[B]{}\hsindent{5}{}\<[5]%
\>[5]{}\mathbf{unlabel}(\Varid{lv}\mkern1mu)\mkern1mu){}\<[E]%
\\[\blanklineskip]%
\>[B]{}{\llangle}\mathbf{case}(\Varid{e},\Varid{x}.\Varid{e}_{1},\Varid{x}.\Varid{e}_{2}\mkern1mu){\rrangle}\mathrel{=}{}\<[E]%
\\
\>[B]{}\hsindent{3}{}\<[3]%
\>[3]{}\mathbf{toLabeled}(\;\mathbf{do}{}\<[E]%
\\
\>[3]{}\hsindent{2}{}\<[5]%
\>[5]{}\Varid{lv}\leftarrow {\llangle}\Varid{e}{\rrangle}{}\<[E]%
\\
\>[3]{}\hsindent{2}{}\<[5]%
\>[5]{}{v}\leftarrow \mathbf{unlabel}(\Varid{lv}\mkern1mu){}\<[E]%
\\
\>[3]{}\hsindent{2}{}\<[5]%
\>[5]{}\Varid{lv'}\leftarrow \mathbf{case}({v},\Varid{x}.{\llangle}\Varid{e}_{1}{\rrangle},\Varid{x}.{\llangle}\Varid{e}_{2}{\rrangle}\mkern1mu){}\<[E]%
\\
\>[3]{}\hsindent{2}{}\<[5]%
\>[5]{}\mathbf{unlabel}(\Varid{lv'}\mkern1mu)\mkern1mu){}\<[E]%
\ColumnHook
\end{hscode}\resethooks
\end{minipage}%
\rulesep
\begin{minipage}[t]{0.5\textwidth}
\setlength{\parindent}{0pt}
\setlength{\mathindent}{6pt}
\begin{hscode}\SaveRestoreHook
\column{B}{@{}>{\hspre}l<{\hspost}@{}}%
\column{3}{@{}>{\hspre}l<{\hspost}@{}}%
\column{5}{@{}>{\hspre}l<{\hspost}@{}}%
\column{7}{@{}>{\hspre}l<{\hspost}@{}}%
\column{8}{@{}>{\hspre}l<{\hspost}@{}}%
\column{E}{@{}>{\hspre}l<{\hspost}@{}}%
\>[B]{}{\llangle}\mathbf{fst}(\Varid{e}\mkern1mu){\rrangle}\mathrel{=}\mathbf{toLabeled}(\;\mathbf{do}{}\<[E]%
\\
\>[B]{}\hsindent{3}{}\<[3]%
\>[3]{}\Varid{lv}\leftarrow {\llangle}\Varid{e}{\rrangle}{}\<[E]%
\\
\>[B]{}\hsindent{3}{}\<[3]%
\>[3]{}{v}\leftarrow \mathbf{unlabel}(\Varid{lv}\mkern1mu){}\<[E]%
\\
\>[B]{}\hsindent{3}{}\<[3]%
\>[3]{}\mathbf{unlabel}(\mathbf{fst}({v}\mkern1mu)\mkern1mu)\mkern1mu){}\<[E]%
\\[\blanklineskip]%
\>[B]{}{\llangle}\mathbf{snd}(\Varid{e}\mkern1mu){\rrangle}\mathrel{=}\mathbf{toLabeled}(\;\mathbf{do}{}\<[E]%
\\
\>[B]{}\hsindent{3}{}\<[3]%
\>[3]{}\Varid{lv}\leftarrow {\llangle}\Varid{e}{\rrangle}{}\<[E]%
\\
\>[B]{}\hsindent{3}{}\<[3]%
\>[3]{}{v}\leftarrow \mathbf{unlabel}(\Varid{lv}\mkern1mu){}\<[E]%
\\
\>[B]{}\hsindent{3}{}\<[3]%
\>[3]{}\mathbf{unlabel}(\mathbf{snd}({v}\mkern1mu)\mkern1mu)\mkern1mu){}\<[E]%
\\[\blanklineskip]%
\>[B]{}{\llangle}\Varid{e}_{1}\;\flows^{?}\;\Varid{e}_{2}{\rrangle}\mathrel{=}\mathbf{toLabeled}(\;\mathbf{do}{}\<[E]%
\\
\>[B]{}\hsindent{3}{}\<[3]%
\>[3]{}\Varid{lv}_{1}\leftarrow {\llangle}\Varid{e}_{1}{\rrangle}{}\<[E]%
\\
\>[B]{}\hsindent{3}{}\<[3]%
\>[3]{}\Varid{lv}_{2}\leftarrow {\llangle}\Varid{e}_{2}{\rrangle}{}\<[E]%
\\
\>[B]{}\hsindent{3}{}\<[3]%
\>[3]{}\Varid{lu}\leftarrow \mathbf{toLabeled}(\mathbf{return}(()\mkern1mu)\mkern1mu){}\<[E]%
\\
\>[B]{}\hsindent{3}{}\<[3]%
\>[3]{}{v}_{1}\leftarrow \mathbf{unlabel}(\Varid{lv}_{1}\mkern1mu){}\<[E]%
\\
\>[B]{}\hsindent{3}{}\<[3]%
\>[3]{}{v}_{2}\leftarrow \mathbf{unlabel}(\Varid{lv}_{2}\mkern1mu){}\<[E]%
\\
\>[B]{}\hsindent{3}{}\<[3]%
\>[3]{}\mathbf{return}(\;\mathbf{if}{}\<[E]%
\\
\>[3]{}\hsindent{4}{}\<[7]%
\>[7]{}{v}_{1}\;\flows^{?}\;{v}_{2}{}\<[E]%
\\
\>[7]{}\hsindent{1}{}\<[8]%
\>[8]{}\mathbf{then}\;\mathbf{inl}(\Varid{lu}\mkern1mu){}\<[E]%
\\
\>[7]{}\hsindent{1}{}\<[8]%
\>[8]{}\mathbf{else}\;\mathbf{inr}(\Varid{lu}\mkern1mu)\mkern1mu){}\<[E]%
\\[\blanklineskip]%
\>[B]{}{\llangle}\mathbf{taint}(\Varid{e}_{1},\Varid{e}_{2}\mkern1mu){\rrangle}\mathrel{=}{}\<[E]%
\\
\>[B]{}\hsindent{3}{}\<[3]%
\>[3]{}\mathbf{toLabeled}(\;\mathbf{do}{}\<[E]%
\\
\>[3]{}\hsindent{2}{}\<[5]%
\>[5]{}\Varid{lv}_{1}\leftarrow {\llangle}\Varid{e}_{1}{\rrangle}{}\<[E]%
\\
\>[3]{}\hsindent{2}{}\<[5]%
\>[5]{}{v}_{1}\leftarrow \mathbf{unlabel}(\Varid{lv}_{1}\mkern1mu){}\<[E]%
\\
\>[3]{}\hsindent{2}{}\<[5]%
\>[5]{}\mathbf{taint}({v}_{1}\mkern1mu){}\<[E]%
\\
\>[3]{}\hsindent{2}{}\<[5]%
\>[5]{}\Varid{lv}_{2}\leftarrow {\llangle}\Varid{e}_{2}{\rrangle}{}\<[E]%
\\
\>[3]{}\hsindent{2}{}\<[5]%
\>[5]{}\mathbf{unlabel}(\Varid{lv}_{2}\mkern1mu)\mkern1mu){}\<[E]%
\\[\blanklineskip]%
\>[B]{}{\llangle}\mathbf{labelOf}(\Varid{e}\mkern1mu){\rrangle}\mathrel{=}{}\<[E]%
\\
\>[B]{}\hsindent{3}{}\<[3]%
\>[3]{}\mathbf{toLabeled}(\;\mathbf{do}{}\<[E]%
\\
\>[3]{}\hsindent{2}{}\<[5]%
\>[5]{}\Varid{lv}\leftarrow {\llangle}\Varid{e}{\rrangle}{}\<[E]%
\\
\>[3]{}\hsindent{2}{}\<[5]%
\>[5]{}\mathbf{labelOf}(\Varid{lv}\mkern1mu)\mkern1mu){}\<[E]%
\\[\blanklineskip]%
\>[B]{}{\llangle}\mathbf{getLabel}{\rrangle}\mathrel{=}{}\<[E]%
\\
\>[B]{}\hsindent{3}{}\<[3]%
\>[3]{}\mathbf{toLabeled}(\mathbf{getLabel}\mkern1mu)\mkern1mu){}\<[E]%
\ColumnHook
\end{hscode}\resethooks
\end{minipage}%
\caption{Translation from \ensuremath{\FG} to \ensuremath{\CG} (expressions).}
\label{fig:fg2cg-expr}
\end{figure}
We show the translation of \ensuremath{\FG} expressions to \ensuremath{\CG} monadic thunks in
Figure \ref{fig:fg2cg-expr}. We use the standard \ensuremath{\mathbf{do}} notation for
readability.\footnote{Syntax \ensuremath{\mathbf{do}\;\Varid{x}\leftarrow \Varid{e}_{1};\Varid{e}_{2}} desugars to \ensuremath{\mathbf{bind}(\Varid{e}_{1},\Varid{x}.\Varid{e}_{2}\mkern1mu)} and syntax \ensuremath{\Varid{e}_{1};\Varid{e}_{2}} to \ensuremath{\mathbf{bind}(\Varid{e}_{1},\anonymous .\Varid{e}_{2}\mkern1mu)}.}
%
First, notice that the translation of all constructs occurs
inside a \ensuremath{\mathbf{toLabeled}(\cdot \mkern1mu)} block.
This achieves two goals, (i) it ensures that the value that results
from a translated expression is \emph{explicitly} labeled and (ii) it
creates an isolated nested context where the translated thunk can
execute without raising the program counter label at the top level.
Inside the \ensuremath{\mathbf{toLabeled}(\cdot \mkern1mu)} block, the program counter label may rise,
e.g., when some intermediate result is unlabeled, and the translation
relies on \ensuremath{\mathbf{LIO}\xspace}'s floating-label mechanism to track dependencies
between data of different security levels.
In particular, we will show later that the value of the program
counter label at the end of each nested block coincides with the label
annotation of the \ensuremath{\FG} value that the original expression evaluates
to.
For example, introduction forms of ground values (unit, labels, and
functions) are simply returned inside the \ensuremath{\mathbf{toLabeled}(\cdot \mkern1mu)} block so that
they get tagged with the current value of the program counter label
just as in the corresponding \ensuremath{\FG} introduction rules
(\srule{Label,Unit,Fun}).
Introduction forms of compounds values such as \ensuremath{\mathbf{inl}(\Varid{e}\mkern1mu)}, \ensuremath{\mathbf{inr}(\Varid{e}\mkern1mu)} and
\ensuremath{(\Varid{e}_{1},\Varid{e}_{2})} follow the same principle.
The translation simply nests the translations of the nested
expressions inside the same constructor, without raising the program
counter label. This matches the behavior of the corresponding \ensuremath{\FG}
rules \srule{Inl,Inr,Pair}.\footnote{We name a variable \ensuremath{\Varid{lv}} if
  it gets bound to a labeled value, i.e., to indicate that the
  variable has type \ensuremath{\mathbf{Labeled}\;\tau}.}
For example, the \ensuremath{\FG} reduction \ensuremath{((),())\;\Downarrow^{\varnothing}_{\Blue{\Conid{L}}}\;{({()}^{\Blue{\Conid{L}}},{()}^{\Blue{\Conid{L}}})}^{\Blue{\Conid{L}}}} maps to a \ensuremath{\CG} term that reduces to
\ensuremath{\mathbf{Labeled}\;\Blue{\Conid{L}}\;(\mathbf{Labeled}\;\Blue{\Conid{L}}\;(),\mathbf{Labeled}\;\Blue{\Conid{L}}\;())} when started with program
counter label \ensuremath{\Blue{\Conid{L}}}.
The translation of variables gives some insight into how the \ensuremath{\CG}
floating-label mechanism can simulate \ensuremath{\FG}'s tainting approach.
First, the type-driven approach set out in Figure \ref{fig:fg2cg-ty}
demands that functions take only labeled values as arguments,
so the variables in the source program are always associated to
a labeled value in the translated program.
The values that correspond to these variables are stored in the
environment \ensuremath{\theta} and translated separately, e.g., if
\ensuremath{\theta(\Varid{x})\mathrel{=}{{r}}^{{\ell}}} in \ensuremath{\FG}, then \ensuremath{\Varid{x}} gets bound to \ensuremath{{\llangle}{{r}}^{{\ell}}{\rrangle}\mathrel{=}\mathbf{Labeled}\;{\ell}\;\thinspace\mkern-3mu{\llangle}{r}{\rrangle}} when translated to \ensuremath{\CG}.
Thus, the translation converts a variable, say \ensuremath{\Varid{x}}, to
\ensuremath{\mathbf{toLabeled}(\mathbf{unlabel}(\Varid{x}\mkern1mu)\mkern1mu)}, so that its label gets tainted with the
current program counter label.
More precisely, \ensuremath{\mathbf{unlabel}(\Varid{x}\mkern1mu)} retrieves the labeled value associated
with the variable, i.e., \ensuremath{\mathbf{Labeled}\;{\ell}\;\thinspace\mkern-3mu{\llangle}{r}{\rrangle}}, taints the program
counter with its label to make it \ensuremath{\Varid{pc}\;\lub\;{\ell}}, and returns the
content, i.e., \ensuremath{{\llangle}{r}{\rrangle}}.
Since \ensuremath{\mathbf{unlabel}(\Varid{x}\mkern1mu)} occurs inside a \ensuremath{\mathbf{toLabeled}(\cdot \mkern1mu)} block, the code
above results in \ensuremath{\mathbf{Labeled}\;(\Varid{pc}\;\lub\;{\ell})\;\thinspace\mkern-3mu{\llangle}{r}{\rrangle}} when evaluated,
matching precisely the tainting behavior of \ensuremath{\FG} rule \srule{Var},
i.e., \ensuremath{\Varid{x}\;\Downarrow^{\update{\theta}{\Varid{x}}{{{r}}^{{\ell}}}}_{\Varid{pc}}\;{{r}}^{\Varid{pc}\;\lub\;{\ell}}}.
The elimination forms for other types (function application, pair
projections and case analysis) follow the same approach.
For example, the code that translates a function application
\ensuremath{\Varid{e}_{1}\;\Varid{e}_{2}} first executes the code that computes the translated
function, i.e., \ensuremath{\Varid{lv}_{1}\leftarrow {\llangle}\Varid{e}_{1}{\rrangle}}, then the code that computes the
argument, i.e., \ensuremath{\Varid{lv}_{2}\leftarrow {\llangle}\Varid{e}_{2}{\rrangle}} and then retrieves the function from
the first labeled value, i.e., \ensuremath{{v}_{1}\leftarrow \mathbf{unlabel}(\Varid{lv}_{1}\mkern1mu)}.\footnote{ Notice
  that it is incorrect to unlabel the function before translating the
  argument, because that would taint the program counter label, which
  would raise at level, say \ensuremath{\Varid{pc}\;\lub\;{\ell}}, and affect the code that
  translates the argument, which was to be evaluated with program
  counter label equal to \ensuremath{\Varid{pc}} by the original \emph{flow-insensitive} \ensuremath{\FG}
  rule \srule{App} for function application.}
%
The function \ensuremath{{v}_{1}} applied to the labeled argument \ensuremath{\Varid{lv}_{2}} gives a
computation that gets executed and returns a labeled value \ensuremath{\Varid{lv}} that
gets unlabeled to expose the final result (the surrounding \ensuremath{\mathbf{toLabeled}(\cdot \mkern1mu)}
wraps it again right away).
The translation of case analysis is analogous.
The translation of pair projections first converts the \ensuremath{\FG} pair into
a computation that gives a \ensuremath{\CG} labeled pair of labeled values, say
\ensuremath{\mathbf{Labeled}\;{\ell}\;(\mathbf{Labeled}\;{\ell}_{1}\;\thinspace\mkern-3mu{\llangle}{r}_{1}{\rrangle},\mathbf{Labeled}\;{\ell}_{2}\;\thinspace\mkern-3mu{\llangle}{r}_{2}{\rrangle})}
and removes the label tag on the pair via \ensuremath{\mathbf{unlabel}}, thus raising the
program counter label to \ensuremath{\Varid{pc}\;\lub\;{\ell}}.
Then, it projects the appropriate component and unlabels it, thus
tainting the program counter label even further with the label of
either the first or the second component.
This coincides with the tainting mechanism of \ensuremath{\FG} for projection
rules, e.g., in rule \srule{Fst} where \ensuremath{\mathbf{fst}(\Varid{e}\mkern1mu)\;\Downarrow^{\theta}_{\Varid{pc}}\;{{r}_{1}}^{\Varid{pc}\;\lub\;{\ell}\;\lub\;{\ell}_{1}}} if \ensuremath{\Varid{e}\;\Downarrow^{\theta}_{\Varid{pc}}\;{({{r}_{1}}^{{\ell}_{1}},{{r}_{2}}^{{\ell}_{2}})}^{{\ell}}}.

Lastly, translating \ensuremath{\mathbf{taint}(\Varid{e}_{1},\Varid{e}_{2}\mkern1mu)} requires (i) translating the
expression \ensuremath{\Varid{e}_{1}} that gives the label, (ii) using \ensuremath{\mathbf{taint}(\cdot \mkern1mu)} from \ensuremath{\CG} to
explicitly taint the program counter label with the label that \ensuremath{\Varid{e}_{1}}
gives, and (iii) translating the second argument \ensuremath{\Varid{e}_{2}} to execute in
the tainted context and unlabeling the result.
The construct \ensuremath{\mathbf{labelOf}(\Varid{e}\mkern1mu)} of \ensuremath{\FG} uses the corresponding \ensuremath{\CG}
primitive applied on the corresponding labeled value, say \ensuremath{\mathbf{Labeled}\;{\ell}\;\thinspace\mkern-3mu{\llangle}{r}{\rrangle}}, obtained from the translated expression.
Notice that \ensuremath{\mathbf{labelOf}(\cdot \mkern1mu)} taints the program counter label in \ensuremath{\CG}, which
rises to \ensuremath{\Varid{pc}\;\lub\;{\ell}}, so the code just described results in \ensuremath{\mathbf{Labeled}\;(\Varid{pc}\;\lub\;{\ell})\;{\ell}}, which corresponds to the translation of the result
in \ensuremath{\FG}, i.e., \ensuremath{{\llangle}{{\ell}}^{{\ell}}{\rrangle}\mathrel{=}\mathbf{Labeled}\;{\ell}\;{\ell}} because \ensuremath{\Varid{pc}\;\lub\;{\ell}\equiv {\ell}},
since \ensuremath{\Varid{pc}\;\flows\;{\ell}} from Property \ref{prop:fg-pc-raise}.
The translation of \ensuremath{\mathbf{getLabel}} follows naturally by simply wrapping
\ensuremath{\CG}'s \ensuremath{\mathbf{getLabel}} inside a \ensuremath{\mathbf{toLabeled}(\cdot \mkern1mu)}, which correctly returns the
program counter label labeled with itself, i.e., \ensuremath{\mathbf{Labeled}\;\Varid{pc}\;\Varid{pc}}.

\subsection{Note on Environments and Weakening}

\begin{figure}[!t]
\begin{mathpar}
\inferrule[(WkenType)]
{\ensuremath{\Gamma\;\setminus\;\overline{x}\vdash\Varid{e}\mathbin{:}\tau}}
{\ensuremath{\Gamma\vdash\textbf{wken}(\overline{x}, \Varid{e})\mathbin{:}\tau}}
\and
\inferrule[(Wken)]
 {\ensuremath{\Varid{e}\;\Downarrow^{\theta\;\setminus\;\overline{x}}\;{v}}}
{\ensuremath{\textbf{wken}(\overline{x}, \Varid{e})\;\Downarrow^{\theta}\;{v}}}
\end{mathpar}
\caption{Typing and semantics rules of \ensuremath{\textbf{wken}} for \ensuremath{\CG}.}
\label{fig:wken}
\end{figure}

The semantics rules of \ensuremath{\FG} and \ensuremath{\CG} feature an environment \ensuremath{\theta}
for input values that gets extended with intermediate values during
program evaluation and that may be captured inside a closure.
%
Unfortunately, this capturing behavior is undesirable for our program
translation.
The program translation defined above introduces temporary
auxiliary variables that carry the value of intermediate results,
e.g., the labeled value obtained from running a computation that
translates some \ensuremath{\FG} expression.
When the translated program is executed, these values end up in the
environment, e.g., by means of rules \srule{App} and \srule{Bind}, and
mix with the input values of the source program and output values as
well, thus complicating the correctness statement of the translation,
which now has to account for those extra variables as well.
In order to avoid this nuisance, we employ a special form of weakening
that allows shrinking the environment at run-time and removing
spurious values that are not needed in the rest of the program.
In particular, expression \ensuremath{\textbf{wken}(\overline{x}, \Varid{e})} has the same type as \ensuremath{\Varid{e}} if
variables \ensuremath{\overline{x}} are not free in \ensuremath{\Varid{e}}, see the formal typing rule
\srule{WkenType} in Figure \ref{fig:wken}.
At run-time, the expression \ensuremath{\textbf{wken}(\overline{x}, \Varid{e})} evaluates \ensuremath{\Varid{e}} in an
environment from which variables \ensuremath{\overline{x}} have been dropped, so that they
do not get captured in any closure created during the execution of
\ensuremath{\Varid{e}}.
Rule \srule{Wken} is part of the pure semantics of \ensuremath{\CG}---the
semantics of \ensuremath{\FG} includes an analogous rule.

%
We remark that this expedient is not essential---we can avoid it by
slightly complicating the correctness statement to explicitly account
for those extra variables. Nor is this expedient particularly
interesting. In fact, we omit \ensuremath{\textbf{wken}} from the code of the program
translations to avoid clutter (our mechanization includes \ensuremath{\textbf{wken}} in
the appropriate places).

\section{References}

\begin{figure}[t]
\centering
\begin{subfigure}[t]{0.35\linewidth}
\setlength{\parindent}{0pt}
\setlength{\mathindent}{0pt}
\begin{hscode}\SaveRestoreHook
\column{B}{@{}>{\hspre}l<{\hspost}@{}}%
\column{3}{@{}>{\hspre}l<{\hspost}@{}}%
\column{5}{@{}>{\hspre}l<{\hspost}@{}}%
\column{11}{@{}>{\hspre}l<{\hspost}@{}}%
\column{E}{@{}>{\hspre}l<{\hspost}@{}}%
\>[B]{}{\llangle}\mathbf{new}(\Varid{e}\mkern1mu){\rrangle}\mathrel{=}{}\<[E]%
\\
\>[B]{}\hsindent{3}{}\<[3]%
\>[3]{}\mathbf{toLabeled}(\;\mathbf{do}{}\<[E]%
\\
\>[3]{}\hsindent{2}{}\<[5]%
\>[5]{}\Varid{lv}\leftarrow {\llangle}\Varid{e}{\rrangle}{}\<[E]%
\\
\>[3]{}\hsindent{2}{}\<[5]%
\>[5]{}\mathbf{new}(\Varid{lv}\mkern1mu)\mkern1mu){}\<[E]%
\\[\blanklineskip]%
\>[B]{}{\llangle}\mathbin{!}\Varid{e}{\rrangle}\mathrel{=}{}\<[E]%
\\
\>[B]{}\hsindent{3}{}\<[3]%
\>[3]{}\mathbf{toLabeled}(\;\mathbf{do}{}\<[E]%
\\
\>[3]{}\hsindent{2}{}\<[5]%
\>[5]{}\Varid{lr}\leftarrow {\llangle}\Varid{e}{\rrangle}{}\<[E]%
\\
\>[3]{}\hsindent{2}{}\<[5]%
\>[5]{}{r}\leftarrow {}\<[11]%
\>[11]{}\mathbf{unlabel}(\Varid{lr}\mkern1mu){}\<[E]%
\\
\>[3]{}\hsindent{2}{}\<[5]%
\>[5]{}\mathbin{!}{r}\mkern1mu){}\<[E]%
\ColumnHook
\end{hscode}\resethooks
\end{subfigure}%
\begin{subfigure}[t]{0.35\linewidth}
\setlength{\parindent}{0pt}
\setlength{\mathindent}{0pt}
\begin{hscode}\SaveRestoreHook
\column{B}{@{}>{\hspre}l<{\hspost}@{}}%
\column{3}{@{}>{\hspre}l<{\hspost}@{}}%
\column{5}{@{}>{\hspre}l<{\hspost}@{}}%
\column{11}{@{}>{\hspre}l<{\hspost}@{}}%
\column{12}{@{}>{\hspre}l<{\hspost}@{}}%
\column{E}{@{}>{\hspre}l<{\hspost}@{}}%
\>[B]{}{\llangle}\Varid{e}_{1}\mathbin{:=}\Varid{e}_{2}{\rrangle}\mathrel{=}{}\<[E]%
\\
\>[B]{}\hsindent{3}{}\<[3]%
\>[3]{}\mathbf{toLabeled}(\;\mathbf{do}{}\<[E]%
\\
\>[3]{}\hsindent{2}{}\<[5]%
\>[5]{}\Varid{lr}\leftarrow {\llangle}\Varid{e}_{1}{\rrangle}{}\<[E]%
\\
\>[3]{}\hsindent{2}{}\<[5]%
\>[5]{}\Varid{lv}\leftarrow {\llangle}\Varid{e}_{2}{\rrangle}{}\<[E]%
\\
\>[3]{}\hsindent{2}{}\<[5]%
\>[5]{}{r}\leftarrow \mathbf{unlabel}(\Varid{lr}\mkern1mu){}\<[E]%
\\
\>[3]{}\hsindent{2}{}\<[5]%
\>[5]{}{r}\mathbin{:=}\Varid{lv}\mkern1mu){}\<[E]%
\\
\>[B]{}\hsindent{3}{}\<[3]%
\>[3]{}\mathbf{toLabeled}(\mathbf{return}(\mkern1mu)\mkern1mu){}\<[E]%
\\[\blanklineskip]%
\>[B]{}{\llangle}\mathbf{labelOfRef}(\Varid{e}\mkern1mu){\rrangle}\mathrel{=}{}\<[E]%
\\
\>[B]{}\hsindent{3}{}\<[3]%
\>[3]{}\mathbf{toLabeled}(\;\mathbf{do}{}\<[E]%
\\
\>[3]{}\hsindent{2}{}\<[5]%
\>[5]{}\Varid{lr}\leftarrow {}\<[12]%
\>[12]{}{\llangle}\Varid{e}{\rrangle}{}\<[E]%
\\
\>[3]{}\hsindent{2}{}\<[5]%
\>[5]{}{r}\leftarrow {}\<[11]%
\>[11]{}\mathbf{unlabel}(\Varid{lr}\mkern1mu){}\<[E]%
\\
\>[3]{}\hsindent{2}{}\<[5]%
\>[5]{}\mathbf{labelOfRef}({r}\mkern1mu)\mkern1mu){}\<[E]%
\ColumnHook
\end{hscode}\resethooks
\end{subfigure}%
\caption{Translation \ensuremath{\FG} to \ensuremath{\CG} (references).}
\label{fig:fg2cg-mem}
\end{figure}

Figure \ref{fig:fg2cg-mem} shows the program translation of \ensuremath{\FG}
primitives that access the store and the heap via references.
Notice that these translations are the same for flow-insensitive and
flow-sensitive references: they replicate the behavior of \ensuremath{\FG}
primitives for each kind of reference using the corresponding
primitives of \ensuremath{\CG}.
Below we explain the translation for flow-insensitive references, the
discussion for flow-sensitive references is analogous.
The translation of \ensuremath{\FG} values wraps references in \ensuremath{\CG} labeled values
(Figure \ref{fig:fg2cg-value}), so the translations of Figure
\ref{fig:fg2cg-mem} take care of boxing and unboxing
references.
%
%
%
The translation of \ensuremath{\mathbf{new}(\Varid{e}\mkern1mu)} has a top-level \ensuremath{\mathbf{toLabeled}(\cdot \mkern1mu)} block that simply
translates the content (\ensuremath{\Varid{lv}\leftarrow {\llangle}\Varid{e}{\rrangle}}) and puts it in a new
reference (\ensuremath{\mathbf{new}(\Varid{lv}\mkern1mu)}).
The \ensuremath{\CG} rule \srule{New} (Figure \ref{fig:cg-memory}) assigns the
label of the translated content to the new reference, which also gets
labeled with the original program counter label\footnote{The nested
  block does not execute any \ensuremath{\mathbf{unlabel}(\cdot \mkern1mu)} nor \ensuremath{\mathbf{taint}(\cdot \mkern1mu)}.}, just as in the
\ensuremath{\FG} rule \srule{New} (Figure \ref{fig:fg-memory}).
In \ensuremath{\FG}, rule \srule{Read} reads from a reference \ensuremath{{{\Varid{n}}_{{\ell}}}^{{\ell}'}} at security level
\ensuremath{{\ell}'} that points to the \ensuremath{{\ell}}-labeled memory, and returns the content \ensuremath{{\Sigma({\ell}){[}\Varid{n}{]}}^{{\ell}\;\lub\;{\ell}'}} at level \ensuremath{{\ell}\;\lub\;{\ell}'}.
Similarly, the translation creates a \ensuremath{\mathbf{toLabeled}(\cdot \mkern1mu)} block that executes to get a labeled
reference \ensuremath{\Varid{lr}\mathrel{=}\mathbf{Labeled}\;{\ell}'\;{\Varid{n}}_{{\ell}}}, extracts the reference \ensuremath{{\Varid{n}}_{{\ell}}}
(\ensuremath{{r}\leftarrow \mathbf{unlabel}(\Varid{lr}\mkern1mu)}) tainting the program counter
label with \ensuremath{{\ell}'}, and then reads the reference's content further tainting the
program counter label with \ensuremath{{\ell}} as well.
The code that translates and updates a reference consists of two
\ensuremath{\mathbf{toLabeled}(\cdot \mkern1mu)} blocks.
The first block is responsible for the update: it extracts the labeled
reference and the labeled new content (\ensuremath{\Varid{lr}} and \ensuremath{\Varid{lv}} resp.), extracts
the reference from the labeled value (\ensuremath{{r}\leftarrow \mathbf{unlabel}(\Varid{lr}\mkern1mu)}) and updates
it (\ensuremath{{r}\mathbin{:=}\Varid{lv}}).
The second block, \ensuremath{\mathbf{toLabeled}(\mathbf{return}(\mkern1mu)\mkern1mu)}, returns unit at
security level \ensuremath{\Varid{pc}}, i.e., \ensuremath{\mathbf{Labeled}\;\Varid{pc}\;()}, similar to the \ensuremath{\FG} rule
\srule{Write}.
The translation of \ensuremath{\mathbf{labelOfRef}(\Varid{e}\mkern1mu)} extracts the reference and projects its label via
the \ensuremath{\CG} primitive \ensuremath{\mathbf{labelOfRef}(\cdot \mkern1mu)}, which additionally taints the program
counter with the label itself, similar to the \ensuremath{\FG} rule \srule{LabelOfRef}.


\section{Correctness}
\label{subsec:fg2cg-correct}
In this section, we establish some desirable properties of the
\ensuremath{\FG}-to-\ensuremath{\CG} translation defined above.
These properties include type and semantics preservation as well as
recovery of non-interference---a meta criterion that rules out a class
of semantically correct (semantics preserving), yet elusive
translations that do not preserve the meaning of security labels
\parencite{Rajani:2018,BartheRB07}.

We start by showing that the program translation preserves typing.
The type translation for typing contexts \ensuremath{\Gamma} is pointwise,
i.e., \ensuremath{{\llangle}\Gamma{\rrangle}\mathrel{=}\lambda \Varid{x}.{\llangle}\Gamma(\Varid{x}){\rrangle}}.

\begin{lemma}[Type Preservation]
  Given a well-typed \ensuremath{\FG} expression, i.e., \ensuremath{\Gamma\vdash\Varid{e}\mathbin{:}\tau},
  the translated \ensuremath{\CG} expression is also well-typed, i.e., \ensuremath{{\llangle}\Gamma{\rrangle}\vdash{\llangle}\Varid{e}{\rrangle}\mathbin{:}\mathbf{LIO}\xspace\;\thinspace\mkern-3mu{\llangle}\tau{\rrangle}}.
\end{lemma}
\begin{proof} By induction on the given typing derivation.
\end{proof}

The main correctness criterion for the translation is semantics preservation.
Intuitively, proving this theorem ensures that the program
translation preserves the meaning of secure \ensuremath{\FG} programs when
translated and executed with \ensuremath{\CG} semantics (under a translated
environment).
In the theorem below, the translation of stores, memories, and heaps
is pointwise, i.e., \ensuremath{{\llangle}\Sigma{\rrangle}\mathrel{=}\lambda {\ell}.{\llangle}\Sigma({\ell}){\rrangle}}, and \ensuremath{{\llangle}[\mskip1.5mu \mskip1.5mu]{\rrangle}\mathrel{=}[\mskip1.5mu \mskip1.5mu]} and \ensuremath{{\llangle}{r}\mathbin{:}\Conid{M}{\rrangle}\mathrel{=}{\llangle}{r}{\rrangle}\mathbin{:}{\llangle}\Conid{M}{\rrangle}} for each
\ensuremath{{\ell}}-labeled memory \ensuremath{\Conid{M}}, and \ensuremath{{\llangle}[\mskip1.5mu \mskip1.5mu]{\rrangle}\mathrel{=}[\mskip1.5mu \mskip1.5mu]} and \ensuremath{{\llangle}{{r}}^{{\ell}}\mathbin{:}\mu{\rrangle}\mathrel{=}{\llangle}{{r}}^{{\ell}}{\rrangle}\mathbin{:}{\llangle}\mu{\rrangle}} for heaps \ensuremath{\mu}.
Furthermore, notice that in the translation, the initial and final
program counter labels are the same. This establishes that the program
translation preserves the flow-insensitive program counter label of
\ensuremath{\FG} (by means of primitive \ensuremath{\mathbf{toLabeled}(\cdot \mkern1mu)}).

\begin{thm}[Semantics Preservation of \ensuremath{{\llangle}\cdot {\rrangle}\mathbin{:}\FG\to \CG} ]\hfill
\label{thm:fg2cg-correct}
For all well-typed \ensuremath{\FG} programs \ensuremath{\Varid{e}}, if \ensuremath{\langle\Sigma,\mu,\Varid{e}\rangle\Downarrow^{\theta}_{\Varid{pc}}\langle\Sigma',\mu',{v}\rangle}, then \ensuremath{\langle{\llangle}\Sigma{\rrangle},{\llangle}\mu{\rrangle},\Varid{pc},{\llangle}\Varid{e}{\rrangle}\rangle\Downarrow^{{\llangle}\theta{\rrangle}}\langle{\llangle}\Sigma'{\rrangle},{\llangle}\mu'{\rrangle},\Varid{pc},{\llangle}{v}{\rrangle}\rangle}.
\end{thm}
\begin{proof} By induction on the given evaluation derivation using
  basic properties of the security lattice and of the translation
  function.\footnote{In our mechanized proofs, this proof also
    requires the (often used) axiom of functional extensionality.}
\end{proof}

\section{Recovery of Non-Interference}
We conclude this section by constructing a proof of
termination-insensitive non-interference for \ensuremath{\FG}
(Theorem~\ref{thm:fg-tini-bij}) from the corresponding theorem for
\ensuremath{\CG} (Theorem~\ref{thm:cg-tini-bij}), using the semantics preserving
translation (Theorem~\ref{thm:fg2cg-correct}), together with a
property that the translation preserves \ensuremath{\Blue{\Conid{L}}}-equivalence
(Lemma~\ref{lm:fg2cg-lift-low-eq-iff}), the validity of references
(Lemma~\ref{lemma:lift-valid-fg2cg}), as well as a property to recover
source \ensuremath{\Blue{\Conid{L}}}-equivalence from target \ensuremath{\Blue{\Conid{L}}}-equivalence
(Lemma~\ref{lm:fg2cg-unlift-low-eq}).
This result ensures that the meaning of labels is preserved by the
translation~\parencite{Rajani:2018,BartheRB07}.
In the absence of such an artifact, one could devise a semantics-preserving translation that
simply does not use the security features of the target
language. While technically correct (i.e., semantics preserving), the
translation would not be meaningful from the perspective of security.%
\footnote{Note that such bogus translations are also ruled out due to the need to preserve the outcome of any label introspection. Nonetheless, building this proof artifact increases our confidence in the robustness of our translation. In contrast, if the enforcement of IFC is \emph{static}, then there is no label introspection, and this proof artifact is extremely important, as argued in prior work~\parencite{Rajani:2018,BartheRB07}.}
%
%
%
%
The following lemma shows that the translation of \ensuremath{\FG} \emph{initial}
configurations, defined as \ensuremath{{{\llangle}\Varid{c}{\rrangle}}^{\Varid{pc}}\mathrel{=}\langle{\llangle}\Sigma{\rrangle},{\llangle}\mu{\rrangle},\Varid{pc},{\llangle}\Varid{e}{\rrangle}\rangle} if \ensuremath{\Varid{c}\mathrel{=}\langle\Sigma,\mu,\Varid{e}\rangle}, preserves \ensuremath{\Blue{\Conid{L}}}-equivalence by lifting
\ensuremath{\Blue{\Conid{L}}}-equivalence from source to target and back.
Since the translation preserves the address of references and
the size of the heap, the lemma relates the target configurations
using the same bijection that relates the source configurations.

%
%
\begin{lemma}[\ensuremath{{\llangle}\cdot {\rrangle}} preserves \ensuremath{\approx_{\Blue{\Conid{L}}}^{\beta}}]
  \label{lm:fg2cg-lift-low-eq-iff}
  For all program counter labels \ensuremath{\Varid{pc}} and bijections \ensuremath{\beta}, \ensuremath{\Varid{c}_{1}\;\approx_{\Blue{\Conid{L}}}^{\beta}\;\Varid{c}_{2}} if and only if \ensuremath{{{\llangle}\Varid{c}_{1}{\rrangle}}^{\Varid{pc}}\;\approx_{\Blue{\Conid{L}}}^{\beta}\;{{\llangle}\Varid{c}_{2}{\rrangle}}^{\Varid{pc}}}.
\end{lemma}
\begin{proof} By definition of \ensuremath{\Blue{\Conid{L}}}-equivalence for initial
configurations in both directions (Sections \ref{subsec:fg-security}
and \ref{subsec:cg-security}), using injectivity of the translation
function, i.e., if \ensuremath{{\llangle}\Varid{e}_{1}{\rrangle}\equiv_{\alpha}{\llangle}\Varid{e}_{2}{\rrangle}} then \ensuremath{\Varid{e}_{1}\equiv_{\alpha}\Varid{e}_{2}}, in the \emph{if} direction, and by mutually proving similar
lemmas for all categories.
\end{proof}

The following lemma recovers \ensuremath{\Blue{\Conid{L}}}-equivalence of \emph{source} final
configurations by back-translating \ensuremath{\Blue{\Conid{L}}}-equivalence of \emph{target}
final configurations.
We define the translation for \ensuremath{\FG} \emph{final} configurations as \ensuremath{{{\llangle}\Varid{c}{\rrangle}}^{\Varid{pc}}\mathrel{=}\langle{\llangle}\Sigma{\rrangle},{\llangle}\mu{\rrangle},\Varid{pc},{\llangle}{v}{\rrangle}\rangle} if \ensuremath{\Varid{c}\mathrel{=}\langle\Sigma,\mu,{v}\rangle}.

\begin{lemma}[\ensuremath{\approx_{\Blue{\Conid{L}}}^{\beta}} recovery via \ensuremath{{\llangle}\cdot {\rrangle}}]
\label{lm:fg2cg-unlift-low-eq}
Let \ensuremath{\Varid{c}_{1}\mathrel{=}\langle\Sigma_1,\mu_{1},{{r}_{1}}^{{\ell}_{1}}\rangle}, \ensuremath{\Varid{c}_{2}\mathrel{=}\langle\Sigma_2,\mu_{2},{{r}_{2}}^{{\ell}_{2}}\rangle} be \ensuremath{\FG} final configurations. For all bijections \ensuremath{\beta} and
program counter label \ensuremath{\Varid{pc}}, such that \ensuremath{\Varid{pc}\;\flows\;{\ell}_{1}} and \ensuremath{\Varid{pc}\;\flows\;{\ell}_{2}}, if \ensuremath{{{\llangle}\Varid{c}_{1}{\rrangle}}^{\Varid{pc}}\approx_{\Blue{\Conid{L}}}^{\beta}{{\llangle}\Varid{c}_{2}{\rrangle}}^{\Varid{pc}}} then \ensuremath{\Varid{c}_{1}\approx_{\Blue{\Conid{L}}}^{\beta}\Varid{c}_{2}}.
\end{lemma}
\begin{proof}
  By case analysis on the \ensuremath{\Blue{\Conid{L}}}-equivalence relation of the target
  final configurations, two cases follow. First, through \ensuremath{{\llangle}\Varid{c}_{1}{\rrangle}\approx_{\Blue{\Conid{L}}}^{\beta}{\llangle}\Varid{c}_{2}{\rrangle}} we recover \ensuremath{\Blue{\Conid{L}}}-equivalence of the source stores, i.e.,
  \ensuremath{\Sigma_1\approx_{\Blue{\Conid{L}}}^{\beta}\Sigma_2}, from that of the target stores, i.e., \ensuremath{{\llangle}\Sigma_1{\rrangle}\approx_{\Blue{\Conid{L}}}^{\beta}{\llangle}\Sigma_2{\rrangle}}, and of the source heaps, i.e., \ensuremath{\mu_{1}\approx_{\Blue{\Conid{L}}}^{\beta}\mu_{2}},
  from the target heaps \ensuremath{{\llangle}\mu_{1}{\rrangle}\approx_{\Blue{\Conid{L}}}^{\beta}{\llangle}\mu_{2}{\rrangle}}.
  Then, the program counter in the target configurations is either (i)
  \emph{above} the attacker's level \srule{Pc\ensuremath{{}_{\Red{\Conid{H}}}}}, i.e., \ensuremath{\Varid{pc}\;\not\flows\;\Blue{\Conid{L}}}, and the source values are \ensuremath{\Blue{\Conid{L}}}-equivalent, i.e.,
  \ensuremath{{{r}_{1}}^{{\ell}_{1}}\approx_{\Blue{\Conid{L}}}^{\beta}{{r}_{2}}^{{\ell}_{2}}} by rule \srule{Value\ensuremath{{}_{\Red{\Conid{H}}}}} applied to
  \ensuremath{{\ell}_{1}\;\not\flows\;\Blue{\Conid{L}}} and \ensuremath{{\ell}_{2}\;\not\flows\;\Blue{\Conid{L}}} (from \ensuremath{\Varid{pc}\;\not\flows\;\Blue{\Conid{L}}}
  and, respectively, \ensuremath{\Varid{pc}\;\flows\;{\ell}_{1}} and \ensuremath{\Varid{pc}\;\flows\;{\ell}_{2}}), or
  (ii) \emph{below} the attacker's level \srule{Pc\ensuremath{{}_{\Blue{\Conid{L}}}}}, i.e.,
  \ensuremath{\Varid{pc}\;\flows\;\Blue{\Conid{L}}}, then \ensuremath{{\llangle}{{r}_{1}}^{{\ell}_{1}}{\rrangle}\approx_{\Blue{\Conid{L}}}^{\beta}{\llangle}{{r}_{2}}^{{\ell}_{2}}{\rrangle}} and the source values are \ensuremath{\Blue{\Conid{L}}}-equivalent, i.e., \ensuremath{{{r}_{1}}^{{\ell}_{1}}\approx_{\Blue{\Conid{L}}}^{\beta}{{r}_{2}}^{{\ell}_{2}}}, by Lemma \ref{lm:fg2cg-lift-low-eq-iff} for
  values.
\end{proof}

Recall that non-interference for \ensuremath{\CG} extended with flow-sensitive
(Theorem~\ref{thm:cg-tini-bij}) requires side conditions about the
validity of program inputs, which must contain only valid heap
addresses (see the \ensuremath{\vdash \mathbf{Valid}(\Varid{c},\theta)} judgment in
Figure~\ref{fig:cg-valid}).
Thus, to recover non-interference for \ensuremath{\FG} through non-interference
for \ensuremath{\CG}, we need to instantiate these judgments in the proof.
%
Luckily, non-interference for \ensuremath{\FG} also requires similar side
conditions (Fig.~\ref{fig:fg-valid}), so we can fulfill these
assumptions by lifting the valid judgment for the source inputs into a
valid judgment for the translated inputs.
%

\begin{lemma}[\ensuremath{{\llangle}\cdot {\rrangle}} preserves \ensuremath{\vdash \mathbf{Valid}(\cdot )}]
\label{lemma:lift-valid-fg2cg}
  For all program counters \ensuremath{\Varid{pc}}, initial configurations \ensuremath{\Varid{c}} and environments \ensuremath{\theta}, if
  \ensuremath{\vdash \mathbf{Valid}(\Varid{c},\theta)}, then \ensuremath{\vdash \mathbf{Valid}({{\llangle}\Varid{c}{\rrangle}}^{\Varid{pc}},{\llangle}\theta{\rrangle})}.
\end{lemma}

Finally, we recover termination insensitive non-interference for \ensuremath{\FG}
through our semantics preserving translation to \ensuremath{\CG}.
Notice that the theorem statement below is identical to
Theorem~\ref{thm:fg-tini-bij}, what is new is the proof of the
theorem, which relies on non-interference for \ensuremath{\CG} and our verified
translation.

\begin{thm}[\ensuremath{\FG}-TINI with Bijections via \ensuremath{{\llangle}\cdot {\rrangle}}]
\label{thm:recovery-fg2cg}
For all program counter labels \ensuremath{\Varid{pc}} and valid inputs \ensuremath{\vdash \mathbf{Valid}(\Varid{c}_{1},\theta_{1})} and \ensuremath{\vdash \mathbf{Valid}(\Varid{c}_{2},\theta_{2})}, such that \ensuremath{\Varid{c}_{1}\;\approx_{\Blue{\Conid{L}}}^{\beta}\;\Varid{c}_{2}},
\ensuremath{\theta_{1}\;\approx_{\Blue{\Conid{L}}}^{\beta}\;\theta_{2}}, if \ensuremath{\Varid{c}_{1}\;\Downarrow^{\theta_{1}}_{\Varid{pc}}\;\Varid{c}_{1}'}, and \ensuremath{\Varid{c}_{2}\;\Downarrow^{\theta_{2}}_{\Varid{pc}}\;\Varid{c}_{2}'}, then there exists an extended bijection \ensuremath{\beta'\;\supseteq\;\beta}, such that \ensuremath{\Varid{c}_{1}'\;\approx_{\Blue{\Conid{L}}}^{\beta'}\;\Varid{c}_{2}'}.
\end{thm}
\begin{proof}
We start by applying the fine to coarse grained program translation to
the initial configurations and environments.
By Theorem \ref{thm:fg2cg-correct} (semantics preservation), we derive
the corresponding \ensuremath{\CG} reductions, i.e., \ensuremath{{{\llangle}\Varid{c}_{1}{\rrangle}}^{\Varid{pc}}\;\Downarrow^{{\llangle}\theta_{1}{\rrangle}}\;{{\llangle}\Varid{c}_{1}'{\rrangle}}^{\Varid{pc}}} and \ensuremath{{{\llangle}\Varid{c}_{2}{\rrangle}}^{\Varid{pc}}\;\Downarrow^{{\llangle}\theta_{2}{\rrangle}}\;{{\llangle}\Varid{c}_{2}'{\rrangle}}^{\Varid{pc}}}.
Then, we lift \ensuremath{\Blue{\Conid{L}}}-equivalence of the initial configurations and
environments from \emph{source} to \emph{target}, i.e., from \ensuremath{\Varid{c}_{1}\approx_{\Blue{\Conid{L}}}^{\beta}\Varid{c}_{2}} to \ensuremath{{{\llangle}\Varid{c}_{1}{\rrangle}}^{\Varid{pc}}\approx_{\Blue{\Conid{L}}}^{\beta}{{\llangle}\Varid{c}_{2}{\rrangle}}^{\Varid{pc}}} and from \ensuremath{\theta_{1}\approx_{\Blue{\Conid{L}}}^{\beta}\theta_{2}} to \ensuremath{{\llangle}\theta_{1}{\rrangle}\approx_{\Blue{\Conid{L}}}^{\beta}{\llangle}\theta_{2}{\rrangle}}
(Lemma~\ref{lm:fg2cg-lift-low-eq-iff}), and similarly we lift the
validity judgments (Lemma~\ref{lemma:lift-valid-fg2cg}), i.e., \ensuremath{\vdash \mathbf{Valid}({\llangle}\Varid{c}_{1}{\rrangle},{\llangle}\theta_{1}{\rrangle})} and \ensuremath{\vdash \mathbf{Valid}({\llangle}\Varid{c}_{2}{\rrangle},{\llangle}\theta_{2}{\rrangle})} from \ensuremath{\vdash \mathbf{Valid}(\Varid{c}_{1},\theta_{1})} and \ensuremath{\vdash \mathbf{Valid}(\Varid{c}_{2},\theta_{2})}, respectively.
Then, we apply \ensuremath{\CG}-\emph{TINI with Bijections}
(Theorem~\ref{thm:cg-tini-bij}) and obtain \ensuremath{\Blue{\Conid{L}}}-equivalence of the
\emph{target} final configurations, i.e., \ensuremath{{{\llangle}\Varid{c}_{1}'{\rrangle}}^{\Varid{pc}}\;\approx_{\Blue{\Conid{L}}}^{\beta'}\;{{\llangle}\Varid{c}_{2}'{\rrangle}}^{\Varid{pc}}} for some bijection \ensuremath{\beta'\;\supseteq\;\beta}.
Finally, we recover \ensuremath{\Blue{\Conid{L}}}-equivalence of the final configurations from
\emph{target} to \emph{source}, i.e., from \ensuremath{{{\llangle}\Varid{c}_{1}'{\rrangle}}^{\Varid{pc}}\;\approx_{\Blue{\Conid{L}}}^{\beta'}\;{{\llangle}\Varid{c}_{2}'{\rrangle}}^{\Varid{pc}}} to \ensuremath{\Varid{c}_{1}'\;\approx_{\Blue{\Conid{L}}}^{\beta'}\;\Varid{c}_{2}'}, via Lemma
\ref{lm:fg2cg-unlift-low-eq}, applied to \ensuremath{\Varid{c}_{1}'\mathrel{=}\langle\anonymous ,\anonymous ,{{r}_{1}}^{{\ell}_{1}}\rangle} and \ensuremath{\Varid{c}_{2}'\mathrel{=}\langle\anonymous ,\anonymous ,{{r}_{2}}^{{\ell}_{2}}\rangle}, and where \ensuremath{\Varid{pc}\;\flows\;{\ell}_{1}} and
\ensuremath{\Varid{pc}\;\flows\;{\ell}_{2}} by Property \ref{prop:fg-pc-raise} applied to the
source reductions, i.e., \ensuremath{\Varid{c}_{1}\;\Downarrow^{\theta_{1}}_{\Varid{pc}}\;\Varid{c}_{1}'} and \ensuremath{\Varid{c}_{2}\;\Downarrow^{\theta_{2}}_{\Varid{pc}}\;\Varid{c}_{2}'}.
\end{proof}



\chapter{Coarse- to Fine-Grained Program Translation}
\label{sec:cg2fg}

We now show a verified program translation in the
opposite direction---from the coarse grained calculus \ensuremath{\CG} to the fine
grained calculus \ensuremath{\FG}.
The translation in this direction is more involved---a program in
\ensuremath{\FG} contains strictly more information than its counterpart in \ensuremath{\CG},
namely the extra \emph{intrinsic} label annotations that tag every value.
The challenge in constructing this translation is
two-fold.
On one hand, the translation must come up with labels for all values.
However, it is not always possible to do this statically during the
translation: Often, the labels depend on input values and arise at
run-time with intermediate results since the \ensuremath{\FG} calculus is designed
to compute and attach labels at run-time.
On the other hand, the translation cannot conservatively
under-approximate the values of labels\footnote{In contrast, previous
  work on \emph{static} type-based fine-to-coarse grained translation
  safely under-approximates the label annotations in types with the
  bottom label of the lattice~\parencite{Rajani:2018}. The proof of
  type preservation of the translation recovers the actual labels via
  \emph{subtyping}.}---\ensuremath{\CG} and \ensuremath{\FG} have label introspection so, in
order to get semantics preservation, labels must be preserved
precisely.
Intuitively, we solve this impasse by crafting a program
translation that (i) preserves the labels that can
be inspected by \ensuremath{\CG} and (ii) lets the \ensuremath{\FG} semantics compute the
remaining label annotations automatically---we account for those
labels with a \emph{cross-language} relation that represents semantic
equivalence between \ensuremath{\FG} and \ensuremath{\CG} modulo extra annotations
(Section~\ref{subsec:semantics-eq}).
%
The fact that the source program in \ensuremath{\CG} cannot inspect those
labels---they have no value counterpart in the source \ensuremath{\CG}
program---facilitates this aspect of the translation.
We elaborate more on the technical details later.
%

%
%
%

%
At a high level, an interesting aspect of the translation (that
informally attests that it is indeed semantics-preserving) is that it
encodes the \emph{flow-sensitive} program counter of the source \ensuremath{\CG}
program into the label annotation of the \ensuremath{\FG} value that results from
executing the translated program.
For example, if a \ensuremath{\CG} monadic expression starts with program counter
label \ensuremath{\Varid{pc}} and results in some value, say \ensuremath{\mathbf{true}}, and final program
counter \ensuremath{\Varid{pc}'}, then the translated \ensuremath{\FG} expression, starting with the
same program counter label \ensuremath{\Varid{pc}}, computes the \emph{same} value
(modulo extra label annotations) at the same security level \ensuremath{\Varid{pc}'},
i.e., the value \ensuremath{{\mathbf{true}}^{\Varid{pc}'}}.
This encoding requires keeping the value of the program counter label
in the source program synchronized with the program counter label in
the target program, by loosening the fine-grained precision of \ensuremath{\FG} at
run-time in a controlled way.

\section{Types and Values}
\begin{figure}[t]
\begin{subfigure}[b]{0.45\textwidth}
\setlength{\mathindent}{16pt}
\begin{hscode}\SaveRestoreHook
\column{B}{@{}>{\hspre}l<{\hspost}@{}}%
\column{E}{@{}>{\hspre}l<{\hspost}@{}}%
\>[B]{}{\llbracket\thinspace\mkern-2mu\xspace }\mathscr{L}{\thinspace\mkern-1mu\rrbracket}\mathrel{=}\mathscr{L}{}\<[E]%
\\
\>[B]{}{\llbracket\thinspace\mkern-2mu\xspace }\mathbf{unit}{\thinspace\mkern-1mu\rrbracket}\mathrel{=}\mathbf{unit}{}\<[E]%
\\
\>[B]{}{\llbracket\thinspace\mkern-2mu\xspace }\tau_{1}\to \tau_{2}{\thinspace\mkern-1mu\rrbracket}\mathrel{=}{\llbracket\thinspace\mkern-2mu\xspace }\tau_{1}{\thinspace\mkern-1mu\rrbracket}\to {\llbracket\thinspace\mkern-2mu\xspace }\tau_{2}{\thinspace\mkern-1mu\rrbracket}{}\<[E]%
\\
\>[B]{}{\llbracket\thinspace\mkern-2mu\xspace }\tau_{1}\mathbin{+}\tau_{2}{\thinspace\mkern-1mu\rrbracket}\mathrel{=}{\llbracket\thinspace\mkern-2mu\xspace }\tau_{1}{\thinspace\mkern-1mu\rrbracket}\mathbin{+}{\llbracket\thinspace\mkern-2mu\xspace }\tau_{2}{\thinspace\mkern-1mu\rrbracket}{}\<[E]%
\\
\>[B]{}{\llbracket\thinspace\mkern-2mu\xspace }\tau_{1}\;\times\;\tau_{2}{\thinspace\mkern-1mu\rrbracket}\mathrel{=}{\llbracket\thinspace\mkern-2mu\xspace }\tau_{1}{\thinspace\mkern-1mu\rrbracket}\times{\llbracket\thinspace\mkern-2mu\xspace }\tau_{2}{\thinspace\mkern-1mu\rrbracket}{}\<[E]%
\\
\>[B]{}{\llbracket\thinspace\mkern-2mu\xspace }\mathbf{Ref}\; s\;\tau{\thinspace\mkern-1mu\rrbracket}\mathrel{=}\mathbf{Ref}\; s\;\thinspace\mkern-3mu{\llbracket\thinspace\mkern-2mu\xspace }\tau{\thinspace\mkern-1mu\rrbracket}{}\<[E]%
\\
\>[B]{}{\llbracket\thinspace\mkern-2mu\xspace }\mathbf{Labeled}\;\tau{\thinspace\mkern-1mu\rrbracket}\mathrel{=}\mathscr{L}\;\negthickspace\mkern-5mu\;\times{\llbracket\thinspace\mkern-2mu\xspace }\tau{\thinspace\mkern-1mu\rrbracket}{}\<[E]%
\\
\>[B]{}{\llbracket\thinspace\mkern-2mu\xspace }\mathbf{LIO}\xspace\;\tau{\thinspace\mkern-1mu\rrbracket}\mathrel{=}\mathbf{unit}\to {\llbracket\thinspace\mkern-2mu\xspace }\tau{\thinspace\mkern-1mu\rrbracket}{}\<[E]%
\ColumnHook
\end{hscode}\resethooks
\caption{Types.}
\label{fig:cg2fg-ty}
\end{subfigure}%
\rulesep
\begin{subfigure}[b]{0.49\textwidth}
\setlength{\mathindent}{16pt}
\begin{hscode}\SaveRestoreHook
\column{B}{@{}>{\hspre}l<{\hspost}@{}}%
\column{E}{@{}>{\hspre}l<{\hspost}@{}}%
\>[B]{}\llbracket()\thinspace\mkern-2mu\rrbracket^{\Varid{pc}}\mathrel{=}{()}^{\Varid{pc}}{}\<[E]%
\\
\>[B]{}\llbracket{\ell}\thinspace\mkern-2mu\rrbracket^{\Varid{pc}}\mathrel{=}{{\ell}}^{\Varid{pc}}{}\<[E]%
\\
\>[B]{}\llbracket\mathbf{inl}({v}\mkern1mu)\thinspace\mkern-2mu\rrbracket^{\Varid{pc}}\mathrel{=}{\mathbf{inl}(\llbracket{v}\thinspace\mkern-2mu\rrbracket^{\Varid{pc}}\mkern1mu)}^{\Varid{pc}}{}\<[E]%
\\
\>[B]{}\llbracket\mathbf{inr}({v}\mkern1mu)\thinspace\mkern-2mu\rrbracket^{\Varid{pc}}\mathrel{=}{\mathbf{inr}(\llbracket{v}\thinspace\mkern-2mu\rrbracket^{\Varid{pc}}\mkern1mu)}^{\Varid{pc}}{}\<[E]%
\\
\>[B]{}\llbracket({v}_{1},{v}_{2})\thinspace\mkern-2mu\rrbracket^{\Varid{pc}}\mathrel{=}{(\llbracket{v}_{1}\thinspace\mkern-2mu\rrbracket^{\Varid{pc}},\llbracket{v}_{2}\thinspace\mkern-2mu\rrbracket^{\Varid{pc}})}^{\Varid{pc}}{}\<[E]%
\\
\>[B]{}\llbracket(\Varid{x}.\Varid{e},\theta)\thinspace\mkern-2mu\rrbracket^{\Varid{pc}}\mathrel{=}{(\Varid{x}.{\llbracket\thinspace\mkern-2mu\xspace }\Varid{e}{\thinspace\mkern-1mu\rrbracket},\llbracket\theta\thinspace\mkern-2mu\rrbracket^{\Varid{pc}})}^{\Varid{pc}}{}\<[E]%
\\
\>[B]{}\llbracket(\Varid{t},\theta)\thinspace\mkern-2mu\rrbracket^{\Varid{pc}}\mathrel{=}{(\anonymous .{\llbracket\thinspace\mkern-2mu\xspace }\Varid{t}{\thinspace\mkern-1mu\rrbracket},\llbracket\theta\thinspace\mkern-2mu\rrbracket^{\Varid{pc}})}^{\Varid{pc}}{}\<[E]%
\\
\>[B]{}\llbracket\mathbf{Labeled}\;{\ell}\;{v}\thinspace\mkern-2mu\rrbracket^{\Varid{pc}}\mathrel{=}{({{\ell}}^{{\ell}},\llbracket{v}\thinspace\mkern-2mu\rrbracket^{{\ell}})}^{\Varid{pc}}{}\<[E]%
\\
\>[B]{}\llbracket{\Varid{n}}_{{\ell}}\thinspace\mkern-2mu\rrbracket^{\Varid{pc}}\mathrel{=}{({\Varid{n}}_{{\ell}})}^{\Varid{pc}}{}\<[E]%
\\
\>[B]{}\llbracket\Varid{n}\thinspace\mkern-2mu\rrbracket^{\Varid{pc}}\mathrel{=}{\Varid{n}}^{\Varid{pc}}{}\<[E]%
\ColumnHook
\end{hscode}\resethooks
\caption{Values.}
\label{fig:cg2fg-val}
\end{subfigure}%
\caption{Translation from \ensuremath{\CG} to \ensuremath{\FG} (part I).}
\label{fig:cg2fg-val-ty}
\end{figure}

\paragraph{Types.}
The \ensuremath{\CG}-to-\ensuremath{\FG} translation follows the same type-driven
approach used in the other direction, starting from the function \ensuremath{\llbracket\cdot \thinspace\mkern-2mu\rrbracket}
in Figure \ref{fig:cg2fg-ty}, that translates a \ensuremath{\FG} type \ensuremath{\tau} into
the corresponding \ensuremath{\CG} type \ensuremath{{\llbracket\thinspace\mkern-2mu\xspace }\tau{\thinspace\mkern-1mu\rrbracket}}.
The translation is defined by induction on \ensuremath{\tau} and preserves all the type constructors standard types. Only the cases corresponding to \ensuremath{\CG}-specific types
are interesting.
In particular, it converts \emph{explicitly} labeled types, i.e.,
\ensuremath{\mathbf{Labeled}\;\tau}, to a standard pair type in \ensuremath{\FG}, i.e., \ensuremath{\mathscr{L}\;\negthickspace\mkern-3mu\;\times{\llbracket\thinspace\mkern-2mu\xspace }\tau{\thinspace\mkern-1mu\rrbracket}}, where the first component is the label and the
second component the content of type \ensuremath{\tau}.
Type \ensuremath{\mathbf{LIO}\xspace\;\tau} becomes a \emph{suspension} in \ensuremath{\FG}, i.e., the function
type \ensuremath{\mathbf{unit}\to {\llbracket\thinspace\mkern-2mu\xspace }\tau{\thinspace\mkern-1mu\rrbracket}} that delays a computation and that can be
forced by simply applying it to the unit value \ensuremath{()}.
%

\paragraph{Values.}
The translation of values follows the type translation,
as shown in Figure \ref{fig:cg2fg-val}.
Notice that the translation is indexed by the program counter label (the translation is written \ensuremath{\llbracket{v}\thinspace\mkern-2mu\rrbracket^{\Varid{pc}}}), which converts the \ensuremath{\CG} value
\ensuremath{{v}} in scope of a computation protected by security level \ensuremath{\Varid{pc}} to the
corresponding fully label-annotated \ensuremath{\FG} value.
The translation is pretty straightforward and uses the program counter
label to tag each value, following the \ensuremath{\CG} principle that
the program counter label protects every value in scope that
  is not explicitly labeled.
The translation converts a \ensuremath{\CG} function closure into a corresponding
\ensuremath{\FG} function closure by translating the body of the function to a
\ensuremath{\FG} expression (see below) and translating the environment pointwise, i.e.,
\ensuremath{\llbracket\theta\thinspace\mkern-2mu\rrbracket^{\Varid{pc}}\mathrel{=}\lambda \Varid{x}.\llbracket\theta(\Varid{x})\thinspace\mkern-2mu\rrbracket^{\Varid{pc}}}.
A thunk value or a \emph{thunk closure}, which denotes a
suspended side-effecful computation, is also converted into a
\ensuremath{\FG} function closure.
%
Technically, the translation would need to introduce a \emph{fresh
  variable} that would get bound to unit when the suspension gets
forced.
However, the argument to the suspension does not have any purpose,
so we do not bother with giving a name to it and write \ensuremath{\anonymous .{\llbracket\thinspace\mkern-2mu\xspace }\Varid{t}{\thinspace\mkern-1mu\rrbracket}} instead. (In our mechanized proofs we employ unnamed De Bruijn
indexes and this issue does not arise.)
%
The translation converts an explicitly labeled value \ensuremath{\mathbf{Labeled}\;{\ell}\;{v}}, into a labeled pair at security level \ensuremath{\Varid{pc}}, i.e., \ensuremath{{({{\ell}}^{{\ell}},\llbracket{v}\thinspace\mkern-2mu\rrbracket^{{\ell}})}^{\Varid{pc}}}. The pair consists of the label \ensuremath{{\ell}} tagged with itself, and the value translated at a security level equal to the
label annotation, i.e., \ensuremath{\llbracket{v}\thinspace\mkern-2mu\rrbracket^{{\ell}}}.
Notice that tagging the label with itself allows us to translate the
\ensuremath{\CG} (label introspection) primitive \ensuremath{\mathbf{labelOf}(\cdot \mkern1mu)} by simply projecting
the first component, thus preserving the label and its security level
across the translation.

\section{Expressions and Thunks}
\begin{figure}[t]
\begin{subfigure}[b]{0.5\textwidth}
\setlength{\parindent}{0pt}
\setlength{\mathindent}{16pt}
\begin{hscode}\SaveRestoreHook
\column{B}{@{}>{\hspre}l<{\hspost}@{}}%
\column{3}{@{}>{\hspre}l<{\hspost}@{}}%
\column{E}{@{}>{\hspre}l<{\hspost}@{}}%
\>[B]{}{\llbracket\thinspace\mkern-2mu\xspace }(){\thinspace\mkern-1mu\rrbracket}\mathrel{=}(){}\<[E]%
\\
\>[B]{}{\llbracket\thinspace\mkern-2mu\xspace }{\ell}{\thinspace\mkern-1mu\rrbracket}\mathrel{=}{\ell}{}\<[E]%
\\
\>[B]{}{\llbracket\thinspace\mkern-2mu\xspace }\Varid{x}{\thinspace\mkern-1mu\rrbracket}\mathrel{=}\Varid{x}{}\<[E]%
\\
\>[B]{}{\llbracket\thinspace\mkern-2mu\xspace }\lambda \Varid{x}.\Varid{e}{\thinspace\mkern-1mu\rrbracket}\mathrel{=}\lambda \Varid{x}.{\llbracket\thinspace\mkern-2mu\xspace }\Varid{e}{\thinspace\mkern-1mu\rrbracket}{}\<[E]%
\\
\>[B]{}{\llbracket\thinspace\mkern-2mu\xspace }\Varid{e}_{1}\;\Varid{e}_{2}{\thinspace\mkern-1mu\rrbracket}\mathrel{=}{\llbracket\thinspace\mkern-2mu\xspace }\Varid{e}_{1}{\thinspace\mkern-1mu\rrbracket}{\llbracket\thinspace\mkern-2mu\xspace }\Varid{e}_{2}{\thinspace\mkern-1mu\rrbracket}{}\<[E]%
\\
\>[B]{}{\llbracket\thinspace\mkern-2mu\xspace }(\Varid{e}_{1},\Varid{e}_{2}){\thinspace\mkern-1mu\rrbracket}\mathrel{=}({\llbracket\thinspace\mkern-2mu\xspace }\Varid{e}_{1}{\thinspace\mkern-1mu\rrbracket},{\llbracket\thinspace\mkern-2mu\xspace }\Varid{e}_{2}{\thinspace\mkern-1mu\rrbracket}){}\<[E]%
\\
\>[B]{}{\llbracket\thinspace\mkern-2mu\xspace }\mathbf{fst}(\Varid{e}\mkern1mu){\thinspace\mkern-1mu\rrbracket}\mathrel{=}\mathbf{fst}({\llbracket\thinspace\mkern-2mu\xspace }\Varid{e}{\thinspace\mkern-1mu\rrbracket}\mkern1mu){}\<[E]%
\\
\>[B]{}{\llbracket\thinspace\mkern-2mu\xspace }\mathbf{snd}(\Varid{e}\mkern1mu){\thinspace\mkern-1mu\rrbracket}\mathrel{=}\mathbf{snd}({\llbracket\thinspace\mkern-2mu\xspace }\Varid{e}{\thinspace\mkern-1mu\rrbracket}\mkern1mu){}\<[E]%
\\
\>[B]{}{\llbracket\thinspace\mkern-2mu\xspace }\mathbf{inl}(\Varid{e}\mkern1mu){\thinspace\mkern-1mu\rrbracket}\mathrel{=}\mathbf{inl}({\llbracket\thinspace\mkern-2mu\xspace }\Varid{e}{\thinspace\mkern-1mu\rrbracket}\mkern1mu){}\<[E]%
\\
\>[B]{}{\llbracket\thinspace\mkern-2mu\xspace }\mathbf{inr}(\Varid{e}\mkern1mu){\thinspace\mkern-1mu\rrbracket}\mathrel{=}\mathbf{inr}({\llbracket\thinspace\mkern-2mu\xspace }\Varid{e}{\thinspace\mkern-1mu\rrbracket}\mkern1mu){}\<[E]%
\\
\>[B]{}{\llbracket\thinspace\mkern-2mu\xspace }\mathbf{case}\;(\Varid{e},\Varid{x}.\Varid{e}_{1},\Varid{x}.\Varid{e}_{2}){\thinspace\mkern-1mu\rrbracket}{}\<[E]%
\\
\>[B]{}\hsindent{3}{}\<[3]%
\>[3]{}\mathrel{=}\mathbf{case}\;({\llbracket\thinspace\mkern-2mu\xspace }\Varid{e}{\thinspace\mkern-1mu\rrbracket},\Varid{x}.{\llbracket\thinspace\mkern-2mu\xspace }\Varid{e}_{1}{\thinspace\mkern-1mu\rrbracket},\Varid{x}.{\llbracket\thinspace\mkern-2mu\xspace }\Varid{e}_{2}{\thinspace\mkern-1mu\rrbracket}){}\<[E]%
\\
\>[B]{}{\llbracket\thinspace\mkern-2mu\xspace }\Varid{e}_{1}\;\flows^{?}\;\Varid{e}_{2}{\thinspace\mkern-1mu\rrbracket}\mathrel{=}{\llbracket\thinspace\mkern-2mu\xspace }\Varid{e}_{1}{\thinspace\mkern-1mu\rrbracket}\flows^{?}{\llbracket\thinspace\mkern-2mu\xspace }\Varid{e}_{2}{\thinspace\mkern-1mu\rrbracket}{}\<[E]%
\\
\>[B]{}{\llbracket\thinspace\mkern-2mu\xspace }\Varid{t}{\thinspace\mkern-1mu\rrbracket}\mathrel{=}\lambda \anonymous .{\llbracket\thinspace\mkern-2mu\xspace }\Varid{t}{\thinspace\mkern-1mu\rrbracket}{}\<[E]%
\ColumnHook
\end{hscode}\resethooks
\caption{Expressions.}
\label{fig:cg2fg-expr}
\end{subfigure}%
\rulesep
\begin{subfigure}[b]{0.44\textwidth}
\setlength{\parindent}{0pt}
\setlength{\mathindent}{12pt}
\begin{hscode}\SaveRestoreHook
\column{B}{@{}>{\hspre}l<{\hspost}@{}}%
\column{3}{@{}>{\hspre}l<{\hspost}@{}}%
\column{5}{@{}>{\hspre}l<{\hspost}@{}}%
\column{6}{@{}>{\hspre}l<{\hspost}@{}}%
\column{E}{@{}>{\hspre}l<{\hspost}@{}}%
\>[B]{}{\llbracket\thinspace\mkern-2mu\xspace }\mathbf{return}(\Varid{e}\mkern1mu){\thinspace\mkern-1mu\rrbracket}\mathrel{=}{\llbracket\thinspace\mkern-2mu\xspace }\Varid{e}{\thinspace\mkern-1mu\rrbracket}{}\<[E]%
\\
\>[B]{}{\llbracket\thinspace\mkern-2mu\xspace }\mathbf{bind}(\Varid{e}_{1},\Varid{x}.\Varid{e}_{2}\mkern1mu){\thinspace\mkern-1mu\rrbracket}\mathrel{=}{}\<[E]%
\\
\>[B]{}\hsindent{3}{}\<[3]%
\>[3]{}\mathbf{let}\;\Varid{x}\mathrel{=}{\llbracket\thinspace\mkern-2mu\xspace }\Varid{e}_{1}{\thinspace\mkern-1mu\rrbracket}\thinspace\mkern-3mu\;()\;\mathbf{in}{}\<[E]%
\\
\>[3]{}\hsindent{2}{}\<[5]%
\>[5]{}\mathbf{taint}(\mathbf{labelOf}(\Varid{x}\mkern1mu),{\llbracket\thinspace\mkern-2mu\xspace }\Varid{e}_{2}{\thinspace\mkern-1mu\rrbracket}()\mkern1mu){}\<[E]%
\\
\>[B]{}{\llbracket\thinspace\mkern-2mu\xspace }\mathbf{unlabel}(\Varid{e}\mkern1mu){\thinspace\mkern-1mu\rrbracket}\mathrel{=}{}\<[E]%
\\
\>[B]{}\hsindent{3}{}\<[3]%
\>[3]{}\mathbf{let}\;\Varid{x}\mathrel{=}{\llbracket\thinspace\mkern-2mu\xspace }\Varid{e}{\thinspace\mkern-1mu\rrbracket}\thinspace\mkern-3mu\;\mathbf{in}{}\<[E]%
\\
\>[3]{}\hsindent{3}{}\<[6]%
\>[6]{}\mathbf{taint}(\mathbf{fst}(\Varid{x}\mkern1mu),\mathbf{snd}(\Varid{x}\mkern1mu)\mkern1mu){}\<[E]%
\\
\>[B]{}{\llbracket\thinspace\mkern-2mu\xspace }\mathbf{toLabeled}(\Varid{e}\mkern1mu){\thinspace\mkern-1mu\rrbracket}\mathrel{=}{}\<[E]%
\\
\>[B]{}\hsindent{3}{}\<[3]%
\>[3]{}\mathbf{let}\;\Varid{x}\mathrel{=}{\llbracket\thinspace\mkern-2mu\xspace }\Varid{e}{\thinspace\mkern-1mu\rrbracket}\thinspace\mkern-3mu\;()\;\mathbf{in}{}\<[E]%
\\
\>[3]{}\hsindent{2}{}\<[5]%
\>[5]{}(\mathbf{labelOf}(\Varid{x}\mkern1mu),\Varid{x}){}\<[E]%
\\
\>[B]{}{\llbracket\thinspace\mkern-2mu\xspace }\mathbf{labelOf}(\Varid{e}\mkern1mu){\thinspace\mkern-1mu\rrbracket}\mathrel{=}\mathbf{fst}({\llbracket\thinspace\mkern-2mu\xspace }\Varid{e}{\thinspace\mkern-1mu\rrbracket}\mkern1mu){}\<[E]%
\\
\>[B]{}{\llbracket\thinspace\mkern-2mu\xspace }\mathbf{getLabel}{\thinspace\mkern-1mu\rrbracket}\mathrel{=}\mathbf{getLabel}{}\<[E]%
\\
\>[B]{}{\llbracket\thinspace\mkern-2mu\xspace }\mathbf{taint}(\Varid{e}\mkern1mu){\thinspace\mkern-1mu\rrbracket}\mathrel{=}\mathbf{taint}({\llbracket\thinspace\mkern-2mu\xspace }\Varid{e}{\thinspace\mkern-1mu\rrbracket},()\mkern1mu){}\<[E]%
\ColumnHook
\end{hscode}\resethooks

\caption{Thunks.}
\label{fig:cg2fg-thunk}
\end{subfigure}%
\caption{Translation from \ensuremath{\CG} to \ensuremath{\FG} (part II).}
\label{fig:cg2fg-expr-thunk}
\end{figure}
The translation of pure expressions (Figure~\ref{fig:cg2fg-expr}) is
trivial: it is homomorphic in all constructs, mirroring the type
translation.
The translation of a thunk expression \ensuremath{\Varid{t}} builds a suspension
explicitly with a \ensuremath{\lambda}-abstraction, i.e., \ensuremath{\lambda \anonymous .{\llbracket\thinspace\mkern-2mu\xspace }\Varid{t}{\thinspace\mkern-1mu\rrbracket}}, (the
name of the variable is again irrelevant, thus we omit it as explained
above), and translates the thunk itself according to the definition in
Figure~\ref{fig:cg2fg-thunk}.
%
The thunk \ensuremath{\mathbf{return}(\Varid{e}\mkern1mu)} becomes \ensuremath{{\llbracket\thinspace\mkern-2mu\xspace }\Varid{e}{\thinspace\mkern-1mu\rrbracket}}, since
\ensuremath{\mathbf{return}(\cdot \mkern1mu)} does not have any side-effect.
When two monadic computations are combined via \ensuremath{\mathbf{bind}(\Varid{e}_{1},\Varid{x}.\Varid{e}_{2}\mkern1mu)}, the translation (i) converts the first computation to a
suspension and forces it by applying unit (\ensuremath{{\llbracket\thinspace\mkern-2mu\xspace }\Varid{e}_{1}{\thinspace\mkern-1mu\rrbracket}\thinspace\mkern-3mu\;()}), (ii)
binds the result to \ensuremath{\Varid{x}} and passes it to the second
computation\footnote{Syntax \ensuremath{\mathbf{let}\;\Varid{x}\mathrel{=}\Varid{e}_{1}\;\mathbf{in}\;\Varid{e}_{2}} where \ensuremath{\Varid{x}} is free in
  \ensuremath{\Varid{e}_{2}} is a shorthand for \ensuremath{(\lambda \Varid{x}.\Varid{e}_{2})\;\Varid{e}_{1}}.}, which is also converted,
forced, and, \emph{importantly}, iii) executed with a program counter label
tainted with the security level of the result of the first
computation (\ensuremath{\mathbf{taint}(\mathbf{labelOf}(\Varid{x}\mkern1mu),{\llbracket\thinspace\mkern-2mu\xspace }\Varid{e}_{2}{\thinspace\mkern-1mu\rrbracket}\thinspace\mkern-3mu\;()\mkern1mu)}).
Notice that \ensuremath{\mathbf{taint}(\cdot \mkern1mu)} is essential to ensure that the second computation
executes with the program counter label set to the correct value---the
precision of the fine-grained system would otherwise retain the
initial lower program counter label according to rule \srule{App}
(Fig.~\ref{fig:fg-semantics}) and the value of the program counter
labels in the source and target programs would differ in the remaining
execution.
%

Similarly, the translation of \ensuremath{\mathbf{unlabel}(\Varid{e}\mkern1mu)} first translates the
labeled expression \ensuremath{\Varid{e}} (the translated expression does not need to be
forced because it is not of a monadic type), binds its result to \ensuremath{\Varid{x}}
and then projects the content in a context tainted with its label, as
in \ensuremath{\mathbf{taint}(\mathbf{fst}(\Varid{x}\mkern1mu),\mathbf{snd}(\Varid{x}\mkern1mu)\mkern1mu)}. This closely follows \ensuremath{\CG}'s
\srule{Unlabel} rule in Figure~\ref{fig:cg-thunk}.
The translation of \ensuremath{\mathbf{toLabeled}(\Varid{e}\mkern1mu)} forces the nested computation
with \ensuremath{{\llbracket\thinspace\mkern-2mu\xspace }\Varid{e}{\thinspace\mkern-1mu\rrbracket}\thinspace\;()}, binds its result to \ensuremath{\Varid{x}} and creates the pair
\ensuremath{(\mathbf{labelOf}(\Varid{x}\mkern1mu),\Varid{x})}, which corresponds to the labeled value obtained in
\ensuremath{\CG} via rule \srule{ToLabeled}.
Intuitively, the translation guarantees that the value of the final
program counter label in the nested computation coincides with the
security level of the translated result (bound to \ensuremath{\Varid{x}}). Therefore, the
first component contains the correct label and it is furthermore at
the right security level, because \ensuremath{\mathbf{labelOf}(\cdot \mkern1mu)} protects the projected
label with the label itself in \ensuremath{\FG}.
Primitive \ensuremath{\mathbf{labelOf}(\Varid{e}\mkern1mu)} simply projects the first component of the pair
that encodes the labeled value in \ensuremath{\FG} as explained above.
Lastly, \ensuremath{\mathbf{getLabel}} in \ensuremath{\CG} maps directly to \ensuremath{\mathbf{getLabel}} in \ensuremath{\FG}---rule
\srule{GetLabel} in \ensuremath{\CG} simply returns the program counter label and
does not raise its value, so it corresponds exactly to rule
\srule{GetLabel} in \ensuremath{\FG}, which returns label \ensuremath{\Varid{pc}} at security level
\ensuremath{\Varid{pc}}.
Similarly, \ensuremath{\mathbf{taint}(\Varid{e}\mkern1mu)} translates to \ensuremath{\mathbf{taint}({\llbracket\thinspace\mkern-2mu\xspace }\Varid{e}{\thinspace\mkern-1mu\rrbracket},()\mkern1mu)}, since rule
\srule{Taint} in \ensuremath{\CG} taints the program counter with the label that
\ensuremath{\Varid{e}} evaluates to, say \ensuremath{{\ell}} and returns unit with program counter label
equal to \ensuremath{\Varid{pc}\;\lub\;{\ell}}, which corresponds to the result of the
translated program, i.e., \ensuremath{{()}^{\Varid{pc}\;\lub\;{\ell}}}.

\section{References}
\begin{figure}[t]
\setlength{\parindent}{0pt}
\setlength{\mathindent}{0pt}
\centering
\begin{hscode}\SaveRestoreHook
\column{B}{@{}>{\hspre}l<{\hspost}@{}}%
\column{3}{@{}>{\hspre}l<{\hspost}@{}}%
\column{5}{@{}>{\hspre}l<{\hspost}@{}}%
\column{E}{@{}>{\hspre}l<{\hspost}@{}}%
\>[B]{}{\llbracket\thinspace\mkern-2mu\xspace }\mathbf{new}(\Varid{e}\mkern1mu){\thinspace\mkern-1mu\rrbracket}\mathrel{=}{}\<[E]%
\\
\>[B]{}\hsindent{3}{}\<[3]%
\>[3]{}\mathbf{let}\;\Varid{x}\mathrel{=}{\llbracket\thinspace\mkern-2mu\xspace }\Varid{e}{\thinspace\mkern-1mu\rrbracket}\thinspace\mkern-3mu\;\mathbf{in}{}\<[E]%
\\
\>[3]{}\hsindent{2}{}\<[5]%
\>[5]{}\mathbf{new}(\mathbf{taint}(\mathbf{fst}(\Varid{x}\mkern1mu),\mathbf{snd}(\Varid{x}\mkern1mu)\mkern1mu)\mkern1mu){}\<[E]%
\\[\blanklineskip]%
\>[B]{}{\llbracket\thinspace\mkern-2mu\xspace }\Varid{e}_{1}\mathbin{:=}\Varid{e}_{2}{\thinspace\mkern-1mu\rrbracket}\mathrel{=}{}\<[E]%
\\
\>[B]{}\hsindent{3}{}\<[3]%
\>[3]{}{\llbracket\thinspace\mkern-2mu\xspace }\Varid{e}_{1}{\thinspace\mkern-1mu\rrbracket}\mathbin{:=}(\mathbf{let}\;\Varid{x}\mathrel{=}{\llbracket\thinspace\mkern-2mu\xspace }\Varid{e}_{2}{\thinspace\mkern-1mu\rrbracket}\thinspace\mkern-3mu\;\mathbf{in}\;\mathbf{taint}(\mathbf{fst}(\Varid{x}\mkern1mu),\mathbf{snd}(\Varid{x}\mkern1mu)\mkern1mu)){}\<[E]%
\\[\blanklineskip]%
\>[B]{}{\llbracket\thinspace\mkern-2mu\xspace }\mathbin{!}\Varid{e}{\thinspace\mkern-1mu\rrbracket}\mathrel{=}\mathbin{!}{\llbracket\thinspace\mkern-2mu\xspace }\Varid{e}{\thinspace\mkern-1mu\rrbracket}{}\<[E]%
\\[\blanklineskip]%
\>[B]{}{\llbracket\thinspace\mkern-2mu\xspace }\mathbf{labelOfRef}(\Varid{e}\mkern1mu){\thinspace\mkern-1mu\rrbracket}\mathrel{=}\mathbf{labelOfRef}({\llbracket\thinspace\mkern-2mu\xspace }\Varid{e}{\thinspace\mkern-1mu\rrbracket}\mkern1mu){}\<[E]%
\ColumnHook
\end{hscode}\resethooks
\caption{Translation from \ensuremath{\CG} to \ensuremath{\FG} (references).}
\label{fig:cg2fg-mem}
\end{figure}
Figure \ref{fig:cg2fg-mem} shows the translation of primitives that
access the store and the heap via references.
Like the translation in the opposite direction, these translations
work homogeneously for flow-insensitive and flow-sensitive references,
therefore we focus our discussion on flow-insensitive references.
Since \ensuremath{\CG}'s rule \srule{New} in Figure \ref{fig:cg-memory} creates a
new reference labeled with the label of the argument (which must be a
labeled value), the translation converts \ensuremath{\mathbf{new}(\Varid{e}\mkern1mu)} to an expression
that first binds \ensuremath{{\llbracket\thinspace\mkern-2mu\xspace }\Varid{e}{\thinspace\mkern-1mu\rrbracket}} to \ensuremath{\Varid{x}} and then creates a new reference
with the same content as the source, i.e., \ensuremath{\mathbf{snd}(\Varid{x}\mkern1mu)}, but tainted with
the label in \ensuremath{\Varid{x}}, i.e., \ensuremath{\mathbf{fst}(\Varid{x}\mkern1mu)}.
Notice that the use of \ensuremath{\mathbf{taint}(\cdot \mkern1mu)} is essential to ensure that \ensuremath{\FG}'s rule
\srule{New} in Figure \ref{fig:fg-memory} assigns the correct label to
the new reference. Due to its \emph{fine-grained} precision, \ensuremath{\FG}
might have labeled the content with a different label that is less
sensitive than the explicit label that \emph{coarsely} approximates
the security level in \ensuremath{\CG}.
Similarly, when updating a reference we translate the new labeled
value into a pair, i.e., \ensuremath{\mathbf{let}\;\Varid{x}\mathrel{=}{\llbracket\thinspace\mkern-2mu\xspace }\Varid{e}_{2}{\thinspace\mkern-1mu\rrbracket}}, and taint the content
with the label of the labeled value projected from the pair, i.e.,
\ensuremath{\mathbf{taint}(\mathbf{fst}(\Varid{x}\mkern1mu),\mathbf{snd}(\Varid{x}\mkern1mu)\mkern1mu)}.\footnote{Technically, tainting is redundant
  for flow-insensitive references, i.e., the translation \ensuremath{{\llbracket\thinspace\mkern-2mu\xspace }\Varid{e}_{1}{\thinspace\mkern-1mu\rrbracket}\mathbin{:=}\mathbf{snd}({\llbracket\thinspace\mkern-2mu\xspace }\Varid{e}_{2}{\thinspace\mkern-1mu\rrbracket}\mkern1mu)} from the conference version of this article
  would preserve the semantics for references \ensuremath{\Gamma\vdash\Varid{e}_{1}\mathbin{:}\mathbf{Ref}\;\Conid{I}\;\tau}.
  Since flow-insensitive references have a \emph{fixed} label, rule
  \srule{Write} in both \ensuremath{\FG} and \ensuremath{\CG} accepts values less sensitive
  than the reference and stores them in the memory labeled like the
  reference.
  In contrast, the label of flow-sensitive references is not fixed and
  tainting is necessary to set the correct label for the value written
  in the heap.  }
%
%
The translation of the primitives that read and query the label of a
reference is trivial.

\section{Cross-Language Equivalence Relation}
\label{subsec:semantics-eq}
When a \ensuremath{\CG} program is translated to \ensuremath{\FG} via the program
translation described above, the \ensuremath{\FG} result contains strictly more
information than the original \ensuremath{\CG} result.
This happens because the semantics of \ensuremath{\FG} tracks flows of information
at fine granularity, in contrast with \ensuremath{\CG}, which instead
coarsely approximates the security level of all values in scope
of a computation with the program counter label.
When translating a \ensuremath{\CG} program, we can capture this condition
precisely for input values \ensuremath{\theta} by \emph{homogeneously} tagging
all standard (unlabeled) values with the initial program counter
label, i.e., \ensuremath{\llbracket\theta\thinspace\mkern-2mu\rrbracket^{\Varid{pc}}}.
However, a \ensuremath{\CG} program handles, creates and mixes unlabeled data that
originated at different security levels at run-time, e.g., when
a secret is unlabeled and combined with previously public (unlabeled)
data.
Crucially, when the translated program executes, the fine-grained
semantics of \ensuremath{\FG} tracks those flows of information precisely and thus
new labels appear (these labels do not correspond to the label of any
labeled value in the source value nor to the program counter label).
Intuitively, this implies that the \ensuremath{\FG} result will \emph{not} be
homogeneously labeled with the final program counter label of the \ensuremath{\CG}
computation, i.e., if a \ensuremath{\CG} program terminates with value \ensuremath{{v}} and
program counter label \ensuremath{\Varid{pc}'}, the translated \ensuremath{\FG} program does
\emph{not} necessarily result in \ensuremath{\llbracket{v}\thinspace\mkern-2mu\rrbracket^{\Varid{pc}'}}.
%
The following example illustrates this issue, which we then address
via a \emph{cross-language} equivalence relation that correctly
approximates the additional labels computed by \ensuremath{\FG}.

%


\begin{example}
\label{example:naive-cg2fg-thm}
Consider the execution of \ensuremath{\CG} program \ensuremath{\Varid{e}\mathrel{=}} \ensuremath{\mathbf{taint}(\Red{\Conid{H}}\mkern1mu);}  \ensuremath{\mathbf{return}(\Varid{x}\mkern1mu)}, i.e.,
\ensuremath{\langle\Sigma,\Blue{\Conid{L}},\mathbf{taint}(\Red{\Conid{H}}\mkern1mu);\mathbf{return}(\Varid{x}\mkern1mu)\rangle\Downarrow^{[\mskip1.5mu \Varid{x}\ \mapsto\ \mathbf{true}\mskip1.5mu]}\langle\Sigma,\Red{\Conid{H}},\mathbf{true}\rangle}, which taints the program counter label with \ensuremath{\Red{\Conid{H}}}, and
then returns \ensuremath{\mathbf{true}\mathrel{=}\mathbf{inl}(()\mkern1mu)} and the store \ensuremath{\Sigma} unchanged.

%
Let \ensuremath{{\llbracket\thinspace\mkern-2mu\xspace }\Varid{e}{\thinspace\mkern-1mu\rrbracket}} be the expression obtained by applying the program translation
from Figure \ref{fig:cg2fg-expr-thunk} to the example program:
\begin{center}
\setlength{\parindent}{0pt}
\setlength{\mathindent}{0pt}
\begin{hscode}\SaveRestoreHook
\column{B}{@{}>{\hspre}l<{\hspost}@{}}%
\column{3}{@{}>{\hspre}l<{\hspost}@{}}%
\column{5}{@{}>{\hspre}l<{\hspost}@{}}%
\column{E}{@{}>{\hspre}l<{\hspost}@{}}%
\>[B]{}{\llbracket\thinspace\mkern-2mu\xspace }\Varid{e}{\thinspace\mkern-1mu\rrbracket}\mathrel{=}\lambda \anonymous .{}\<[E]%
\\
\>[B]{}\hsindent{3}{}\<[3]%
\>[3]{}\mathbf{let}\;\Varid{y}\mathrel{=}\mathbf{taint}(\Red{\Conid{H}},()\mkern1mu)\;\mathbf{in}{}\<[E]%
\\
\>[3]{}\hsindent{2}{}\<[5]%
\>[5]{}\mathbf{taint}(\mathbf{labelOf}(\Varid{y}\mkern1mu),\Varid{x}\mkern1mu){}\<[E]%
\ColumnHook
\end{hscode}\resethooks
\end{center}
When we force the program \ensuremath{{\llbracket\thinspace\mkern-2mu\xspace }\Varid{e}{\thinspace\mkern-1mu\rrbracket}} and execute it starting from
program counter label equal to \ensuremath{\Blue{\Conid{L}}}, and an input environment
translated according to the initial program counter label, i.e.,
\ensuremath{\Varid{x}\ \mapsto\ \llbracket\mathbf{true}\thinspace\mkern-2mu\rrbracket^{\Blue{\Conid{L}}}\mathrel{=}{\mathbf{inl}({()}^{\Blue{\Conid{L}}}\mkern1mu)}^{\Blue{\Conid{L}}}\mathrel{=}{\mathbf{true}}^{\Blue{\Conid{L}}}}, we do \emph{not} obtain the translated result homogeneously
labeled with \ensuremath{\Red{\Conid{H}}}:
\begin{center}
\begin{tabular}{r l}
\ensuremath{\langle{\llbracket\thinspace\mkern-2mu\xspace }\Sigma{\thinspace\mkern-1mu\rrbracket},{\llbracket\thinspace\mkern-2mu\xspace }\Varid{e}{\thinspace\mkern-1mu\rrbracket}\thinspace\mkern-3mu\;()\rangle\Downarrow^{\Varid{x}\ \mapsto\ [\mskip1.5mu {\mathbf{true}}^{\Blue{\Conid{L}}}\mskip1.5mu]}_{\Blue{\Conid{L}}}} & \ensuremath{\langle{\llbracket\thinspace\mkern-2mu\xspace }\Sigma{\thinspace\mkern-1mu\rrbracket},{\mathbf{true}}^{\Red{\Conid{H}}}\rangle} \\
\ensuremath{\mathrel{=}} & \ensuremath{\langle{\llbracket\thinspace\mkern-2mu\xspace }\Sigma{\thinspace\mkern-1mu\rrbracket},{\mathbf{inl}({()}^{\Blue{\Conid{L}}}\mkern1mu)}^{\Red{\Conid{H}}}\rangle} \\
\ensuremath{\neq} & \ensuremath{\langle{\llbracket\thinspace\mkern-2mu\xspace }\Sigma{\thinspace\mkern-1mu\rrbracket},{\mathbf{inl}({()}^{\Red{\Conid{H}}}\mkern1mu)}^{\Red{\Conid{H}}}\rangle} \\
\ensuremath{\mathrel{=}} & \ensuremath{\langle{\llbracket\thinspace\mkern-2mu\xspace }\Sigma{\thinspace\mkern-1mu\rrbracket},\llbracket\mathbf{true}\thinspace\mkern-2mu\rrbracket^{\Red{\Conid{H}}}\rangle}
\end{tabular}
\end{center}
In particular, \ensuremath{\FG} preserves the public label tag on data nested
inside the left injection, i.e., \ensuremath{{()}^{\Blue{\Conid{L}}}} in \ensuremath{{\mathbf{inl}({()}^{\Blue{\Conid{L}}}\mkern1mu)}^{\Red{\Conid{H}}}} above. This happens because \ensuremath{\FG}'s rule \srule{Var} taints
only the \emph{outer} label of the value \ensuremath{{\mathbf{true}}^{\Blue{\Conid{L}}}} when it looks
up variable \ensuremath{\Varid{x}} in program counter label \ensuremath{\Red{\Conid{H}}}.
\end{example}

\paragraph{Solution.}
\begin{figure}[t]
\begin{mathpar}
\inferrule[(Value)]
{\ensuremath{{\ell}_{1}\;\flows\;\Varid{pc}} \\ \ensuremath{{r}_{1}\ceqN{\Varid{pc}}{v}_{2}} }
{\ensuremath{{{r}_{1}}^{{\ell}_{1}}\ceqN{\Varid{pc}}{v}_{2}}}
\and
\inferrule[(Unit)]
{}
{\ensuremath{()\ceqN{\Varid{pc}}()}}
\and
\inferrule[(Label)]
{}
{\ensuremath{{\ell}\ceqN{\Varid{pc}}{\ell}}}
\and
\inferrule[(Ref)]
{}
{\ensuremath{{\Varid{n}}_{{\ell}}\ceqN{\Varid{pc}}{\Varid{n}}_{{\ell}}}}
\and
\inferrule[(Ref-FS)]
{}
{\ensuremath{\Varid{n}\ceqN{\Varid{pc}}\Varid{n}}}
\and
\inferrule[(Inl)]
{\ensuremath{{v}_{1}\ceqN{\Varid{pc}}{v}_{1}'}}
{\ensuremath{\mathbf{inl}({v}_{1}\mkern1mu)\ceqN{\Varid{pc}}\mathbf{inl}({v}_{1}'\mkern1mu)}}
\and
\inferrule[(Inr)]
{\ensuremath{{v}_{2}\ceqN{\Varid{pc}}{v}_{2}'}}
{\ensuremath{\mathbf{inr}({v}_{2}\mkern1mu)\ceqN{\Varid{pc}}\mathbf{inr}({v}_{2}'\mkern1mu)}}
\and
\inferrule[(Pair)]
{\ensuremath{{v}_{1}\ceqN{\Varid{pc}}{v}_{1}'} \\ \ensuremath{{v}_{2}\ceqN{\Varid{pc}}{v}_{2}'}}
{\ensuremath{({v}_{1},{v}_{2})\ceqN{\Varid{pc}}({v}_{1}',{v}_{2}')}}
\and
\inferrule[(Fun)]
{\ensuremath{\theta_{1}\ceqN{\Varid{pc}}\theta_{2}}}
{\ensuremath{(\Varid{x}.{\llbracket\thinspace\mkern-2mu\xspace }\Varid{e}{\thinspace\mkern-1mu\rrbracket},\theta_{1})\ceqN{\Varid{pc}}(\Varid{x}.\Varid{e},\theta_{2})}}
\and
\inferrule[(Thunk)]
{\ensuremath{\theta_{1}\ceqN{\Varid{pc}}\theta_{2}}}
{\ensuremath{(\anonymous .{\llbracket\thinspace\mkern-2mu\xspace }\Varid{t}{\thinspace\mkern-1mu\rrbracket},\theta_{1})\ceqN{\Varid{pc}}(\Varid{t},\theta_{2})}}
\and
\inferrule[(Labeled)]
{\ensuremath{{v}_{1}\ceqN{{\ell}}{v}_{2}}}
{\ensuremath{({{\ell}}^{{\ell}},{v}_{1})\ceqN{\Varid{pc}}(\mathbf{Labeled}\;{\ell}\;{v}_{2})}}
\end{mathpar}
\caption{Cross-language value equivalence modulo label annotations.}
\label{fig:cross-lang-eq}
\end{figure}

In order to recover a notion of semantics preservation, we
introduce a key contribution of this work, a \emph{cross-language}
binary relation that associates values of the two calculi that, in the
scope of a computation at a given security level, are semantically
equivalent up to the extra annotations present in the \ensuremath{\FG}
value.\footnote{This relation is conceptually similar to the logical
  relation developed by~\citet{Rajani:2018} for their translations with \emph{static} IFC enforcement, but technically
  different in the treatment of labeled values.}
Technically, we use this equivalence in the semantics preservation
theorem in Section~\ref{subsec:cg2fg-correct}, which \emph{existentially}
quantifies over the result of the translated \ensuremath{\FG} program, but
guarantees that it is semantically equivalent to the result obtained
in the source program.
Concretely, for a \ensuremath{\FG} value \ensuremath{{v}_{1}} and a \ensuremath{\CG} value \ensuremath{{v}_{2}}, we write \ensuremath{{v}_{1}\ceqN{\Varid{pc}}{v}_{2}} if the label annotations (including those nested inside
compound values) of \ensuremath{{v}_{1}} are no more sensitive than label
\ensuremath{\Varid{pc}} and its raw value corresponds to \ensuremath{{v}_{2}}.
Figure \ref{fig:cross-lang-eq} formalizes this intuition by means of
two mutually inductive relations, one for \ensuremath{\FG} values and one for \ensuremath{\FG}
raw values.
In particular, rule \srule{Value} relates \ensuremath{\FG} value \ensuremath{{{r}_{1}}^{{\ell}_{1}}} and
\ensuremath{\CG} value \ensuremath{{v}_{2}} at security level \ensuremath{\Varid{pc}} if the label annotation on the
raw value \ensuremath{{r}_{1}} flows to the program counter label, i.e., \ensuremath{{\ell}_{1}\;\flows\;\Varid{pc}},
and if the raw value is in relation with the standard value, i.e.,
\ensuremath{{r}_{1}\ceqN{\Varid{pc}}{v}_{2}}.
The relation between raw values and standard values relates only
semantically equivalent values, i.e., it is syntactic equality for
ground types (\srule{Unit,Label,Ref,Ref-FS}), requires the same
injection for values of the sum type (\srule{Inl,Inr}) and requires
related components for pairs (\srule{Pair}).

Rules \srule{Fun} (resp.\ \srule{Thunk}) relates function (resp.\
thunk) closures only when environments are related pointwise, i.e.,
\ensuremath{\theta_{1}\ceqN{\Varid{pc}}\theta_{2}} iff \ensuremath{dom(\theta_{1})\equiv dom(\theta_{2})}
and \ensuremath{\forall\Varid{x}.\theta_{1}(\Varid{x})\ceqN{\Varid{pc}}\theta_{2}(\Varid{x})}, and the
\ensuremath{\FG} function body \ensuremath{\Varid{x}.{\llbracket\thinspace\mkern-2mu\xspace }\Varid{e}{\thinspace\mkern-1mu\rrbracket}} (resp.\ thunk body \ensuremath{\anonymous .{\llbracket\thinspace\mkern-2mu\xspace }\Varid{t}{\thinspace\mkern-1mu\rrbracket}})
is obtained from the \ensuremath{\CG} function body \ensuremath{\Varid{e}} (resp.\ thunk \ensuremath{\Varid{t}}) via the
program translation defined above.
Lastly, rule \srule{Labeled} relates a \ensuremath{\CG} labeled value
\ensuremath{\mathbf{Labeled}\;{\ell}\;{v}_{1}} to a pair \ensuremath{({{\ell}}^{{\ell}},{v}_{2})}, consisting of the label
\ensuremath{{\ell}} protected by itself in the first component and value \ensuremath{{v}_{2}} related with
the content \ensuremath{{v}_{1}} at security level \ensuremath{{\ell}} (\ensuremath{{v}_{1}\ceqN{{\ell}}{v}_{2}}) in
the second component.
This rule follows the principle of \lio\ that for explicitly
labeled values, the label annotation represents an upper bound on
the sensitivity of the content.
Stores are related pointwise, i.e., \ensuremath{\Sigma_1\shortdownarrow\approx\Sigma_2} iff
\ensuremath{\Sigma_1({\ell})\shortdownarrow\approx\Sigma_2({\ell})} for \ensuremath{{\ell}\;\in\;\mathscr{L}}, and
\ensuremath{{\ell}}-labeled memories relate their contents respectively at security
level \ensuremath{{\ell}}, i.e., \ensuremath{[\mskip1.5mu \mskip1.5mu]\shortdownarrow\approx[\mskip1.5mu \mskip1.5mu]} and \ensuremath{({r}\mathbin{:}\Conid{M}_1)\shortdownarrow\approx({v}\mathbin{:}\Conid{M}_2)} iff \ensuremath{{r}\ceqN{{\ell}}{v}} and \ensuremath{\Conid{M}_1\shortdownarrow\approx\Conid{M}_2} for \ensuremath{\FG} and \ensuremath{\CG} memories \ensuremath{\Conid{M}_1,\Conid{M}_2\mathbin{:}\Conid{Memory}\;{\ell}}.
Heaps are also related pointwise, i.e., \ensuremath{[\mskip1.5mu \mskip1.5mu]\shortdownarrow\approx[\mskip1.5mu \mskip1.5mu]} and \ensuremath{({{r}}^{{\ell}}\mathbin{:}\mu_{1})\shortdownarrow\approx(\mathbf{Labeled}\;{\ell}\;{v}\mathbin{:}\mu_{2})} iff \ensuremath{{r}\ceqN{{\ell}}{v}} and \ensuremath{\mu_{1}\shortdownarrow\approx\mu_{2}}, where
values at corresponding positions must additionally have the same
label.
Lastly, we lift the relation
to initial and final configurations.

\begin{defn}[Cross-Language Equivalence of Configurations]
\label{defn:eq-conf}
For all initial and final configurations:
\begin{itemize}[leftmargin=12pt,label=$\triangleright$]
\item \ensuremath{\langle\Sigma_1,\mu_{1},{\llbracket\thinspace\mkern-2mu\xspace }\Varid{e}{\thinspace\mkern-1mu\rrbracket}\thinspace\mkern-3mu\;()\rangle\shortdownarrow\approx\langle\Sigma_2,\mu_{2},\Varid{pc},\Varid{e}\rangle} iff \ensuremath{\Sigma_1\shortdownarrow\approx\Sigma_2} and \ensuremath{\mu_{1}\shortdownarrow\approx\mu_{2}},
\item \ensuremath{\langle\Sigma_1,\mu_{1},{\llbracket\thinspace\mkern-2mu\xspace }\Varid{t}{\thinspace\mkern-1mu\rrbracket}\rangle\shortdownarrow\approx\langle\Sigma_2,\mu_{2},\Varid{pc},\Varid{t}\rangle} iff \ensuremath{\Sigma_1\shortdownarrow\approx\Sigma_2} and \ensuremath{\mu_{1}\shortdownarrow\approx\mu_{2}}.
\item \ensuremath{\langle\Sigma_1,\mu_{1},{{r}}^{\Varid{pc}}\rangle\shortdownarrow\approx\langle\Sigma_2,\mu_{2},\Varid{pc},{v}\rangle} iff
  \ensuremath{\Sigma_1\shortdownarrow\approx\Sigma_2}, \ensuremath{\mu_{1}\shortdownarrow\approx\mu_{2}}, and \ensuremath{{r}\ceqN{\Varid{pc}}{v}}.
\end{itemize}
\end{defn}

First, the relation requires the stores and heaps of initial and final
configurations to be related.
Additionally, for initial configurations, the relation requires the
\ensuremath{\FG} code to be obtained from the \ensuremath{\CG} expression (resp. thunk) via
the program translation function \ensuremath{\llbracket\cdot \thinspace\mkern-2mu\rrbracket} defined above (similar to
rules \srule{Fun} and \srule{Thunk} in Figure
\ref{fig:cross-lang-eq}).
Furthermore, in the first case (expressions), the relation
additionally forces the translated suspension \ensuremath{{\llbracket\thinspace\mkern-2mu\xspace }\Varid{e}{\thinspace\mkern-1mu\rrbracket}} by applying
it to \ensuremath{()}, so that when the \ensuremath{\FG} security monitor executes the
translated program, it obtains the result that corresponds to the
\ensuremath{\CG} monadic program \ensuremath{\Varid{e}}.
Finally, in the definition for final configurations, the security
level of the final \ensuremath{\FG} result must match the program counter label
\ensuremath{\Varid{pc}} of the final \ensuremath{\CG} configuration, and the final \ensuremath{\CG} result must
correspond to the \ensuremath{\FG} result up to extra annotations at security
level \ensuremath{\Varid{pc}}, i.e., \ensuremath{{r}\ceqN{\Varid{pc}}{v}}.

Before showing semantics preservation, we prove some basic properties
of the cross-language equivalence relation that will be useful later.
The following property allows instantiating the semantics preservation
theorem with the \ensuremath{\CG} initial configuration.
The translation for initial configurations is per-component, i.e.,
\ensuremath{{\llbracket\thinspace\mkern-2mu\xspace }\langle\Sigma,\mu,\Varid{pc},\Varid{t}\rangle{\thinspace\mkern-1mu\rrbracket}\mathrel{=}\langle{\llbracket\thinspace\mkern-2mu\xspace }\Sigma{\thinspace\mkern-1mu\rrbracket},{\llbracket\thinspace\mkern-2mu\xspace }\mu{\thinspace\mkern-1mu\rrbracket},{\llbracket\thinspace\mkern-2mu\xspace }\Varid{t}{\thinspace\mkern-1mu\rrbracket}\rangle} and forcing for suspensions, i.e., \ensuremath{{\llbracket\thinspace\mkern-2mu\xspace }\langle\Sigma,\mu,\Varid{pc},\Varid{e}\rangle{\thinspace\mkern-1mu\rrbracket}\mathrel{=}\langle{\llbracket\thinspace\mkern-2mu\xspace }\Sigma{\thinspace\mkern-1mu\rrbracket},{\llbracket\thinspace\mkern-2mu\xspace }\mu{\thinspace\mkern-1mu\rrbracket},{\llbracket\thinspace\mkern-2mu\xspace }\Varid{e}{\thinspace\mkern-1mu\rrbracket}\thinspace\;()\rangle}, pointwise for
stores, i.e., \ensuremath{{\llbracket\thinspace\mkern-2mu\xspace }\Sigma{\thinspace\mkern-1mu\rrbracket}\mathrel{=}\lambda {\ell}.{\llbracket\thinspace\mkern-2mu\xspace }\Sigma({\ell}){\thinspace\mkern-1mu\rrbracket}}, memories,
i.e., \ensuremath{{\llbracket\thinspace\mkern-2mu\xspace }[\mskip1.5mu \mskip1.5mu]{\thinspace\mkern-1mu\rrbracket}\mathrel{=}[\mskip1.5mu \mskip1.5mu]} and \ensuremath{{\llbracket\thinspace\mkern-2mu\xspace }{v}\mathbin{:}\Conid{M}{\thinspace\mkern-1mu\rrbracket}\mathrel{=}\llbracket{v}\thinspace\mkern-2mu\rrbracket^{{\ell}}\mathbin{:}{\llbracket\thinspace\mkern-2mu\xspace }\Conid{M}{\thinspace\mkern-1mu\rrbracket}}
for \ensuremath{{\ell}}-labeled memory \ensuremath{\Conid{M}}, and heaps, i.e., \ensuremath{{\llbracket\thinspace\mkern-2mu\xspace }[\mskip1.5mu \mskip1.5mu]{\thinspace\mkern-1mu\rrbracket}\mathrel{=}{\llbracket\thinspace\mkern-2mu\xspace }[\mskip1.5mu \mskip1.5mu]{\thinspace\mkern-1mu\rrbracket}}
and \ensuremath{{\llbracket\thinspace\mkern-2mu\xspace }\mathbf{Labeled}\;{\ell}\;{v}\mathbin{:}\mu{\thinspace\mkern-1mu\rrbracket}\mathrel{=}\llbracket{v}\thinspace\mkern-2mu\rrbracket^{{\ell}}\mathbin{:}{\llbracket\thinspace\mkern-2mu\xspace }\mu{\thinspace\mkern-1mu\rrbracket}}.


\begin{property}[Reflexivity of \ensuremath{\shortdownarrow\approx}]
\label{prop:ceq-refl}
For all \ensuremath{\CG} initial configurations \ensuremath{\Varid{c}}, \ensuremath{{\llbracket\thinspace\mkern-2mu\xspace }\Varid{c}{\thinspace\mkern-1mu\rrbracket}\shortdownarrow\approx\Varid{c}}.
\end{property}
\begin{proof} The proof is by induction and relies on analogous
properties for all syntactic categories: for stores, \ensuremath{{\llbracket\thinspace\mkern-2mu\xspace }\Sigma{\thinspace\mkern-1mu\rrbracket}\shortdownarrow\approx\Sigma}, for memories, \ensuremath{{\llbracket\thinspace\mkern-2mu\xspace }\Conid{M}{\thinspace\mkern-1mu\rrbracket}\shortdownarrow\approx\Conid{M}}, for heaps \ensuremath{{\llbracket\thinspace\mkern-2mu\xspace }\mu{\thinspace\mkern-1mu\rrbracket}\shortdownarrow\approx\mu}, for environments \ensuremath{\llbracket\theta\thinspace\mkern-2mu\rrbracket^{\Varid{pc}}\ceqN{\Varid{pc}}\theta}, for values
\ensuremath{\llbracket{v}\thinspace\mkern-2mu\rrbracket^{\Varid{pc}}\ceqN{\Varid{pc}}{v}}, for any label \ensuremath{\Varid{pc}}.
\end{proof}

The next property guarantees that values and environments remain in
the relation when the program counter label rises.
\begin{property}[Weakening]
\label{prop:wken-ceq}
For all labels \ensuremath{\Varid{pc}} and \ensuremath{\Varid{pc}'} such that \ensuremath{\Varid{pc}\;\flows\;\Varid{pc}'}, and for
all \ensuremath{\FG} raw values \ensuremath{{r}_{1}}, values \ensuremath{{v}_{1}} and environments \ensuremath{\theta_{1}}, and
\ensuremath{\CG} values \ensuremath{{v}_{2}} and environments \ensuremath{\theta_{2}}:
\begin{CompactItemize}
 \item If \ensuremath{{r}_{1}\ceqN{\Varid{pc}}{v}_{2}} then \ensuremath{{r}_{1}\ceqN{\Varid{pc}'}{v}_{2}}
 \item If \ensuremath{{v}_{1}\ceqN{\Varid{pc}}{v}_{2}} then \ensuremath{{v}_{1}\ceqN{\Varid{pc}'}{v}_{2}}
 \item If \ensuremath{\theta_{1}\ceqN{\Varid{pc}}\theta_{2}} then \ensuremath{\theta_{1}\ceqN{\Varid{pc}'}\theta_{2}}
\end{CompactItemize}
\end{property}
\begin{proof} By mutual induction on the cross-language
equivalence relation.
\end{proof}
\section{Correctness}
\label{subsec:cg2fg-correct}

With the help of the cross-language relation defined above, we can now
state and prove that the \ensuremath{\CG}-to-\ensuremath{\FG} translation is correct, i.e., it
satisfies a semantics-preservation theorem analogous to that
proved for the translation in the opposite direction.
At a high level, the theorem ensures that the translation preserves
the meaning of a secure terminating \ensuremath{\CG} program when executed under
\ensuremath{\FG} semantics, with the same program counter label and translated
input values.
Since the translated \ensuremath{\FG} program computes strictly more information
than the original \ensuremath{\CG} program, the theorem existentially quantify
over the \ensuremath{\FG} result, but insists that it is semantically equivalent
to the original \ensuremath{\CG} result at a security level equal to the final value
of the program counter label, using the cross-language
relation just defined.

We start by proving that the program translation preserves typing.

\begin{lemma}[Type Preservation]
  If \ensuremath{\Gamma\vdash\Varid{e}\mathbin{:}\tau} then \ensuremath{{\llbracket\thinspace\mkern-2mu\xspace }\Gamma{\thinspace\mkern-1mu\rrbracket}\vdash{\llbracket\thinspace\mkern-2mu\xspace }\Varid{e}{\thinspace\mkern-1mu\rrbracket}\mathbin{:}{\llbracket\thinspace\mkern-2mu\xspace }\tau{\thinspace\mkern-1mu\rrbracket}}.
\end{lemma}
\begin{proof}
By induction on the typing judgment.
\end{proof}

Next, we prove semantics preservation of \ensuremath{\CG} pure reductions. Since
these reductions do not perform any security-relevant operation (they
do not read or write state), they can be executed with \emph{any}
program counter label in \ensuremath{\FG} and do not change the state in \ensuremath{\FG}.

\begin{lemma}[\ensuremath{\llbracket\cdot \thinspace\mkern-2mu\rrbracket\mathbin{:}\CG\to \FG} preserves Pure Semantics]
  If \ensuremath{\Varid{e}\;\Downarrow^{\theta}\;{v}} then for any program counter label \ensuremath{\Varid{pc}}, \ensuremath{\FG}
  store \ensuremath{\Sigma}, heap \ensuremath{\mu}, and environment \ensuremath{\theta'} such that \ensuremath{\theta'\ceqN{\Varid{pc}}\theta}, there exists a raw value \ensuremath{{r}}, such that \ensuremath{\langle\Sigma,\mu,{\llbracket\thinspace\mkern-2mu\xspace }\Varid{e}{\thinspace\mkern-1mu\rrbracket}\rangle\Downarrow^{\theta'}_{\Varid{pc}}\langle\Sigma,\mu,{{r}}^{\Varid{pc}}\rangle} and \ensuremath{{r}\ceqN{\Varid{pc}}{v}}.
 \label{lemma:cg2fg-correct}
\end{lemma}
\begin{proof}
By induction on the given evaluation derivation and using basic
properties of the lattice.
\end{proof}

Notice that the lemma holds for \emph{any} input target environment
\ensuremath{\theta'} in relation with the source environment \ensuremath{\theta} at security
level \ensuremath{\Varid{pc}} rather than just for the translated environment \ensuremath{\llbracket\theta\thinspace\mkern-2mu\rrbracket^{\Varid{pc}}}.
Intuitively, we needed to generalize the lemma so that the induction
principle is strong enough to discharge cases where (i) we need to
prove reductions that use an existentially quantified environment,
e.g., \srule{App}, and (ii) when some intermediate result at a
security level other than \ensuremath{\Varid{pc}} gets added to the environment, so the
environment is no longer homogenously labeled with \ensuremath{\Varid{pc}}.
While the second condition does not arise in pure reductions, it does
occur in the reduction of monadic expressions considered in the
following theorem.

\begin{thm}[Thunk and Forcing Semantics Preservation via \ensuremath{\shortdownarrow\approx}]
  For all \ensuremath{\FG} environments \ensuremath{\theta_{1}}, initial configurations \ensuremath{\Varid{c}_{1}} and
  \ensuremath{\CG} environments \ensuremath{\theta_{2}} and initial configuration \ensuremath{\Varid{c}_{2}}, such that
  \ensuremath{\theta_{1}\ceqN{\Varid{c}_{2}.\Varid{pc}}\theta_{2}} and \ensuremath{\Varid{c}_{1}\shortdownarrow\approx\Varid{c}_{2}}, if \ensuremath{\Varid{c}_{2}\;\Downarrow^{\theta_{2}}\;\Varid{c}_{2}'}, then there exists a final configuration \ensuremath{\Varid{c}_{1}'}, such that \ensuremath{\Varid{c}_{1}\;\Downarrow^{\theta_{1}}_{\Varid{c}_{2}.\Varid{pc}}\;\Varid{c}_{1}'} and \ensuremath{\Varid{c}_{1}'\shortdownarrow\approx\Varid{c}_{2}'}.
%
%
\label{thm:cg2fg-forcing-correct}
\end{thm}
\begin{proof} By mutual induction on the given derivation for
  expressions and thunks, using Lemma \ref{lemma:cg2fg-correct} for
  pure reductions, \emph{Weakening} (Property
  \ref{prop:wken-ceq}),
  and basic properties for operations on related stores and heaps.
\end{proof}

We finally instantiate the semantics-preservation theorem
(Theorem~\ref{thm:cg2fg-forcing-correct}) with the translated input
environment at security level \ensuremath{\Varid{pc}}, via \emph{reflexivity} of the
cross-language relation (Property~\ref{prop:ceq-refl}).

\begin{cor}[Semantics Preservation of \ensuremath{\llbracket\cdot \thinspace\mkern-2mu\rrbracket\mathbin{:}\CG\to \FG}]
\label{cor:cg2fg-correct}
For all \ensuremath{\CG} well-typed initial configurations \ensuremath{\Varid{c}_{2}} and environments
\ensuremath{\theta}, if \ensuremath{\Varid{c}_{2}\;\Downarrow^{\theta}\;\Varid{c}_{2}'}, then there exists a final \ensuremath{\FG}
configuration \ensuremath{\Varid{c}_{1}'} such that \ensuremath{\Varid{c}_{1}'\shortdownarrow\approx\Varid{c}_{2}'} and \ensuremath{{\llbracket\thinspace\mkern-2mu\xspace }\Varid{c}_{2}{\thinspace\mkern-1mu\rrbracket}\Downarrow^{\llbracket\theta\thinspace\mkern-2mu\rrbracket^{\Varid{pc}}}_{\Varid{pc}}\;\Varid{c}_{1}'}, where \ensuremath{\Varid{pc}\mathrel{=}\Varid{c}_{2}.\Varid{pc}}.
\end{cor}

In the corollary above, the flow-sensitive program counter of the
source \ensuremath{\CG} program gets encoded in the security level of the result
of the \ensuremath{\FG} translated program.
For example, if \ensuremath{\langle\Sigma_2,\mu_{2},\Varid{pc},\Varid{e}\rangle\Downarrow^{\theta}\langle\Sigma_{2}',\mu_{2}',\Varid{pc}',{v}\rangle} then, by Corollary \ref{cor:cg2fg-correct} and unrolling
Definition \ref{defn:eq-conf}, there exists a raw value \ensuremath{{r}} at
security level \ensuremath{\Varid{pc}'}, a store \ensuremath{\Sigma_{1}'}, and a heap \ensuremath{\mu_{1}'} such that
\ensuremath{\langle{\llbracket\thinspace\mkern-2mu\xspace }\Sigma_2{\thinspace\mkern-1mu\rrbracket},{\llbracket\thinspace\mkern-2mu\xspace }\mu_{2}{\thinspace\mkern-1mu\rrbracket},{\llbracket\thinspace\mkern-2mu\xspace }\Varid{e}{\thinspace\mkern-1mu\rrbracket}\thinspace\mkern-3mu\;()\rangle\Downarrow^{\llbracket\theta\thinspace\mkern-2mu\rrbracket^{\Varid{pc}}}_{\Varid{pc}}\langle\Sigma_{1}',\mu_{1}',{{r}}^{\Varid{pc}'}\rangle}, \ensuremath{{r}\ceqN{\Varid{pc}'}{v}}, \ensuremath{\Sigma_{1}'\shortdownarrow\approx\Sigma_{2}'}, and \ensuremath{\mu_{1}'\shortdownarrow\approx\mu_{2}'}.

\section{Recovery of Non-Interference}
Similarly to our presentation of Theorem~\ref{thm:recovery-fg2cg} for
the translation in the opposite direction, we conclude this section
with a sanity check---recovering a proof of termination-insensitive
non-interference (\emph{TINI}) for \ensuremath{\CG} through the program
translation defined above, semantics preservation
(Corollary~\ref{cor:cg2fg-correct}), and \ensuremath{\FG} non-interference
(Theorem~\ref{thm:fg-tini-bij}).
By reproving non-interference of the source language from the target
language, we show that our program translation is authentic.

To prove this result, we first need to prove that the translation
preserves \ensuremath{\Blue{\Conid{L}}}-equivalence (Lemma~\ref{lm:cg2fg-lift-low-eq}) and the
validity of references (Lemma~\ref{lemma:lift-valid-cg2fg}), as well
as a property for recovering source \ensuremath{\Blue{\Conid{L}}}-equivalence from target
\ensuremath{\Blue{\Conid{L}}}-equivalence through the cross-language relation
(Lemma~\ref{lm:cg2fg-unlift-low-eq}).
%
%
%
%
%
%
The following lemma ensures that the translation of initial
configurations preserves \ensuremath{\Blue{\Conid{L}}}-equivalence.

\begin{lemma}[\ensuremath{\llbracket\cdot \thinspace\mkern-2mu\rrbracket} preserves \ensuremath{\approx_{\Blue{\Conid{L}}}^{\beta}}]
\label{lm:cg2fg-lift-low-eq}
For all bijections \ensuremath{\beta} and initial configurations \ensuremath{\Varid{c}_{1}} and \ensuremath{\Varid{c}_{2}}, if
\ensuremath{\Varid{c}_{1}\approx_{\Blue{\Conid{L}}}^{\beta}\Varid{c}_{2}}, then \ensuremath{{\llbracket\thinspace\mkern-2mu\xspace }\Varid{c}_{1}{\thinspace\mkern-1mu\rrbracket}\approx_{\Blue{\Conid{L}}}^{\beta}{\llbracket\thinspace\mkern-2mu\xspace }\Varid{c}_{2}{\thinspace\mkern-1mu\rrbracket}}.
\end{lemma}
\begin{proof} By induction on the \ensuremath{\Blue{\Conid{L}}}-equivalence judgment and
proving similar lemmas for all syntactic categories.
\end{proof}

The following lemmas recovers \ensuremath{\CG} \ensuremath{\Blue{\Conid{L}}}-equivalence from \ensuremath{\FG}
\ensuremath{\Blue{\Conid{L}}}-equivalence using the cross-language equivalence relation for
values and environments in public contexts.
\begin{lemma}[\ensuremath{\approx_{\Blue{\Conid{L}}}^{\beta}} recovery from $\ceqN{\ensuremath{\Blue{\Conid{L}}}}$ for values and envs]
\label{lm:cg2fg-unlift-low-eq-syntax}
For all bijections \ensuremath{\beta} and public program counter labels \ensuremath{\Varid{pc}\;\flows\;\Blue{\Conid{L}}}, for all \ensuremath{\FG} values \ensuremath{{v}_{1}}, \ensuremath{{v}_{2}}, raw values \ensuremath{{r}_{1}}, \ensuremath{{r}_{2}},
environments \ensuremath{\theta_{1}}, \ensuremath{\theta_{2}}, and corresponding \ensuremath{\CG} values \ensuremath{{v}_{1}'},
\ensuremath{{v}_{2}'}, and environments \ensuremath{\theta_{1}'}, \ensuremath{\theta_{2}'}:
\begin{CompactItemize}
\item If \ensuremath{{v}_{1}\approx_{\Blue{\Conid{L}}}^{\beta}{v}_{2}}, \ensuremath{{v}_{1}\ceqN{\Varid{pc}}{v}_{1}'} and \ensuremath{{v}_{2}\ceqN{\Varid{pc}}{v}_{2}'}, then \ensuremath{{v}_{1}'\approx_{\Blue{\Conid{L}}}^{\beta}{v}_{2}'},
\item If \ensuremath{{r}_{1}\approx_{\Blue{\Conid{L}}}^{\beta}{r}_{2}}, \ensuremath{{r}_{1}\ceqN{\Varid{pc}}{v}_{1}'} and \ensuremath{{r}_{2}\ceqN{\Varid{pc}}{v}_{2}'}, then \ensuremath{{v}_{1}'\approx_{\Blue{\Conid{L}}}^{\beta}{v}_{2}'},
\item If \ensuremath{\theta_{1}\approx_{\Blue{\Conid{L}}}^{\beta}\theta_{2}}, \ensuremath{\theta_{1}\ceqN{\Varid{pc}}\theta_{1}'} and \ensuremath{\theta_{2}\ceqN{\Varid{pc}}\theta_{2}'}, then \ensuremath{\theta_{1}'\approx_{\Blue{\Conid{L}}}^{\beta}\theta_{2}'}.
\end{CompactItemize}
\end{lemma}
\begin{proof} The lemmas are proved mutually, by induction on the
\ensuremath{\Blue{\Conid{L}}}-equivalence relation and the cross-language equivalence relations
and using injectivity of the translation function \ensuremath{\llbracket\cdot \thinspace\mkern-2mu\rrbracket} for closure
values.\footnote{Technically, the function \ensuremath{\llbracket\cdot \thinspace\mkern-2mu\rrbracket} presented in
  Section \ref{sec:cg2fg} is not injective.  For example, consider the
  type translation function from Figure \ref{fig:cg2fg-ty}: \ensuremath{{\llbracket\thinspace\mkern-2mu\xspace }\mathbf{Labeled}\;\mathbf{unit}{\thinspace\mkern-1mu\rrbracket}\mathrel{=}\mathscr{L}\;\times\;\mathbf{unit}\mathrel{=}{\llbracket\thinspace\mkern-2mu\xspace }\mathscr{L}\;\times\;\mathbf{unit}{\thinspace\mkern-1mu\rrbracket}} but \ensuremath{\mathbf{Labeled}\;\mathbf{unit}\neq\mathscr{L}\;\times\;\mathbf{unit}}, and \ensuremath{{\llbracket\thinspace\mkern-2mu\xspace }\mathbf{LIO}\xspace\;\mathbf{unit}{\thinspace\mkern-1mu\rrbracket}\mathrel{=}\mathbf{unit}\to \mathbf{unit}\mathrel{=}{\llbracket\thinspace\mkern-2mu\xspace }\mathbf{unit}\to \mathbf{unit}{\thinspace\mkern-1mu\rrbracket}} but \ensuremath{\mathbf{LIO}\xspace\;\mathbf{unit}\neq\mathbf{unit}\to \mathbf{unit}}. We make the translation injective by
  (i) adding a wrapper type \ensuremath{\mathbf{Id}\;\tau} to \ensuremath{\FG}, together with
  constructor \ensuremath{\mathbf{Id}(\Varid{e})}, a deconstructor \ensuremath{\mathbf{unId}(\Varid{e})} and raw value \ensuremath{\mathbf{Id}({v})},
  and (ii) tagging security-relevant types and terms with the wrapper,
  i.e., \ensuremath{{\llbracket\thinspace\mkern-2mu\xspace }\mathbf{Labeled}\;\tau{\thinspace\mkern-1mu\rrbracket}\mathrel{=}\mathbf{Id}\;(\mathscr{L}\;\times{\llbracket\thinspace\mkern-2mu\xspace }\tau{\thinspace\mkern-1mu\rrbracket})} and
  \ensuremath{\mathbf{LIO}\xspace\;\tau\mathrel{=}\mathbf{Id}\;\mathbf{unit}\to {\llbracket\thinspace\mkern-2mu\xspace }\tau{\thinspace\mkern-1mu\rrbracket}}. Adapting the translations
  in both directions is tedious but straightforward; we refer the
  interested reader to our mechanized proofs for details.}
\end{proof}

Next, we extend this result to program state, i.e., stores, memories,
and heaps.

\begin{lemma}[\ensuremath{\approx_{\Blue{\Conid{L}}}^{\beta}} recovery from \ensuremath{\shortdownarrow\approx} for state]
\label{lm:cg2fg-unlift-low-eq-state}
For all bijections \ensuremath{\beta}, \ensuremath{\FG} memories \ensuremath{\Conid{M}_1}, \ensuremath{\Conid{M}_2}, stores \ensuremath{\Sigma_1},
\ensuremath{\Sigma_2}, heaps \ensuremath{\mu_{1}}, \ensuremath{\mu_{2}}, and corresponding \ensuremath{\CG} memories \ensuremath{\Conid{M}_{1}'}, \ensuremath{\Conid{M}_{2}'},
stores \ensuremath{\Sigma_{1}'}, \ensuremath{\Sigma_{2}'}, heaps \ensuremath{\mu_{1}'} and \ensuremath{\mu_{2}'}:
\begin{CompactItemize}
\item If \ensuremath{\Conid{M}_1\approx_{\Blue{\Conid{L}}}^{\beta}\Conid{M}_2}, \ensuremath{\Conid{M}_1\shortdownarrow\approx\Conid{M}_{1}'} and \ensuremath{\Conid{M}_2\shortdownarrow\approx\Conid{M}_{2}'}, then \ensuremath{\Conid{M}_{1}'\approx_{\Blue{\Conid{L}}}^{\beta}\Conid{M}_{2}'},
\item If \ensuremath{\Sigma_1\approx_{\Blue{\Conid{L}}}^{\beta}\Sigma_2}, \ensuremath{\Sigma_1\shortdownarrow\approx\Sigma_{1}'} and \ensuremath{\Sigma_2\shortdownarrow\approx\Sigma_{2}'}, then \ensuremath{\Sigma_{1}'\approx_{\Blue{\Conid{L}}}^{\beta}\Sigma_{2}'},
\item If \ensuremath{\mu_{1}\approx_{\Blue{\Conid{L}}}^{\beta}\mu_{2}}, \ensuremath{\mu_{1}\shortdownarrow\approx\mu_{1}'} and \ensuremath{\mu_{2}\shortdownarrow\approx\mu_{2}'}, then \ensuremath{\mu_{1}'\approx_{\Blue{\Conid{L}}}^{\beta}\mu_{2}'}.
\end{CompactItemize}
\end{lemma}
\begin{proof}
  By induction on the \ensuremath{\Blue{\Conid{L}}}-equivalence relation and the cross-language
  equivalence relations and using
  Lemma~\ref{lm:cg2fg-unlift-low-eq-syntax}.
\end{proof}

The next lemma lifts the previous lemmas to final configurations.

%

\begin{lemma}[\ensuremath{\approx_{\Blue{\Conid{L}}}^{\beta}} recovery from \ensuremath{\shortdownarrow\approx}]
\label{lm:cg2fg-unlift-low-eq}
Let \ensuremath{\Varid{c}_{1}} and \ensuremath{\Varid{c}_{2}} be \ensuremath{\FG} final configurations, let \ensuremath{\Varid{c}_{1}'} and \ensuremath{\Varid{c}_{2}'} be
\ensuremath{\CG} final configurations. If \ensuremath{\Varid{c}_{1}\approx_{\Blue{\Conid{L}}}^{\beta}\Varid{c}_{2}}, \ensuremath{\Varid{c}_{1}\shortdownarrow\approx\Varid{c}_{1}'} and \ensuremath{\Varid{c}_{2}\shortdownarrow\approx\Varid{c}_{2}'}, then \ensuremath{\Varid{c}_{1}'\approx_{\Blue{\Conid{L}}}^{\beta}\Varid{c}_{2}'}.
\end{lemma}
\begin{proof}
Let \ensuremath{\Varid{c}_{1}\mathrel{=}\langle\Sigma_1,\mu_{1},{v}_{1}\rangle}, \ensuremath{\Varid{c}_{2}\mathrel{=}\langle\Sigma_2,\mu_{2},{v}_{2}\rangle}, \ensuremath{\Varid{c}_{1}'\mathrel{=}\langle\Sigma_{1}',\mu_{1}',\Varid{pc}_{1},{v}_{1}'\rangle}, \ensuremath{\Varid{c}_{2}'\mathrel{=}\langle\Sigma_{2}',\mu_{2}',\Varid{pc}_{2},{v}_{2}'\rangle}.
From \ensuremath{\Blue{\Conid{L}}}-equivalence of the \ensuremath{\FG} final configurations, we have
\ensuremath{\Blue{\Conid{L}}}-equivalence for the stores and the values, i.e., \ensuremath{\Sigma_1\approx_{\Blue{\Conid{L}}}^{\beta}\Sigma_2} and \ensuremath{{v}_{1}\approx_{\Blue{\Conid{L}}}^{\beta}{v}_{2}} from \ensuremath{\Varid{c}_{1}\approx_{\Blue{\Conid{L}}}^{\beta}\Varid{c}_{2}} (Section
\ref{subsec:fg-security}).
Similarly, we derive cross-language equivalence relations for the
components of the final configurations, i.e., respectively \ensuremath{\Sigma_1\shortdownarrow\approx\Sigma_{1}'}, \ensuremath{\mu_{1}\shortdownarrow\approx\mu_{1}'}, and \ensuremath{{v}_{1}\ceqN{\Varid{pc}_{1}}{v}_{1}'} from \ensuremath{\Varid{c}_{1}\shortdownarrow\approx\Varid{c}_{2}}, and
\ensuremath{\Sigma_2\shortdownarrow\approx\Sigma_{2}'}, \ensuremath{\mu_{2}\shortdownarrow\approx\mu_{2}'}, and \ensuremath{{v}_{2}\ceqN{\Varid{pc}_{2}}{v}_{2}'} from \ensuremath{\Varid{c}_{2}\shortdownarrow\approx\Varid{c}_{2}'} (Definition \ref{defn:eq-conf}).
First, the \ensuremath{\CG} stores and heaps are \ensuremath{\Blue{\Conid{L}}}-equivalent,
i.e., \ensuremath{\Sigma_{1}'\approx_{\Blue{\Conid{L}}}^{\beta}\Sigma_{2}'} and \ensuremath{\mu_{1}'\approx_{\Blue{\Conid{L}}}^{\beta}\mu_{2}'} by Lemma
\ref{lm:cg2fg-unlift-low-eq-state} for stores and heaps,
respectively.
Then, two cases follow by case split on \ensuremath{{v}_{1}\approx_{\Blue{\Conid{L}}}^{\beta}{v}_{2}}.
Either (i) both label annotations on the values are not observable
(\srule{Value\ensuremath{{}_{\Red{\Conid{H}}}}}), then the program counter labels are also
not observable (\ensuremath{\Varid{pc}_{1}\;\not\flows\;\Blue{\Conid{L}}} and \ensuremath{\Varid{pc}_{2}\;\not\flows\;\Blue{\Conid{L}}} from \ensuremath{{v}_{1}\ceqN{\Varid{pc}_{1}}{v}_{1}'} and \ensuremath{{v}_{2}\ceqN{\Varid{pc}_{2}}{v}_{2}'}) and \ensuremath{\Varid{c}_{1}'\approx_{\Blue{\Conid{L}}}^{\beta}\Varid{c}_{2}'} by rule
\srule{Pc\ensuremath{{}_{\Red{\Conid{H}}}}} or (ii) the label annotations are equal and
observable by the attacker (\srule{Value\ensuremath{{}_{\Blue{\Conid{L}}}}}), i.e., \ensuremath{\Varid{pc}_{1}\equiv \Varid{pc}_{2}\;\flows\;\Blue{\Conid{L}}}, then \ensuremath{{v}_{1}'\approx_{\Blue{\Conid{L}}}^{\beta}{v}_{2}'} by Lemma
\ref{lm:cg2fg-unlift-low-eq-syntax} for values and \ensuremath{\Varid{c}_{1}'\approx_{\Blue{\Conid{L}}}^{\beta}\Varid{c}_{2}'} by
rule \srule{Pc\ensuremath{{}_{\Blue{\Conid{L}}}}}.
\end{proof}

Before recovering non-interference, we show that the translation
preserves validity of references, i.e., the assumption of the
\emph{TINI} theorem for \ensuremath{\FG} extended with flow-sensitive references.

\begin{lemma}[\ensuremath{\llbracket\cdot \thinspace\mkern-2mu\rrbracket} preserves \ensuremath{\vdash \mathbf{Valid}(\cdot )}]
\label{lemma:lift-valid-cg2fg}
  For all initial configurations \ensuremath{\Varid{c}} and environments \ensuremath{\theta}, if
  \ensuremath{\vdash \mathbf{Valid}(\Varid{c},\theta)}, then \ensuremath{\vdash \mathbf{Valid}({\llbracket\thinspace\mkern-2mu\xspace }\Varid{c}{\thinspace\mkern-1mu\rrbracket},\llbracket\theta\thinspace\mkern-2mu\rrbracket^{\Varid{c}.\Varid{pc}})}.
\end{lemma}

Finally, we combine these lemmas and prove \emph{TINI} for \ensuremath{\CG}
through our verified translation.

\begin{thm}[\ensuremath{\CG}-TINI with Bijections via \ensuremath{\llbracket\cdot \thinspace\mkern-2mu\rrbracket}]
For all valid inputs \ensuremath{\vdash \mathbf{Valid}(\Varid{c}_{1},\theta_{1})} and \ensuremath{\vdash \mathbf{Valid}(\Varid{c}_{2},\theta_{2})} and
bijections \ensuremath{\beta}, such that \ensuremath{\Varid{c}_{1}\;\approx_{\Blue{\Conid{L}}}^{\beta}\;\Varid{c}_{2}}, \ensuremath{\theta_{1}\;\approx_{\Blue{\Conid{L}}}^{\beta}\;\theta_{2}}, if \ensuremath{\Varid{c}_{1}\;\Downarrow^{\theta_{1}}\;\Varid{c}_{1}'}, and \ensuremath{\Varid{c}_{2}\;\Downarrow^{\theta_{2}}\;\Varid{c}_{2}'}, then
there exists an extended bijection \ensuremath{\beta'\;\supseteq\;\beta}, such that
\ensuremath{\Varid{c}_{1}'\;\approx_{\Blue{\Conid{L}}}^{\beta'}\;\Varid{c}_{2}'}.
\label{thm:recovery-cg2fg}
\end{thm}
\begin{proof}
First, we apply the translation \ensuremath{\llbracket\cdot \thinspace\mkern-2mu\rrbracket\mathbin{:}\CG\to \FG} to the initial
configurations \ensuremath{\Varid{c}_{1}} and \ensuremath{\Varid{c}_{2}} and the respective environments \ensuremath{\theta_{1}}
and \ensuremath{\theta_{2}}.
Let \ensuremath{\Varid{pc}} be the initial program counter label common to configurations
\ensuremath{\Varid{c}_{1}} and \ensuremath{\Varid{c}_{2}} (it is the same because \ensuremath{\Varid{c}_{1}\approx_{\Blue{\Conid{L}}}\Varid{c}_{2}}).
%
Then, \emph{Semantics Preservation of \ensuremath{\llbracket\cdot \thinspace\mkern-2mu\rrbracket}} (Corollary
\ref{cor:cg2fg-correct}) ensures that there exist two \ensuremath{\FG}
configurations \ensuremath{\Varid{c}_{1}''} and \ensuremath{\Varid{c}_{2}''}, such that \ensuremath{{\llbracket\thinspace\mkern-2mu\xspace }\Varid{c}_{1}{\thinspace\mkern-1mu\rrbracket}\Downarrow^{\llbracket\theta_{1}\thinspace\mkern-2mu\rrbracket^{\Varid{pc}}}_{\Varid{pc}}\;\Varid{c}_{1}''} and \ensuremath{\Varid{c}_{1}''\shortdownarrow\approx\Varid{c}_{1}'}, and \ensuremath{{\llbracket\thinspace\mkern-2mu\xspace }\Varid{c}_{2}{\thinspace\mkern-1mu\rrbracket}\Downarrow^{\llbracket\theta_{2}\thinspace\mkern-2mu\rrbracket^{\Varid{pc}}}_{\Varid{pc}}\;\Varid{c}_{2}''} and \ensuremath{\Varid{c}_{2}''\shortdownarrow\approx\Varid{c}_{2}'}.
We then lift \ensuremath{\Blue{\Conid{L}}}-equivalence of source configurations and
environments to \ensuremath{\Blue{\Conid{L}}}-equivalence in the target language via Lemma
\ref{lm:cg2fg-lift-low-eq}, i.e., \ensuremath{\llbracket\theta_{1}\thinspace\mkern-2mu\rrbracket^{\Varid{pc}}\approx_{\Blue{\Conid{L}}}^{\beta}\llbracket\theta_{2}\thinspace\mkern-2mu\rrbracket^{\Varid{pc}}}
from \ensuremath{\theta_{1}\approx_{\Blue{\Conid{L}}}^{\beta}\theta_{2}} and \ensuremath{{\llbracket\thinspace\mkern-2mu\xspace }\Varid{c}_{1}{\thinspace\mkern-1mu\rrbracket}\approx_{\Blue{\Conid{L}}}^{\beta}{\llbracket\thinspace\mkern-2mu\xspace }\Varid{c}_{2}{\thinspace\mkern-1mu\rrbracket}} from \ensuremath{\Varid{c}_{1}\approx_{\Blue{\Conid{L}}}^{\beta}\Varid{c}_{2}}.
Similarly, we lift the valid judgments for \ensuremath{\CG} inputs to \ensuremath{\FG} via
Lemma~\ref{lemma:lift-valid-cg2fg}, i.e., \ensuremath{\vdash \mathbf{Valid}({\llbracket\thinspace\mkern-2mu\xspace }\Varid{c}_{1}{\thinspace\mkern-1mu\rrbracket},{\llbracket\thinspace\mkern-2mu\xspace }\theta_{1}{\thinspace\mkern-1mu\rrbracket})} and \ensuremath{\vdash \mathbf{Valid}({\llbracket\thinspace\mkern-2mu\xspace }\Varid{c}_{2}{\thinspace\mkern-1mu\rrbracket},{\llbracket\thinspace\mkern-2mu\xspace }\theta_{2}{\thinspace\mkern-1mu\rrbracket})} from \ensuremath{\vdash \mathbf{Valid}(\Varid{c}_{1},\theta_{1})} and \ensuremath{\vdash \mathbf{Valid}(\Varid{c}_{2},\theta_{2})}, respectively.
Then, we apply \ensuremath{\FG}-\emph{TINI with Bijections} (Theorem
\ref{thm:fg-tini-bij}) to the reductions i.e., \ensuremath{{\llbracket\thinspace\mkern-2mu\xspace }\Varid{c}_{1}{\thinspace\mkern-1mu\rrbracket}\Downarrow^{\llbracket\theta_{1}\thinspace\mkern-2mu\rrbracket^{\Varid{pc}}}_{\Varid{pc}}\;\Varid{c}_{1}''} and \ensuremath{{\llbracket\thinspace\mkern-2mu\xspace }\Varid{c}_{2}{\thinspace\mkern-1mu\rrbracket}\Downarrow^{\llbracket\theta_{2}\thinspace\mkern-2mu\rrbracket^{\Varid{pc}}}_{\Varid{pc}}\;\Varid{c}_{2}''}, which gives \ensuremath{\Blue{\Conid{L}}}-equivalence of the resulting
configurations, i.e., \ensuremath{\Varid{c}_{1}''\;\approx_{\Blue{\Conid{L}}}^{\beta'}\;\Varid{c}_{2}''}, up to some bijection
\ensuremath{\beta'\;\supseteq\;\beta}.
Then, we apply Lemma \ref{lm:cg2fg-unlift-low-eq} to \ensuremath{\Varid{c}_{1}''\;\approx_{\Blue{\Conid{L}}}^{\beta'}\;\Varid{c}_{2}''}, \ensuremath{\Varid{c}_{1}''\shortdownarrow\approx\Varid{c}_{1}'}, and \ensuremath{\Varid{c}_{2}''\shortdownarrow\approx\Varid{c}_{2}'}, and recover
\ensuremath{\Blue{\Conid{L}}}-equivalence for the source configurations, i.e., \ensuremath{\Varid{c}_{1}'\;\approx_{\Blue{\Conid{L}}}^{\beta'}\;\Varid{c}_{2}'}, up to the same bijection \ensuremath{\beta'\;\supseteq\;\beta}.
\end{proof}


%


\chapter{Related work}
\label{sec:related}

\section{Relative Expressiveness of IFC Systems}
\paragraph{Fine- and Coarse-Grained IFC.}
Systematic study of the relative expressiveness of fine- and
coarse-grained information flow control (IFC) systems has started only
recently.
\citet{RajaniBRG17} initiated this study in the context of
\emph{static} coarse- and fine-grained IFC, enforced via type
systems. In more recent work, \citet{Rajani:2018} show that a
fine-grained IFC type system, which they call FG, and two variants of
a coarse-grained IFC type system, which they call CG, are equally
expressive. Their approach is based on type-directed translations,
which are type- and semantics-preserving.
For proofs, they develop logical relations models of FG and the
  two variants of CG, as well as cross-language logical relations.
Our work and some of our techniques are directly inspired by their
work, but we examine \emph{dynamic} IFC systems based on runtime
monitors. As a result, our technical development is completely
different. In particular, in our work we handle label introspection,
which has no counterpart in the earlier work on static IFC systems,
and which also requires significant care in translations. Our dynamic
setting also necessitated the use of tainting operators in both the
fine-grained and the coarse-grained systems.
Furthermore, our languages and translations support flow-sensitive
references, which are not considered by \citet{RajaniBRG17}, as the
type-system of FG and CG is only flow-insensitive.

Our coarse-grained system \ensuremath{\CG} is the dynamic
analogue of the second variant of \citet{Rajani:2018}'s CG type
system. This variant is described only briefly in their paper (in
Section 4, paragraph ``Original HLIO'') but covered extensively in
Part-II of the paper's appendix. \citet{Rajani:2018} argue that
translating their fine-grained system FG to this variant of CG is very
difficult and requires significant use of parametric label
polymorphism. The astute reader may wonder why we do not encounter the
same difficulty in translating our fine-grained system \ensuremath{\FG} to
\ensuremath{\CG}. The reason for this is that our fine-grained system \ensuremath{\FG} is not
a direct dynamic analogue of \citet{Rajani:2018}'s FG. In \ensuremath{\FG}, a
value constructed in a context with program counter label \ensuremath{\Varid{pc}}
automatically receives the security label \ensuremath{\Varid{pc}}. In contrast, in
\citet{Rajani:2018}'s FG, all introduction rules create values
(statically) labeled $\bot$. Hence, leaving aside the
static-vs-dynamic difference, FG's labels are more precise than
\ensuremath{\FG}'s, and this makes \citet{Rajani:2018}'s FG to CG translation more
difficult than our \ensuremath{\FG} to \ensuremath{\CG} translation. In fact, in earlier work,
\citet{RajaniBRG17} introduced a different type system called
$\mbox{FG}^{-}$, a static analogue of \ensuremath{\FG} that labels all constructed
values with \ensuremath{\Varid{pc}} (statically), and noted that translating it to the
second variant of CG is much easier (in the static setting).



\paragraph{Flow-Insensitive and Flow-Sensitive IFC.}
Researchers have explored the relative permissiveness of static and
dynamic IFC systems with respect to the flow-sensitivity of the
analysis.
\citet{Hunt:2006} show that flow-sensitive \emph{static} analysis are
a natural generalization of flow-insensitive analysis.
In the dynamic settings, \citet{Buiras:2014} provide a
semantics-preserving translation that embeds flow-sensitive references
into flow-insensitive references in
LIO~\parencite{stefan:2012:addressing}.
Their embedding tracks the mutable label of references through an
extra level of indirection, i.e., they translate each flow-sensitive
references into a flow-insensitive reference, which points to another
flow-insensitive reference that stores the content.
Interestingly, this result seems to suggest that flow-sensitive
references do not fundamentally increase the expressiveness and
permissiveness of dynamic coarse-grained IFC systems.
In contrast to our translations, however, their embedding is not local
(or macro-expressible using the terminology
of~\citet{Felleisen:1991}): it relies on the flow-sensitive heap to
assign a fixed label to flow-insensitive references.
%
%
%

%
Some works also study the relative permissiveness of static and
dynamic IFC systems.
As one would expected, dynamic flow-insensitive analysis are more
permissive than their static
counterpart~\parencite{Sabelfeld:Russo:PSI09}.
\citet{Russo:2010} show that, perhaps surprisingly, this is not the
case for \emph{flow-sensitive} analysis, i.e., purely dynamic and
static flow-sensitive analysis are incomparable in terms of
permissiveness, and propose a more permissive class of hybrid IFC
monitors.

\paragraph{Other IFC Techniques.}
\citet{Balliu2017WeAF} develop a formal framework to study the
soundness and permissiveness trade-offs of dynamic information-flow
trackers, but their analysis is only with respect to explicit and
implicit flows.
\citet{Bielova16} compare dynamic and hybrid information-flow monitors
for imperative languages with respect to soundness and transparency.
\emph{Secure multi-execution} (SME) is a dynamic
IFC mechanisms that enforces security precisely, at the cost of
executing a program multiple times~\parencite{Devriese:2010}.
\emph{Multiple facets} (MF) simulates SME through a single execution
that maintains multiple views on data, but provides weaker security
guarantees than SME~\parencite{AustinF12}.
\citet{BielovaR16} show that MF is not equivalent to SME (even when
SME is relaxed to enforce the same security condition as MF) and adapt
MF to provide the same guarantees of SME.
\citet{Schmitz:2018} unify MF and SME in a single framework that
combines their best features by allowing programs to switch between
the two mechanisms at run-time.

%
\section{Coarse-Grained Dynamic IFC}
Coarse-grained dynamic IFC systems are prevalent in security research in operating systems
\parencite{Efstathopoulos:2005,krohn:flume,Zeldovich:2006}.
These ideas have also been successfully applied to other domains,
e.g., the web
\parencite{GiffinLSTMMR12,stefan:2014:protecting,Yip:2009,browser-info-flow:ndss15},
mobile applications \parencite{android:esorics13,
  nadkarni2016practical}, IoT~\parencite{fernandes2016flowfence}, and
distributed systems~\parencite{zeldovich:dstar,Pedersen19,Cheng12}.
Our \ensuremath{\CG} calculus is based on LIO, a domain-specific language embedded
in Haskell that rephrases OS-like IFC enforcement into a
language-based setting \parencite{stefan:lio,stefan:2012:addressing}.
\citet{heule:2015:ifc-inside} introduce a general framework for
retrofitting coarse-grained IFC in any programming language in which
external effects can be controlled.
Co-Inflow \parencite{XiangC2021} extends Java with coarse-grained
dynamic IFC, which is implemented via compilation, similar to our
\ensuremath{\FG}-to-\ensuremath{\CG} translation.
Laminar \parencite{roy2009laminar} unifies mechanisms for IFC in
programming languages and operating systems, resulting in a mix of
dynamic fine- and coarse-grained enforcement.

\section{Fine-Grained Dynamic IFC}

%
The dangerous combination of highly dynamic scripting languages and
third-party code in web pages~\parencite{Nikiforakis12} and IoT
platforms~\parencite{Surbatovich17,Mohammad21} has stirred a line of
work on dynamic fine-grained IFC systems for JavaScript
interpreters~\parencite{hedin2014jsflow},
engines~\parencite{bichhawat:2014,Vineet15}, and IoT
apps~\parencite{Bastys18}.
Our \ensuremath{\FG} calculus is inspired by the calculus of
\citet{Austin:Flanagan:PLAS09}, which we have extended with
flow-insensitive references and label introspection.
In a follow up work, \citet{Austin:Flanagan:PLAS10} replace the
\emph{no-sensitive upgrade} check with a \emph{permissive upgrade}
strategy, which tracks partially leaked data to ensure it is not
completely leaked.
Breeze is conceptually similar to our \ensuremath{\FG}, except for the \ensuremath{\mathbf{taint}(\cdot \mkern1mu)}
primitive~\parencite{hritcu2013all}.
Breeze~\parencite{hritcu2013all},
JSFlow~\parencite{DBLP:conf/csfw/HedinS12}, and other dynamic
fine-grained IFC languages~\parencite{Bichhawat2021,Austin17} feature
exception handling primitives, which allow programs to recover from
IFC violations without leaking data.
Since LIO~\parencite{stefan:2017:flexible} features similar
primitives, we believe that our results extend also to IFC languages
with exceptions.

\section{Label Introspection and Flow-Sensitive References}
In general, dynamic fine-grained IFC systems often do not support
label introspection, with Breeze \parencite{hritcu2013all} as notable
exception.
\citet{stefan:2017:flexible} show that careless label introspection
can leak data and discuss alternative flow-sensitive and
flow-insensitive APIs (see
Footnote~\ref{footnote:flow-insensitive-api}).
\citet{XiangC2021} argue that the flow-sensitive API does not provide
a usable programming model (because inspecting a label can
unpredictably taint the program counter label) and use \emph{opaque
  labeled values} instead.
To support label introspection securely, our calculi protect each
label with the label itself.
\citet{Kozyri19} generalizes this mechanism to chains of labels of
arbitrary length (where each label defines the sensitivity of its
predecessor) and study the trade-offs between permissiveness and
storage.

Several dynamic fine-grained IFC systems support references with
flow-sensitive labels~\parencite{hedin2014jsflow,Austin:Flanagan:PLAS10,Austin:Flanagan:PLAS09,bichhawat:2014}.
This design choice, however, allows label changes to be exploited as a
covert channel for information
leaks~\parencite{Russo:2010,Austin:Flanagan:PLAS10,Austin:Flanagan:PLAS09}.
There are many approaches to preventing such leaks---from using static
analysis techniques~\parencite{Sabelfeld:Myers:JSAC}, to disallowing
label upgrades depending on sensitive data (i.e.,
no-sensitive-upgrades~\parencite{zdancewic:thesis,Austin:Flanagan:PLAS09}),
to avoiding branching on data whose labels have been upgraded (i.e.,
permissive-upgrades~\parencite{Austin:Flanagan:PLAS10,Bichhawat2021,Bichhawat14}).
\citet{Buiras:2014} extend LIO with flow-sensitive references and
explicitly protect the label of these references with
the program counter label at creation time.
The semantics of \ensuremath{\CG} avoids keeping track of this extra label by
using the label of the value itself as a sound approximation (see
rules~\srule{New-FS} and~\srule{Write-FS} in
Fig.~\ref{fig:fs-cg-refs-dynamics}), which corresponds precisely to
the semantics of \ensuremath{\FG}.

%
%

\section{Proof Techniques for Termination-Insensitive Non-Interference}
Since \citet{Goguen:Meseguer:Noninterference} introduced the
notion of \emph{non-interference}, different proof techniques for IFC
languages have emerged.
Our proof technique based on \emph{\ensuremath{\Blue{\Conid{L}}}-equivalence preservation} and
\emph{confinement} dates back to the seminal work
by~\citet{Volpano:Smith:Irvine:Sound} and is similar to the proof by
\citet{Austin:Flanagan:PLAS09}, although we do not make any
assumptions about the heap allocator.
Instead, our treatment of heap addresses is inspired
by~\citet{Banerjee:2005}: we extend the \ensuremath{\Blue{\Conid{L}}}-equivalence relation with
a bijection, which accounts precisely for different, yet
indistinguishable addresses.
Although bijections can complicate the formal analysis,
\citet{VASSENA2017} argue that they can be avoided by partitioning
data structures per security level (as we do here for the store).
Moreover, the separation between pure computations and side-effects
further simplifies the security analysis of monadic languages like
\ensuremath{\CG} and, as we explain in Section~\ref{sec:cg-discussion}, it leads
to shorter proofs than in impure languages like \ensuremath{\FG}.
\citet{Hirsch21} generalize this insight to other effects
(non-termination and exceptions) through a new proof technique for
pure languages that provide effects through a monad.
In their fine-grained static IFC \ensuremath{\lambda}-calculus,
\citet{Pottier:Simonet:POPL02} represent secret values and expressions
explicitly, through a syntactic bracketed pair construct.
This representation simplifies the non-interference proof (which is
derived from subject reduction), but requires reasoning about a
non-standard semantics.
%


%
\paragraph{Logical Relations.}
The first proofs of non interference that use logical relations are
for the pure fragment of the fine-grained static IFC languages by
\citet{zdancewic:thesis} and \citet{Heintze:Riecke:Slam}.
\citet{RajaniBRG17} extend this proof technique for FG and CG, but
their logical relation require step-indexed Kripke
worlds~\parencite{Birkedal11} to avoid circular arguments when
reasoning about state.
Recently, \citet{Gregersen21} develop a mechanized semantic model
based on logical relations on top of the Iris
framework~\parencite{Iris:18} for an expressive fine-grained static
IFC language.
Proofs based on logical relations for stateful languages feature two
types of logical relations: a \emph{binary} relation for observable
values (similar to \ensuremath{\Blue{\Conid{L}}}-equivalence), and a \emph{unary} relation for
secret values, which provides a semantics interpretation of the
\emph{confinement} lemma.

\paragraph{PER and Parametricity.}
\citet{Abadi+:Core} suggests a connection between parametricty and
non-interference, which is formalized through a domain-theoretic
semantics for the Dependency Core Calculus (DCC).
\citet{Sabelfeld01} develop this idea further with a general semantics
model of information flow based on partial equivalence relations
(PER).
\citet{BowmanA15} prove non-interference for DCC from parametricity by
proving that their translation from DCC into System F is fully
abstract.
\citet{Algehed19} simplifies this proof by embedding DCC into the
Calculus of Construction and applies the same technique to derive a
shorter proof for the core of LIO~\parencite{stefan:2017:flexible}.
%



\chapter{Conclusion}
\label{sec:conclusions}
This tutorial presents a detailed and homogeneous account of dynamic
fine- and coarse-grained IFC security and unifies these paradigms,
which were considered fundamentally at odds with respect to precision
and permissiveness.
To this end, we formalized two representative IFC languages that track
information flows with fine and coarse granularity, established their
security guarantees using standard proof techniques, and devised
verified semantics- and security-preserving translations between them.
%
%
%
These results formally establish a connection between dynamic fine-
and coarse-grained enforcement for IFC, showing that these paradigms
are equally expressive under reasonable assumptions.
Indeed, this work provides a systematic way to bridging the gap
between a wide range of dynamic IFC techniques often proposed by the
programming languages (fine-grained) and operating systems (coarse-grained)
communities.
As consequence, this allows future designs of dynamic IFC to choose a
coarse-grained approach, which is easier to implement and use, without
giving up on the precision of fine-grained IFC.\looseness=-1





\clearpage
\newpage




\backmatter  

\printbibliography

\end{document}
